\documentclass[11pt]{article}
\usepackage{xr}
\makeatletter
\newcommand*{\addFileDependency}[1]{
  \typeout{(#1)}
  \@addtofilelist{#1}
  \IfFileExists{#1}{}{\typeout{No file #1.}}
}
\makeatother

\newcommand*{\myexternaldocument}[1]{
    \externaldocument{#1}
    \addFileDependency{#1.tex}
    \addFileDependency{#1.aux}
}
\myexternaldocument{Appendix-Round-01}

\usepackage{color, amsmath, amssymb, amsthm, graphicx, bbm, footnote}
\usepackage[top=1in, bottom=1in, left=1in, right=1in]{geometry}

\usepackage[toc,page]{appendix}
\usepackage{tikz}
\usepackage{titling}
\allowdisplaybreaks

\usepackage{setspace}
\usepackage{adjustbox}
\usepackage[colorlinks,allcolors=blue]{hyperref}
\usepackage[round]{natbib}  

\newtheorem{theorem}{Theorem}
\newtheorem{definition}{Definition}

\newtheorem{corollary}{Corollary}
\newtheorem{lemma}{Lemma}
\newtheorem{remark}{Remark}
\newtheorem{example}{Example}
\newtheorem{condition}{Condition}
\newtheorem{assumption}{Assumption}

\newcommand{\vertiii}[1]{{\left\vert\kern-0.25ex\left\vert\kern-0.25ex\left\vert #1 
		\right\vert\kern-0.25ex\right\vert\kern-0.25ex\right\vert}}

\newenvironment{condition-prime}[1]
{%
	\addtocounter{condition}{-1}%
	\begin{condition}}
	{\end{condition}}

\newenvironment{condition-rep}[1]
{%
	\addtocounter{condition}{-1}%
	\begin{condition}}
	{\end{condition}}

\newenvironment{assumption-prime}[1]
{%
	\addtocounter{assumption}{-1}%
	\begin{assumption}}
	{\end{assumption}}

\newenvironment{condition-2prime}[1]
{%
	\addtocounter{condition}{-1}%
	\begin{condition}}
	{\end{condition}}

\newenvironment{assumption-2prime}[1]
{%
	\addtocounter{assumption}{-1}%
	\begin{assumption}}
	{\end{assumption}}

\newcommand\independent{\protect\mathpalette{\protect\independenT}{\perp}}
\def\independenT#1#2{\mathrel{\rlap{$#1#2$}\mkern2mu{#1#2}}}
	\title{Debiased Inference on Heterogeneous Quantile Treatment Effects with Regression Rank-Scores}
\author{Alexander Giessing\thanks{Department of Statistics, University of Washington, Seattle. E-mail: giessing@uw.edu.} \and Jingshen Wang\thanks{Division of Biostatistics, University of California, Berkeley. E-mail: jingshenwang@berkeley.edu.} }

\date{\today}

\begin{document}

\maketitle

\begin{abstract}
	Understanding treatment effect heterogeneity is vital to many scientific fields because the same treatment may affect different individuals differently. Quantile regression provides a natural framework for modeling such heterogeneity. We propose a new method for inference on heterogeneous quantile treatment effects in the presence of high-dimensional covariates. Our estimator combines an $\ell_1$-penalized regression adjustment with a quantile-specific bias correction scheme based on rank scores. We study the theoretical properties of this estimator, including weak convergence and semiparametric efficiency of the estimated heterogeneous quantile treatment effect process. We illustrate the finite-sample performance of our approach through simulations and an empirical example, dealing with the differential effect of statin usage for lowering low-density lipoprotein cholesterol levels for the Alzheimer's disease patients who participated in the UK Biobank study.
	
\end{abstract}

\section{Introduction}
\subsection{Motivation}

Understanding treatment effect heterogeneity in observational studies is vital to many scientific fields because often the same treatment affects different individuals differently.
For instance, in modern drug development, it is important to test for the existence (or the lack) of treatment effect heterogeneity and to identify subpopulations for which a treatment is most beneficial (or harmful) \citep{lipkovich2017tutorial, ma2017concave}. Similarly, in precision medicine, it is essential to be able to generalize causal effect estimates from a small experimental sample to a target population \citep{kern2016assessing, coppock2018generalizability}.

Quantile regression \citep{koenker2005quantile} models the effect of covariates on the conditional distribution of the response variable and thus provides a natural framework for studying treatment heterogeneity. In this paper, we propose a new method for inference on the heterogeneous quantile treatment effects (HQTE) curve in the presence of high-dimensional covariates. The HQTE curve is defined as the difference between the quantiles of the conditional distributions of treatment and control group:
\begin{align}\label{eq:subsec:Motivation-1}
\alpha(\tau; z) := Q_1(\tau; z) - Q_0(\tau;z),
\end{align}
where $Q_1(\tau;z)$ ($Q_0(\tau;z)$) is the conditional quantile curve of the potential outcome of the treated  group (the control group) evaluated at a quantile level $\tau \in (0,1)$ and covariate $z \in \mathbb{R}^p$.

The HQTE curve provides information about the treatment effect at every quantile level. Unlike the average treatment effect, which only gives the mean effect of a treatment, the HQTE curve offers a more nuanced analysis by examining the treatment effects at different points in the distribution. For instance, the bio-medical literature documents that maternal hypertension is a risk factor for low infant birth weight and this effect is more pronounced in the lower quantiles of the birth weight distribution~\citep{bowers2011birth,mhanna2015weight}. By utilizing a statistical procedure that focuses on detecting treatment effects at the lower quantiles, researchers can gain more useful insights than relying solely on estimating the average treatment effect. Another scenario where the HQTE curve proves beneficial is when the outcome variable exhibits a skewed distribution, such as survival times.~\cite{wang2018quantile} demonstrate that treatment regimes aimed at maximizing the average treatment effect, or the mean-optimal treatment regimes, may not be optimal for individuals who significantly differ from the typical sample population. In such cases, an adaptive quantile-optimal treatment regime based on the HQTE becomes preferable as it considers the effects across different quantiles and can be tailored to individuals' unique characteristics.

\subsection{Contribution and outline of the paper}
The primary contribution of this article is the novel \emph{rank-score debiased estimator} of the heterogeneous quantile treatment effects (HQTE) curve and a comprehensive study of its theoretical properties. We break summarize our contributions as follows: 

\begin{itemize}
	\item \emph{Statistical methodology:} We show how to use inverse-density weighted regression rank-scores to debias estimates of the conditional quantile function when these estimates are obtained from solving an $\ell_1$-penalized quantile regression problem. We rationalize this idea in two different ways: a bias-variance trade-off and an approximate Neyman orthogonalization procedure (Section~\ref{sec:Methodology}).
	
	\item \emph{Statistical theory:} Our main theoretical result is the weak convergence of the rank-score debiased HQTE curve to a Gaussian process in $\ell^\infty(\mathcal{T})$. The large sample properties of this process are needed whenever one would like to conduct simultaneous inference on the HQTE curve on several (or a continuum) of quantile levels $\mathcal{T} \subset (0,1)$. We propose two uniformly consistent estimators for the covariance functions of the Gaussian limit process. Moreover, for fixed dimensions, we prove that the rank-score debiased estimator is semiparametric efficient (Section~\ref{sec:Theory}).
	
	\item \emph{Algorithmic implementation:} We propose a systematic way of selecting the tuning parameters in the proposed estimation procedure. Our procedure is similar to optimization problems adopted for covariate balancing in causal inference \citep{zubizarreta2015stable, wang2017minimal, athey2018approximate}. While conventional covariate balancing procedures are rather sensitive to the choice of tuning parameters, our systematic procedure makes use of the dual formulation of the rank-score debiasing program and is fully automatic (Section~\ref{sec:Implementation}). We illustrate the finite-sample performance of our approach through Monte Carlo experiments (Section~\ref{sec:SimulationStudies}) and an empirical example, dealing with the differential effect of statin usage on lowering the low-density lipoprotein cholesterol (LDL) levels for the Alzheimer's disease patients (Section~\ref{sec:Application}).
	
	\item \emph{Technical results:} To analyze the theoretical properties of the quantile rank-score debiasing problem, we develop new technical tools that complement existing results on the consistency of $\ell_1$-penalized quantile regression~\citep{wang2021analysis, belloni2019valid, belloni2011L1penalized} and the weak convergence of quantile regression processes in growing dimension~\citep{ belloni2019conditional,chao2017quantile}. Two new results are particularly interesting: the dual formulation of the rank-score debiasing program and the Bahadur-type representation for the rank-score debiased estimator (Sections~\ref{sec:Preamble}--\ref{sec:Proofs} of the Supplementary Materials).
\end{itemize}

\subsection{Prior and related work}

Treatment effect heterogeneity is of significant interest in causal inference and is analyzed from many different angles. \cite{imai2013estimating} formulate the estimation of heterogeneous mean treatment effects as a variable selection problem. \cite{angrist2004treatment} studies mean treatment effect heterogeneity through instrumental variables. In recent publications \cite{semenova2021debiased}, \cite{kunzel2019metalearners} and \cite{nie2017quasi} propose several new meta-learners to estimate conditional average treatment effects. \cite{firpo2007efficient, frolich2013unconditional, cattaneo2010efficient} study (marginal) quantile treatment effects through modeling inverse propensity scores. \cite{chernozhukov2005iv} and \cite{abadie2002instrumental} show how instrumental variables can be helpful in identifying conditional quantile treatment effects in the presence of unmeasured confounding variables. Our paper contributes to this thriving field by introducing a novel quantile estimator to address treatment effect heterogeneity.

Three recent articles specifically study the problem of debiased inference for high-dimensional quantile regression: \cite{belloni2019valid} propose an efficient debiased estimator of a single quantile regression coefficient using Neyman orthogonal scores. \cite{bradic2017} consider the problem of debiasing the $\ell_1$-penalized estimate of the quantile regression vector when the response is homoscedastic. \cite{zhao2019debiasing} consider the same problem as \cite{bradic2017} but propose a different estimator that can deal with heteroscedastic responses. 
Allowing for heteroscedastic responses is of great practical importance since the ability to model heteroscedasticity is a key reason for using quantile regression in the first place. We provide a detailed (mathematical) comparison of our approach with the ones by~\cite{belloni2019valid} and~\cite{zhao2019debiasing} in Section~\ref{sec:Comparisons} of the Supplementary Materials. The following are the three key points of this comparison:

First, the crucial conceptual difference between the approaches by~\cite{belloni2019valid} and~\cite{zhao2019debiasing} and ours is that we treat the solution of the $\ell_1$-penalized quantile regression problem as a nuisance parameter and directly debias the scalar estimate of the conditional quantile function $Q_d(\tau;z)$. Unlike them, we do not debias a low-dimensional or coordinate-wise projection of a high-dimensional regression vector.
	
Second, when the goal is to debias a single regression coefficient, our estimator is asymptotically equivalent to the one proposed by \cite{belloni2019valid}. However, our estimator is more flexible as it can debias arbitrarily many linear combinations of regression coefficients.
	
Third, in principle, the estimator by~\cite{zhao2019debiasing} can also be used to construct a debiased estimate of the conditional quantile function. However, our approach has the following three advantages: First, our estimator is statistically more efficient, in theory and simulation studies. Second, to debias the quantile regression coefficient vector we do not need to estimate the inverse of a high-dimensional covariance matrix. Therefore, our estimator is also computationally more efficient. Third, our estimator is asymptotically normal even in growing dimensions.

\section{Causal framework and identification}\label{sec:Framework}

Throughout this paper, $Y \in \mathbb{R}$ denotes the response variable, $D \in \{0,1\}$ a binary treatment variable, and $X \in \mathbb{R}^p$ a vector of covariates. Following the framework of~\cite{rubin1974estimating}, we define the causal effect of interest in terms of so-called potential outcomes: Potential outcomes describe counterfactual states of the world, i.e. possible responses if certain treatments were administered. More formally, we index the outcomes of the response variable $Y$ by the treatment variable $D$ and write $Y_D$ for the potential outcomes of $Y$. With this notation, the potential outcome $Y_d$ corresponds to the response that we would observe if treatment $D=d$ was assigned. The causal quantity of interest in this paper is the heterogeneous quantile treatment effect (HQTE) curve evaluated at covariates $z \in \mathbb{R}^p$, 
\begin{align}\label{eq:sec:Framework-1}
\alpha(\tau; z) := Q_1(\tau; z) - Q_0(\tau;z),
\end{align}
where $Q_d(\tau;z) = \inf \left\{ y \in \mathbb{R} : F_{Y_d|X}(y|z) \geq \tau\right\}$ is the conditional quantile function (CQF) of the potential outcome $Y_d \mid X=z$ at a quantile level $\tau \in (0,1)$ and $F_{Y_d|X}$ denotes the corresponding conditional distribution function.

The key challenge in causal inference is that for each individual we only observe its potential outcome $Y_D$ under one of the two possible treatment assignments $D \in \{0,1\}$ but never under both. In other words, the observed response variable is given as $Y = D Y_1 + (1- D) Y_0$. Since the potential outcomes $Y_0$ and $Y_1$ are not observed, a priori, it is unclear how to estimate $Q_1(\tau;z)$ and $Q_0(\tau;z)$. To make headway, we introduce the following condition:
\begin{condition}[Unconfoundedness]\label{condition:Unconfoundedness}
	$(Y_0, Y_1)$ is independent of $D$ given $X$, i.e.	$(Y_0, Y_1) \independent D \mid X$.
\end{condition}
Colloquially speaking, this condition guarantees that after controlling for relevant covariates the treatment assignment is completely randomized. Under this condition, $Q_d(\tau;z)$ is identifiable and can be recast as the solution to the following program:
\begin{align}\label{eq:sec:Framework-2}
Q_d(\tau; \cdot)  \in \arg \min_{q( \cdot)} \mathbb{E}\left[\rho_\tau(Y - q(X)) - \rho_\tau(Y) \mid D= d\right],
\end{align}
where $\rho_\tau(u) = u(\tau - \mathbf{1}\{u \leq 0\})$ is the so-called check-loss and the minimum is taken over all measurable functions $q(\cdot)$ of $X$ \citep{koenker2005quantile, angrist2006quantile}.
While unconfoundedness of treatment assignments is a standard condition in the literature on causal inference, it cannot be verified from the data alone. \cite{rubin2009should} argues that unconfoundedness is more plausible when $X$ is a rich set of covariates. This motivates us to frame our problem as a high-dimensional statistical problem with predictors $X \in \mathbb{R}^p$ whose dimension $p$ exceeds the sample size $n$.

The convex optimization program \eqref{eq:sec:Framework-2} poses already a formidable challenge in low dimensions and to make it tractable in high dimensions we need to impose further structural constraints:
\begin{condition}[Sparse linear quantile regression function]\label{condition:LinearSparseQR}
	Let $\mathcal{T}$ be a compact subset of $(0,1)$. The CQF of $Y_d \mid X = z$ is given by $Q_d(\tau ; z) = z'\theta_d(\tau)$ and $\sup_{\tau \in \mathcal{T}}\left\|\theta_d(\tau)\right\|_0 \ll  p \wedge n$.
\end{condition}
In principle, this condition can be relaxed to approximate linearity and approximate sparsity similar to \cite{belloni2019valid}, but we do not pursue the technical refinements in this direction. Under Conditions~\ref{condition:Unconfoundedness} and~\ref{condition:LinearSparseQR}, the program \eqref{eq:sec:Framework-2} reduces to the linear quantile regression program
\begin{align}\label{eq:sec:Framework-4}
\theta_d(\tau) \in \arg \min_{\theta \in \mathbb{R}^p} \mathbb{E}\left[\rho_\tau(Y - X'\theta) - \rho_\tau(Y) \mid D= d\right],
\end{align}
and the HQTE curve is identified as
\begin{align}\label{eq:sec:Framework-5}
\alpha(\tau; z) = z'\theta_1(\tau) - z'\theta_0(\tau).
\end{align}

Despite the linearity condition, the HQTE curve in~\eqref{eq:sec:Framework-5} is flexible and can capture three different aspects of treatment heterogeneity. First, by keeping $z \in \mathbb{R}^p$ fixed and varying only the quantile levels $\tau \in \mathcal{T}$ we can investigate treatment effect heterogeneity across different quantile levels. Second, by keeping $\tau \in \mathcal{T}$ fixed and varying $z \in \mathbb{R}^p$ we can analyze individual treatment effects for individuals characterized by different covariates $z$. Third, by keeping $\tau \in \mathcal{T}$ fixed and letting $z \in \mathbb{R}^p$ be a sparse contrast we can identify differential effects of treatments in different sub-populations characterized by a few pre-treatment covariates (e.g. race, marriage status, gender, socioeconomic status, etc.).

\section{Methodology}\label{sec:Methodology}

In this section, we introduce the rank-score debiasing procedure for estimating the HQTE curve. We show that the estimator solves a bias-variance trade-off problem and discuss its relation to Neyman orthogonalization \citep{neyman1959optimal, belloni2019valid}. 

\subsection{The rank-score debiasing procedure }\label{subsec:RankScoreBalancedEstimator}
Let $\{(Y_i, D_i, X_i)\}_{i=1}^n$ be a random sample of response variable $Y$, treatment indicator $D$, and covariates $X$. Denote by $f_{Y_d|X}$ the conditional density of $Y_d \mid X$, $d\in \{0,1\}$. To simplify notation, write $f_i(\tau) = f_{Y_{D_i}|X}(X_i'\theta_{D_i}(\tau)|X_i)$, $i=1, \ldots, n$. Moreover, assume that the first $n_0$ observations belong to the control group and the remaining $n_1 = n- n_0$ observations to the treatment group.

\textbf{Step 1.}  For $d \in \{0,1\}$, compute pilot estimates of $\theta_d(\tau)$ as the solution of the $\ell_1$-penalized quantile regression program,
\begin{align}\label{eq:subsec:RankScoreBalancedEstimator-1}
\hat{\theta}_d(\tau) \in  \arg\min_{\theta \in \mathbb{R}^p} \left\{ \sum_{i: D_i = d} \rho_\tau(Y_i - X_i'\theta) + \lambda_d \|\theta\|_1 \right\},
\end{align}
where $\lambda_d > 0$ is a regularization parameter. Use the pilot estimates $\hat{\theta}_d(\tau)$ to estimate the conditional densities $f_i(\tau)$ as
\begin{align}\label{eq:subsec:RankScoreBalancedEstimator-2}
\hat{f}_i(\tau) := 
\begin{cases}
\frac{2h}{X_i'\hat{\theta}_1(\tau + h) - X_i'\hat{\theta}_1(\tau - h)}, & i \in \{j : D_j = 1\}\\
\frac{2h}{X_i'\hat{\theta}_0(\tau + h) - X_i'\hat{\theta}_0(\tau - h)}, & i \in \{j : D_j = 0\},
\end{cases}
\end{align}
where $h  > 0$ is a bandwidth parameter.  We discuss the choice of $\lambda_d$ and $h$ in Sections~\ref{subsec:TuningParameters} and~\ref{sec:BandwidthSelection}.

\textbf{Step 2.} Solve the \emph{rank-score debiasing program} with plug-in estimates of the conditional densities from Step 1,
\begin{align}\label{eq:subsec:RankScoreBalancedEstimator-3}
\widehat{w}(\tau; z) \in \arg\min_{w \in \mathbb{R}^n} \left\{ \sum_{i=1}^n w_i^2 \hat{f}_i^{-2}(\tau) : \:  \left\|z - \frac{1}{\sqrt{n}} \sum_{i: D_i = d} w_iX_i\right\|_\infty \leq \frac{\gamma_d}{n}, \:\: d\in \{0,1\}\right\},
\end{align}
where the $\gamma_d > 0$ are tuning parameters. We discuss the choice of $\gamma_d$ in Section~\ref{Sec:Rank-score-balncing-implement}.

\textbf{Step 3.} Define the \emph{rank-score debiased estimator of the CQF} as
\begin{align}\label{eq:subsec:RankScoreBalancedEstimator-4}
\widehat{Q}_d(\tau;z) : = z'\hat{\theta}_d(\tau) + \frac{1}{\sqrt{n}}\sum_{i:D_i =d} \widehat{w}_i(\tau;z)\hat{f}_i^{-1}(\tau)\big(\tau - \mathbf{1}\{Y_i \leq X_i'\hat{\theta}_d(\tau)\}\big).
\end{align}

\textbf{Step 4.} Define the \emph{rank-score debiased estimator of the HQTE curve} as 
\begin{align*}
\widehat{\alpha}(\tau;z) := \widehat{Q}_1(\tau;z) - \widehat{Q}_0(\tau;z).
\end{align*}
and construct an asymptotic $95\%$ confidence interval of $\alpha(\tau; z)$ as
\begin{align*}
\left[ \widehat{\alpha}(\tau;z) \pm 1.96\times \sqrt{\frac{\tau(1 - \tau)}{n} \sum_{i = 1}^n \widehat{w}_i^2(\tau;z)\hat{f}_i^{-2}(\tau)}\: \right].
\end{align*}

Steps 2 and 3 constitute the core of the rank-score debiasing procedure. In Step 2 we compute quantile-specific debiasing weights and in Step 3 we augment the estimated conditional quantile function $z'\hat{\theta}_d(\tau)$ with a bias correction based on these weights. This bias correction addresses the penalization bias in $z'\hat{\theta}_d(\tau)$, because the $\ell_1$-penalty introduces a regularization bias by shrinking coefficients in $\hat{\theta}_d(\tau)$ towards zero. Also, since the quantile regression vector $\hat{\theta}_d(\tau)$ is based on the observed covariates $\{X_i : D_i = d\}$ alone, estimating $Q_d(\tau;z)$ as $z'\hat{\theta}_d(\tau)$ introduces a sort of mismatch bias. The more $z$ differs from a typical covariate in $\{X_i : D_i = d\}$ the larger is this bias.
We refer to our estimator as the rank-score debiased estimator, because its key component is a weighted sum of quantile regression rank scores with weights that approximately match the covariates.

\subsection{Heuristic explanation in terms of a bias-variance trade-off}\label{subsec:Heuristics-BiasVariance}

The rank-score debiased estimator can be motivated in terms of a bias-variance trade-off. This perspective offers a first glimpse at its theoretical properties.

Let $\theta \in \mathbb{R}^p$ and $w \in \mathbb{R}^n$ be arbitrary. To simplify notation, write $f_i(\tau) = f_{Y_{D_i}|X}(X_i'\theta_{D_i}(\tau)|X_i)$ and $F_i(\tau) = F_{Y_{D_i}|X}(X_i'\theta|X_i)$ for $i=1, \ldots, n$. Define $\varphi_i(\theta) = \mathbf{1}\{Y_i \leq X_i'\theta\} - \mathbf{1}\{Y_i \leq X_i'\theta_d(\tau)\}$ and note that $\mathbb{E}[f_i^{-1}(\tau)\varphi_i(\theta)\mid X_i] =  f_i^{-1}(\tau)\big(F_{Y_{D_i}|X}(X_i'\theta|X_i) - F_i(\tau)\big)$. Thus, a first-order Taylor approximation at $\theta = \theta_{D_i}(\tau)$ yields
\begin{align*}
\frac{1}{\sqrt{n}} \sum_{i: D_i =d} w_i\mathbb{E}\left[ f_i^{-1}(\tau) \varphi_i(\tau)  \mid X_i\right] = \frac{1}{\sqrt{n}} \sum_{i: D_i =d} w_i X_i' \big(\theta - \theta_d(\tau)\big) +  a_n(\theta),
\end{align*}
where $|a_n(\theta)| \leq \big\|\frac{1}{\sqrt{n}} \sum_{i: D_i=d} w_if_i^{-1}(\tau)\xi_{i,\tau} X_iX_i\big\|_{op} \big\|\theta - \theta_d(\tau)\big\|_2^2$ and $\xi_{i,\tau} = f_{Y_{D_i}|X}'(X_i'\xi|X_i)$ with $\xi$ a point on the line connecting $\theta$ and $\theta_{D_i}(\tau)$. Suppose that this identity remains (approximately) true for $\theta = \hat{\theta}_d(\tau)$. Then, re-arranging this expansion leads to
\begin{align}\label{eq:subsec:RankScoreBalancedEstimator-8}
&z'\hat{\theta}_d(\tau)+ \frac{1}{\sqrt{n}}\sum_{i:D_i =d} w_if_i^{-1}(\tau)\big(\tau - \mathbf{1}\{Y_i \leq X_i'\hat{\theta}_d(\tau)\}\big) \nonumber\\
\begin{split}
&\quad{}\quad{}\quad{}= z'\theta_d(\tau) + \frac{1}{\sqrt{n}}\sum_{i:D_i =d} w_if_i^{-1}(\tau)\big(\tau - \mathbf{1}\{Y_i \leq X_i'\theta_d(\tau)\}\big)\\
&\quad{} \quad{}\quad{}\quad{} \quad{}\quad{}+ \left(z - \frac{1}{\sqrt{n}} \sum_{i: D_i=d} w_i X_i\right)'\big(\hat{\theta}_d(\tau) - \theta_d(\tau)\big) + a_n\big(\hat{\theta}_d(\tau)\big) + b_n\big(\hat{\theta}_d(\tau)\big),
\end{split} 
\end{align}
where $b_n(\theta) = -\frac{1}{\sqrt{n}} \sum_{i: D_i =d} w_i \big(f_i^{-1}(\tau) \varphi_i(\theta) - \mathbb{E}[f_i^{-1}(\tau) \varphi_i(\theta) \mid X_i]\big)$.

If $\hat{\theta}_d(\tau)$ is consistent for $\theta_d(\tau)$ and if the remainder terms $a_n\big(\hat{\theta}_d(\tau)\big)$ and $b_n\big(\hat{\theta}_d(\tau)\big)$ can be shown to be asymptotically negligible, then the statistical behavior of the left hand side of eq.~\eqref{eq:subsec:RankScoreBalancedEstimator-8} is governed by the first three terms on the right hand side. In particular, the first term on the right hand side, $z'\theta_d(\tau)$, is deterministic, the second term has mean zero and variance $\tau (1- \tau)n^{-1}\sum_{i:D_i=d} w_i^2 f_i^{-2}(\tau)$ (expectations taken conditionally on the $X_i$'s), and the third term can be upper bounded by $\big\|z - \frac{1}{\sqrt{n}} \sum_{i:D_i=d} w_i X_i \big\|_\infty \big\|\hat{\theta}_d(\tau) - \theta_d(\tau)\big\|_1$. Since the weights $w$ are arbitrary, we can choose them to fine-tune the statistical behavior of the left hand side of eq.~\eqref{eq:subsec:RankScoreBalancedEstimator-8}. Given above observations, it is natural to seek weights $w$ that minimize the variance $\tau (1- \tau)n^{-1}\sum_{i:D_i=d} w_i^2 f_i^{-2}(\tau)$ while controlling the bias term $\big\|z - \frac{1}{\sqrt{n}} \sum_{i:D_i=d} w_i X_i \big\|_\infty$. The rank-score debiasing program~\eqref{eq:subsec:RankScoreBalancedEstimator-3} with plug-in estimates $\hat{f}_i(\tau)$ can be viewed as a feasible sample version of this constrained minimization problem.
Since the weights are chosen to minimize the variance of the right hand side, we expect that the rank-score balanced estimator can be more efficient than other debiasing procedures. 
We emphasize that the theoretical analysis the rank-score debiased estimator does not rely on this Taylor expansion because it is impossible to bound the remainder terms uniformly in $w \in \mathbb{R}^n$ as $n$ diverges.

\subsection{Connection to Neyman orthogonalization}\label{subsec:Heuristics-Causal}

Our algorithm can also be rationalized as an approximate Neyman orthogonalization procedure~\citep{neyman1959optimal, belloni2019valid, chernozhukov2018double}.

Given the target $Q_d(\tau;z)$, one may interpret the true quantile regression coefficient $\theta_d(\tau)$ as a nuisance parameter, say $\eta_0 \equiv \theta_d(\tau)$. To carry out valid inference on $Q_d(\tau;z)$ when the high-dimensional nuisance parameter $\eta_0$ cannot be estimated at $\sqrt{n}$-rate, one then seeks a score function $\psi(q , \eta  )$ such that for all $\eta$ in a (shrinking) neighborhood $\mathcal{N}_n$ of $\eta_0$ and a null sequence $(\delta_n)_{n \geq 1}$,
\begin{align}\label{eq:near-orthogonal-score-conditions}	
\mathbb{E}\big[ \psi(Q_d(\tau;z), \eta_0  ) \mid X \big] = 0 \quad{} \quad{} \mathrm{and} \quad{}\quad{} \sup_{\eta \in \mathcal{N}_n} \left| \frac{\partial}{\partial \eta } \mathbb{E}\big[  \psi(Q_d(\tau;z) , \eta_0) \mid X \big](\eta - \eta_0) \right| \leq \delta_n n^{-1/2}.
\end{align}
These equations are known as Neyman near-orthogonality conditions~\citep[][Section 3.2]{chernozhukov2018double}. The neighborhood $\mathcal{N}_n$ is also called the nuisance realization set and chosen such that it contains the estimated nuisance parameter $\hat{\eta}$ with high probability. Given the definition of the rank-score debiased estimator in eq.~\eqref{eq:subsec:RankScoreBalancedEstimator-4}, a natural choice for the score function is 
\begin{align*} 
	\psi_w(q, \eta ) := q - z'\eta - \frac{1}{\sqrt{n}}\sum_{i:D_i =d}  w_i f_i^{-1}(\tau) (\tau - \mathbf{1}\big(\{Y_i \leq X_i'\eta \}\big),
\end{align*}
where $w \in \mathbb{R}^n$ is a tuning parameter to be chosen later. One easily verifies that the score function $\psi_w$ satisfies the first equality in eq.~\eqref{eq:near-orthogonal-score-conditions} for all $w \in \mathbb{R}^n$. Furthermore, provided that $w \in \mathbb{R}^n$ satisfies the box-constraint in program~\eqref{eq:subsec:RankScoreBalancedEstimator-3} and that the nuisance realization set can be chosen as $\mathcal{N}_n = \big\{\eta \in \mathbb{R}^p: \|\eta - \eta_0\|_1 \leq \frac{\delta_n}{\gamma_d} n^{1/2}\big\}$, the second inequality in eq.~\eqref{eq:near-orthogonal-score-conditions} holds as well: Indeed, for all $\eta \in \mathcal{N}_n$, by H{\"o}lder's inequality,
\begin{align*}
\left| \frac{\partial}{\partial \eta } \mathbb{E}\big[  \psi_w(Q_d(\tau;z) , \eta_0) \mid X\big](\eta - \eta_0) \right| = \left| \left(z - \frac{1}{\sqrt{n}} \sum_{i:D_i = d} w_i X_i \right)' (\eta - \eta_0) \right| \leq \delta_n n^{-1/2}.
\end{align*}
Next, denote by $\widehat{Q}_d(\tau;z, w)$ the generalized method of moment estimator that solves $\sum_{i:D_i=d}\psi_w \big(\\ \widehat{Q}_d(\tau;z, w),\hat{\eta}\big) = 0$. Conditionally on the $X_i$'s, $\widehat{Q}_d(\tau;z, w)$ has asymptotic variance $\tau (1- \tau) n^{-1}\sum_{i:D_i=d}\\ w_i^2 f_i^{-2}(\tau)$ \citep[e.g.][Section 3.2]{chernozhukov2018double}. Since $w \in \mathbb{R}^n$ is arbitrary, it is sensible to choose $w$ to minimize this asymptotic variance. Hence, the rank-score debiasing algorithm with plug-in estimates $\hat{f}_i(\tau)$ can be viewed as a feasible sample version of this approximate Neyman orthogonalization procedure. Intuitively, the box-constraint in program~\eqref{eq:subsec:RankScoreBalancedEstimator-3} relaxes the strict Neyman orthogonality condition since in high dimensions one can not hope to match $z$ exactly with a linear combination of the $X_i$'s. Furthermore, the inverse-density weighting of the weights in the expression $\sum_{i:D_i=d} w_i^2 f_i^{-2}(\tau)$ ensures that observations associated with low density at the $\tau$th quantile are given smaller debiasing weights.

\section{Theoretical analysis}\label{sec:Theory}
In this section we establish joint asymptotic normality of the HQTE process, propose consistent estimators of its asymptotic covariance function, and discuss the duality theory of the rank-score debiasing program which underlies the theoretical results.

\subsection{Regularity conditions}\label{subsec:RegularityConditions}
Throughout, we assume that $\{(Y_i, D_i, X_i)\}_{i=1}^n$ are i.i.d. copies of $(Y,D,X)$. Recall that $Y = D Y_1 + (1 - D)Y_0 \in \mathbb{R}$, where $Y_1$ and $Y_0$ are potential outcomes, $D \in\{0,1\}$, and $X \in \mathbb{R}^p$.
For examples of quantile regression models that satisfy below conditions, we refer to Section~\ref{subsec:ExampleSufficientConditions}.

\begin{condition}[Sub-Gaussian predictors]\label{condition:SubGaussianity} $X \in \mathbb{R}^p$ is a sub-Gaussian vector, i.e. $\|X - \mathbb{E}[X]\|_{\psi_2} \lesssim \big(\mathbb{E}[(X'u)^2] \big)^{1/2}$ for all $u \in \mathbb{R}$.
\end{condition}

Condition~\ref{condition:SubGaussianity} is standard in high-dimensional statistics. We introduce it to analyze the rank-score debiasing program~\eqref{eq:subsec:RankScoreBalancedEstimator-3}, but it also simplifies the theoretical analysis of the quantile regression program~\eqref{eq:subsec:RankScoreBalancedEstimator-1}. The specific formulation of sub-Gaussianity is convenient because it allows us to relate higher moments of (sparse) linear combinations $X'u$ to (sparse) eigenvalues of their covariance and second moment matrix (i.e. design matrix).

We require the following conditions on the conditional quantiles and density of $Y_d$ given $X$:

\begin{condition}[Sparsity and Lipschitz continuity of $\tau \mapsto \theta_d(\tau)$]\label{condition:SparsityLipschitzQRVector} Let $\mathcal{T}$ be compact subset of $(0,1)$.
	\begin{itemize}
		\item[(i)] There exists $s_\theta \geq 1$ such that $\sup_{d \in \{0,1\}} \sup_{\tau \in \mathcal{T}} \big|T_{\theta_d}(\tau)\big| \leq s_\theta$ for $T_{\theta_d}(\tau) =\mathrm{support}\big(\theta_d(\tau)\big)$;
		\item[(ii)] There exists $L_\theta \geq 1$ such that $\sup_{d \in \{0,1\}}\|\theta_d(\tau) - \theta_d(\tau')\|_2 \leq L_\theta |\tau - \tau'|$ for all $\tau, \tau' \in \mathcal{T}$.
	\end{itemize}
\end{condition}

\begin{condition}[Boundedness and Lipschitz continuity of $f_{Y_d|X}$]\label{condition:LipschitzBoundednessDensity} Let $a,b, x \in \mathbb{R}^p$ be arbitrary.
	\begin{itemize}
		\item[(i)] There exists $\bar{f} \geq 1$ such that $\sup_{d \in \{0,1\}} f_{Y_d|X}(a| x) \leq \bar{f}$;
		\item[(ii)] There exists $\underline{f} > 0$ such that $\inf_{d \in \{0,1\}} \inf_{\tau \in \mathcal{T}} f_{Y_d|X}(x'\theta_d(\tau)| x) \geq \underline{f}$;
		\item[(iii)] There exists $L_f \geq 1$ such that $\sup_{d \in \{0,1\}} \left|f_{Y_d|X}(x'a|x) - f_{Y_d|X}(x'b|x)\right| \leq L_f|x'a - x'b|$.
	\end{itemize}
\end{condition}

\begin{condition}[Differentiability of $\tau \mapsto Q_d(\tau; X)$]\label{condition:DifferentiabilityCQF}
	Let $\mathcal{T}$ be a compact subset of $(0,1)$. The CQF $Q_d(\tau; X)$ is three times boundedly differentiable on $\mathcal{T}$, i.e. there exists $C_Q \geq 1$ such that $\sup_{d \in \{0,1\}}\left|Q_d'''(\tau; x)\right| \leq C_Q$ for all $x \in \mathbb{R}^p$ and $\tau \in \mathcal{T}$.
\end{condition}

Conditions~\ref{condition:SparsityLipschitzQRVector} and~\ref{condition:LipschitzBoundednessDensity} are common in the literature on high-dimensional quantile regression \citep{belloni2011L1penalized, chao2017quantile, belloni2019valid, wang2021analysis}. They are relevant for establishing weak convergence of the rank-score debiased HQTE process to a Gaussian process in $\ell^\infty(\mathcal{T})$. Conditions~\ref{condition:LipschitzBoundednessDensity} $(i)$ and $(ii)$ are only needed for the theoretical analysis of program~\eqref{eq:subsec:RankScoreBalancedEstimator-3} and for establishing uniform (in $\tau \in \mathcal{T}$) consistency of the non-parametric estimates of the conditional densities in~\eqref{eq:subsec:RankScoreBalancedEstimator-2}; for all other purposes they can be dropped. If one is only interested in consistency and asymptotic normality of a single (or finitely many) quantile level(s), one can also drop Conditions~\ref{condition:SparsityLipschitzQRVector} $(ii)$ and Condition~\ref{condition:LipschitzBoundednessDensity} $(iii)$. Condition~\ref{condition:DifferentiabilityCQF} was introduced recently in~\cite{belloni2019valid} as part of the sufficient conditions for establishing consistency of the non-parametric estimates of the conditional densities in~\eqref{eq:subsec:RankScoreBalancedEstimator-2}. It might be possible to relax this condition to $Q_d(\tau;x)$ belonging to a H{\"o}lder class of functions, which is a common assumption in non-parametric (quantile) spline estimation~\citep{he1994convergence, he2013quantile}.

The next two definitions and conditions are variations of canonical assumptions for high-dimensional regression models.

\begin{definition}[$s$-sparse maximum eigenvalues] We define the $s$-sparse maximum eigenvalues of the population and sample design matrices by
	\begin{align*}
	\varphi_{\max,d}(s) := \sup_{u : \|u\|_0 \leq s} \frac{\mathbb{E}[(X'u)^2 \mathbf{1}\{D= d\}]}{\|u\|_2^2} \hspace{20pt} \mathrm{and} \hspace{20pt} \widehat{\varphi}_{\max,d}(s) := \sup_{u : \|u\|_0 \leq s} \frac{n^{-1}\sum_{i: D_i = d}(X_i'u)^2}{\|u\|_2^2}.
	\end{align*}
\end{definition}

\begin{condition}[Bounds on maximum eigenvalues]\label{condition:UpperBoundEigenvalues} There exists an absolute constant $\varphi_{\max} \geq 1$ such that
	\begin{align*}
		\varphi_{\max,d}\big(n_d/\log(n_dp)\big) \vee \widehat{\varphi}_{\max,d}\big(n_d/\log(n_dp)\big) \leq \varphi_{\mathrm{max}}, \quad{}\quad{} d \in \{0,1\}.
	\end{align*}
\end{condition}
Under Condition~\ref{condition:SubGaussianity} and for $\log p = o(n_d)$ one can upper bound the empirical maximal eigenvalue $\widehat{\varphi}_{\max,d}\big(n_d/\log(n_dp)\big)$ by a constant multiple of $\varphi_{\max,d}\big(n_d/\log(n_dp)\big)$ with probability tending to 1 (e.g. apply Lemma~\ref{lemma:MaxInequalityCovarianceCone} in the Appendix). Hence, Condition~\ref{condition:UpperBoundEigenvalues} is first and foremost a condition on the maximum eigenvalue of the population design matrix. 

To state the next definition recall that for $J\subseteq \{1, \ldots, p\}$, $q \geq 1$, and $\vartheta \in [0, \infty]$ the cone of $(J, \vartheta)$-dominant coordinate is defined as $C^p_q(J, \vartheta) := \left\{ u\in \mathbb{R}^{p}: \|u_{J^c}\|_q \leq \vartheta\|u_J\|_q  \right\}$. 

\begin{definition}[$(\omega, \vartheta, \varrho)$-restricted minimum eigenvalue of the design matrix] Let $\mathcal{T}$ be a compact subset of $(0,1)$ and $\omega, \vartheta, \varrho \geq 0$. We define the $(\omega, \vartheta, \varrho)$-restricted minimum eigenvalue of the design matrix as
	\begin{align*}
		\kappa_\omega(\vartheta, \varrho) : = \min_{d \in \{0,1\}}\inf_{\tau \in \mathcal{T}} \inf_{\|\zeta\|_2 \leq \varrho\:} \inf_{u \in C^p_1(T_\theta(\tau), \vartheta) \cap \partial B^p_2(0,1) } \mathbb{E} \left[f_{Y|X}^\omega\big(X'\theta_0(\tau) + X'\zeta |X\big)(X'u)^2\mathbf{1}\{D= d\}\right].
	\end{align*}
	To simplify notation we write $\kappa_\omega(\vartheta) : = \kappa_\omega(\vartheta, 0)$.
\end{definition}


\begin{condition}[$\varrho_n$-restricted identifiability of $\theta_d(\tau)$]\label{condition:RestrictedIdentifiability} 
	Let $\mathcal{T}$ be a compact subset of $(0,1)$ and $(\varrho_n)_{n\geq 1}$ a null sequence. The quantile regression vectors $\theta_d(\tau)$ with $d \in \{0,1\}$ and $\tau \in \mathcal{T}$ are $\varrho_n$-restricted identifiable, i.e. $	\kappa_1(2) > 0 \ \mathrm{and} \ \kappa_1(2, \varrho_n) \gtrsim \kappa_1(2).$
\end{condition}

Condition~\ref{condition:RestrictedIdentifiability} guarantees that the objective function of the $\ell_1$-penalized quantile regression program~\eqref{eq:subsec:RankScoreBalancedEstimator-1} can be locally minorized by a quadratic function. To the best of our knowledge this identifiability condition for high-dimensional quantile regression vectors is new. We use it with $\varrho_n \asymp \sqrt{s_\theta (\log np)/n}$. For this choice of $\varrho_n$, Condition~\ref{condition:RestrictedIdentifiability} is milder than the restricted identifiability and nonlinearity condition D.5 in~\cite{belloni2011L1penalized} and also slightly less restrictive than Condition (C1) in~\cite{wang2021analysis}. For a comparison of these conditions, see Remark 1 in~\cite{wang2021analysis} and Section~\ref{subsec:Assumptions-L1} in the Supplementary Materials.

The last set of definitions and conditions concern the dual of the rank-score debiasing program~\ref{eq:subsec:RankScoreBalancedEstimator-3}. Readers may skip over these conditions and return to them after having read Section~\ref{subsec:RankScoreDual}.

\begin{definition}[$\epsilon$-approximation]\label{def:EpsApprox}  Let $\epsilon \geq 0$. We call a vector $\tilde{v} \in \mathbb{R}^p$ an $\epsilon$-approximation of $v \in \mathbb{R}^p$ if $\|v- \tilde{v}\|_2 \leq \epsilon \|v\|_2$.
\end{definition}

\begin{condition}[Sparse $\epsilon_n$-approximate solution to the population dual]~\label{condition:SparsityDual}
	For $z \in \mathbb{R}^p$, $\tau \in \mathcal{T}$ and $d \in \{0,1\}$ define
	\begin{align*}
		v_d(\tau;z) := -2 \mathbb{E}\left[ f_{Y_d|X}^2\big(X'\theta_d(\tau)|X\big)XX'\mathbf{1}\{D= d\} \right]^{-1}z.
	\end{align*}
	Let $(\epsilon_n)_{n \geq 1}$ be a null sequence and $\big(\tilde{v}_{d,n}(\tau;z)\big)_{n\geq 1}$ the associated collection of $\epsilon_n$-approximations of $v_d(\tau;z)$. We assume that there exists $(s_{v,n})_{n\geq 1}$ such that
	\begin{align*}
		\sup_{d \in \{0,1\}} \sup_{\tau \in \mathcal{T}} \big|T_{v_{d,n}}(\tau)\big| \leq s_{v,n} \ll n \wedge p, \quad{}\quad{} \mathrm{where} \quad{} \quad{}T_{v_{d,n}}(\tau) =\mathrm{support}\big(\tilde{v}_{d,n}(\tau; z)\big).
	\end{align*}
	We drop the subscript $n$ on $s_{v,n}$ and $v_{d,n}(\tau;z)$ if this does not cause confusion.
\end{condition}

Condition~\ref{condition:SparsityDual} is a technical condition that allows us to analyze the rank-score debiasing weights. The plausibility of Condition~\ref{condition:SparsityDual} depends crucially on the choice of $(\epsilon_n)_{n\geq1} $. Intuitively, the larger $\epsilon_n \geq 0$, the easier it is to find a sparse $\epsilon_n$-approximation $\tilde{v}_d(\tau;z)$ of $v_d(\tau; z)$. Indeed, if $\epsilon_n \geq 1$, then one may take $\tilde{v}_d(\tau;z) \equiv 0$ with $s_v = 0$. In contrast, if $\epsilon_n = 0$, then, necessarily, $\tilde{v}_d(\tau;z) = v_d(\tau; z)$ and $s_v = \|v_d(\tau; z)\|_0$, which may or may not be less than $n \wedge p$. Our theoretical results hold for any null sequence $\epsilon_n \lesssim 1/\sqrt{s_v}$. Typically, we choose $s_v \asymp \log n$, and, hence, Definition~\ref{def:EpsApprox} and Condition~\ref{condition:SparsityDual} combine the notion of sieve estimators from classical statistics~\citep[e.g.][]{chen2007large} with the concept of compressibility from the literature on compressive sensing~\citep[e.g.][]{foucart2013mathematical}. We provide concrete examples and high-level conditions under which Condition~\ref{condition:SparsityDual} holds in Section~\ref{subsec:ExampleSufficientConditions}. 
To simplify the presentation, above definition and condition are stated somewhat informal. The rigorous formulations can be found in Section~\ref{subsec:Assumptions-RankScore} in the Supplementary Materials.

\begin{condition}[Identifiability of $\tilde{v}_d(\tau;z)$]\label{condition:LowerBoundEigenvalues} The sparse $\epsilon_n$-approximate solution to the population dual $\tilde{v}_d(\tau;z)$ is identifiable, i.e. $\kappa_2(\infty)  > 0$.
\end{condition}
Condition~\ref{condition:LowerBoundEigenvalues} guarantees that the objective function of the dual of program~\eqref{eq:subsec:RankScoreBalancedEstimator-3} can be locally minorized by a quadratic function. 

\subsection{Examples of simple sufficient conditions}\label{subsec:ExampleSufficientConditions}
We illustrate the general Conditions~\ref{condition:SubGaussianity}--\ref{condition:LowerBoundEigenvalues} with some simple sufficient conditions. We emphasize that the conditions of Section~\ref{subsec:RegularityConditions} are significantly more general than the examples discussed here.

\begin{example}[Location model with Gaussian predictors and autoregressive covariance structure]\label{example:LocationGaussianAR}
Consider the location model
	\begin{align*}
		Y_d = \alpha_d + X'\beta_d + \varepsilon, \quad{}\quad{} X \independent \varepsilon, \quad{}\quad{} d \in \{0,1\},
	\end{align*}
where $\varepsilon \sim N(0, \sigma_\varepsilon^2)$, $\sigma_\varepsilon > 0$ fixed, $X \sim N(0, \Sigma)$, and smallest and largest eigenvalues of $\Sigma \in \mathbb{R}^{p \times p}$ bounded from below by $\underline{\kappa} > 0$ and from above by $\bar{\varphi} < \infty$. Moreover, suppose that the precision matrix $\Sigma^{-1} \equiv \Omega = (\omega_{jk})_{j,k=1}^p$ has bandwidth $1 \leq q < p$, i.e. $\omega_{jk} = 0$ if $ k < j - q$ or $k > j + q$.
\end{example}

\begin{lemma}\label{lemma:example:LocationGaussianAR}
	Let $z \in \mathbb{R}^p$ be sparse with $\|z\|_0 \leq s_z$ and $\mathcal{T} = [\xi, 1-\xi]$. Under the design in Example~\ref{example:LocationGaussianAR}, Condition~\ref{condition:SubGaussianity}--\ref{condition:LowerBoundEigenvalues} are satisfied with
	\begin{align*}
		&s_\theta \leq \max_{d \in \{0,1\}}\|\beta_d\|_0 + 1, \quad{} L_\theta = \sigma_\varepsilon/\xi \vee 1, \quad{} \bar{f} = 1/\sqrt{2\pi \sigma^2_\varepsilon} \vee 1, \quad{} \underline{f} = \sqrt{\xi}/\sqrt{2\pi\sigma^2_\varepsilon},  \quad{} s_v \leq (q +1)s_z,\\
		& L_f = \sqrt{e/(2\pi\sigma^4_\varepsilon)} \vee 1,  \quad{}C_Q = 4\sigma_\varepsilon/\xi^4,\quad{} \varphi_{\max} = \bar{\varphi}, \quad{}\kappa_1(2) \geq \sqrt{\xi}\underline{\kappa}/\sqrt{2\pi \sigma^2_\varepsilon} , \quad{} \kappa_2(\infty) \geq \xi \underline{\kappa}/(2\pi \sigma^2_\varepsilon),\\
		&\varrho_n = o(1), \quad{} \epsilon_n = 0.
	\end{align*}
\end{lemma}

In Example~\ref{example:LocationGaussianAR} the covariance structure and the sparsity of $z$ guarantee that $v_d(\tau;z)$ is sparse. Hence, Condition~\ref{condition:SparsityDual} is trivially satisfied. In the next two examples we only require $v_d(\tau;z)$ to lie in some cone of dominant coordinates. This is a mild assumption and allows $v_d(\tau;z)$ to be dense and/ or weakly sparse (see also Lemma~\ref{lemma:HighLevelDualSolution} below).

\begin{example}[Location model with Gaussian predictors]\label{example:LocationGaussianGeneral}
Consider the location model
\begin{align*}
	Y_d = \alpha_d + X'\beta_d + \varepsilon, \quad{}\quad{} X \independent \varepsilon, \quad{}\quad{} d \in \{0,1\},
\end{align*}
where $\varepsilon \sim N(0, \sigma_\varepsilon^2)$, $\sigma_\varepsilon > 0$ fixed, $X \sim N(0, \Sigma)$, and smallest and largest eigenvalues of $\Sigma \in \mathbb{R}^{p \times p}$ bounded from below by $\underline{\kappa} > 0$ and from above by $\bar{\varphi} < \infty$.
\end{example}

\begin{lemma}\label{lemma:example:LocationGaussianGeneral}
	Let $\mathcal{T} = [\xi, 1-\xi]$, $c_0 \in (0, \infty]$, and $J \subseteq \{1, \ldots, p\}$ with $|J| \leq s$. Suppose that $v_d(\tau; z) \in C^p_1(J, c_0)$ for $d \in \{0,1\}$ and $\tau \in \mathcal{T}$. Under the design in Example~\ref{example:LocationGaussianGeneral}, Condition~\ref{condition:SubGaussianity}--\ref{condition:LowerBoundEigenvalues} are satisfied with
	\begin{align*}
		&s_\theta \leq \max_{d \in \{0,1\}}\|\beta_d\|_0 + 1, \quad{} L_\theta = \sigma_\varepsilon/\xi \vee 1, \quad{} \bar{f} = 1/\sqrt{2\pi \sigma^2_\varepsilon} \vee 1, \quad{} \underline{f} = \sqrt{\xi }/\sqrt{2\pi \sigma^2_\varepsilon},  \quad{} s_v = s \log n\\
		& L_f = \sqrt{e/(2\pi\sigma^4_\varepsilon)} \vee 1,  \quad{}C_Q = 4\sigma_\varepsilon/\xi^4,\quad{} \varphi_{\max} = \bar{\varphi}, \quad{}\kappa_1(2) \geq \sqrt{\xi}\underline{\kappa}/\sqrt{2\pi \sigma^2_\varepsilon} , \quad{} \kappa_2(\infty) \geq \xi \underline{\kappa}/(2\pi \sigma^2_\varepsilon),\\
		&\varrho_n = o(1), \quad{} \epsilon_n = o\left(c_0/\sqrt{\log n}\right).
	\end{align*}
\end{lemma}

\begin{example}[Location-scale model with bounded predictors]\label{example:LocationScaleGeneral}
	Consider the location-scale model
	\begin{align*}
		Y_d =  X'\beta_d + \varepsilon \cdot X'\eta_d, \quad{}\quad{} X \independent \varepsilon, \quad{}\quad{} d \in \{0,1\},
	\end{align*}
	where $\varepsilon \sim F$ with twice boundedly differentiable density $f$. Suppose that the smallest and largest eigenvalues of $\mathbb{E}[XX'] \in \mathbb{R}^{p \times p}$ are bounded from below by $\underline{\kappa} > 0$ and from above by $\bar{\varphi} < \infty$. Furthermore, suppose that there exit absolute constants $K, \upsilon, \Upsilon >0$ such that $\max_{1 \leq k \leq p} |x^{(k)}| \leq K$ and $0 < \upsilon \leq x'\eta \leq \Upsilon < \infty$ for all $x = (x^{(1)}, \ldots, x^{(p)})'$ in the range of $X$.	
\end{example}

\begin{lemma}\label{lemma:example:LocationScaleGeneral}
	Let $\mathcal{T} = [\xi, 1-\xi]$, $c_0 \in (0, \infty]$, and $J \subseteq \{1, \ldots, p\}$ with $|J| \leq s$. Suppose that $v_d(\tau; z) \in C^p_1(J, c_0)$ for $d \in \{0,1\}$ and $\tau \in \mathcal{T}$. Under the design in Example~\ref{example:LocationScaleGeneral}, Condition~\ref{condition:SubGaussianity}--\ref{condition:LowerBoundEigenvalues} are satisfied with
	\begin{align*}
		&s_\theta \leq \max_d \|\beta_d\|_0 + \|\eta_d\|_0, \quad{} L_\theta = \max_d\|\eta_d\|_2 \underline{f}, \quad{} \bar{f} = \max_{y} f(y)/\upsilon \vee 1,\\
		&\underline{f} = \min_{\tau \in \mathcal{T}} f(F^{-1}(\tau))/\Upsilon,  \quad{} s_v = s \log n, \quad{}L_f = \max_ff'(y)/\upsilon^2 \vee 1, \\
		&C_Q = \max_y \left(f''(y)/\underline{f}^4 + 3 L_f^2 \upsilon^4/\underline{f}^5 \right)\Upsilon, \quad{} \varphi_{\max} = \bar{\varphi}, \quad{}\kappa_1(2) \geq \underline{f}\underline{\kappa},\\
		&\kappa_2(\infty) \geq \underline{f}^2 \underline{\kappa}, \quad{} \varrho_n = o(1), \quad{} \epsilon_n = o\left(c_0/\sqrt{\log n }\right).
	\end{align*}
\end{lemma}

In above three examples we have imposed high-level assumptions on $v_0(\tau;d)$ which guarantee that Condition~\ref{condition:SparsityDual} holds. The next lemma provides more specific and (to some extent) testable sufficient conditions under which Condition~\ref{condition:SparsityDual} is met.

\begin{lemma}[Sufficient conditions for sparse $\epsilon_n$-approximate solutions to the population dual]\label{lemma:HighLevelDualSolution} 
	To simplify notation, write $A = [A_1, \ldots, A_p] : =\mathbb{E}\left[ f_{Y_d|X}^2\big(X'\theta_d(\tau)|X\big)XX' \mathbf{1} \{D = d \} \right]^{-1} \in \mathbb{R}^{p \times p}$. For subsets $S, T \subseteq\{1, \ldots, p\}$ let $A_{S,T} \in \mathbb{R}^{|S| \times |T|}$ be the sub-matrix obtained from $A$ by deleting all rows in $S^c$ and columns in $T^c$. Denote by $\sigma_{\min}(A_{S,T})$ the smallest singular value of $A_{S,T}$ and set $\kappa_{\min}(S, c_0) : = \inf_{u \in C^p_1(S, c_0)}\|A_Su\|_2/\|u\|_2$ for $c_0 \geq 0$.
	\begin{itemize}
		\item[(i)] If each column of $A$ has at most $q \geq 1$ non-zero entries and $z \in \mathbb{R}^p$ has at most $s_z \geq 1$ non-zero entries, then Condition~\ref{condition:SparsityDual} holds with $s_v = q s_z$ and $\epsilon_n \equiv 0$ for all $n \geq 1$.
		\item[(ii)] Suppose that there exists $\vartheta \in (0, \infty)$ such that $A_k \in C^p_1(J_k, \vartheta)$, $J_k \subseteq\{1, \ldots, p\}$, for all $1 \leq k \leq p$ and $z \in \mathbb{R}^p$ has support set $\mathrm{support}(z) = T_z$ of size at most $s_z \geq 1$. Let $J \subseteq \{1, \ldots, p\}$ be such that $\sigma_{\min}(A_{J, T_z}) > 0$. Then Condition~\ref{condition:SparsityDual} holds with $s_v = |J| \log n$ and $\epsilon_n = O\big( (1 + \vartheta) K(J, z)/\sqrt{\log n}\big)$, where $K(J, z) = \max_{k \in T_z} \sqrt{s_z}  \left\| A_{J_k, k}\right\|_1/ \sigma_{\min}(A_{J, T_z})$.
		\item[(iii)] Suppose that there exists $\vartheta \in (0, \infty)$ such that $A_k \in C^p_1(J_k, \vartheta)$, $J_k \subseteq\{1, \ldots, p\}$, for all $1 \leq k \leq p$. Let $J \subseteq \{1, \ldots, p\}$ be such that $z_J \neq 0$ and $\kappa_{\min}(J, c_0) > 0$ with  $c_0 =  \|z_{J^c}\|_1/\|z_J\|_1$. Then Condition~\ref{condition:SparsityDual} holds with $s_v = |J| \log n$ and $\epsilon_n = O\big((1 + \vartheta) K(J, z)/\sqrt{\log n}\big)$, where $K(J, z) = (1 + c_0) \max_{ 1 \leq k \leq p} \sqrt{|J|} \left\| A_{J_k, k}\right\|_1/ \kappa_{\min}(J,c_0)$.
		\item[(iv)] Suppose that there exists $\vartheta \in (0, \infty)$ such that $A_k \in C^p_1(J_k, \vartheta)$, $J_k \subseteq\{1, \ldots, p\}$, for all $1 \leq k \leq p$ and $z \in U_z \subseteq \mathbb{R}^p$, $\dim(U_z) \leq s_z$. Let $J \subseteq \{1, \ldots, p\}$ be such that $z_J \neq 0$ and $\min_{u \in U_z \cap S^{p-1}} \|A_J u\|_2 > 0$. Then Condition~\ref{condition:SparsityDual} holds with $s_v = |J| \log n$ and $\epsilon_n = O\big( (1 + \vartheta) K(J, z)/\sqrt{\log n}\big)$, where $K(J, z) = \|z\|_1/\|z_J\|_1 \max_{ 1 \leq k \leq p} \sqrt{|J|} \left\| A_{J_k, k}\right\|_1/ \min_{u \in U_z \cap S^{p-1}} \|A_J u\|_2$.
	\end{itemize}
\end{lemma}

From this lemma we infer that Condition~\ref{condition:SparsityDual} holds whenever the columns of $\mathbb{E}\big[ f_{Y_d|X}^2\big(X'\theta_d(\tau)|X\big)\\XX' \mathbf{1} \{D = d \} \big]^{-1}\in \mathbb{R}^{p \times p}$ are (weakly) sparse. Typically, this is the case if most predictors are only weakly correlated. Moreover, sparsity of $z \in \mathbb{R}^p$ is not necessary; in particular, by part $(iii)$ and $(iv)$, $\|z\|_1 = O(1)$ is sufficient. We illustrate these facts in the following example:

\begin{example}[Homoscedastic quantile regression model]\label{example:HomoscedasticQR} Suppose that $Y_d = \alpha_d + X'\beta_d + \varepsilon$ with $X, D, \varepsilon$ independent of each other for all $d \in \{0,1\}$. Let $Q_\varepsilon(\tau)$ be the $\tau$th quantile of the error $\varepsilon$ and $0 < \mathbb{P}\{D = 1\} = \pi_1 = 1- \pi_0 = 1- \mathbb{P}\{D = 0\} < 1$. Then,
	\begin{align*}
		v_d(\tau; z) = -2\pi_d^{-1} f_\varepsilon(Q_\varepsilon(\tau))^{-2} \mathbb{E}[XX']^{-1}z.
	\end{align*}
	From this expression we easily read off the following:
	\begin{itemize}
		\item[(i)] If $z\in \mathbb{R}^p$ has at most $s_z \geq 1$ non-zero entries and at least $p - q \geq 1$ entries in $X \sim N(0,\Sigma)$ are independent or $X$ follows an AR(q) process, $q \geq 1$, then Lemma~\ref{lemma:HighLevelDualSolution} (i) applies.
		\item[(ii)] If $z\in \mathbb{R}^p$ has at most $s_z  \geq 1$ non-zero entries and $X$ follows an MA(q) process, $q \geq 1$, then there exist a set $J \subseteq\{1, \ldots, p\}$ with $|J| \leq s_z$ and $\vartheta \in [0, \infty)$ such that Lemma~\ref{lemma:HighLevelDualSolution} (ii) applies.
		\item[(iii)] Suppose that $z \in \mathbb{R}^p$ has $p$ non-zero entries and $\|z\|_1 = O(1)$. If at least $p - q \geq 1$ entries in $X \sim N(0,\Sigma)$ are independent or $X$ follows an AR(q) or MA(q) process, then there exist a set $J \subseteq\{1, \ldots, p\}$ with $|J|= 1$, $\vartheta \in [0, \infty)$, and $U_z \subset \mathbb{R}^d$ with $\mathrm{dim}(U_z) = 1$ such that Lemma~\ref{lemma:HighLevelDualSolution} (iv) applies.
	\end{itemize}
\end{example}

\subsection{Weak convergence results}\label{subsec:WeakConvergence}

In this section we establish weak convergence of the rank-score debiased CQF and the HQTE processes,
\begin{align*}
\left\{\sqrt{n}\big(\widehat{Q}_d(\tau;z) - Q_d(\tau;z)\big) : \tau \in \mathcal{T}\right\} \quad{}\quad{} \mathrm{and} \quad{} \quad{} \left\{\sqrt{n}\big(\widehat{\alpha}(\tau;z) - \alpha(\tau;z)\big) : \tau \in \mathcal{T}\right\}.
\end{align*}
The large sample properties of these processes are needed whenever one would like to conduct inference on the HQTE curve on more than just one quantile at a time. For example, statistical comparisons of the HQTE across different quantiles require uniform confidence bands that hold for all quantiles under consideration. Similarly, testing hypotheses about subsets of quantiles requires constructing rejection regions that hold across these quantiles. In both cases, process methods provide a natural way of addressing these problems. We provide concrete examples below.

To formulate the theoretical results we introduce the following operator:
\begin{align*} 
H_d^{(n)}(\tau_1, \tau_2; z) := v_d'(\tau_1; z)\mathbb{E}[f_{Y_d|X}(X'\theta_d(\tau_1)|X)f_{Y_d|X}(X'\theta_d(\tau_2)|X) XX' \mathbf{1}\{D= d\}]v_d(\tau_2; z),
\end{align*}
where $v_d(\tau;z) = -2 \big(\mathbb{E}[ f_{Y_d|X}^2\big(X'\theta_d(\tau)|X\big)XX'\mathbf{1}\{D= d\}]\big)^{-1}z$. Since the dimension $n$ may grow with the sample size $n$, we make the dependence of $H_d^{(n)}(\tau_1, \tau_2; z)$ on $n$ explicit.

The following theorem establishes joint asymptotic normality of the rank-score balanced CQF process.

\begin{theorem}[Weak convergence of the rank-score debiased CQF process]\label{theorem:WeakConvergence-MainPaper-CQF}
	Let $\mathcal{T}$ be a compact subset of $(0,1)$. Suppose that Conditions~\ref{condition:Unconfoundedness}--\ref{condition:LowerBoundEigenvalues} hold with $\varrho_n = \sqrt{ (s_v + s_\theta)\log(np) / n}$ and $\epsilon_n^2 = O(\sqrt{n} h^{-1} \varrho_n^2 + h^2)$. In addition, suppose that $(s_v + s_\theta)^3\log^3( np)\log^3(n)= o(nh^3)$, $h^2 s_v = o(1)$, and $\|z\|_2 = O(1)$. If $\lambda_d \asymp \sqrt{\varphi_{\max}}\sqrt{n \log(np)}$ and $\gamma_d \asymp \|z\|_2 \left(h^{-1} s_\theta\log(np) +\sqrt{n}h^2\right)\sqrt{n}$, then
	\begin{align*}
	\sqrt{n} \big(\widehat{Q}_d(\cdot;z) - Q_d(\cdot; z) \big) \leadsto \mathbb{G}_d(\cdot\:; z) \quad{} \mathrm{in} \quad{} \ell^\infty(\mathcal{T}),
	\end{align*}
	where $\mathbb{G}_d(\cdot\:;z)$ is a centered Gaussian process with covariance function $(\tau_1, \tau_2) \mapsto H_d(\tau_1, \tau_2; z) := \lim_{n \rightarrow \infty} \frac{\tau_1 \wedge \tau_2 - \tau_1 \tau_2}{4}  H_d^{(n)}(\tau_1, \tau_2; z)$ provided this limit exists pointwise for all $\tau_1, \tau_2 \in \mathcal{T}$.
\end{theorem}
\begin{remark}[On the existence of the covariance function]
	It is easy to verify that the limit $H_d(\tau_1, \tau_2; z)$ is finite for all $\tau_1, \tau_2 \in \mathcal{T}$ whenever Condition~\ref{condition:LowerBoundEigenvalues} holds and $\|z\|_2 = O(1)$. 
	However, this alone does not imply existence of the limit, since $ H_d^{(n)}(\tau_1, \tau_2; z)$ may oscillate with the sample size $n$. Hence, we impose pointwise convergence of $H_d^{(n)}(\tau_1, \tau_2; z)$ for all $\tau_1, \tau_2 \in \mathcal{T}$ as an additional assumption. In the context of abstract weak convergence results for classes of functions that may change with the sample size $n$ this assumption is standard~\citep[e.g.][ch. 2.11.3]{vaartwellner1996weak}; in the context of high-dimensional quantile regression this assumption also appears in~\cite{chao2017quantile}.
	If $H_d^{(n)}(\tau_1, \tau_2; z)$ does not converge pointwise for all $\tau_1, \tau_2 \in \mathcal{T}$ weak process convergence fails, but we still have asymptotic normality of the studentized rank-score debiased CQF: Indeed, for all (fixed) $\tau \in \mathcal{T}$, Lemma~\ref{lemma:ConsistencyCovarianceFunction} implies that $\widehat{\sigma}_2^{-1}(\tau;z)\sqrt{n} \big(\widehat{Q}_d(\tau;z) - Q_d(\tau; z) \big) \leadsto N(0,1)$, where $\widehat{\sigma}_2(\tau;z)$ is defined in eq.~\eqref{eq:subsec:ConsistecyCovariance-3}.
\end{remark}

Assume, for a moment, that dimension $p$ is fixed. Then, Theorem~\ref{theorem:WeakConvergence-MainPaper-CQF} implies that
\begin{align*}
\sqrt{n}\big(\widehat{Q}_d(\tau;z) - Q_d(\tau;z)\big) \leadsto N\left(0, \tau(1-\tau)z'\left( \mathbb{E}[f_{Y_d|X}^2(X'\theta_d(\tau)|X) XX'\mathbf{1}\{D= d\}]\right)^{-1}z\right).
\end{align*}
What is of interest here is that the asymptotic variance $\tau(1-\tau)z'(\mathbb{E}[f_{Y_d|X}^2(X'\theta_d(\tau)|X) XX' \mathbf{1}\{D= d\}])^{-1}z$ is known to be the semi-parametric efficiency bound for all estimators of the linear conditional quantile function~\citep{newey1990efficient}. In particular, the rank-score balanced estimator of the CQF is as efficient as the estimate of the CQF based on the weighted quantile regression program~\citep{koenker2005quantile, koenker1994estimatton, zhao2001asymptotically}. This lends further support to the heuristic arguments made in Section~\ref{subsec:Heuristics-BiasVariance}. Though we note that as the conditional densities can be hard to estimate, the weighted quantile regression problems can be less popular in practice. 

Since the rank-score debiased estimates of $Q_1(\tau;z)$ and $Q_0(\tau;z)$ are asympototically independent, Theorem~\ref{theorem:WeakConvergence-MainPaper-CQF} and the Continuous Mapping Theorem yield the following result for the HQTE process. 

\begin{theorem}[Weak convergence of the rank-score debiased HQTE process]\label{theorem:WeakConvergence-MainPaper-HQTE}
	Let $\mathcal{T}$ be a compact subset of $(0,1)$. Under the conditions of Theorem~\ref{theorem:WeakConvergence-MainPaper-CQF},
	\begin{align*}
	\sqrt{n} \big(\widehat{\alpha}(\cdot;z) - \alpha(\cdot; z) \big)  \leadsto  \mathbb{G}_1(\cdot\:; z) +  \mathbb{G}_0(\cdot\:; z) \quad{} \mathrm{in} \quad{} \ell^\infty(\mathcal{T}),
	\end{align*}
	where $\mathbb{G}_1(\cdot\:;z)$,  $\mathbb{G}_0(\cdot\:;z)$ are independent, centered Gaussian processes with covariance functions $(\tau_1, \tau_2) \mapsto H_d(\tau_1, \tau_2; z)$ with $d \in\{0,1\}$.
\end{theorem}

The takeaway from Theorem~\ref{theorem:WeakConvergence-MainPaper-HQTE} is that the HQTE process converges weakly to the sum of two independent centered Gaussian processes. We illustrate Theorem~\ref{theorem:WeakConvergence-MainPaper-HQTE} with four examples; for more elaborate applications of process weak convergence in the context of quantile regression we refer to~\cite{belloni2019conditional, chao2017quantile, angrist2006quantile, chernozhuov2005subsampling}.

\begin{example}[Asymptotic normality of the HQTE estimator] \label{example:AsymptoticNormalityHQTE}
	For fixed quantile $\tau \in \mathcal{T}$, Theorem~\ref{theorem:WeakConvergence-MainPaper-HQTE} implies that  $\sqrt{n}\big(\widehat{\alpha}(\tau;z) - \alpha(\tau; z) \big)$ is asymptotically normal with mean zero and variance $\sigma^2(\tau;z) := \lim_{n \rightarrow \infty}\sigma_{(n)}^2(\tau;z)$, where
	\begin{align*}
	\sigma_{(n)}^2(\tau;z) &:= \tau(1- \tau) z'\left[\Big(\pi_1\mathbb{E}[f_{Y_1|X}^2(X'\theta_1(\tau)|X) XX'\mid D = 1]\Big)^{-1} \right.\\
	&\left. \hspace{60pt}\quad{} + \Big(\pi_0\mathbb{E}[f_{Y_0|X}^2(X'\theta_0(\tau)|X) XX' \mid D = 0]\Big)^{-1} \right]z,
	\end{align*}
	where $0 <\pi_1 = 1- \pi_0 = \mathbb{P}\left\{D= 1\right\} < 1$. 
\end{example}

\begin{example}[Joint asymptotic normality of the HQTE estimator at finitely many quantiles] \label{example:JointAsymptoticNormalityHQTE}
	Consider a finite collection of quantile levels $\{\tau_1, \ldots, \tau_K\} \subset \mathcal{T}$. Theorem~\ref{theorem:WeakConvergence-MainPaper-HQTE} implies that the collection $\sqrt{n}\big(\widehat{\alpha}(\tau_j;z) - \alpha(\tau_j; z) \big)$, $j=1, \ldots, K$, is jointly asymptotically normal with mean zero and covariance matrix $\Sigma = (H_1(\tau_j, \tau_k; z)+H_0(\tau_j, \tau_k; z))_{j,k=1}^K$.
\end{example}

\begin{example}[Uniform confidence bands for the HQTE curve]\label{example:UniformConfidenceBand}
	Define $K(z) := \sup_{\tau \in \mathcal{T}}|\mathbb{G}_1(\tau; z)/\sigma(\tau;z) + \mathbb{G}_0(\tau; z)/\sigma(\tau;z)|$, where $\sigma^2(\tau;z)$ is the variance from Example~\ref{example:AsymptoticNormalityHQTE}. Let $\hat{\kappa}(\alpha;z)$  and $\widehat{\sigma}_n^2(\tau;z)$ be (uniformly) consistent estimates of the $\alpha$ quantile of $K(z)$ and $\sigma^2(\tau;z)$, respectively. Then,
	\begin{align*}
	\lim_{n \rightarrow \infty}\mathbb{P} \left\{ \alpha(\tau;z) \in \left[\hat{\alpha}(\tau;z) \pm \hat{\kappa}(\alpha;z)\frac{\widehat{\sigma}_n(\tau;z)}{\sqrt{n}} \right], \: \tau \in \mathcal{T}\right\} = \alpha.
	\end{align*}
	A consistent estimate $\hat{\kappa}(\alpha;z)$ can be obtained via simulation based bootstrap, i.e. sampling from $\widehat{K}(z) = \sup_{\tau \in \mathcal{T}}|\widetilde{\mathbb{G}}_1(\tau; z) + \widetilde{\mathbb{G}}_0(\tau; z)|$, where $\widetilde{\mathbb{G}}_1(\tau; z)$ and $\widetilde{\mathbb{G}}_0(\tau; z)$ are independent centered Gaussian processes with covariance functions based on uniformly consistent plug-in estimates of the operators $(\tau_1, \tau_2) \mapsto H_d(\tau_1, \tau_2;z)/(\sigma(\tau_1;z)\sigma(\tau_2;z))$, $d \in \{0,1\}$.
\end{example}

\begin{example}[Asymptotic theory for the integrated HQTE curve]
	Assessing the HQTE on a specific quantile is often less relevant than assessing the average HQTE over a continuum of quantile levels $\mathcal{T}$ (e.g., lower, middle, or upper quantiles). In such cases, it is natural to consider the integrated HQTE. Theorem~\ref{theorem:WeakConvergence-MainPaper-HQTE} and the continuous mapping theorem imply that $\sqrt{n}\int_\mathcal{T} \big(\widehat{\alpha}(\tau;z) - \alpha(\tau; z)\big)d\tau   \leadsto I(z)$, where $I(z) :=  \int_\mathcal{T} \mathbb{G}_1(\tau, z)d\tau + \int_\mathcal{T} \mathbb{G}_0(\tau;z) d \tau.$ While the random variable $I(z)$ is not distribution-free, its distribution can be approximated via re-sampling techniques~\citep{chernozhuov2005subsampling}.
\end{example}

\subsection{Duality theory for the rank-score debiasing program}\label{subsec:RankScoreDual}
In this section we introduce the dual to the rank-score debiasing program~\eqref{eq:subsec:RankScoreBalancedEstimator-3} and explain its pivotal role in the proofs of the weak convergence results in Sections~\ref{subsec:WeakConvergence}. The dual program is also important for constructing uniformly consistent estimates of the covariance function in Sections~\ref{subsec:ConsistencyCovariance}.

Observe that the solution to the rank-score debiasing program~\eqref{eq:subsec:RankScoreBalancedEstimator-3} can be written as $\widehat{w}(\tau;z) = \widehat{w}_0(\tau;z) + \widehat{w}_1(\tau;z)$, with the $\widehat{w}_d(\tau;z)$'s being the solutions to two independent optimization problems: 
\begin{align}\label{eq:BalancingWeights-Primal}
\widehat{w}_d(\tau;z) \in  \arg \min_{w \in \mathbb{R}^n} \left\{ \sum_{i=1}^n w_i^2\hat{f}_i^{-2}(\tau) : \:  \left\|z - \frac{1}{\sqrt{n}} \sum_{i: D_i = d} w_iX_i\right\|_\infty \leq \frac{\gamma_d}{n}\right\}, \quad{}\quad{} d \in \{0,1\}.
\end{align}
These two optimization problems have the following two duals:
\begin{align}\label{eq:BalancingWeights-Dual}
\hat{v}_d(\tau;z) \in  \arg \min_{v \in \mathbb{R}^p} \:\:\:&\left\{ \frac{1}{4n}\sum_{i:D_i = d }\hat{f}_i^2(\tau) (X_i'v)^2 + z' v + \frac{\gamma_d}{n} \| v\|_1\right\}, \quad{}\quad{} d \in \{0,1\}.
\end{align}
Provided that strong duality holds, we can estimate the rank-score debiasing weights $\widehat{w}_d(\tau;z)$ by either solving the primal problems~\eqref{eq:BalancingWeights-Primal} or by solving the dual problems~\eqref{eq:BalancingWeights-Dual} and exploiting the explicit relationship between primal and dual solutions. To be precise, we have the following result:

\begin{lemma}[Dual characterization of the rank-score debiasing program]\label{lemma:CharacterizationBalancingEstimator-MainPaper}
	\noindent
	
	\begin{itemize}
		\item[(i)] Programs~\eqref{eq:BalancingWeights-Primal} and~\eqref{eq:BalancingWeights-Dual} form a primal-dual pair.
		\item[(ii)] Let $\delta \in (0,1)$ and $d \in \{0,1\}$. Suppose that Conditions~\ref{condition:SubGaussianity},~\ref{condition:LipschitzBoundednessDensity} (i), and~\ref{condition:UpperBoundEigenvalues} hold. There exists an absolute constant $c_1 > 1$ such that for all $\gamma_d > 0$ that satisfy $\gamma_d \geq c_1 \varphi_{\max} \kappa_2^{-1}(\infty) \bar{f}^2\|z\|_2\sqrt{n\log(p/\delta)}$, we have with probability at least $1 - \delta$, for all $1 \leq i \leq n$ and $\tau \in \mathcal{T}$,
		\begin{align*}
		\widehat{w}_{d, i}(\tau; z)= \begin{cases}
		-\frac{\hat{f}_i^2(\tau)}{2\sqrt{n}}X_i'\hat{v}_d(\tau; z), & i \in \{j: D_j = d\}\\
		0, &i \notin \{j: D_j = d\},
		\end{cases}
		\end{align*}
		where $\widehat{w}_d(\tau;z)$ and $\hat{v}_d(\tau;z)$ are the solutions to the programs~\eqref{eq:BalancingWeights-Primal} and~\eqref{eq:BalancingWeights-Dual}, respectively.
	\end{itemize}
\end{lemma}

The important takeaway from Lemma~\ref{lemma:CharacterizationBalancingEstimator-MainPaper} is that, with high probability, for $\gamma_d > 0$ large enough, the rank-score balanced estimator~\eqref{eq:subsec:RankScoreBalancedEstimator-4} has the following equivalent dual formulation:
\begin{align}\label{eq:RankScoreBalancedEstimator-Dual}
\widehat{Q}_d(\tau;z) = z'\hat{\theta}_d(\tau) - \frac{1}{2n}\sum_{i:D_i =d} \hat{f}_i(\tau)\big(\tau - \mathbf{1}\{Y_i \leq X_i'\hat{\theta}_d(\tau)\}\big)X_i'\hat{v}_d(\tau;z).
\end{align}

Thus, while the original formulation of the rank-score balanced estimator involves a complicated sum over the rank-score debiasing weights $\widehat{w}_1(\tau;z), \ldots, \widehat{w}_n(\tau;z)$, the dual formulation is a simple linear function of the dual solution $\hat{v}_d(\tau;z) \in \mathbb{R}^p$. Therefore, we can expect that (at least for fixed $\tau$ and $p$) the rank-score debiased estimator can be approximated by a sum of $n$ independent and identically distributed random variables. The following non-asymptotic Bahadur-type representation is a significantly refined version of this statement (holding uniformly in $\tau \in \mathcal{T}$ and for $p \geq n$). It is key to the weak convergence results in Section~\ref{subsec:WeakConvergence}.

\begin{lemma}[Bahadur-type representation]\label{lemma:BahadurTypeRep-MainPaper}
	Let $\mathcal{T}$ be a compact subset of $(0,1)$ and $\delta \in (0,1)$. Suppose that Conditions~\ref{condition:Unconfoundedness}--\ref{condition:LowerBoundEigenvalues} hold with $\varrho_n = \sqrt{ (s_v + s_\theta)\log(np/\delta) / n}$ and $\epsilon_n^2 = O(\sqrt{n} h^{-1} \varrho_n^2 + h^2)$. In addition, suppose that $(s_v + s_\theta)^2\log^2( np/\delta)\log^2(n)= o(nh^2)$, $h^2 s_v = o(1)$, and $\|z\|_2 = O(1)$. If $\lambda_d \asymp \sqrt{\varphi_{\max}}\sqrt{n \log(np/\delta)}$ and $\gamma_d \asymp \|z\|_2 \left(h^{-1} s_\theta\log(np/\delta) +\sqrt{n}h^2\right)\sqrt{n}$, then
	\begin{align*} 
	&\widehat{Q}_d(\tau;z) - Q_d(\tau; z)\\
	&\quad{}\quad{}\quad{} = - \frac{1}{2n}\sum_{i:D_i = d} f_{Y_d|X}(X_i'\theta_d(\tau)|X_i)\big(\tau - \mathbf{1}\{Y_i \leq X_i'\theta_d(\tau)\}\big)X_i'v_d(\tau; z) + e_d(\tau; z),
	\end{align*}
	where $v_d(\tau;z) = -2 \big(\mathbb{E}[ f_{Y_d|X}^2\big(X'\theta_d(\tau)|X\big)XX'\mathbf{1}\{D= d\}]\big)^{-1}z$, and, with probability at least $1- \delta$,
	\begin{align*}
		\sup_{\tau \in \mathcal{T}} |e_n(\tau; z)| &\lesssim c_2 \left( \varrho_n^{3/2} (\log n)^{3/4} +  h^2 \varrho_n + h^{-1}\varrho_n^2\right),
	\end{align*}
	where $c_2 > 0$ depends on $\bar{f}, \underline{f}, L_f, L_\theta, C_Q, \kappa_1(2), \kappa_2(\infty), \varphi_{\max}, \|z\|_2$.
\end{lemma}

The upper bound (or: rate) on the remainder term $e_n(\tau; z)$ comprises a parametric and a non-parametric part. The parametric part is $\varrho_n^{3/2} (\log n)^{3/4} =  (s_v + s_\theta)^{3/4}\log^{3/4}(np/\delta)\log^{3/4}(n)/n^{3/4}$. Up to the $\log$-factors this rate matches the optimal rate of the residuals of the Bahadur representation for classical estimators of the quantile function~\citep{bahadur1966anote, kiefer1967bahadur} as well as quantile regression estimators in low dimensions~\citep{zhou1996direct}. The non-parametric part $h^2 \varrho_n + h^{-1}\varrho_n^2$ depends on the bandwidth $h > 0$. The particular dependence of the bandwidth is the result of the two-fold dependence of the rank-score debiased estimator on the non-parametric density estimates: a direct dependence via $\hat{f}_i(\tau)$ and an indirect dependence via $\widehat{v}_d(\tau;z)$.

\subsection{Consistent estimates of the covariance function}\label{subsec:ConsistencyCovariance}
The weak convergence results and examples from Section~\ref{subsec:WeakConvergence} are only practically relevant together with an estimator of the asymptotic covariance function that is uniformly consistent in $\tau_1, \tau_2 \in \mathcal{T}$. Here, we show how to exploit the duality formalism from Section~\ref{subsec:RankScoreDual} to construct such estimators.

An estimate for the covariance function $(\tau_1, \tau_2) \mapsto H_d(\tau_1, \tau_2; z)$ is given by
\begin{align*} 
\widehat{H}_d(\tau_1, \tau_2;z) :=(\tau_1 \wedge \tau_2 - \tau_1 \tau_2) \hat{v}_d' (\tau_1; z) \left(\frac{1}{4n}\sum_{i :D_i=d} \hat{f}_i(\tau_1) \hat{f}_i(\tau_2) X_iX_i' \right)\hat{v}_d (\tau_2; z),
\end{align*}
where $\hat{v}_d(\tau; z)$ is the solution to the dual program~\eqref{eq:BalancingWeights-Dual} (see Section~\ref{subsec:RankScoreDual}). By the following lemma this estimate is uniformly consistent in $\tau_1, \tau_2 \in \mathcal{T}$.
\begin{lemma}\label{lemma:ConsistencyCovarianceFunction}
	Recall the setup of Theorem~\ref{theorem:WeakConvergence-MainPaper-CQF} and let $\varrho_n = \sqrt{(s_v + s_\theta)\log(np)/ n}$. The following holds:
	\begin{align*}
	\sup_{\tau_1, \tau_2 \in \mathcal{T}}\left|\widehat{H}_d(\tau_1, \tau_2;z) -  H_d(\tau_1, \tau_2;z)  \right| = O_p\left(c_3\Big(\varrho_n + \sqrt{n} \sqrt{s_v} h^{-1} \varrho_n^2  + \sqrt{s_v}h^2\Big)\right),
	\end{align*}
	where $c_3 > 0$ depends on $\bar{f}, \underline{f}, L_f, L_\theta, C_Q, \kappa_1(2), \kappa_2(\infty), \varphi_{\max}, \|z\|_2$.
\end{lemma}
As a consequence, a uniformly consistent estimate of the asymptotic variance $\sigma^2(\tau; z)$ of the HQTE process at a single quantile $\tau \in \mathcal{T}$ (see Example~\ref{example:AsymptoticNormalityHQTE}) is given by
\begin{align}\label{eq:subsec:ConsistecyCovariance-2}
\widehat{\sigma}_1^2(\tau;z) :=  \tau(1 - \tau)\left( \frac{1}{4n}\sum_{i: D_i = 1} \hat{f}_i^2(\tau) \big(X_i'\hat{v}_1(\tau; z)\big)^2 +\frac{1}{4n}\sum_{i: D_i = 0} \hat{f}_i^2(\tau) \big(X_i'\hat{v}_0(\tau; z)\big)^2 \right),
\end{align}
where $\hat{v}_1(\tau; z)$ and $\hat{v}_0(\tau; z)$ are the solutions to the dual problems~\eqref{eq:BalancingWeights-Dual}. The duality formalism from Section~\ref{subsec:RankScoreDual} implies that another uniformly consistent estimate for $\sigma^2(\tau;z)$ is given by 
\begin{align}\label{eq:subsec:ConsistecyCovariance-3}
\widehat{\sigma}_2^2(\tau;z) := \tau(1 - \tau) \sum_{i = 1}^n \widehat{w}_i^2(\tau;z) \hat{f}_i^{-2}(\tau),
\end{align}	
where the $\widehat{w}_i(\tau;z)$'s are the rank-score debiasing weights. Neither of the two estimates requires inverting a (high-dimensional) matrix, which may be surprising given the form of the target $\sigma^2(\tau; z)$.

\section{A practical guide to the rank-score debiasing procedure}\label{sec:Implementation}
In the following, we explain how we implement the rank-score debiasing procedure with the help of the dual problem. As the rank-score debiasing estimator of the HQTE depends on the four regularization parameters $\lambda_0$, $\lambda_1$, $\gamma_0$, and $\gamma_1 > 0$ and the bandwidth $h > 0$ of the non-parametric density estimator, we also explain how to choose these parameters in robust and data-dependent ways. 

\subsection{Implementing the $\ell_1$-penalized quantile regression program}\label{subsec:TuningParameters}

To select $\lambda_d > 0$ in a data dependent way, we substantially deviate from the vanilla quantile regression program~\eqref{eq:subsec:RankScoreBalancedEstimator-1} and instead implement the weighted $\ell_1$-penalized quantile regression problem by~\cite{belloni2011L1penalized}. That is, we compute the pilot estimate of $\theta_d(\tau)$ as
\begin{align}\label{eq:weighted-lasso-belloni}
\hat{\theta}_d(\tau) \in \arg\min_{\theta \in \mathbb{R}^p} \left\{\sum_{i: D_i = d} \rho_\tau(Y_i - X_i'\theta) + \lambda_d\sqrt{\tau(1-\tau)} \sum_{k=1}^p\widehat{\sigma}_{d,k} |\theta_k| \right\},
\end{align}
with $\widehat{\sigma}_{d,k}^2 = n^{-1}\sum_{i; D_i = d} X_{ik}^2$ and $\lambda_d = 1.5 \cdot \Lambda_d(0.9|X_1, \ldots, X_n)$, where $\Lambda_d(0.9|X_1, \ldots, X_n)$ is the 90\%-quantile of $\Lambda_d| X_1, \ldots, X_n $ and
\begin{align*}
\Lambda_d :=  \sup_{\tau \in \mathcal{T}}\max_{1 \leq k \leq p} \left|\sum_{i: D_i = d}\frac{ (\tau - 1\{U_i \leq \tau\})X_{ik}}{\widehat{\sigma}_{d,k} \sqrt{\tau(1- \tau)} }\right|,
\end{align*}
with $U_1, \ldots, U_n$ be i.i.d. Uniform(0,1) random variables, independent of $X_1, \ldots, X_n$

\subsection{Implementing the rank-score debiasing program}\label{Sec:Rank-score-balncing-implement}

Recall the primal and dual programs \eqref{eq:BalancingWeights-Primal} and \eqref{eq:BalancingWeights-Dual}, respectively. Provided that strong duality holds, we can estimate the rank-score balancing weights $\widehat{w}(\tau;z)$ by solving either of the two problems. However, from a statistical and computational point of view, it is preferable to solve the dual problems.

First, since the dual programs \eqref{eq:BalancingWeights-Dual} are unconstrained optimization problems they allow us to choose the tuning parameter $\gamma_d > 0$ systematically via cross-validation. In contrast, the primal problems are constrained optimization problems which do not naturally lend themselves to cross-validation procedures. In the simulation study we therefore implement a 10-fold cross validation procedure on the dual problems and choose $\gamma_d > 0$ as the smallest tuning parameter which yields a risk that is at most one standard deviation away from the smallest cross-validated risk. The main point of this one-standard-deviation (1SE) rule is to estimate debiasing weights with small bias $\|z - \frac{1}{\sqrt{n}}\sum_{i :D_i = d} w_i X_i\|_\infty$ whose risk is comparable to the one of the optimal weights. A smaller $\gamma_d > 0$ produces a less biased estimate, which leads to a better coverage probability of the confidence interval.
It is instructive to compare our 1SE rule with the 1SE rule popularized by~\cite{breiman1984classification}.~\cite{breiman1984classification} aim to improve the out-of-sample (classification) accuracy of their estimator and hence advocate choosing the least variable model whose risk is comparable to the model with the smallest cross-validated risk. In contrast, we aim to improve statistical inferential validity and hence are less concerned about the variability of our estimate than its bias.

Second, since the primal problems \eqref{eq:BalancingWeights-Primal} are constrained optimization programs, finding feasible points can be difficult. In contrast, the dual programs are unconstrained convex optimization problems and therefore can be easily solved by off-shelf optimization packages. In our simulation studies, we solve the primal problem using \texttt{R} package \texttt{CVXR} \citep{fu2017cvxr}, and the dual problem using Alternating Direction Method of Multipliers (ADMM) to the $l_1$-regularized quadratic program \citep{wahlberg2012admm} via \texttt{R} package \texttt{accSDA} \citep{admmpackage}.

Third, since the dual programs do not involve the inverses of the estimated densities $\hat{f}_i(\tau)$ they are numerically more stable than the primal problems.  Therefore, in the simulation studies and the real data analysis we only report results obtained via the dual problem.

\subsection{Selecting bandwidth $h$ for the non-parametric density estimator}\label{sec:BandwidthSelection}

To stabilize the density estimator~\eqref{eq:subsec:RankScoreBalancedEstimator-2}, we replace the $\ell_1$-penalized quantile regression estimates with refitted quantile regression estimates. The refitted estimates are obtained by fitting a quantile regression model to the data using only the covariates in the support set of $\hat{\theta}_d(\tau)$. As this density estimator takes a similar form as the one in~\cite{belloni2019valid}, we follow their advice and set bandwidth $h = \min\{ n^{-1/6}, \tau(1-\tau)/2 \}$.

\section{Simulation study}\label{sec:SimulationStudies}
We carry out simulation studies to investigate the performance of the rank-score debiased estimator. The goal of the simulation studies is to: (1) illustrate our rank-score debiased estimator provides consistent estimate of HQTE with nominal-level coverage probabilities, (2) showcase the rank-score debiased estimator is more efficient than the unweighted quantile regression estimator, and (3) provide numerical evidence supporting the theoretical results from Section \ref{subsec:WeakConvergence}.

\subsection{Simulation design}

Our simulation design mimics high-dimensional observation studies where treatments are assigned based on covariates. We consider the following generative model:
\begin{align*}
&Y_1 = X'\theta_1 + \varepsilon\sigma_{1}(X), \qquad Y_0 =  X'\theta_0+ \varepsilon\sigma_{0}(X) , \qquad X \independent \varepsilon, \qquad \varepsilon \sim N(0,1),\\
& D \mid X\sim \text{Bernoulli}\left( \frac{ e^{1 - X_{7} +X_{8} } }{1 + e^{1 - X_{7} +X_{8} }} \right), \qquad
Y = D Y_1  + (1-D) Y_0. 
\end{align*}
For the noise level $\sigma_d(X)$ and the covariates $X$, we consider two sets of covariate designs for the homoscedastic case and the heteroscedastic case. We first generate $W \sim N( 0, \Sigma)$ where $\Sigma = (\Sigma_{jk})_{j,k=1}^{p-1}$ and $\Sigma_{jk} = 0.5^{|j-k|}$. Then, in the homoscedastic case, we set $\sigma_1(X) = \sigma_0(X)=1$ and generate the covariates with $X_1 = 1$ and $X_j = W_j$, for $2 \leq j\leq p$. In the heteroscedastic case, we set $X_1 = 1$, $X_2 = |W_2| + 0.1$, $X_3  = W_3^2 + 0.5$, $X_j = W_j$ for $ 4 \leq j \leq p$, and $\sigma_d(X) =(1- d)X_2 +d X_3$ for $d \in \{0,1\}$. In both cases, we set $\theta_0 = (0.5, 0, 1, -1, 0, \ldots, 0)' \in \mathbb{R}^p$ and consider the following three scenarios for $\theta_1$: sparse ($\theta_1 \propto (1,1,1, 1, 1, 1, 0, \ldots, 0)'$), dense ($\theta_1 \propto(1, 1/\sqrt{2}, \ldots, 1/\sqrt{p})'$) and pseudo dense ($\theta_1 \propto(1, 1/{2}, \ldots, 1/{p})'$). We consider three different signal strengths $||\theta_1||_2\in \{1,2,4\}$. We choose the sample size $n$ and the dimension of the covariates $p$ from $ (n,p)\in \{ (600, 400), (1000, 600) \}$. As we estimate the CQF separately by using the observed data in the treated and control groups, the effective sample size for our rank-score debiasing program $n_d$ is approximately half of the sample size. Thus, the effective sample size is always less than $p$. Lastly, we set $z = (0, 1/\sqrt{2}, 1/\sqrt{2}, 0, \ldots, 0)'$ or $z = (1, 1, 1/\sqrt{2}, \ldots, 1/\sqrt{p})'$. Under this data generating process, the HQTE at $z$ is the linear function $\alpha(\tau ;z ) = z'(\theta_1(\tau) -\theta_0(\tau))$. 

We implement the rank-score debiased estimator as discussed in Section~\ref{sec:Implementation}. In particular, this means that even in the case of homoscedastic noise we do not use a specialized density estimator that could exploit this extra information. Since in practice homoscedasticity may be difficult to detect, we do not to want rely on the validity of the homoscedasticity assumption.
To illustrate the bias-variance trade-off that underlies the tuning parameter $\gamma_d$ we report results not just for the ``1SE'' rule (``Rank-1SE") but also for a ``2SE" rule (``Rank-2SE"). The ``2SE'' rule chooses the smallest $\gamma_d > 0$ that is less than two standard errors away from the tuning parameter with the lowest dual loss function.

To showcase the merit of the rank-score debiased estimator, we compare it with the following four methods: ``Unweighted Oracle", ``Refit'', ``Lasso'' and ``Debiased. The ``Unweighted Oracle'' method fits a quantile regression model based on the true model, i.e. based on the covariates in the support set of $\theta_d(\tau)$ only. The (unweighted) oracle estimate of the HQTE is $\widehat{\alpha}^{\mathrm{oracle}}(\tau ;z ) = z'(\hat{\theta}_1^{\mathrm{oracle}}(\tau) - \hat{\theta}_0^{\mathrm{oracle}}(\tau))$. We compute this (unweighted) oracle estimate only in the scenario with sparse $\theta_d(\tau)$. The ``Refit" method is the following two-step procedure: We first obtain estimates $\hat{\theta}_d(\tau)$ by solving the weighted $\ell_1$-penalized quantile regression program \eqref{eq:weighted-lasso-belloni}. Then, we compute a refitted estimate $\hat{\theta}_d^{\mathrm{refit}}(\tau)$ by fitting a quantile regression model based only on the covariates in the support set of $\hat{\theta}_d(\tau)$. The refitted estimate of the HQTE is $\widehat{\alpha}^{\mathrm{refit}}(\tau ;z ) = z'(\hat{\theta}_1^{\mathrm{refit}}(\tau) - \hat{\theta}_0^{\mathrm{refit}}(\tau))$. The ``Lasso" method refers to simply using the estimates $\hat{\theta}_d(\tau)$ from the $\ell_1$-penalized quantile regression program \eqref{eq:weighted-lasso-belloni} without any adjustments. The Lasso estimate of the HQTE is thus $\widehat{\alpha}^{\mathrm{lasso}}(\tau ;z ) = z'(\hat{\theta}_1^{\mathrm{lasso}}(\tau) - \hat{\theta}_0^{\mathrm{lasso}}(\tau))$. The ``Debiased" method refers to the debiased $\ell_1$-penalized quantile regression coefficient estimate proposed by \cite{zhao2019debiasing}. 
 We denote the debiased estimate of the HQTE as $\widehat{\alpha}^{\mathrm{debias}}(\tau ;z ) = z'(\hat{\theta}_1^{\mathrm{debias}}(\tau) - \hat{\theta}_0^{\mathrm{debias}}(\tau))$ with
\begin{align*}
		& \hat{\theta}_d^{\mathrm{debias}} (\tau)= \hat{\theta}_d^{\mathrm{Lasso}} (\tau)+ \hat{\Theta}_d(\tau)\cdot \frac{1}{n_d}\sum_{i:D_i = d} X_i ( \tau - \mathbf{1}\{Y_i \leq X_i'\hat{\theta}_d(\tau) \} ) ,
\end{align*}
where $\hat{\Theta}_d(\tau) $ is an estimate of the inverse covariance matrix $\big[ \mathbb{E}[f_{Y_d|X}( X'\theta_d(\tau)|X ) XX' \mathbf{1}\{D = d\} ]\big]^{-1}$. Following the recommendation by \cite{zhao2019debiasing}, we use the \texttt{R} package \texttt{clime} to obtain $\hat{\Theta}_d (\tau)$.

Confidence intervals for the rank-score debiased estimator are based on the asymptotic normality results in Section~\ref{subsec:WeakConvergence} and hold under the mild regularity conditions stated in Section~\ref{subsec:RegularityConditions}. Confidence intervals for the (unweighted) oracle method are constructed using standard large sample theory~\citep{angrist2006quantile}. Confidence intervals for the Lasso and the Refit method are constructed assuming that the selected models equal the true model, i.e. the support sets of $\hat{\theta}_d^\mathrm{lasso}(\tau)$, $\hat{\theta}_d^\mathrm{refit}(\tau)$ equal the support set of $\theta_d(\tau)$. This assumption is satisfied under strong oracle conditions \citep{fan2001variable}. As \cite{zhao2019debiasing} focus on providing accurate point estimate of the quantile regression coefficients, they do not construct confidence interval for $\theta_d(\tau)$. Based on our conjecture provided in the Supplementary Materials, we construct confidence intervals based on normal approximation with an estimated asymptotic variance equal to $\pi_1	z'\hat{\Theta}_1(\tau) \frac{1}{n_1}\sum_{i:D_i = 1} X_i X_i' \hat{\Theta}_1(\tau) z+\pi_0	z'\hat{\Theta}_0(\tau) \frac{1}{n_0}\sum_{i:D_i = 0} X_i X_i' \hat{\Theta}_0(\tau) z$.

\begin{figure}[h]
	\includegraphics[width=\linewidth]{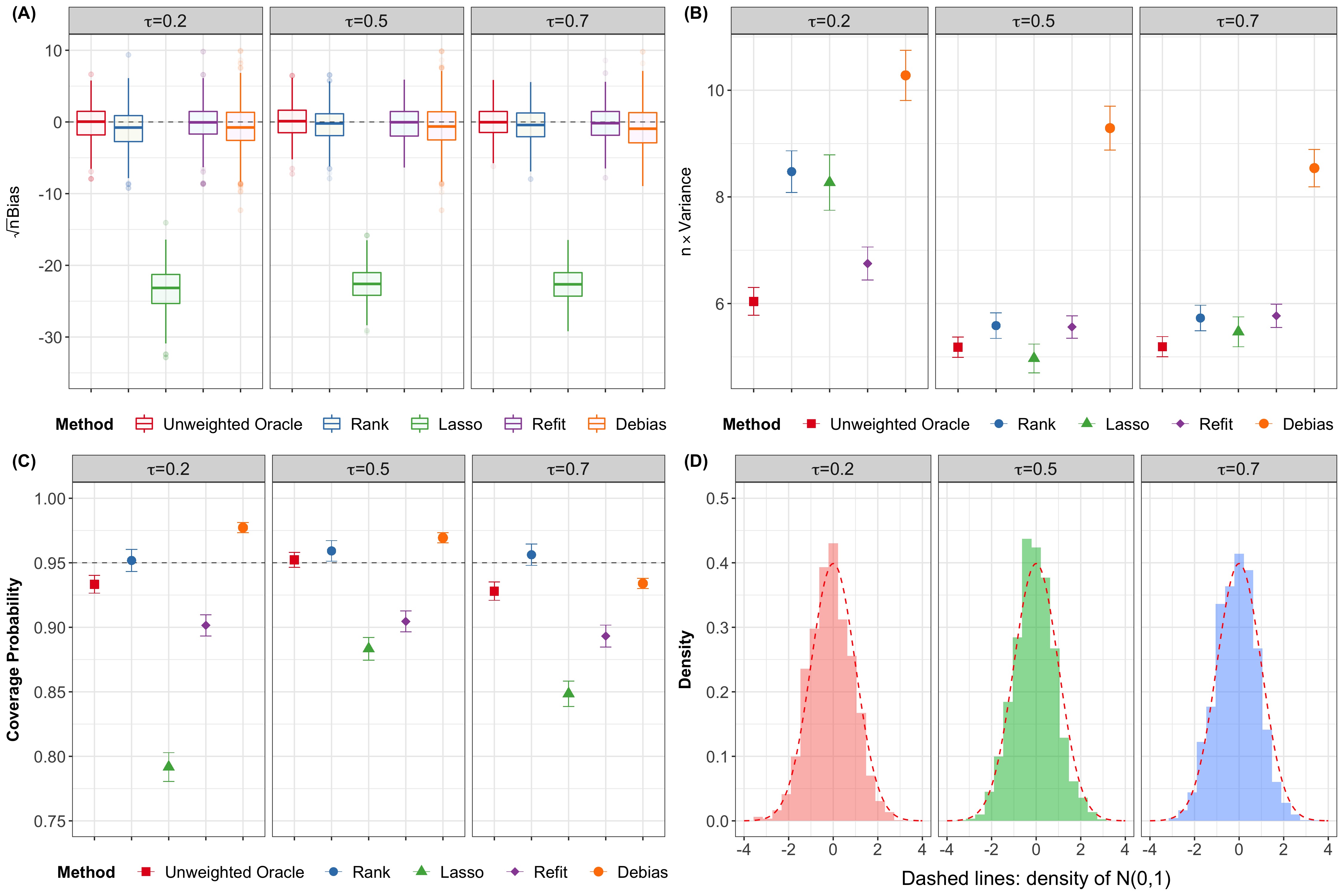}
	\caption{Simulation results for homoscedastic data with $(n, p) = (1000, 600)$, sparse $\theta_1$ with $||\theta_1||_2 = 4$ and sparse $z$. Panel (A) Bias comparison. Panel (B) Variance comparison. Panel (C) Coverage probability of the confidence interval while the nominal coverage probability is 95\%. Panel (D) Histograms of the standardized estimates of the rank-score debiased estimator displayed in \eqref{eq:standardized-rank}, density of $N(0,1)$ in red.}
	\label{figure-homo}
\end{figure}

\begin{figure}[h]
	\includegraphics[width=\linewidth]{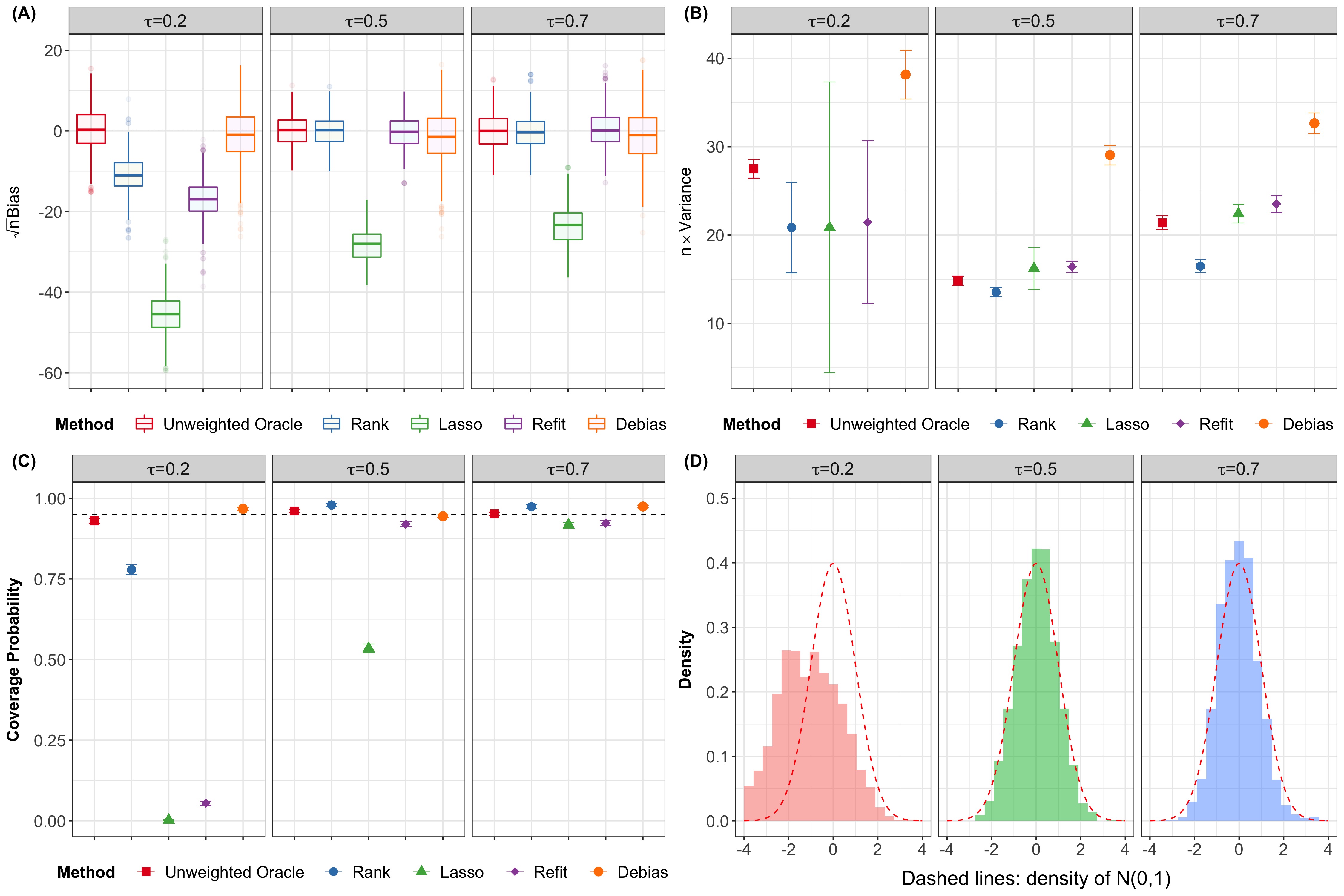}
	\caption{Simulation results for heteroscedastic data with $(n, p) = (1000, 600)$, sparse $\theta_1$ with $||\theta_1||_2 = 4$, and sparse $z$. Panel (A) Bias comparison. Panel (B) Variance comparison. Panel (C) Coverage probability of the confidence interval while the nominal coverage probability is 95\%. Panel (D) Histograms of the standardized estimates of the rank-score debiased estimator displayed in \eqref{eq:standardized-rank}, density of $N(0,1)$ in red. }
	\label{figure-heter}
\end{figure}

\subsection{Simulation results}

We measure the performance of the estimators in terms of their biases (computed as the differences between the mean of the Monte Carlo estimates of $\alpha(\tau;z)$ and the true HQTE), variances (computed as the variances of the Monte Carlo estimates of $\alpha(\tau;z)$) and coverage probabilities of the confidence intervals with the nominal coverage probability of 95\%. We provide finite sample comparisons through Table \ref{table-homo} and Figure \ref{figure-homo} for homoscedastic data, and Table \ref{table-heter} and Figure \ref{figure-heter} for heteroscedastic data. Details about the model parameters are given in the captions of these tables and figures. Our simulation results are evaluated through 2,000 Monte Carlo samples.

The main takeaway from the simulation study is that the rank-score debiased estimator with $\gamma_d$ selected by the 1SE rule outperforms the Refitted and Lasso estimators in terms of bias, variance and the validity of inference in most scenarios. In the following we highlight three conclusions. First, the rank-score debiased estimator performs better in sparse than in dense models. Second, the rank-score debiased estimator can have a smaller variance than the unweighted Oracle estimator in the heteroscedastic cases when $\theta_1$ is sparse. Third, the asymptotic normality results from Section~\ref{subsec:WeakConvergence} continue to hold reasonably well in finite samples. This can be deduced from Figures \ref{figure-homo}(D) and \ref{figure-heter}(D), in which we provide histograms of the standardized estimates of the rank-score debiased estimator, 
\begin{align}\label{eq:standardized-rank}
\widehat{\sigma}_2^{-1}(\tau;z)\cdot\sqrt{n}\big( \widehat{\alpha}(\tau;z) - \alpha(\tau;z) \big),
\end{align}
where $ \widehat{\sigma}_2(\tau;z)$ is the estimate defined in eq.~\eqref{eq:subsec:ConsistecyCovariance-2}. These histograms fit the overlaid N(0,1) densities. In contrast, the Lasso estimator is clearly biased (Figures~\ref{figure-homo}(A) and~\ref{figure-heter}(B)) and so is the Refitting estimator in scenarios with smaller signal to noise ratio, i.e. scenarios with small $\|\theta_1(\tau)\|_2$  (Tables \ref{table-homo} and \ref{table-heter}). These biases suggest that the oracle condition is violated and hence the finite sample distributions of these estimators may not be approximated by a standard normal distribution. The debiased quantile Lasso estimator has small biases, but it often has larger variance compared to the rank-score debiased estimator. This observation is in-line with our conjecture based on the derivation provided in the Supplementary Materials.

\begin{table}[h!]\clearpage
\begin{center}
\resizebox{\textwidth}{!}{	\begin{tabular}{cccccccc}
		\hline \hline 
		& $\tau$  & Unweighted Oracle & Refit & Lasso & Debias & Rank-1SE &  Rank-2SE  \\\cline{2-8}
& & \multicolumn{6}{c}{$n=600, p = 400$, sparse $\theta_1$ with $||\theta_1||_2 = 1$, sparse $z$ } \\\cline{2-8}
$\sqrt{n}$Bias & $ 0.2 $ & $ 0.12 ( 0.08 )$ & $ -0.03 ( 0.08 )$ & $ -1.25 ( 0.08 )$ & $ -0.37 ( 0.10 )$ &  $ -0.50 ( 0.09 )$ & $ -0.47 ( 0.09 )$  \\  
& $ 0.5 $ & $ 0.14 ( 0.07 )$ & $ -0.1 ( 0.07 )$ & $ -1.15 ( 0.08 )$ & $ -0.41 ( 0.09 )$ & $ -0.27 ( 0.08 )$ & $ -0.23 ( 0.09 )$ \\  
& $ 0.7 $ & $ 0.10 ( 0.07 )$ & $ -0.11 ( 0.07 )$ & $ -1.16 ( 0.07 )$ & $ -0.54 ( 0.08 )$ &  $ -0.26 ( 0.08 )$ & $ -0.25 ( 0.08 )$ \\\cline{3-8} 
$n$Var & $ 0.2 $ & $ 6.58 ( 0.31 )$ & $ 7.10 ( 0.33 )$ & $ 6.48 ( 0.35 )$ & $ 9.58 ( 0.47 )$ & $ 7.97 ( 0.37 )$ & $ 8.06 ( 0.38 )$  \\  
& $ 0.5 $ & $ 5.13 ( 0.21 )$ & $ 5.60 ( 0.24 )$ & $ 6.13 ( 0.34 )$ & $ 7.37 ( 0.35 )$ &  $ 6.53 ( 0.33 )$ & $ 6.77 ( 0.33 )$  \\ 
& $ 0.7$ & $ 5.14 ( 0.21 )$ & $ 5.56 ( 0.25 )$ & $ 5.13 ( 0.28 )$ & $ 7.14 ( 0.29 )$ &  $ 5.9 ( 0.25 )$ & $ 6.01 ( 0.26 )$ \\\cline{3-8}
Coverage & $0.2 $ & $ 0.94 $ & $ 0.85 $ & $ 0.84 $ & $ 0.97 $ & $ 0.91 $ & $ 0.93 $ \\  & $ 0.5 $ & $ 0.95 $ & $ 0.91 $ & $ 0.85 $ & $ 0.98 $ & $ 0.95 $ & $ 0.96  $\\  & $0.7$ & $ 0.95 $ & $ 0.89 $ & $ 0.88 $ & $ 0.96 $ & $ 0.95 $ & $ 0.95 $ \\\cline{3-8}

	& & \multicolumn{6}{c}{$n=600, p = 400$, pseudo sparse $\theta_1$ with $||\theta_1||_2 = 1$, sparse $z$ } \\\cline{2-8}
$\sqrt{n}$Bias & $ 0.2 $ & $ -$ & $ -0.34 ( 0.09 )$ & $ -2.15 ( 0.08 )$ & $ -1.20 ( 0.10 )$ & $ -0.90 ( 0.09 )$ & $ -0.86 ( 0.09 )$ \\  
& $ 0.5 $ & $ -$ & $ -0.33 ( 0.08 )$ & $ -1.60 ( 0.08 )$ & $ -0.93 ( 0.09 )$ & $ -0.64 ( 0.09 )$ & $ -0.58 ( 0.09 )$\\ 
& $ 0.7 $ & $-$ & $ -0.27 ( 0.08 )$ & $ -1.54 ( 0.08 )$ & $ -1.09 ( 0.09 )$ & $ -0.75 ( 0.08 )$ & $ -0.72 ( 0.09 )$  \\\cline{3-8} 
$n$Var & $ 0.2 $ & $ -$ & $ 7.55 ( 0.32 )$ & $ 6.75 ( 0.51 )$ & $ 9.15 ( 0.52 )$ & $ 8.03 ( 0.41 )$ & $ 8.05 ( 0.4 )$ \\  
& $ 0.5 $ & $-$ & $ 6.39 ( 0.27 )$ & $ 6.46 ( 0.41 )$ & $ 8.61 ( 0.49 )$ & $ 8.36 ( 0.43 )$ & $ 8.96 ( 0.46 )$ \\  
& $ 0.7$ & $-$ & $ 6.15 ( 0.31 )$ & $ 5.85 ( 0.37 )$ & $ 8.21 ( 0.42 )$ & $ 7.01 ( 0.32 )$ & $ 7.46 ( 0.34 )$  \\\cline{3-8} 
Coverage & $0.2 $ & $ - $ & $ 0.82 $ & $ 0.73 $ & $ 0.97 $ &$ 0.97 $ & $ 0.98 $ \\  & $ 0.5 $ & $ -$ & $ 0.88 $ & $ 0.82 $ & $ 0.96 $ &   $ 0.94 $ & $ 0.96  $ \\  & $0.7$ & $-$ & $ 0.86 $ & $ 0.77 $ & $ 0.97 $ & $ 0.96 $ & $ 0.98 $\\\cline{3-8}

	& & \multicolumn{6}{c}{$n=600, p = 400$, dense $\theta_1$ with $||\theta_1||_2 = 1$, sparse $z$ } \\\cline{3-8}
$\sqrt{n}$Bias & $ 0.2 $ & $-$ & $ -2.02 ( 0.12 )$ & $ -3.06 ( 0.09 )$ & $ -2.15 ( 0.12 )$ & $ -1.91 ( 0.10 )$ & $ -1.79 ( 0.1 )$ \\  
& $ 0.5 $ & $-$ & $ -1.76 ( 0.11 )$ & $ -3.04 ( 0.09 )$ & $ -1.73 ( 0.12 )$ & $ -1.55 ( 0.10 )$ & $ -1.46 ( 0.10 )$ \\  
& $ 0.7 $ & $ -$ & $ -1.64 ( 0.11 )$ & $ -2.94 ( 0.09 )$ & $ -2.14 ( 0.12 )$ & $ -1.83 ( 0.09 )$ & $ -1.73 ( 0.10 )$  \\\cline{3-8} 
$n$Var & $ 0.2 $ & $ -$ & $ 15.30 ( 0.91 )$ & $ 7.39 ( 0.74 )$ & $ 13.80 ( 0.83 )$ & $ 9.72 ( 0.60 )$ & $ 9.96 ( 0.59 )$ \\ 
& $ 0.5 $ & $ -$ & $ 12.55 ( 0.75 )$ & $ 8.45 ( 0.75 )$ & $ 14.6 ( 0.79 )$ & $ 9.99 ( 0.54 )$ & $ 10.38 ( 0.54 )$ \\  
& $ 0.7$ & $ -$ & $ 12.45 ( 0.77 )$ & $ 7.59 ( 0.70 )$ & $ 14.09 ( 0.83 )$ & $ 9.06 ( 0.55 )$ & $ 9.14 ( 0.53 )$  \\\cline{3-8} 
Coverage & $0.2 $ & $-$ & $ 0.65 $ & $ 0.63 $ & $ 0.96 $ & $ 0.93 $ & $ 0.94 $ \\ 
& $ 0.5 $ & $ - $ & $ 0.70 $ & $ 0.60 $ & $ 0.98 $ & $ 0.96 $ & $ 0.98  $ \\  
& $0.7$ & $ - $ & $ 0.72 $ & $ 0.60 $ & $ 0.96 $ & $ 0.97 $ & $ 0.98 $  \\\cline{3-8}

	& & \multicolumn{6}{c}{$n=600, p = 400$, sparse $\theta_1$ with $||\theta_1||_2 = 2$, dense $z$ } \\\cline{3-8}
$\sqrt{n}$Bias & $ 0.2 $ & $ 0.79 ( 0.12 )$ & $ 0.77 ( 0.13 )$ & $ 4.79 ( 0.11 )$ & $ 2.00 ( 0.15 )$ & $ 2.71 ( 0.12 )$ & $ 2.64 ( 0.12 )$ \\  & $ 0.5 $ & $ 0.44 ( 0.11 )$ & $ 0.35 ( 0.11 )$ & $ -1.32 ( 0.12 )$ & $ -0.81 ( 0.15 )$ & $ -0.26 ( 0.13 )$ & $ -0.24 ( 0.13 )$ \\  & $ 0.7 $ & $ 0.57 ( 0.11 )$ & $ 0.56 ( 0.13 )$ & $ -1.16 ( 0.11 )$ & $ -1.03 ( 0.16 )$ & $ -0.49 ( 0.13 )$ & $ -0.49 ( 0.13 )$  \\\cline{3-8} $n$Var & $ 0.2 $ & $ 14.59 ( 0.69 )$ & $ 16.52 ( 0.73 )$ & $ 11.33 ( 1.49 )$ & $ 23.32 ( 1.27 )$ & $ 14.62 ( 0.99 )$ & $ 14.89 ( 0.98 )$ \\  & $ 0.5 $ & $ 11.22 ( 0.52 )$ & $ 12.55 ( 0.58 )$ & $ 14.97 ( 0.73 )$ & $ 22.43 ( 1.09 )$ & $ 16.69 ( 0.82 )$ & $ 16.97 ( 0.84 )$ \\  & $ 0.7$ & $ 11.66 ( 0.55 )$ & $ 16.95 ( 0.83 )$ & $ 12.61 ( 0.62 )$ & $ 27.38 ( 1.27 )$ & $ 17.63 ( 0.87 )$ & $ 17.97 ( 0.89 )$  \\\cline{3-8} Coverage & $0.2 $ & $ 0.95 $ & $ 0.95 $ & $ 0.78 $ & $ 0.99 $ & $ 0.93 $ & $ 0.94 $ \\ 
& $ 0.5 $ & $ 0.95 $ & $ 0.93 $ & $ 0.88 $ & $ 0.99 $ & $ 0.95 $ & $ 0.97  $ \\ 
& $0.7$ & $ 0.97 $ & $ 0.88 $ & $ 0.94 $ & $0.98 $ & $ 0.96 $ & $ 0.96 $  \\\cline{3-8}

\hline\hline
	\end{tabular}}
	\caption{Homoscedastic data. Standard errors of estimates based on 1,000 Monte Carlo samples are given in parenthesis. All standard errors of the coverage probability are smaller than 0.01 and thus are omitted.  }
	\label{table-homo}
\end{center}
\end{table}

\begin{table}[h!]
	\begin{center}
\resizebox{\textwidth}{!}{	\begin{tabular}{cccccccc}
		\hline \hline 
		& $\tau$  & Unweighted Oracle & Refit & Lasso  & Debias & Rank-1SE &  Rank-2SE  \\\cline{2-8}
	& & \multicolumn{6}{c}{$n=600, p = 400$,  sparse $\theta_1$ with $||\theta_1||_2 = 1$, sparse $z$} \\\cline{3-8}
$\sqrt{n}$Bias & $ 0.2 $ & $ 0.21 ( 0.18 )$ & $ -10.65 ( 0.13 )$ & $ -7.77 ( 0.07 )$ & $ -1.14 ( 0.15 )$ & $ -3.71 ( 0.1 )$ & $ -3.51 ( 0.1 )$ \\  
& $ 0.5 $ & $ -0.29 ( 0.12 )$ & $ -2.1 ( 0.15 )$ & $ -5.76 ( 0.11 )$ & $ -0.97 ( 0.16 )$ & $ -1.71 ( 0.14 )$ & $ -1.22 ( 0.15 )$ \\ 
& $ 0.7 $ & $ 0.2 ( 0.15 )$ & $ 0.24 ( 0.16 )$ & $ -2.59 ( 0.14 )$ & $ -1.22 ( 0.15 )$ & $ -0.83 ( 0.13 )$ & $ -0.73 ( 0.13 )$ \\\cline{3-8}
$n$Var & $ 0.2 $ & $ 31.10 ( 1.33 )$ & $ 16.39 ( 4.63 )$ & $ 4.42 ( 2.26 )$ & $ 24.18 ( 1.16 )$ &  $ 10.26 ( 0.98 )$ & $ 10.81 ( 0.95 )$ \\
& $ 0.5 $ & $ 15.54 ( 0.61 )$ & $ 23.07 ( 1.02 )$ & $ 11.26 ( 1.82 )$ & $ 24.29 ( 1.10 )$ & $ 17.34 ( 0.91 )$ & $ 17.80 ( 0.92 )$ \\  
& $ 0.7$ & $ 22.25 ( 0.85 )$ & $ 26.19 ( 1.09 )$ & $ 21.13 ( 1.32 )$ & $ 23.79 ( 1.09 )$ &  $ 17.66 ( 0.81 )$ & $ 18.01 ( 0.82 )$  \\\cline{3-8} 
Coverage & $0.2 $ & $ 0.92 $ & $ 0.17 $ & $ 0.12 $ & $ 0.97 $ & $ 0.88 $ & $ 0.90 $  \\  
& $ 0.5 $ & $ 0.94 $ & $ 0.64 $ & $ 0.50 $ & $ 0.96 $ & $ 0.90 $ & $ 0.92  $ \\  
& $0.7$ & $ 0.94 $ & $ 0.91 $ & $ 0.90 $ & $ 0.99 $ & $ 0.94 $ & $ 0.98 $  \\\cline{3-8}

	& & \multicolumn{6}{c}{$n=600, p = 400$, pseudo sparse $\theta_1$ with $||\theta_1||_2 = 1$, sparse $z$ } \\\cline{3-8}
$\sqrt{n}$Bias & $ 0.2 $ & $ -$ & $ -7.75 ( 0.13 )$ & $ -4.55 ( 0.08 )$ & $ 1.00 ( 0.17 )$ & $ -0.90 ( 0.11 )$ & $ -0.46 ( 0.12 )$ \\ 
& $ 0.5 $ & $-$ & $ -2.79 ( 0.14 )$ & $ -4.69 ( 0.09 )$ & $ -1.44 ( 0.14 )$ & $ -1.81 ( 0.12 )$ & $ -1.38 ( 0.13 )$ \\ 
& $ 0.7 $ & $-$ & $ -0.63 ( 0.17 )$ & $ -3.00 ( 0.16 )$ & $ -2.92 ( 0.17 )$ & $ -1.76 ( 0.15 )$ & $ -1.50 ( 0.16 )$  \\\cline{3-8} 
$n$Var & $ 0.2 $ & $ -$ & $ 16.79 ( 2.82 )$ & $ 5.84 ( 1.02 )$ & $ 29.20 ( 1.49 )$ & $ 14.53 ( 0.49 )$ & $ 17.20 ( 0.59 )$ \\ 
& $ 0.5 $ & $-$ & $ 18.50 ( 1.16 )$ & $ 8.92 ( 1.31 )$ & $ 19.34 ( 1.10 )$ & $ 13.91 ( 0.88 )$ & $ 15.44 ( 0.95 )$ \\ 
& $ 0.7$ & $ -$ & $ 30.67 ( 1.43 )$ & $ 25.12 ( 1.55 )$ & $ 29.61 ( 1.78 )$ & $ 23.08 ( 1.28 )$ & $ 24.71 ( 1.25 )$  \\\cline{3-8} 
Coverage & $0.2 $ & $ -$ & $ 0.26 $ & $ 0.39 $ & $ 0.96$ & $ 0.96 $ & $ 0.97 $ \\ 
& $ 0.5 $ & $ -$ & $ 0.63 $ & $ 0.54 $ & $ 0.96 $ & $ 0.93 $ & $ 0.94  $ \\ 
& $0.7$ & $ -$ & $ 0.84 $ & $ 0.85 $ & $ 0.98 $ & $ 0.95 $ & $ 0.96 $  \\\cline{3-8}

	& & \multicolumn{6}{c}{$n=600, p = 400$, dense $\theta_1$ with $||\theta_1||_2 = 1$, sparse $z$ } \\\cline{3-8}
$\sqrt{n}$Bias & $ 0.2 $ & $ -$ & $ -10.94 ( 0.16 )$ & $ -3.59 ( 0.07 )$ & $ -0.28 ( 0.19 )$ & $ -1.38 ( 0.13 )$ & $ -1.31 ( 0.14 )$ \\ 
& $ 0.5 $ & $-$ & $ -3.53 ( 0.15 )$ & $ -4.60 ( 0.10 )$ & $ -1.10 ( 0.17 )$ & $ -1.49 ( 0.14 )$ & $ -1.43 ( 0.14 )$ \\ 
& $ 0.7 $ & $ -$ & $ -3.61 ( 0.20 )$ & $ -4.39 ( 0.16 )$ & $ -4.41 ( 0.21 )$ & $ -4.02 ( 0.18 )$ & $ -3.93 ( 0.18 )$  \\\cline{3-8} 
$n$Var & $ 0.2 $ & $-$ & $ 25.23 ( 6.01 )$ & $ 4.87 ( 0.77 )$ & $ 35.73 ( 1.44 )$ & $ 18.35 ( 0.85 )$ & $ 19.27 ( 0.86 )$ \\ 
& $ 0.5 $ & $ -$ & $ 24.13 ( 1.45 )$ & $ 10.84 ( 1.24 )$ & $ 29.81 ( 1.60 )$ & $ 19.78 ( 0.95 )$ & $ 20.12 ( 0.95 )$ \\  
& $ 0.7$ & $ -$ & $ 41.7 ( 2.54 )$ & $ 27.39 ( 2.00 )$ & $ 43.87 ( 2.91 )$ & $ 31.38 ( 2.24 )$ & $ 31.77 ( 2.23 )$  \\\cline{3-8} 
Coverage & $0.2 $ & $ -$ & $ 0.03 $ & $ 0.61 $ & $ 0.96 $ & $0.93$ & $ 0.94 $ \\  
& $ 0.5 $ & $- $ & $ 0.60 $ & $ 0.57 $ & $ 0.99 $ & $ 0.95 $ & $ 0.97  $ \\ 
& $0.7$ & $ - $ & $ 0.72 $ & $ 0.72 $ & $ 0.98 $ & $ 0.93 $ & $ 0.94 $  \\\cline{3-8}

	& & \multicolumn{6}{c}{$n=600, p = 400$, sparse $\theta_1$ with $||\theta_1||_2 = 2$, dense $z$ } \\\cline{3-8}
$\sqrt{n}$Bias & $ 0.2 $ & $ 0.32 ( 0.10 )$ & $ -6.47 ( 0.12 )$ & $ -8.68 ( 0.20 )$ & $ -7.63 ( 0.19 )$ &  $ -6.37 ( 0.19 )$ & $ -6.23 ( 0.19 )$\\  & $ 0.5 $ & $ 0.36 ( 0.07 )$ & $ 0.08 ( 0.08 )$ & $ -1.71 ( 0.08 )$ & $ 2.08 ( 0.10 )$ & $ -0.55 ( 0.08 )$ & $ -0.46 ( 0.08 )$ \\  & $ 0.7 $ & $ 0.64 ( 0.1 )$ & $ 0.96 ( 0.11 )$ & $ 1.78 ( 0.10 )$ & $ 2.82 ( 0.12 )$ & $ 0.43 ( 0.10 )$ & $ 0.47 ( 0.10 )$  \\\cline{3-8}   
$n$Var & $ 0.2 $ & $ 13.06 ( 0.58 )$ & $ 15.93 ( 1.03 )$ & $ 38.42 ( 4.95 )$ & $ 38.25 ( 4.39 )$ & $ 34.45 ( 4.04 )$ & $ 34.39 ( 3.97 )$  \\  & $ 0.5 $ & $ 5.85 ( 0.22 )$ & $ 5.74 ( 0.28 )$ & $ 6.69 ( 0.46 )$ & $ 9.11 ( 0.64 )$ & $ 6.27 ( 0.33 )$ & $ 6.43 ( 0.33 )$ \\  & $ 0.7$ & $ 11.23 ( 0.48 )$ & $ 13.24 ( 0.64 )$ & $ 10.16 ( 0.62 )$ & $ 13.7 ( 1.00 )$ & $ 10.71 ( 0.51 )$ & $ 10.78 ( 0.52 )$  \\\cline{3-8} 

Coverage & $0.2 $ & $ 0.96 $ & $ 0.60 $ & $ 0.39 $ & $ 0.66 $ & $ 0.82 $ & $ 0.85 $\\  & $ 0.5 $ & $ 0.96 $ & $ 0.95 $ & $ 0.86 $ & $ 0.99 $ & $ 0.94 $ & $ 0.94  $ \\  & $0.7$ & $ 0.95 $ & $ 0.90 $ & $ 0.92 $ & $ 0.94$ & $ 0.95 $ & $ 0.95 $  \\\cline{3-8}

\hline\hline
	\end{tabular}}
	\caption{Heteroscedastic data. Standard errors of estimates based on 1,000 Monte Carlo samples are given in parenthesis. All standard errors of the coverage probability are smaller than 0.01 and thus are omitted. }
	\label{table-heter}
	\end{center}\clearpage
\end{table}

\section{A case study}\label{sec:Application}

\subsection{Study design}

To illustrate the advantages of considering HQTE, we apply the proposed method to study the heterogeneous effect of statin usage, especially when combined with a healthy lifestyle, in lowering the low--density lipoprotein (LDL) chlestrol concentration levels for older Alzheimer's disease (AD) patients enrolled in the UK Biobank study. 

Alzheimer’s disease (AD) is the sixth leading cause of death in the United States, directly affecting an estimated 5.8 million Americans and incurring nearly \$236 billion of total healthcare costs \citep{alzheimer20192019}. While there is no disease-modifying treatment available for AD, several studies have reported a reduced risk for progression of AD in statin-treated populations \citep{jick2000statins, rockwood2002use, geifman2017evidence}\footnote{Statin is a commonly prescribed drug due to its clear benefits in reducing the level of LDL cholesterol through 3-hydroxy-3-methylglutaryl-coenzyme A reductase (HMGCR) inhibition \citep{nissen2005statin}.}. This slowed progression of AD might be linked to the reduced cholesterol generation after statin usage suggested by a substantial body of cellular and molecular mechanistic evidence \citep{di2011linking,mcguinness2010statins}. Thus, our study may provide some additional evidence for the conjecture that statins are helpful for AD patients as they lower the LDL cholesterol levels.

On the top of management of diseases related to high LDL cholesterol concentrations, there has been increased global attention on prevention and risk reduction of AD by maintenance of healthy lifestyle patterns \citep{barthold2020association, world2019risk, lourida2019association}. In this case study, we first adopt our proposed method to examine if the combined effect of healthy dietary patterns, increased physical activities, reduced alcohol intake, and reduced smoking on lowering LDL cholesterol concentration in the statin-treated group is different from the statin-controlled group in AD patients. Since that the Mediterranean diet is one of the dietary patterns most commonly investigated \citep{kivipelto2018lifestyle}, we define the healthy dietary pattern on the basis of adherence to the following characteristics: consumption of an increased amount of fruits, vegetables, fish, and a reduced amount of processed meats and unprocessed red meats. We then move on to examine if the statin usage takes heterogeneous effects across different individuals with different lifestyles in the study cohorts. Detailed descriptions of our data structure and scientific questions are provided in the next section.

Since existing clinical studies suggest that investigating the benefit of statin usage can be susceptible to unmeasured confounding factors which induce potential selection bias, we adopt a genetic variant rs12916-T as a surrogate treatment variable. This means that if the subject carries the variant rs12916-T, the treatment indicator variable is set to be one $D=1$, otherwise is set to be zero. We adopt this genetic surrogate biomarker as the treatment because the rs12916-T allele only affects the LDL cholesterol concentration through HMGCR inhibition and is functionally equivalent to statin usage \citep{swerdlow2015hmg}. Moreover, given that genetic variants are randomly inherited from parents and are fixed at conception, our treatment variable (whether the individual carrying rs12916-T) is thus independent of unmeasured factors such as lifestyle modification after statin usage \citep{swerdlow2015hmg, wurtz2016metabolomic}. To further avoid potential confounding issues introduced by genetic pleiotropy and linkage disequilibrium, our approach includes $p=637$ additional SNPs and lifestyle factors associated with LDL cholesterol concentration as covariates. We hope that such a study design makes Condition \ref{condition:Unconfoundedness} more plausible to believe in this case study. See Supplementary Materials for our data pre-processing steps. 

Lastly, we recognize that our study design has some potential limitations. Since the treatment variable is defined as carrying rs12916-T allele or not, it is a surrogate measurement of statin usage. This suggests that generalizing the current study findings still warrants further confirmation from clinical trials.

\begin{figure}[h!]
	\centering
	\includegraphics[width=\linewidth]{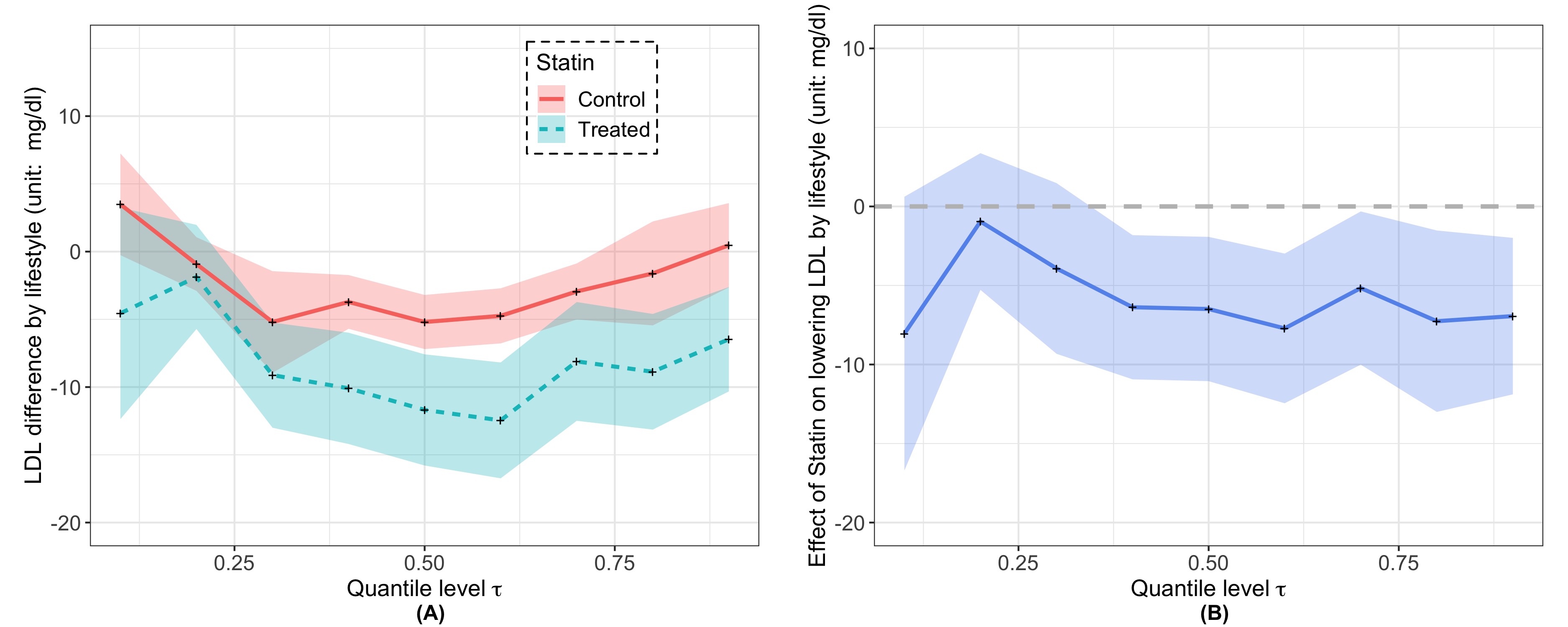}
	\caption{Panel (A) LDL plasma concentration for the treated and control groups. Panel (B) HQTE of statin usage by healthy lifestyles. Uniform 95\% confidence bands discussed in Example \ref{example:UniformConfidenceBand} are given by the shaded regions.  }
	\label{fig:realdata-intro}
\end{figure}

\subsection{Data structure and analysis}

Our data source is the UK Biobank study. The UK Biobank study cohort is a prospective cohort that enrolled about 500,000 individuals aged from 40 to 69 in the United Kingdom, started in 2006. We focus on AD (and AD proxy) patients older than 65 years since the majority of AD patients experience their first symptoms in their mid-60 \citep{jack2010hypothetical}. To avoid complications due to missing data, we only include patients with complete covariates. 
This results in a cohort of $n_0 = \sum_{i=1}^n(1-D_i) = 563$ subjects who do not carry the variant rs12916-T, and $n_1 =\sum_{i=1}^nD_i= 3150$ subjects who carry the variant rs12916-T. In this dataset, the sample size in the controlled group $n_0$ is less than the dimension $p$. Our response variable $Y_i$ is the individual plasma LDL-cholesterol concentration measured in mg/dl. 

As for the covariates $X_i$ for the subject $i$, we include the following variables: $X_{i1}$ is the intercept, $X_{i2}$ represents age, $X_{i3}$ represents number of days of moderate physical activity, $X_{i4}$ represents number of days of vigorous physical activity, $X_{i5}$ represents cooked vegetable intake, $X_{i6}$ represents salad/raw vegetable intake, $X_{i7}$ represents fresh fruit intake, $X_{i8}$ represents dried fruit intake, $X_{i9}$ represents oily fish intake, $X_{i,10}$ represents non-oily fish intake, $X_{i,11}$ represents processed meat intake, $X_{i,12}$ represents poultry intake, $X_{i,13}$ represents beef intake, $X_{i,14}$ represents lamb/mutton intake, $X_{i,15}$ represents pork intake, $X_{i, 16}$ represents alcohol intake frequency per week, $X_{i, 17}$ represents smoking status, $X_{i,19}$ represents insulin medication usage, $X_{i,18}$ represents gender, and $X_{i,20}, \ldots, X_{i,p}$ contain additional 619 SNPs associated with the LDL cholesterol concentration. The unit measurement of the included dietary variables is tablespoons/day. We have provided detailed data pre-processing steps in the Supplementary Materials. 

Since our goal is to investigate whether statin usage has differential effects on the study cohort, we estimate the HQTE $\alpha(\tau;z) =  z'(\theta_{1}(\tau) -\theta_{0}(\tau) )$ for two different sets of the vector $z$: 

In the first design, we study whether the combined effect of healthy dietary patterns, physical activities, and reduced smoking differs in the statin-treated and control groups for lowering LDL levels. We thus set  
\begin{align*}
	z = (0, 0, \underbrace{1, \ldots, 1}_{8}, \underbrace{-1, \ldots, -1}_{6}, 0 \ldots, 0)'\in \mathbb{R}^{p}.
\end{align*}
Figure~\ref{fig:realdata-intro}(A) shows the estimated linear combination of quantile regression coefficients  $ z'\hat{\theta}_{1}(\tau)$ (green curve) and $z' \hat{\theta}_{0}(\tau) $ (red curve) along with estimated uniform 95\% confidence bands (see Example \ref{example:UniformConfidenceBand} for details on how they were constructed). We observe that the effect of statin usage, moderate to vigorous physical activity combined with healthier dietary patterns on reducing the plasma LDL cholesterol concentration is the largest among those patients whose cholesterol are in the upper quantiles (i.e. whose cholesterol levels are high relative to the population).
For subjects who do not take statins, the effect of increased moderate and vigorous physical activity combined with healthier dietary patterns on reducing the plasma LDL cholesterol concentration is roughly a quadratic function of the quantile $\tau$, but the effect overall seems to be marginal. Figure~\ref{fig:realdata-intro}(B) shows the estimate of $\alpha(\tau;z) $. We see that the effect of statin usage is heterogeneous across different quantiles of the LDL cholesterol concentration -- its influence is more significant at the right tail of the distribution. This suggests that statin usage may help further reduce the LDL cholesterol level when combined with healthy lifestyles for AD patients with rather high LDL cholesterol concentration. Our findings might be helpful for researchers to design future clinical trials to study the effect of statin usage on patients with high LDL concentrations at baseline.

\begin{figure}
	\centering
	\includegraphics[width=0.33\linewidth]{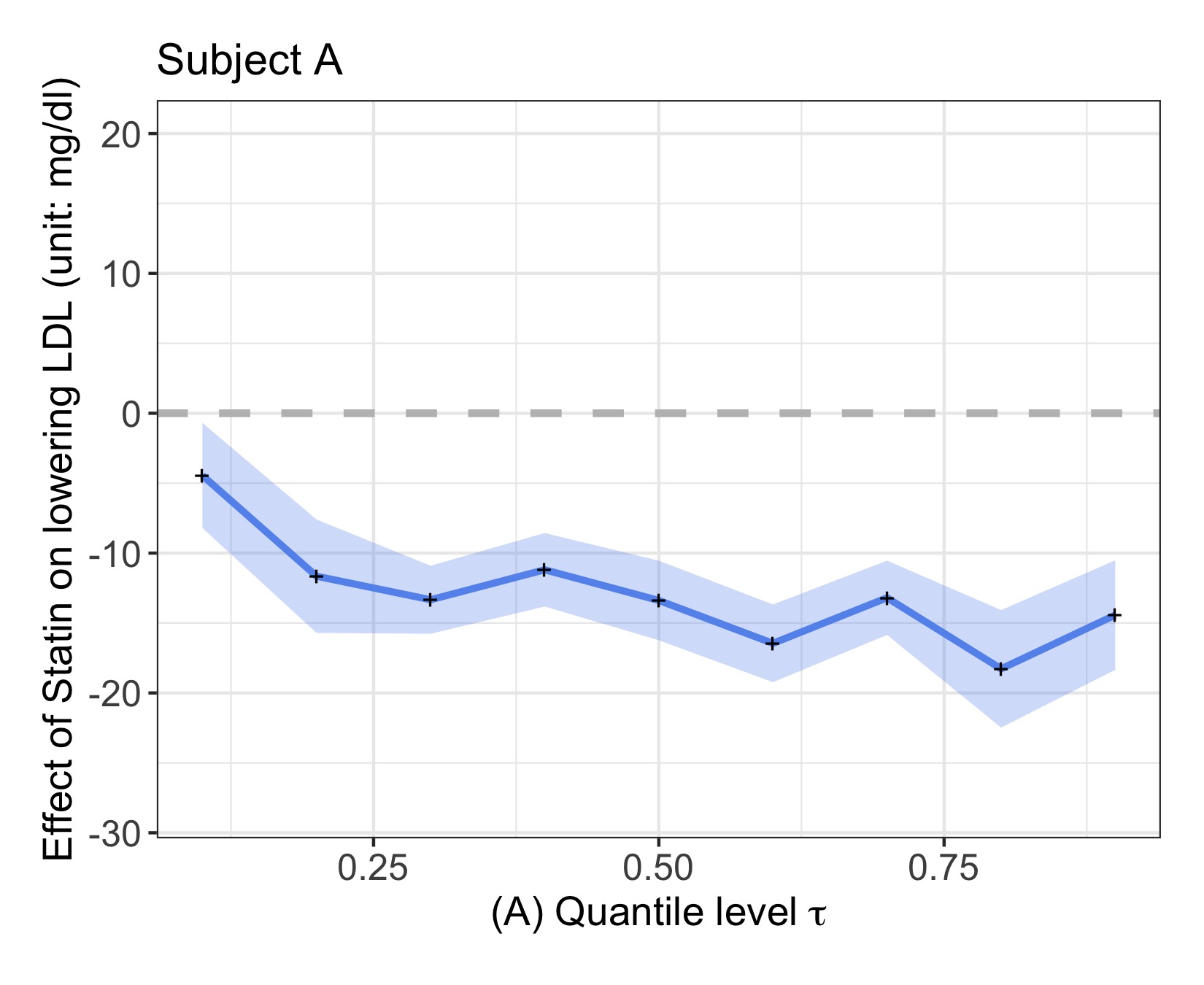}\includegraphics[width=0.33\linewidth]{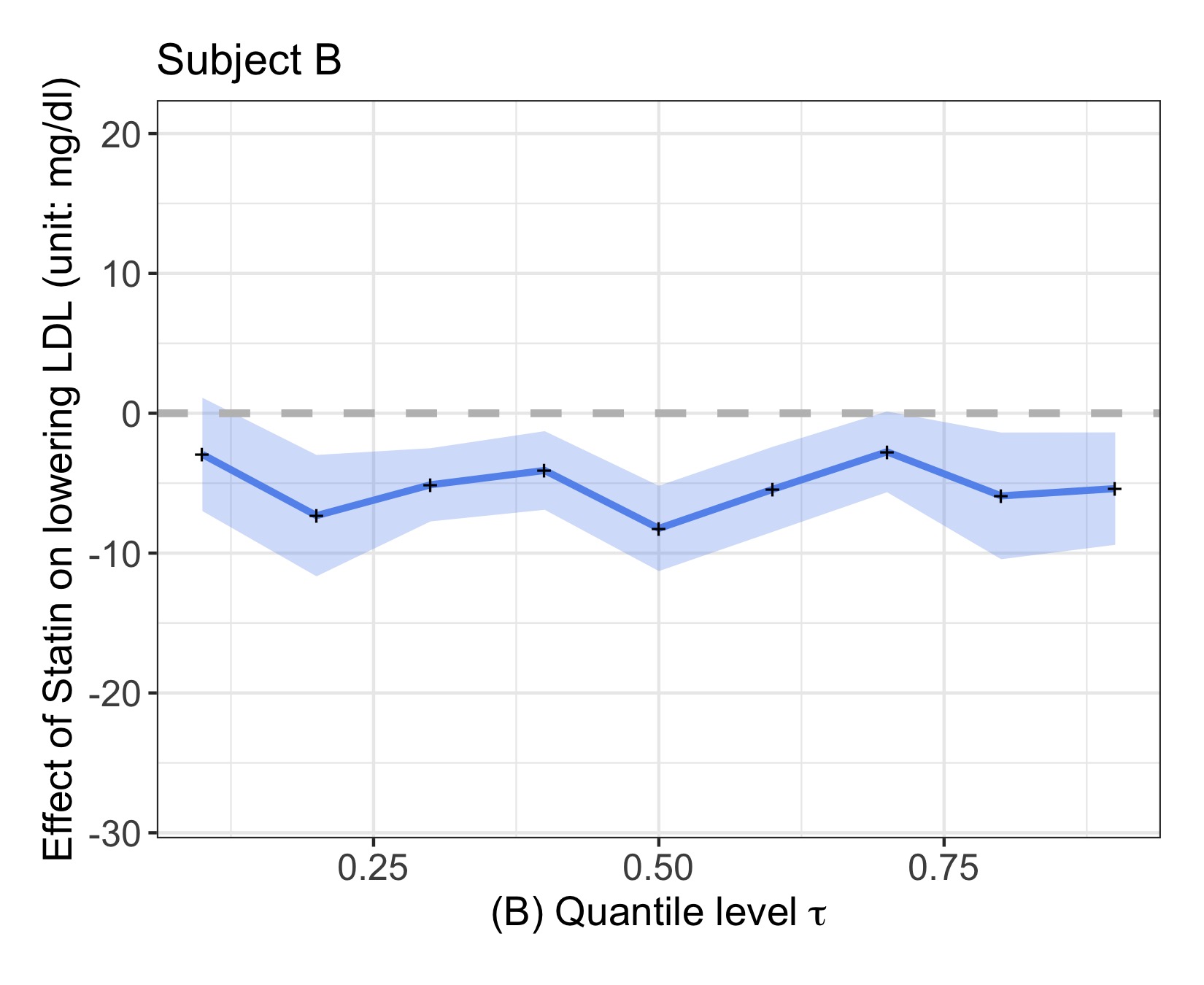}
	\includegraphics[width=0.33\linewidth]{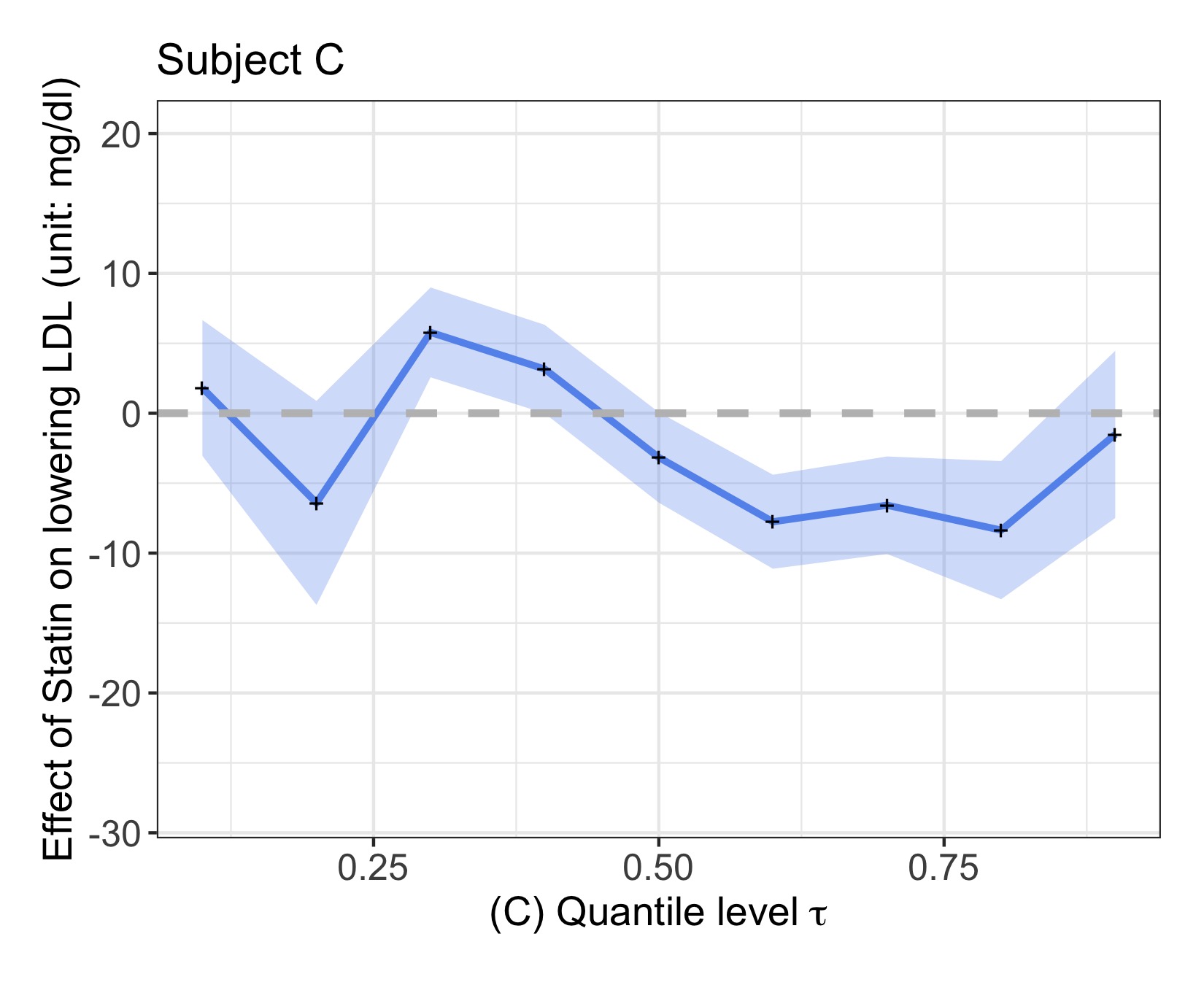}
	\caption{Heterogeneous quantile treatment effects of statin usage for three subjects in the considered study sample with AD from the UK Biobank study. Uniform 95\% confidence bands discussed in Example \ref{example:UniformConfidenceBand} are given by the shaded regions. }
	\label{fig:my_label}
\end{figure}

In the second design, we estimate HQTEs for three study participants in the UK Biobank cohorts with different lifestyles: The first patient is a 65-year-old subject (A) who exercises 4 times a week, has sufficient cooked and raw vegetable intake (4 tablespoons per day), has one tablespoon of fruit intake per day, has 2 drinks per week and with a recent smoking history. The second patient is a 70-year-old subject (B) who exercises every day, has sufficient vegetable and fruit intake (more than 3 tablespoons per day), has 2 drinks per week and with a recent smoking history. The third patient is a 66-year-old subject (C) who has no moderate/vigorous physical activity, less than 2 tablespoons vegetable and fruit intake per day, 5 drinks per week and no recent smoking history. We do not observe a notable difference for meat intake among these three subjects. The point estimates and their corresponding 95\% uniform confidence bands are reported in Figure \ref{fig:my_label}. We have excluded these three individuals from implementing our rank-score debiasing procedure. Although the observed differential effects across the above three subjects might be ascribed to the fact that study participants have different genotypes, our study results suggest that the benefit of statin usage can be heterogeneous across study participants. 

\section{Discussion}

In this article, we have introduced a new procedure to study treatment effect heterogeneity based on quantile regression modeling and rank-score debiasing. While our rank-score debiased estimator is easy to implement and enjoys strong theoretical guarantees, the following points merit future research: 
First, it is worthwhile to relax the unconfoundedness assumption, simply because unmeasured confounding presents a critical challenge to causal inference from observational studies. A classical approach to mitigate the confounding bias is to include instrumental variable methods. In this context, the identification condition \ref{condition:Unconfoundedness} can be modified similar to that of \cite{chernozhukov2005iv}. In future work, we therefore plan to investigate how to combine instrumental variables with our proposed debiasing procedure.
Second, it is desirable to further study the asymptotic efficiency of the rank-score debiasing procedure. 
The existing semiparametric efficiency bounds for quantile regression apply only to fixed dimensional settings when the quantile regression vector is independent of the sample size~\citep{newey1990efficient, zhao2001asymptotically}. The treatment of high-dimensional quantile regression models requires a more elaborate analysis since the quantile regression vector may change with sample size.
In future work we intend to develop a concept of semi-parametric efficiency of high-dimensional processes indexed by changing function classes.

\section*{Acknowledgement}
We thank the associate editor and two anonymous reviewers for their constructive suggestions that greatly helped to improve presentation and technical accuracy of this work. We thank the individuals involved in the UK Biobank for their participation and the research teams for their work on collecting, processing, and sharing these datasets. This research has been conducted using the UK Biobank Resource (application number 48240), subject to a data transfer agreement. Alexander Giessing's research was supported by NSF grant DMS-2310578, Jingshen Wang's research by NSF grant DMS-2015325 and NIH grant R01MH125746.

\newpage
\normalsize
\setcounter{page}{1}

\bibliography{High-dim-QR-Theory-Final-Ref-Complete}
\bibliographystyle{apalike}

\newpage

\title{Supplementary Materials for ``Debiased Inference on Heterogeneous Quantile Treatment Effects with Regression Rank-Scores''}
\author{Alexander Giessing\protect\footnotemark[1] \and  Jingshen Wang\protect\footnotemark[2] }
\date{	\today}

\maketitle

\small
\appendix

\setcounter{page}{1}

\section{Connection with existing literature}\label{sec:Comparisons}

In this section, we briefly discuss the differences between our approach and the ones propose by \cite{belloni2019valid} and \cite{zhao2019debiasing}. To simplify the discussion, we introduce the following simple quantile regression model: Let $\{(Y_i, X_i)\}_{i=1}^n$ be a random sample of response variable $Y$ and covariates $X$.  For a qantile level $\tau$, suppose that the conditional quantile function of a response $Y$ conditional on the covariates $X$ is given by 
\begin{align*}
	Q(\tau ; x) = 	x'\beta(\tau), \quad \quad{}\beta(\tau) = \big(\beta_1(\tau), \ldots, \beta_p(\tau)    \big)' \quad{} \quad{} \sup_{\tau \in \mathcal{T}}\left\|\beta(\tau)\right\|_0 \ll  p \wedge n.
\end{align*}
We define the weighted covariance matrix
\begin{align*}
	H(\tau) = \mathbb{E}\big[f_{i}^2(\tau) X_iX_i'\big] = \mathbb{E} \begin{bmatrix}
		f_{i}^2(\tau) X_{i1}^2 &    f_{i}^2(\tau) X_{i1} X_{i,-1}' \\
		f_{i}^2(\tau) X_{i1} X_{i,-1} &  f_{i}^2(\tau) X_{i,-1} X_{i,-1}'
	\end{bmatrix} \equiv \begin{bmatrix}
		H_{11}(\tau) &   H_{12}(\tau) \\
		H_{21}(\tau) &  H_{22}(\tau)
	\end{bmatrix},
\end{align*}
where $f_i(\tau)$ is the conditional density  of $Y|X$ evaluated at $X_i'\beta(\tau)$. 

\cite{belloni2019valid} provide a procedure for inference on a single regression coefficient, say $\beta_1(\tau)$. When $p$ is larger than the sample size $n$ and model selection is adopted to carry out inference on $\beta_1(\tau)$, the authors propose a method that is robust to model selection mistakes by constructing orthogonal score functions. They hypothesize the auxiliary regression problem
\begin{align*}
	& f_i(\tau) X_{i1} = f_i(\tau) X_{i,-1}'\gamma(\tau) + \nu_i, \quad{} \quad{} \mathbb{E} [	f_i(\tau) X_{i1} \nu_i]=0,\quad{}\quad{} i=1, \ldots, n,\\
	&\gamma(\tau) = \arg\min_\gamma \frac{1}{n}\sum_{i=1}^n\mathbb{E}[f_i^2(\tau) ( X_{i1}  - X_{i,-1}'\gamma )^2 ],
\end{align*}
where $X_{i,-1} = ( X_{i2}, \ldots, X_{ip} )'$ with regression vector $\gamma(\tau) = H_{22}^{-1}(\tau)H_{21}(\tau)$ and residuals
$	\nu_i = f_i(\tau) X_{i1} - f_i(\tau) X_{i,-1}'H_{22}^{-1}(\tau)H_{21}(\tau)$.
This auxiliary regression motivates the authors to define the orthogonal score function $\psi_\tau(Y, X'b, \nu) = (\tau -  \mathbf{1}\{Y \leq X'b\})\nu$. This score function not only satisfies $\mathbb{E}[\psi_\tau(Y_i, X_i'\beta(\tau), \nu_i) ]=0$ but (by the very definition of the regression vector $\gamma(\tau)$ and the residuals $\nu_i$) also the orthogonality conditions
\begin{align*}
	& \frac{\partial }{\partial b } \frac{1}{n}\sum_{i=1}^n\mathbb{E} \big[ \big(\tau - \mathbf{1}\{ Y_i \leq X_{i,1}\beta_1(\tau) +X_{i, -1}'b \} \big) \cdot \nu_i \big]\bigg|_{ b = \beta_{-1}(\tau) }= 0, \\
	&  \frac{\partial }{\partial \gamma } \frac{1}{n}\sum_{i=1}^n\mathbb{E} \big[ (\tau - \mathbf{1}\{ Y_i \leq X_i'\beta(\tau) \}\big)\cdot  f_i(\tau) \cdot \big(X_{i1} -X_{i,-1}'\gamma \big) \big]\bigg|_{ \gamma = \gamma(\tau) }= 0.
\end{align*}
The authors proceed to use this score function to construct an estimator $\hat{\beta}_1(\tau)$ which, by virtue of above orthogonality conditions, is protected against the estimation biases of the high-dimensional nuisance parameters $\gamma(\tau)$ and $\big(\beta_2(\tau), \ldots, \beta_p(\tau)    \big)'$~\citep[][Section 2]{belloni2019valid}. The theoretical investigation of \cite{belloni2019valid} shows that the asymptotic variance of their estimator equals 
\begin{align*}
	\tau(1-\tau) \bar{\mathbb{E}}[\nu_i^{2}]^{-1} = \tau(1-\tau)  \cdot \big[ H_{11}(\tau) - H_{12}(\tau)H_{22}^{-1}(\tau)H_{21}(\tau)  \big]^{-1} =  \tau(1-\tau)  [H(\tau)^{-1}]_{11}. 
\end{align*}

Comparing this result with our Theorem 1, we see that when the goal is to estimate a single regression coefficient, our rank-score debiased estimator is asymptotically equivalent to the one proposed by \cite{belloni2019valid}. 
The key difference between \cite{belloni2019valid} and our approach is that we directly debias the linear combination of the high-dimensional quantile regression coefficient $z'\beta(\tau)$. We do not consider a low dimensional projection of the high-dimensional regression vector.

\cite{zhao2019debiasing} propose a debiased penalized quantile regression coefficient estimator by directly estimating the weighted inverse covariance matrix $\mathbb{E}\big[f_{i}(\tau) X_iX_i'\big]$ based on ideas from \cite{vandegeer2014on} and \cite{zhang2014confidence}. Concretely, their estimator for $\beta(\tau)$ is of the form 
\begin{align*}
	& \hat{\beta}_{\mathrm{Lasso}}(\tau) =\arg\min_{b\in\mathbb{R}^p} \left\{ \frac{1}{n}\sum_{i=1}^n \rho_{\tau}\big( Y_i  - X_i'b \big) + p_{\lambda}(b)\right\}, \\
	& \hat{\beta} (\tau)= \hat{\beta}_{\mathrm{Lasso}} (\tau)+ \widehat{\Theta} \left( \frac{1}{n}\sum_{i=1}^n  \big(\tau - \mathbf{1}\{Y_i \leq X_i'\hat{\beta}_{\texttt{Lasso}}(\tau)\} \big) X_i\right),
\end{align*}
where $\widehat{\Theta} $ is an estimate of the inverse covariance matrix $D_1(\tau)^{-1} := \Big[ \mathbb{E}\big[f_{i}(\tau) X_iX_i'\big]\Big]^{-1}$ (the authors use the \texttt{R} package \texttt{clime} to estimate the inverse covariance matrix). The authors establish consistency of this estimator in the $\ell_\infty$-norm but do not derive weak convergence results. Based on the theoretical derivation given in their Appendix A, their estimator satisfies that 
\begin{align*}
	\hat{\beta}_1(\tau) - \beta_1(\tau) = e_1'\widehat{\Theta} \left(\frac{1}{n}\sum_{i=1}^n \big(\tau - \mathbf{1}\{Y_i \leq X_i'\hat{\beta}_{\texttt{Lasso}}(\tau)\} \big) X_i\right) + r_n, 
\end{align*}
where $e_1 = (1, 0, \ldots, 0)'$ and $r_n$ is a remainder term such that $\|r_n\|_\infty \downarrow 0$. The rate at which $\|r_n\|_\infty$ vanishes is too slow for any weak convergence guarantees, but if it was fast enough, the asymptotic variance of their estimator would be of the typical sandwich-formula
\begin{align*}
	\tau(1-\tau)\cdot 	e_1' D_1^{-1}(\tau) D_0 D_1^{-1}(\tau)e_1,
\end{align*}
where $D_0 := \mathbb{E}\big[ X_iX_i'\big]$. This indicates that the estimator in \cite{zhao2019debiasing} is not as efficient as the rank-score debiased estimator~\citep[see also][Section 5.3.1]{koenker2005quantile}. As an aside, this asymptotic variance can obviously be estimated as $\frac{1}{n}\sum_{i=1}^n (e_1'\widehat{\Theta} X_i)^2$.

\section{Additional implementation details for the case study}~\label{sec:data-preproc}

We first provide a pipeline for cleaning the UK Biobank data for our case study. We obtain phenotype and genotype data from the UK Biobank study with following steps. For the phenotype data, we download encoded data in \textit{.enc} format from  the UK Biobank's Access Management System (AMS). To decrypt the encoded data, we download three helper programs: \texttt{ukb\_md5}, \texttt{ukb\_unpack}, and \texttt{ukb\_conv}. Note that the helper programs are only supported by Windows and Linux systems. Second, we verify the integrity of the encoded data via \texttt{ukb\_md5} and unpack them into \textit{.enc\_ukb} format with \texttt{ukb\_unpack}. To convert the data into a readible format, we use \texttt{ukb\_conv} to convert the \textit{.enc\_ukb} data into \textit{.csv} format (other options include txt, r, sas stata or bulk format). The data dictionary can be obtained using \texttt{ukb\_conv} with the \textit{docs} option. After decrypting and converting the encoded data, we obtain a dataset with sample size n = $502,481$ and $20,502$ features. In our study, we work with the phenotypes at the baseline to avoid confounding issues, and we extract baseline variables with the suffix ``-0.0". The phenotype data we extract include gender, age at recruitment, AD family history, International Statistical Classification of Disease 9th revision (ICD 9) and 10th revision (ICD 10) codes and  self-report for T2D and AD.

For the genotype data, we download imputed genotypes and associated sample information for 23 chromosomes. \textit{Imputation BGEN} and \textit{Imputation sample} can be obtained via \texttt{ukb\_gene} program.   \textit{Imputation BGI} and \textit{Imputation MAF+info} can be downloaded directly from UK Biobank resources $1965$ and $1967$. Second, we use the \textit{Imputation sample} file to remove individuals without genotype information, which yields sample size $n=407,057$.  Finally, we read in BGEN files with \texttt{snp\_readBGEN} function in \texttt{R} package \texttt{bigsnpr} using \textit{HapMap3} as the reference genomes. \texttt{snp\_readBGEN} converts the BGEN files into an \texttt{R} object comprising of two elements: \textit{genotype} and \textit{map}, where \textit{genotype} represents the imputed genotypes in a matrix format and \textit{map} contains the features of SNPs (chromosome, rsid, physical position, major and minor alleles and allele frequency). We only extract the genotype matrix from \textit{genotype} and \textit{rsid} from \textit{map} as our genotype data. We restrict our sample to subjects used for the principle components (PCs) computation, since those individuals are unrelated. From the extracted genotype data, we obtain the treatment variable: rs12916 (on chromosome 5), a functionally equivalent SNP of statins. 

Since rs12916 and LDL cholestrol concertration are associated with other SNPs due to genetic pleiotropy and linkage disequilibrium, we adjust for low-density lipoprotein (LDL) and rs12916 related SNPs in our study. To find disease-associated SNPs, we rely on published GWAS studies from the GWAS catalogue. In our study, we define the disease-associated SNPs as SNPs associated with LDL with $p$-values less than $5\times 10^{-5}$. To determine the $p-$values for multiple correlated SNPs in the same locus, we use the linkage disequilibrium clumping procedure with $R^2 < 0.01$.  Our filtration criteria yield 619 disease-associated SNPs. 

Lastly, we note that as the \texttt{R} package \texttt{quantreg} with the ``Lasso" option tends to be numerically unstable in our real data analyses (it produces error message ``singular design matrix"), we switch to the the \texttt{R} package \texttt{conquer} for the implementation of the quantile regression problems. 

\section{Overview of technical results}\label{sec:Preamble}
In the the remainder of these supplementary materials we develop the theoretical backbone of the results in the main text. The key to the weak convergence results in the main text is the Bahadur-type representation of the rank-score balanced estimate of the CQF. Establishing this Bahadur-type representation with a reasonable non-asymptotic bound on the remainder term is very involved. In particular, the proof requires the following auxiliary results:
\begin{itemize}
	\item uniform consistency of the \emph{unweighted} $\ell_1$-penalized quantile regression vector;
	\item uniform control over the empirical sparsity of the \emph{unweighted} $\ell_1$-penalized quantile regression vector;
	\item uniform consistency of the solution to the dual program of the rank-score balancing program.
\end{itemize}
These results are new. Existing results and proofs concerning uniform consistency and empirical sparsity of the \emph{weighted} $\ell_1$-penalized quantile regression vector~\citep{belloni2011L1penalized, belloni2019valid} do not apply to the case of \emph{unweighted} $\ell_1$-penalized quantile regression. The reason for this is that the empirical processes associated with the weighted $\ell_1$-penalty are self-normalized whereas those corresponding to the unweighted $\ell_1$-penalty are unbounded. The self-normalized processes can be analyzed via conventional tools in empirical process theory, while the unbounded empirical processes require new tools developed in~\cite{giessing2020maximal}. Similarly, the dual of the rank-score balancing process is a non-standard problem that cannot be recast as a standard regression problem. Hence, its analysis requires a new approach as well.

In order to make these auxiliary results as transparent and accessible as possible we prove them under slightly weaker conditions than stated in the main paper and provide explicit constants whenever possible. The results from the main text are therefore simple corollaries of the more refined results provided in Sections~\ref{sec:L1-QR} and~\ref{sec:RankScore-QR}.

\section{Notation}\label{sec:Notation}
In addition to the notation of the main text we introduce the following conventions.

Unless otherwise stated, we denote by $\{X_i\}_{i \in \mathbb{N}}$ a sequence of independent $S$-valued random elements with common law $P$, i.e. $X_i : S^{\mathbb{N}} \rightarrow S$, $i \in \mathbb{N}$, are the coordinate projections of the infinite product probability space $(\Omega, \mathcal{A}, \mathbb{P}) =(S^{\mathbb{N}}, \mathcal{S}^{\mathbb{N}}, P^{\mathbb{N}})$. If auxiliary variables independent of the $X_i$'s are involved, the underlying probability space is assumed to be of the form $(\Omega, \mathcal{A}, \mathbb{P}) =(S^{\mathbb{N}}, \mathcal{S}^{\mathbb{N}}, P^{\mathbb{N}}) \times (Z, \mathcal{Z}, Q)$. We write $\mathbb{E}$ for the expectation with respect to $\mathbb{P}$, and $\mathbb{P}_X$, $\mathbb{E}_X$ for the partial integration with respect to the joint law of $\{X_i\}_{i \in \mathbb{N}}$ only. For events $A \in \mathcal{A}$ with $\mathbb{P}\{A\} > 0$ and a random element $Y$ on $(\Omega, \mathcal{A}, \mathbb{P})$ we define the conditional expectation of $Y$ given $A$ as $\mathbb{E}\left[Y\mid A\right] := (\int_A Yd \mathbb{P}) /\mathbb{P}\{A\}$.
For any measure $Q$ and any real-valued $Q$-integrable function $f$ on $(S, \mathcal{S})$ we write $Qf = Q(f) := \int f dQ$. We denote by $L_p(S, \mathcal{S}, Q)$, $p \in [1, \infty)$, the space of all real-valued measurable functions $f$ on $(S, \mathcal{S})$ with finite $L_p(Q)$-norm, i.e. $\|f\|_{Q, p} := (\int |f|^p dQ)^{1/p} < \infty$. For a random variable $\xi$ on $(\Omega, \mathcal{A}, \mathbb{P})$ we set $\|\xi\|_p := (\mathbb{E}|\xi|^p)^{1/p}$.
We define the empirical measures $P_n$ associated with observations $\{X_i,\}_{i=1}^n$, $n \in \mathbb{N}$, as random measures on $(S, \mathcal{S})$ given by $P_n(\omega) := n^{-1}\sum_{i=1}^n \delta_{X_i(\omega)}$, $\omega \in \mathcal{S}$, where $\delta_{x}$ is the Dirac measure at $x$. We denote the empirical processes indexed by a class $\mathcal{F}\subset L_1(S, \mathcal{S}, P)$ by $\mathbb{G}_n(f) := \sqrt{n}(P_n - P)(f) := n^{-1/2}\sum_{i=1}^n (f(X_i) - Pf)$, $f \in \mathcal{F}$, and the corresponding symmetrized empirical processes by $\mathbb{G}_n^\circ(f) := n^{-1/2}\sum_{i=1}^n \varepsilon_i f(X_i)$, $f \in \mathcal{F}$, where $\{\varepsilon_i\}_{i=1}^n$ is a sequence of i.i.d. Rademacher random variables independent of $\{X_i\}_{i=1}^n$. 
For probability measures $Q$ on $(S, \mathcal{S})$ we set $\|Q\|_\mathcal{F} = \sup\{|Qf| : f \in \mathcal{F}\}$. To avoid distracting measurablity questions regarding the quantity $\|Q\|_\mathcal{F}$, we assume that the function class $\mathcal{F}$ is either countable or that the (symmetrized) empirical process index by $\mathcal{F}$ is separable. The latter is true for all instances in this paper, since all function classes considered are indexed by vectors that live in $\mathbb{R}^p$, $p \in [1, \infty)$. Given a pseudometric space $(T, d)$ and $\varepsilon > 0$, $N(\varepsilon, T, d)$ denotes the $\varepsilon$-covering number of $T$ with respect to $d$. If $d$ is induced by a norm $n$ on $T$, we also write $N(\varepsilon, T, n)$.

For non-negative real-valued sequences $\{a_n\}_{n\geq 1}$ and $\{b_n\}_{n \geq 1}$, the relation $a_n \lesssim b_n$ means that there exists an absolute constant $c > 0$ independent of $n, d, p$ and an integer $n_0 \in \mathbb{N}$ such that $a_n \leq c b_n$ for all $n \geq n_0$. We write $a_n \asymp b_n$ if  $a_n \lesssim b_n$ and $b_n \lesssim a_n$. We define $a_n \vee b_n = \max\{a_n, b_n\}$  and $a_n \wedge b_n = \min\{a_n, b_n\}$. For a vector $a \in \mathbb{R}^d$ and $p \in [1, \infty)$ we write $\|a\|_p = (\sum_{k=1}^d |a_k|^p)^{1/p}$. Also, we write $\|a\|_\infty = \max_{1 \leq k \leq d} |a_k|$. For a scalar random variable $\xi$ and $\alpha \in (0, 2]$ we define the $\psi_\alpha$-Orlicz norm by $\|\xi \|_{\psi_\alpha} = \inf\{t > 0 : \mathrm{E}[\exp(|\xi|^\alpha/t^\alpha)] \leq 2\}$. For a sequence of scalar random variables $\{\xi_n\}_{n \geq 1}$ we write $\xi_n = O_p(a_n)$ if $\xi_n/a_n$ is stochastically bounded. For any matrix $M \in \mathbb{R}^{d \times d}$ we denote its operator norm by $\|M\|_{op}$ (its largest singular value).

\section{Proofs of the results in the main text}\label{sec:ProofsMainText}
\begin{proof}[\textbf{Proof of Lemma 1}]
	The proof is modeled after a similar result in~\cite{belloni2011L1penalized}. Denote cdf and df of the standard normal distribution by $\Phi$ and $\phi$. The model implies a linear quantile regression model with coefficients $\theta_d^{(1)}(\tau) = \alpha_d + \sigma_\varepsilon \Phi^{-1}(\tau)$ and $\theta_d^{(k)}(\tau) = \beta_d^{(k-1)}$ for $k= 2, \ldots, p+1$. Hence, $s_\theta \leq \max_{d \in \{0,1\}} \|\beta_d\|_0 + 1$, $\bar{f} = 1/\sqrt{2\pi \sigma^2_\varepsilon} \vee 1$, $\underline{f} = 1/\sqrt{2\pi \xi}$, and $\kappa_1(2) \geq \sqrt{2\pi \xi \sigma^2_\varepsilon} \underline{\kappa}$ and $\kappa_2(\infty) \geq  2\pi \xi \sigma^2_\varepsilon \underline{\kappa}$. Moreover, $L_\theta = \sup_{\tau \in \mathcal{T}} \sigma_\varepsilon/ \phi(\Phi^{-1}(\tau)) \leq \sigma_\varepsilon/\xi$ and for all $x \in \mathbb{R}^p$ and $\tau \in [\xi, 1-\xi]$,	
	\begin{align*}
		|Q_d'''(\tau; x)|  &= \big|\sigma_\varepsilon \frac{d^3}{d\tau^3}\Phi^{-1}(\tau) \big| \\
		&= \big|- \sigma_\varepsilon \phi''(\Phi^{-1}(\tau))/\phi(\Phi^{-1}(\tau))^4 + 3\sigma_\varepsilon \phi'(\Phi^{-1}(\tau))^2/\phi(\Phi^{-1}(\tau))^5\big| \\
		&\leq \sigma_\varepsilon/\xi^4 + 3 \sigma_\varepsilon/\xi^3 \leq  4\sigma_\varepsilon/\xi^4.
	\end{align*}
	It is easy to see that under the sparsity of $z$ and the sparsity of the precision matrix, $v_d(\tau, z)$ is sparse as well with at most $(2q-1)s$ non-zero entries. Hence, we may set $\epsilon_n = 0$ and take $\tilde{v}_d(\tau;z) = v_d(\tau;z)$. By the Lipschitz continuity of $\phi$, we have
	\begin{align*}
		\kappa_1(2, \varrho) &\geq \kappa_1(2) - \varrho_n L_f\max_{d \in \{0,1\}}\sup_{\tau \in \mathcal{T}} \sup_{\|\zeta\|_2 \leq \varrho\:} \sup_{u \in C^p_1(T_\theta(\tau), \vartheta) \cap \partial B^p_2(0,1) } \mathbb{E} \left[|X'u|^3\mathbf{1}\{D= d\}\right]\\
		&\overset{(a)}{\geq} \kappa_1(2) - \varrho_n L_f\bar{\varphi}^{3/2}\\
		&\gtrsim \kappa_1(2),
	\end{align*}
	where (a) holds by Corollary 3.2 in~\cite{ledoux1996probability}. This completes the proof.
\end{proof}

\begin{proof}[\textbf{Proof of Lemma 2}]
	By Theorem 2.5 in~\cite{foucart2013mathematical} there exists $\tilde{v}_d(\tau;z) \in \mathbb{R}^p$ such that $\|\tilde{v}_d(\tau;z) \|_0 = s_v$ and 
	\begin{align*}
		\|\tilde{v}_d(\tau;z) - v_d(\tau;z)\|_2 \leq \frac{1}{2\sqrt{s_v}}\| v_d(\tau;z)\|_1.
	\end{align*}
	Since $v_d(\tau; z) \in C^p_1(J, c_0)$ it follows that for $s_v = s \log(np)$,
	\begin{align*}
		\|\tilde{v}_d(\tau;z) - v_d(\tau;z)\|_2 &\leq \frac{1 + c_0}{2\sqrt{s_v}}\big\| \big( v_d(\tau;z)\big)_J\big\|_1 \\
		&\leq \frac{1 + c_0}{2} \sqrt{\frac{s}{s_v}} \|v_d(\tau; z)\|_2  = \frac{(1 + c_0)/2}{\sqrt{\log n}} \|v_d(\tau; z)\|_2. 
	\end{align*}
	Thus, $\epsilon_n = o\big(c_0/\sqrt{\log n}\big)$. The other quantities follow as in the proof of Example 1.
\end{proof}

\begin{proof}[\textbf{Proof of Lemma 3}]
	The model implies a linear quantile regression model with coefficient vector $\theta_d(\tau) = \beta_d + F^{-1}(\tau) \eta_d$.  Hence, $s_\theta \leq \max_{d \in \{0,1\}} \|\beta_d\|_0 + \|\eta_d\|_0 $, $\bar{f} = \max_{y} f(y)/\upsilon \vee 1$, $\underline{f} = \min_{\tau \in \mathcal{T}} f(F^{-1}(\tau))/\Upsilon$, $L_f = \max_ff'(y)/\upsilon^2$, $\kappa_1(2) \geq \underline{f}\underline{\kappa}$,  $\kappa_2(\infty) \geq \underline{f}^2 \underline{\kappa}$, $L_\theta = \max_d\|\eta_d\|_2\underline{f}$, and for all $x \in \mathbb{R}^p$ and $\tau \in [\xi, 1-\xi]$,	
	\begin{align*}
		|Q_d'''(\tau; x)|  &= \big| \frac{d^3}{d\tau^3}F^{-1}(\tau) \big| \Upsilon \\
		&= \big|- f''(F^{-1}(\tau))/f(F^{-1}(\tau))^4 + 3f'(F^{-1}(\tau))^2/f(F^{-1}(\tau))^5\big|\Upsilon \\
		&\leq \max_y \left(f''(y)/\underline{f}^4 + 3 L_f^2 \upsilon^4/\underline{f}^5 \right)\Upsilon.
	\end{align*}
	The remainder follows from the proof of Lemma 2. The $\rho_n$-identifiability follows as in the proof of Example 1. (Note that Corollary 3.2 in~\cite{ledoux1996probability} is not needed when the predictors are bounded.)
\end{proof}

\begin{proof}[\textbf{Proof of Lemma 4}]
	Define $A = [A_1, \ldots, A_p] : =\mathbb{E}\left[ f_{Y_d|X}^2\big(X'\theta_d(\tau)|X\big)XX' \mathbf{1} \{D = d \} \right]^{-1} \in \mathbb{R}^{p \times p}$. For subsets $S, T \subseteq\{1, \ldots, p\}$ let $A_{S,T} \in \mathbb{R}^{|S| \times |T|}$ be the sub-matrix obtained from $A$ by deleting all rows in $S^c$ and columns in $T^c$. Denote by $\sigma_{\min}(A_{S,T})$ the smallest positive singular value of $A_{S,T}$.
	
	Proof of claim (i). By assumption, $\max_{1 \leq k \leq p} \|A_k\|_0 \leq q$. Let $T_z \subseteq \{1, \ldots, p\}$ be the support set of $z \in \mathbb{R}^p$. Then, $\|z\|_0 = |T_z| \leq s_z$. Now, compute
	\begin{align*}
		\|v_d(\tau; z)\|_0 = \|Az\|_0 = \left\|  \sum_{k =1}^p A_k z_k\right\|_0 = \left\|  \sum_{k \in T_z} A_k z_k\right\|_0 \leq \sum_{k \in T_z}\left\| A_k \right\|_0 \leq qs_z.
	\end{align*}
	Thus, $v_d(\tau; z) \in \mathbb{R}^p$ is sparse and we can take $\tilde{v}_{d,n}(\tau; z) = v_d(\tau;z)$ to meet Condition 9 with $s_v = q s_z$ and $\epsilon_n \equiv 0$ for all $n \geq 1$.
	
	Proof of claim (ii). As in the proof of claim (i) let $T_z$ be the support set of $z \in \mathbb{R}^p$, hence, by assumption $\|z\|_0 = |T_z| \leq s_z$. Since by assumption $s_z < |J|$, it follows that $A_{J, T_z}'A_{J, T_z} = \sum_{k \in J} A_{k, T_z} A_{k,T_z}'\in \mathbb{R}^{|T_z| \times |T_z|}$ is positive definite. Thus, $\sigma_{\min}(A_{J,T_z}) > 0$. By assumption, $\|A_{J^c,k}\|_1 \leq \vartheta \|A_{J,k}\|_1$ for all $1 \leq k \leq p$. 
	
	Now, compute
	\begin{align*}
		\|(Az)_{J^c}\|_1  &= \left\| \sum_{k =1}^p A_{J^c, k} z_k\right\|_1 \\
		&\leq \sum_{k \in T_z} \left\| A_{J^c, k}\right\|_1 |z_k|\\
		& \leq \sum_{k \in T_z} \vartheta \left\| A_{J, k}\right\|_1 |z_k| \\
		& \leq  \vartheta \sqrt{s_z} \max_{k \in T_z}\left\| A_{J, k}\right\|_1 \|z\|_2\\
		& \leq \vartheta \sqrt{s_z} \max_{k \in T_z}\left\| A_{J, k}\right\|_1 \sup_{ \mathrm{support}(u)=T_z} \frac{\|u\|_2}{ \|A_J u\|_2}\|A_J z\|_2\\
		& \leq \max_{k \in T_z}\frac{\vartheta \sqrt{s_z}  \left\| A_{J, k}\right\|_1}{ \sigma_{\min}(A_{J,T_z}) }\|(A z)_J\|_1\\
		&\equiv \vartheta K(J, z)\|(A z)_J\|_1.
	\end{align*}
	Thus, $v_d(\tau;z) \in C^p_1\left(J, \:  \vartheta K(J, z) \right)$ and, as in the proof of Lemma 2, we conclude that there exists $\tilde{v}_d(\tau;z) \in \mathbb{R}^p$ with $\|\tilde{v}_d(\tau;z) \|_0 = |J| \log n$ and 
	\begin{align*}
		\|\tilde{v}_d(\tau;z) - v_d(\tau;z)\|_2 &\leq \frac{1 + \vartheta K(J, z)}{2\sqrt{|J|\log n}}\big\| \big( v_d(\tau;z)\big)_J\big\|_1 \leq \frac{1 + \vartheta K(J, z)}{2 \sqrt{\log n}}  \|v_d(\tau; z)\|_2.
	\end{align*}
	Thus, Condition 9 holds with $s_v = |J| \log n$ and $\epsilon_n = O\big(\vartheta K(J, z)/\sqrt{\log n}\big)$.
	
	Proof of claim (iii). Since $z_J \neq 0$, it follows that $\|z_{J^c}\|_1 \leq c_0 \|z_J\|_1$ for $c_0 = \|z_{J^c}\|_1/\|z_J\|_1$. Define $ \kappa_{\min}(J, c_0) : = \inf_{u \in C^p_1(J, c_0)}\|A_Ju\|_2$. Thus,
	\begin{align*}
		\|(Az)_{J^c}\|_1  &= \left\| \sum_{k =1}^p A_{J^c, k} z_k\right\|_1 \\
		&\leq \sum_{k \in J} \left\| A_{J^c, k}\right\|_1 |z_k| + \sum_{k \in J^c} \left\| A_{J^c, k}\right\|_1 |z_k| \\
		&\leq \max_{k \in J} \left\| A_{J^c, k}\right\|_1 \|z_J\|_1 + \max_{k \in J^c} \left\| A_{J^c, k}\right\|_1  \|z_{J^c}\|_1\\
		&\leq (1 + c_0) \max_{1 \leq k \leq p} \left\| A_{J^c, k}\right\|_1 \|z_J\|_1\\
		&\leq (1 + c_0) \max_{1 \leq k \leq p} \left\| A_{J^c, k}\right\|_1 \sqrt{|J|} \|z_J\|_2\\
		&\leq \vartheta (1 + c_0) \sqrt{|J|} \max_{1 \leq k \leq p} \left\| A_{J, k}\right\|_1 \sup_{u \in C^P_1(J, c_0)} \frac{\|u\|_2}{ \|A_J u\|_2} \|A_Jz\|_2\\
		&\leq \vartheta (1 + c_0) \max_{1 \leq k \leq p}\frac{ \sqrt{|J|} \left\| A_{J, k}\right\|_1 }{ \kappa_{\min}(J, c_0) } \|A_J z\|_2\\
		&\equiv \vartheta K(J, z) \|(A z)_J\|_1.
	\end{align*}
	Therefore, as in the proof of claim (ii), we conclude that Condition 9 holds with $s_v = |J| \log n$ and $\epsilon_n = O\big(\vartheta K(J, z)/\sqrt{\log n}\big)$.
\end{proof}

\begin{proof}[\textbf{Proof of Example 4}]
	
	Throughout, $A =  \mathbb{E}[XX']^{-1}$. The proof of claim (i) is immediate and hence omitted. 
	
	Proof of claim (ii). By Theorem 2.2 in~\cite{demko1977inverses} $A = (A_{jk})_{j,k=1}^p$ satisfies
	\begin{align}\label{eq:example:HomoscedasticQR-1}
		|A_{jk}| \leq K \|A^{-1}\|_{op} r^{|j-k|}, \quad\quad 1 \leq j,k \leq p,
	\end{align}
	where $K > 0$ and $r \in (0,1)$ depend on $q \geq 1$ and $\|A\|_{op}$ only. Thus, by~\eqref{eq:example:HomoscedasticQR-1} for all $1 \leq k \leq p$,
	\begin{align}\label{eq:example:HomoscedasticQR-2}
		\sum_{j=1}^p |A_{jk}| \leq  \frac{K \|A^{-1}\|_{op} }{1-r} < \infty.
	\end{align}
	Since $A$ is positive definite each row of $A_k$ has at least one non-zero entry. Thus, by~\eqref{eq:example:HomoscedasticQR-2} there exist $\vartheta \in (0, \infty)$ such that for all $1 \leq k \leq p$ we can find $J_k \subseteq \{1 , \ldots, p\}$ , $|J_k| = 1$, with
	\begin{align*}
		\sum_{j \in J_k^c} |A_{jk}| \leq \vartheta \sum_{j \in J_k} |A_{jk}| \quad \Leftrightarrow \quad A_k \in C^p_1(J_k, \vartheta).
	\end{align*}
	In particular, we can choose $J_k = k$ for all $1 \leq k \leq p$.
	
	Since $A$ is positive definite, every leading principle minor of $A$ is positive definite. In fact, permuting rows and columns, we conclude that every principle minor is positive definite. Therefore, $\sigma_{\min}(A_{T_z, T_z}) > 0$. Hence, choose $J = T_z$. This completes the proof of claim (ii).
	
	Proof of claim (iii). Since $A \in \mathbb{R}^{p \times p}$ is positive definite, $Az \neq 0 \in \mathbb{R}^d$. Since $z_k \neq 0$ for all $1 \leq k \leq p$, there exists at least one index $k^*$ such that $A_{k^*}'z \neq 0$ and $z_{k^*} \neq 0$. We set $J = \{k^*\}$ and $U_z = \mathrm{span}\{z\}$. By construction of $J$ and $U_z$, $K(J, z) = \|z\|_1/\|z_J\|_1 \max_{ 1 \leq k \leq p} \sqrt{|J|} \left\| A_{J_k, k}\right\|_1/ \min_{u \in U_z \cap S^{p-1}}$ $\|A_J u\|_2 < \infty$. Existence of $\vartheta \in [0, \infty]$ follows as in the proofs of claims (i) and (ii). 
\end{proof}

\begin{proof}[\textbf{Proof of Theorem 1}]
	The claim follows from Theorem~\ref{theorem:WeakConvergenceRankScoreBalancedEstimator} upon noting hat the conditions Conditions 1-10 with $\varrho_n = \sqrt{s_v + s_\theta} \times \sqrt{\log(np)/n}$ guarantee that $\lambda_d, \gamma_d > 0$ satisfy, with high probability, eq.~\eqref{eq:theorem:EmpiricalSparsity-0} and eq.~\eqref{eq:corollary:theorem:ConsistencyDual-00}, respectively.
\end{proof}	

\begin{proof}[\textbf{Proof of Theorem 2}]
	By Theorem 1, $\sqrt{n} \big(\widehat{Q}_d(\cdot;z) - Q_d(\cdot; z) \big) \leadsto \mathbb{G}(\cdot\:; z)$ in $\ell^\infty(\mathcal{T})$, where $\mathbb{G}(\cdot\:;z)$ is a centered Gaussian process with covariance function $(\tau_1, \tau_2) \mapsto H_d(\tau_1, \tau_2; z)$. The processes $\{\sqrt{n} \big(\widehat{Q}_1(\cdot;z) - Q_1(\cdot; z) \big) : \tau \in \mathcal{T}\} $ and $\{\sqrt{n} \big(\widehat{Q}_0(\cdot;z) - Q_0(\cdot; z) \big) : \tau \in \mathcal{T}\}$ are asymptotically independent. Hence, by Corollary 1.4.5 (Example 1.4.6) in~\cite{vaartwellner1996weak}, $\big(\sqrt{n} \big(\widehat{Q}_1(\cdot;z) - Q_1(\cdot; z) \big), \: \sqrt{n} \big(\widehat{Q}_0(\cdot;z) - Q_0(\cdot; z) \big) \big) \leadsto \big(\mathbb{G}_1, \mathbb{G}_0\big)$. Hence, the Continuous Mapping Theorem~\citep[e.g.][Theorem 1.11.1]{vaartwellner1996weak} yields the claim of the theorem. 
\end{proof}

\begin{proof}[\textbf{Proof of Lemma 5}]
	Apply Lemma~\ref{lemma:CharacterizationBalancingEstimator}.
\end{proof}

\begin{proof}[\textbf{Proof of Lemma 6}]
	The claim follows from Corollary~\ref{corollary:theorem:BahadurTypeRep}. As in the proof of Theorem 1 the conditions Conditions 1-10 with $\varrho_n = \sqrt{s_v + s_\theta) \log(np/\delta)/n}$ guarantee that $\lambda_d, \gamma_d > 0$ satisfy, with high probability, eq.~\eqref{eq:theorem:EmpiricalSparsity-0} and eq.~\eqref{eq:corollary:theorem:ConsistencyDual-00}, respectively.
\end{proof}

\begin{proof}[\textbf{Proof of Lemma 7}]
	Apply Lemma~\ref{lemma:ConsistencyCovarianceFunction-2}.
\end{proof}

\section{The $\ell_1$-penalized quantile regression problem}\label{sec:L1-QR}
\subsection{Setting}\label{subsec:Setting-L1}
We consider a high-dimensional parametric quantile regression model with continuous response $Y \in \mathbb{R}$ and predictors $X \in \mathbb{R}^p$, where the number of parameters $p$ diverges with (and possibly exceeds) the sample size $n$. Let $F$ be the joint distribution of $(Y,X)$, and let $F_{Y|X}(\cdot|x)$ and $f_{Y|X}(\cdot|x)$ be the conditional distribution and conditional density of $Y \mid X=x$. Recall that the $\tau$th conditional quantile function (CQF) of $Y$ given $X$ is 
\begin{align*}
	Q(\tau ; X) = \inf\left\{y : F_{Y|X}(y|X) \geq \tau \right\}.
\end{align*}
We assume that, at least over a compact subset $\mathcal{T} \subset (0,1)$ of quantile levels, the true CQF is linear function of only a few predictor variables, i.e. $Q(\tau ; X) = X'\theta(\tau)$. Given a random sample $\{(Y_i, X_i)\}_{i=1}^n$ we therefore estimate $\theta_0(\tau)$ as the solution $\hat{\theta}_\lambda(\tau)$ to the (unweighted) $\ell_1$-penalized quantile regression problem,
\begin{align}\label{eq:subsec:SettingNotation-1}
	\min_{\theta \in \mathbb{R}^p} \left\{\sum_{i=1}^n \rho_\tau(Y_i - X_i'\theta) + \lambda \|\theta\|_1 \right\}.
\end{align}

\subsection{Assumptions}\label{subsec:Assumptions-L1}
The following assumptions partially refine (and weaken) the conditions of the main text. Throughout, we assume that $\{(Y_i, X_i)\}_{i=1}^n$ is a random sample of independent and identically distributed random variables with joint distribution $F$.

\begin{assumption}[Sub-Gaussian predictors]\label{assumption:SubGaussianity}
	The random vector $X \in \mathbb{R}^p$ has positive definite covariance matrix $\Sigma$ and satisfies
	\begin{align*}
		\left\|\big(X - \mathbb{E}[X]\big)'u\right\|_{\psi_2}^2 \lesssim u'\Sigma u \quad{} \quad{} \forall \: u \in \mathbb{R}^p.
	\end{align*}
\end{assumption}

\begin{assumption}[Linear conditional quantile function]\label{assumption:LinearCQF}
	Let $\mathcal{T}$ be a compact subset of $(0,1)$. For all $\tau \in \mathcal{T}$, the $\tau$th conditional quantile of $Y$ given $X$ is a linear function in $X$, i.e. $Q_Y(\tau |X) = X'\theta_0(\tau)$, $\theta_0(\tau) \in \mathbb{R}^p$.
\end{assumption}

\begin{assumption}[Sparsity of $\tau \mapsto \theta_0(\tau)$]\label{assumption:SparsityQRVector}
	Let $\mathcal{T}$ be compact subset of $(0,1)$.
	\begin{align*}
		&T_\theta(\tau) :=\mathrm{support}\big(\theta_0(\tau)\big)  \hspace{20pt} \mathrm{and} \hspace{20pt} s_\theta := \sup_{\tau \in \mathcal{T}}\|\theta_0(\tau)\|_0 < n\wedge p.
	\end{align*}
\end{assumption}

\begin{assumption}[Lipschitz continuity of $\tau \mapsto \theta_0(\tau)$]\label{assumption:LipschitzQRVector}
	Let $\mathcal{T}$ be compact subset of $(0,1)$. There exists a constant $L_\theta \geq 1$ such that for all $s, \tau \in \mathcal{T}$,
	\begin{align*}
		\|\theta_0(\tau) - \theta_0(s)\|_2 \leq L_\theta |\tau - s|.
	\end{align*}
\end{assumption}

\begin{assumption}[Lipschitz continuity of the conditional density]\label{assumption:LipschitzDensity}
	The conditional density of $Y$ given $X$, $f_{Y|X}$, exists and is Lipschitz continuous, i.e. there exists a constant $L_f  \geq 1$ such that for all $a, b, x \in \mathbb{R}^p$,
	\begin{align*}
		\left|f_{Y|X}(x'a|x) - f_{Y|X}(x'b|x)\right| \leq L_f|x'a - x'b|.
	\end{align*}
\end{assumption}

\begin{assumption}[Growth condition for consistency]\label{assumption:GrowthCondition}
	The parameters $\delta \in (0,1)$ and $s_\theta, p, n \geq 1$ satisfy
	\begin{align*}
		s_\theta \log(ep/s_\theta) =o(n), \quad{}  \log(1/\delta) = O\big(\log(ep/s_\theta) \wedge \log n\big), \quad{} \mathrm{and} \quad{} s_\theta + 2  < p.
	\end{align*}
\end{assumption}

Next we introduce several definitions and assumptions pertaining the population and sample covariance and gram matrix of the predictors.

\begin{definition}[$s$-sparse maximum eigenvalues of covariance matrix] Let $s \in \{1, \ldots, p\}$ and define the $s$-sparse maximum eigenvalue the population and sample the covariance matrices $\Sigma$ and $\widehat{\Sigma}$, respectively, by
	\begin{align*}
		\phi_{\max}(s) := \sup_{u : \|u\|_0 \leq s} \frac{u' \Sigma u}{\|u\|_2^2} \hspace{20pt} \mathrm{and} \hspace{20pt} \widehat{\phi}_{\max}(s) := \sup_{u : \|u\|_0 \leq s} \frac{u'\widehat{\Sigma}u}{\|u\|_2^2}.
	\end{align*}
\end{definition}

\begin{definition}[$\ell_q$-cone of $(J, \vartheta)$-dominant coordinates]
	For $J\subseteq \{1, \ldots, p\}$, $\vartheta \in [0, \infty]$ and $q \geq 1$ define the $\ell_q$-cone of $(J, \vartheta)$-dominant coordinates by
	\begin{align*}
		C^p_q(J, \vartheta) := \left\{ u\in \mathbb{R}^{p}: \|u_{J^c}\|_q \leq \vartheta\|u_J\|_q  \right\}.
	\end{align*}
\end{definition}
\begin{remark}
	Observe that $\vartheta \in [0, \infty]$ controls the (approximate) sparsity level of the vectors in $C^p_q\big(J, \vartheta\big)$. Indeed, if $\vartheta= 0$, then the vectors in $C^p_q\big(J, 0\big)$ are $|J|$-sparse and only entries with index in $J$ are non-zero. In contrast, if $\vartheta = \infty$, then $C^p_q\big(J, \infty\big) = \mathbb{R}^p$.
\end{remark}

\begin{definition}[$\ell_q$-ball]
	For $q \geq 1$ define the $\ell_q$-ball with center $a \in \mathbb{R}^p$ and radius $r  > 0$ as $B^p_q(a, r)= \{u \in \mathbb{R}^p: \|u-a\|_q \leq r\}$.
\end{definition}

\begin{definition}[$(\omega, \vartheta, \varrho)$-restricted minimum eigenvalue of the design matrix] Let $\mathcal{T}$ be a compact subset of $(0,1)$ and $\omega, \vartheta, \varrho \geq 0$. Define the $(\omega, \vartheta, \varrho)$-restricted minimum eigenvalue of the design matrix as
	\begin{align*}
		\kappa_\omega(\vartheta, \varrho) : = \inf_{\tau \in \mathcal{T}} \inf_{\|\zeta\|_2 \leq \varrho} \inf_{u \in C^p_1(T_\theta(\tau), \vartheta) \cap \partial B^p_2(0,1)} \mathbb{E} \left[f_{Y|X}^\omega(X'\theta_0(\tau) + X'\zeta |X)(X'u)^2\right].
	\end{align*}
	To simplify notation we write $\kappa_\omega(\vartheta) : = \kappa_\omega(\vartheta, 0)$.
\end{definition}
\begin{remark}
	If $\mathcal{T} = \{\tau\}$ then $\kappa_\omega(\infty)$ is simply the minimum eigenvalue of $\mathbb{E} \left[f_{Y|X}^\omega(X'\theta_0(\tau) |X)XX'\right]$
\end{remark}

\begin{definition}[$(\omega, \vartheta)$-restricted nonlinearity coefficient] Let $\mathcal{T}$ be a compact subset of $(0,1)$ and $\omega, \vartheta \geq 0$.
	Define the $(\omega, \vartheta)$-restricted nonlinearity coefficient as
	\begin{align*}
		q_\omega(\vartheta) : = \sup_{\tau \in \mathcal{T}}\sup_{u \in C^p_1(T_\theta(\tau),\vartheta) \cap \partial B^p(0,1)} \frac{L_f^\omega\mathbb{E}[ |X'u|^3]}{\mathbb{E}[f_{Y|X}^\omega (X'\theta_0(\tau)|X)(X'u)^2]}.
	\end{align*}
\end{definition}

\begin{assumption}[$(\vartheta, \varrho)$-restricted identifiability]\label{assumption:RestrictedIdentifiability} 
	Let $\mathcal{T}$ be a compact subset of $(0,1)$ and $\vartheta, \varrho > 0$. The quantile regression vector $\theta_0(\tau)$ is $(\vartheta, \varrho)$-restricted identifiable for all $\tau \in \mathcal{T}$, i.e.
	\begin{align*}
		\kappa_1(\vartheta, \varrho) > 0 \quad{}\quad{} \mathrm{and} \quad{}\quad{} \kappa_1(\vartheta, \varrho) \gtrsim \kappa_1(\vartheta).
	\end{align*}
\end{assumption}

\begin{assumption-prime}{assumption:RestrictedIdentifiability}[$(\vartheta, \varrho)'$-restricted identifiability]\label{assumption:RestrictedPrimeIdentifiability} 
	Let $\mathcal{T}$ be a compact subset of $(0,1)$ and $\vartheta, \varrho > 0$. The quantile regression vector $\theta_0(\tau)$ is $(\vartheta, \varrho)'$-restricted identifiable for all $\tau \in \mathcal{T}$, i.e.
	\begin{align*}
		\kappa_1(\vartheta) > 0 \quad{}\quad{} \mathrm{and} \quad{}\quad{} q_1(\vartheta) \lesssim \frac{1}{\varrho}.
	\end{align*}
\end{assumption-prime}

\begin{assumption-2prime}{assumption:RestrictedIdentifiability}[$(\vartheta, \varrho)''$-restricted identifiability]\label{assumption:Restricted2PrimeIdentifiability} 
	Let $\mathcal{T}$ be a compact subset of $(0,1)$ and $\vartheta, \varrho > 0$. The quantile regression vector $\theta_0(\tau)$ is $(\vartheta, \varrho)''$-restricted identifiable for all $\tau \in \mathcal{T}$, i.e. there exist an absolute constant $\underline{f} > 0$ such that
	\begin{align*}
		\kappa_0(\vartheta) > 0 \quad{} \quad{} \mathrm{and} \quad{}\quad{}  \inf_{x \in \mathbb{R}^p}\inf_{\tau \in \mathcal{T}} \inf_{ |a| \leq 2 q_0(\vartheta)\varrho } f_{Y|X}\left(x'\theta_0(\tau) + a |x\right) \geq \underline{f}.
	\end{align*}
\end{assumption-2prime}

\begin{remark}
	Assumptions~\ref{assumption:RestrictedIdentifiability},~\ref{assumption:RestrictedPrimeIdentifiability}, and~\ref{assumption:Restricted2PrimeIdentifiability} can be used interchangeably. All three assumptions guarantee that the expected value of the quantile regression loss function can be (locally) minorized by a quadratic function (see Lemma~\ref{lemma:LocallyQuadraticMinorization}).  Assumption~\ref{assumption:RestrictedPrimeIdentifiability} is a version of a condition introduced by~\cite{belloni2011L1penalized}; versions of Assumption~\ref{assumption:Restricted2PrimeIdentifiability} are frequently used in the literature on quantile regression~\cite{chao2017quantile, wang2021analysis}. To the best of our knowledge, we are the first to introduce  Assumption~\ref{assumption:RestrictedIdentifiability}.
\end{remark}
\begin{remark}
	Assumption~\ref{assumption:RestrictedIdentifiability} appears to be the mildest assumption among all three in the following sense: Since we will use $1/ \varrho \asymp \sqrt{\frac{n}{s_\theta \log p}}$ and since $q_1(\vartheta)$ and $q_0(\vartheta)$ both grow in the dimension $p$, Assumptions~\ref{assumption:RestrictedPrimeIdentifiability} and~\ref{assumption:Restricted2PrimeIdentifiability} (implicitly) restrict the growth rate of $p$ relative to the sample size $n$. (If the quantity  $q_0(\vartheta)\varrho$ diverges, Assumption~\ref{assumption:Restricted2PrimeIdentifiability} cannot hold.) In contrast, Assumption~\ref{assumption:RestrictedIdentifiability} does not impose restrictions on the growth rates of $p$ and $n$. It is worth discussing the special case in which Assumption~\ref{assumption:SubGaussianity} holds and $\mathrm{E}[X] = 0$. In this case, $q_1(\vartheta) \vee q_0(\vartheta) = O(1)$ and Assumptions~\ref{assumption:RestrictedPrimeIdentifiability} and~\ref{assumption:Restricted2PrimeIdentifiability} do not restrict the growth rate of $p$ relative to $n$. However, both assumptions are still more restrictive than Assumption~\ref{assumption:RestrictedIdentifiability} because of the required Lipschitz-continuity and the lower bound on the density $f_{Y|X}$.
\end{remark}
\begin{remark}
	Sufficient conditions for Assumption~\ref{assumption:RestrictedIdentifiability}: If $f_{Y|X}$ is continuous, then $\kappa_1(\vartheta, \varrho)/ \kappa_1(\vartheta) \rightarrow 1$ as $\varrho \rightarrow 0$ and~\ref{assumption:RestrictedIdentifiability} holds for $\varrho > 0$ sufficiently small. If $f_{Y|X}$ is integrable with countably many discontinuity points, and the law of $X$ has a continuous density, then we still have $\kappa_1(\vartheta, \varrho)/ \kappa_1(\vartheta) \rightarrow 1$ as $\varrho \rightarrow 0$ and ~\ref{assumption:RestrictedIdentifiability} holds for $\varrho > 0$ sufficiently small. Unlike Assumption~\ref{assumption:Restricted2PrimeIdentifiability}, Assumption~\ref{assumption:RestrictedIdentifiability} may also hold for densities $f_{Y|X}$ that vanish on a non-null set of the real line.
\end{remark}
\subsection{Consistency}\label{subsec:Consistency-L1}
\begin{theorem}[Consistency]\label{theorem:Consistency}
	Let $\mathcal{T}$ be a compact subset of $(0,1)$, $\delta \in (0, 1)$, $c_0 > 1$, and $\lambda > 0$. Set $\bar{c} := (c_0 + 1)/(c_0 -1)$ and
	\begin{align}\label{eq:theorem:Consistency-r}
		r_\theta := \left(\frac{\bar{c}\phi_{\max}^{1/2}(2s_\theta)L_\theta}{\kappa_1(\bar{c})} \vee 1\right)\sqrt{ \frac{s_\theta \log(ep/ s_\theta) + \log n + \log(1/\delta)}{n}} \bigvee \frac{\bar{c}}{\kappa_1(\bar{c})}\frac{\lambda \sqrt{s_\theta}}{n}.
	\end{align}
	Suppose that Assumptions~\ref{assumption:SubGaussianity}--\ref{assumption:GrowthCondition} and Assumption~\ref{assumption:RestrictedIdentifiability} with $(\vartheta, \varrho) = (\bar{c}, r_\theta)$ hold and that $\lambda > 0$ satisfies eq.~\eqref{eq:lemma:RestrictedConeProp-QRVector-1}. With probability at least $1- \delta$, 
	\begin{align*}
		\sup_{\tau \in \mathcal{T}} \|\hat{\theta}_\lambda(\tau) - \theta_0(\tau)\|_2 \lesssim r_\theta.
	\end{align*}
\end{theorem}
\begin{remark}
	Observe that the rate $r_\theta$ depends on the data only through the largest $2s_\theta$-sparse eigenvalue of the covariance matrix $\phi_{\max}(2s_\theta)$, the smallest restricted eigenvalue $\kappa_1(\bar{c})$, and the Lipschitz constant $L_f$. In particular, the conditional density of $Y$ given $X$ does not need to be bounded away from $0$ or $\infty$, and the largest (sparse) eigenvalue of the Gram matrix may diverge.
\end{remark}
\begin{corollary}\label{corollary:theorem:Consistency}
	Let $\mathcal{T}$ be a compact subset of $(0,1)$, $\delta \in (0, 1)$, and $c_0 > 1$. Set $\bar{c} := (c_0 + 1)/(c_0 -1)$ and $\tilde{r}_\theta := \sqrt{ s_\theta \log(p/ \delta)/n + (\log n)/n }$. Suppose that Assumptions~\ref{assumption:SubGaussianity}--\ref{assumption:GrowthCondition} and Assumption~\ref{assumption:RestrictedIdentifiability} with $(\vartheta, \varrho) = (\bar{c}, \tilde{r}_\theta)$ hold. Then there exists an absolute constant $c_1 > c_0$ such that for all $C_\lambda \geq c_1$ and
	\begin{align}
		\lambda := C_\lambda \varphi_{\max}^{1/2}(1)\sqrt{n\log(p/\delta)},
	\end{align}
	with probability at least $1- \delta$, 
	\begin{align*}
		\sup_{\tau \in \mathcal{T}} \|\hat{\theta}_\lambda(\tau) - \theta_0(\tau)\|_2 \lesssim \left(\frac{\bar{c}\phi_{\max}^{1/2}(2s_\theta)L_\theta}{\kappa_1(\bar{c})} \vee \frac{C_\lambda\bar{c} \varphi_{\max}^{1/2}(1)}{\kappa_1(\bar{c})}\right) \sqrt{ \frac{s_\theta \log(p/ \delta) + \log n }{n}}.
	\end{align*}
\end{corollary}

\subsection{Empirical sparsity}\label{subsec:EmpiricalSparsity}
We introduce the following notation:

\begin{definition}[$s$-sparse maximum eigenvalues of gram matrix] Let $s \in \{1, \ldots, p\}$ and define the $s$-sparse maximum eigenvalue the population and sample the gram matrices by
	\begin{align*}
		\varphi_{\max}(s) := \sup_{u : \|u\|_0 \leq s} \frac{ \mathbb{E}[(X'u)^2]}{\|u\|_2^2} \hspace{20pt} \mathrm{and} \hspace{20pt} \widehat{\varphi}_{\max}(s) := \sup_{u : \|u\|_0 \leq s} \frac{n^{-1}\sum_{i=1}^n (X_i'u)^2}{\|u\|_2^2}.
	\end{align*}
\end{definition}

\begin{theorem}[Empirical Sparsity]\label{theorem:EmpiricalSparsity}  Let $\mathcal{T}$ be a compact subset of $(0,1)$.
	\begin{itemize}
		\item[(i)] Let $1 \leq m \leq p$ be arbitrary. If $\lambda \geq n\sqrt{2} \sqrt{\widehat{\varphi}_{\max}(m)/m}$, then $\hat{s}_\lambda := \sup_{\tau \in \mathcal{T}}\|\hat{\theta}_\lambda(\tau)\|_0\leq m \wedge n \wedge p$ with probability one.
		
		\item[(ii)] Let $\delta \in (0, 1)$ and $c_0 > 1$. Set $\bar{c} := (c_0 + 1)/(c_0 -1)$ and $\bar{r} := \sqrt{ s_\theta \log(np/ \delta)/n}$. Suppose that Assumptions~\ref{assumption:SubGaussianity}--\ref{assumption:GrowthCondition} and Assumption~\ref{assumption:RestrictedIdentifiability} with $(\vartheta, \varrho) = (\bar{c}, \bar{r}_\theta)$ hold. There exists an absolute constant $c_1 > c_0 \vee \sqrt{2}$ such that for all $C_\lambda \geq c_1$ and
		\begin{align}\label{eq:theorem:EmpiricalSparsity-0}
			\lambda := C_\lambda \sqrt{\varphi_{\max}\big(n/\log(np/\delta)\big) \vee \widehat{\varphi}_{\max}\big(n/\log(np/\delta)\big)}\sqrt{n \log(np/\delta)},
		\end{align}
		with probability at least $1- \delta$,
		\begin{align*}
			\sup_{\tau \in \mathcal{T}} \|\hat{\theta}_\lambda(\tau) - \theta_0(\tau)\|_2 \lesssim C_1\bar{r}_\theta,
		\end{align*}
		and
		\begin{align*}
			\hat{s}_\lambda := \sup_{\tau \in \mathcal{T}}\|\hat{\theta}_\lambda(\tau)\|_0 &\lesssim C_1^2C_2^2 s_\theta,
		\end{align*}
		where 
		\begin{align*}
			C_1 &: = \left(\frac{\bar{c}\phi_{\max}^{1/2}(2s_\theta)L_\theta}{\kappa_1(\bar{c})} \vee \frac{C_\lambda \bar{c} \varphi_{\max}^{1/2}\big(n/\log(np/\delta)\big)}{\kappa_1(\bar{c})} \vee \frac{C_\lambda \bar{c} \widehat{\varphi}_{\max}^{1/2}\big(n/\log(np/\delta)\big)}{\kappa_1(\bar{c})} \vee 1\right),\\
			C_2 & : = \left(\frac{ \bar{c} (2 + \bar{c})^2\phi_{\max}(s_\theta)L_f}{C_\lambda} \vee \frac{\bar{c}(2 + \bar{c}) \varphi_{\max}^{1/2}(s_\theta)}{C_\lambda} \right).
		\end{align*}
	\end{itemize}
\end{theorem}

\subsection{Auxiliary results}\label{subsec:AuxResults-L1}
\begin{lemma}[Restricted cone property]\label{lemma:RestrictedConeProp-QRVector}
	Let $\mathcal{T}$ be a compact subset of $(0,1)$ and $c_0 > 1$. Set $\bar{c} := (c_0 + 1)/(c_0 -1)$. Suppose that Assumption~\ref{assumption:SparsityQRVector} holds and 
	\begin{align}\label{eq:lemma:RestrictedConeProp-QRVector-1}
		\lambda c_0^{-1} \geq \sup_{\tau \in \mathcal{T}}\left\|\sum_{i=1}^n X_i\big(\tau - 1\{Y_i\leq X_i'\theta_0(\tau)\}\big) \right\|_\infty.
	\end{align}
	Then, for all $\tau \in \mathcal{T}$,
	\begin{align*}
		\widehat{\theta}_\lambda(\tau) - \theta_0(\tau) \in C^p\big(T_\theta(\tau), \bar{c}\big).
	\end{align*}
\end{lemma}
\begin{remark}
	This lemma states that $\widehat{\theta}_\lambda(\tau) - \theta_0(\tau) $ lies in a cone of dominant coordinates. This is instrumental for establishing (uniform) consistency of the $\ell_1$-penalized quantile regression vector.
\end{remark}

\begin{lemma}[A new look at Knight's identity]\label{lemma:NewKnightsIdentity}
	Let $s, \tau \in \mathcal{T}$, $y \in \mathbb{R}$, and $\theta, x \in \mathbb{R}^p$ be arbitrary. Define
	\begin{align*}
		\phi_{\tau,x,y}(z) :=\int_0^{z} 1\{y \leq x'\theta_0(\tau) + u\} du.
	\end{align*}
	The following holds true:
	\begin{enumerate}
		\item[(i)] $\phi_{\tau, x, y}$ is a contraction and $\phi_{\tau, x,y}(0) = 0$;
		\item[(ii)] $\phi_{\tau,x,y}\big(x'\theta - x'\theta_0(\tau)\big) = \phi_{s,x, y}\big(x'\theta - x'\theta_0(s)\big) - \phi_{s,x,y}\big(x'\theta_0(\tau) - x'\theta_0(s)\big)$;
		\item[(iii)] $\rho_\tau(y - x'\theta) - \rho_\tau\big(y - x'\theta_0(\tau)\big) =  - \tau \big(x'\theta - x'\theta_0(\tau)\big) + \phi_{\tau, x, y}\big(x'\theta - x'\theta_0(\tau)\big)$.
	\end{enumerate}
\end{lemma}
\begin{remark}
	The first property allows us to apply the contraction principle for conditional Rademacher averages; the second property helps us when using (simple) chaining arguments over quantile levels $\tau \in \mathcal{T}$. While the first and second properties appear to be new, the third property is a simple consequence of Knight's identity. Implicitly,~\cite{belloni2011L1penalized} use the same properties in their proof of Lemma 5.
\end{remark}

\begin{lemma}[Locally quadratic minorization]\label{lemma:LocallyQuadraticMinorization} Let $\mathcal{T}$ be a compact subset of $(0,1)$, $c_0 \geq 0$, and $r_0 > 0$.
	\begin{itemize}
		\item[(i)] Suppose that Assumption~\ref{assumption:RestrictedIdentifiability} holds with $(\vartheta, \varrho) = (c_0, r_0)$. Then, for all points $(\theta, \tau) \in \mathbb{R}^p \times \mathcal{T}$ with $\theta - \theta_0(\tau) \in C^p(T_\theta(\tau), c_0) \cap \partial B^p(0, r_0)$,
		\begin{align*}
			\mathbb{E}\left[ \rho_\tau(Y-X'\theta) - \rho_\tau\big(Y-X'\theta_0(\tau)\big)\right] \gtrsim \kappa_1(c_0) r_0^2.
		\end{align*}
		\item[(ii)] Suppose that Assumption~\ref{assumption:RestrictedPrimeIdentifiability} holds with $(\vartheta, \varrho) = (c_0, r_0)$. Then, for all points $(\theta, \tau) \in \mathbb{R}^p \times \mathcal{T}$ with $\theta - \theta_0(\tau) \in C^p(T_\theta(\tau), c_0) \cap \partial B^p(0, r_0)$,
		\begin{align*}
			\mathbb{E}\left[ \rho_\tau(Y-X'\theta) - \rho_\tau\big(Y-X'\theta_0(\tau)\big)\right] \gtrsim \kappa_1(c_0) r_0^2.
		\end{align*}
		\item[(iii)] Suppose that Assumption~\ref{assumption:Restricted2PrimeIdentifiability} holds with $(\vartheta, \varrho) = (c_0, r_0)$. Then, for all points $(\theta, \tau) \in \mathbb{R}^p \times \mathcal{T}$ with $\theta - \theta_0(\tau) \in C^p(T_\theta(\tau), c_0) \cap \partial B^p(0, r_0)$,
		\begin{align*}
			\mathbb{E}\left[ \rho_\tau(Y-X'\theta) - \rho_\tau\big(Y-X'\theta_0(\tau)\big)\right] \gtrsim \underline{f}\kappa_0(c_0) r_0^2.
		\end{align*}
	\end{itemize}	
	
\end{lemma}

\begin{lemma}[Size of $\ell_q$-cones of dominant coordinates (A useful version of Lemma 7.1,~\citeauthor{koltchinskii2011oracle},~\citeyear{koltchinskii2011oracle})]\label{lemma:SizeConesDominatedCoordinates}
	Let $\vartheta \in [0,\infty]$, $J \subseteq \{1, \ldots, p\}$, $s = \mathrm{card}(J)$, $q \geq 1$, and $p \geq s + 2$. Define
	\begin{align*}
		\mathcal{M} = \bigcup _{I \subset \{1, \ldots, p\},  \mathrm{card}(I) \leq s } \mathcal{N}_I,
	\end{align*}
	where $\mathcal{N}_I$ is the minimal $1/2$-net of $B_I =\big\{ \{u_i\}_{i \in I} : \sum_{i \in I} |u_i|^2 \leq 1\big\}$. The following holds true:
	\begin{itemize}
		\item[(i)] $C^p_q(J, \vartheta) \cap B^p_2(0,1) \subset 2(2 + \vartheta s^{1/2(1/q-1)}) \mathrm{conv}(\mathcal{M})$;
		\item[(ii)] $ \mathrm{card}(\mathcal{M}) \leq \frac{3}{2}\left(\frac{5ep}{s}\right)^s$;
		\item[(iii)] $\forall u \in \mathcal{M} : \|u\|_0 \leq s$.
	\end{itemize}
\end{lemma}

\begin{lemma}[Maxima of biconvex function]\label{lemma:MaximaBiconvexFunction}
	Let $f : \mathcal{X} \times \mathcal{Y} \rightarrow \mathbb{R}$ be a biconvex function. Then,
	\begin{align*}
		\sup_{x \in \mathrm{conv}(\mathcal{X})} \sup_{y \in \mathrm{conv}(\mathcal{Y})} f(x,y) = \sup_{x \in \mathcal{X}} \sup_{y \in \mathcal{Y}} f(x,y),
	\end{align*}
	Moreover, the identity remains true if $f$ is replaced by $|f|$.
\end{lemma}

\begin{lemma}\label{lemma:LocalizedLoss} Let $\delta \in (0,1)$ be arbitrary, $\mathcal{T}$ be a compact subset of $(0,1)$, $c_0 \geq 0$, $r_0 > 0$. Suppose that Assumptions~\ref{assumption:SubGaussianity}--\ref{assumption:LipschitzQRVector} hold. Define
	\begin{align*}
		\mathcal{G} = \big\{g: \mathbb{R}^{p+1} \rightarrow \mathbb{R} : g(X,Y) = \rho_\tau\big(Y-X'\theta_0(\tau)\big) - \rho_\tau(Y-X'\theta), \\
		\theta - \theta_0(\tau) \in C^p(T_\theta(\tau), c_0) \cap B^p(0, r_0), \: \tau \in \mathcal{T} \big\}.
	\end{align*}
	With probability at least $1 - \delta$, 
	\begin{align*}
		\|\mathbb{G}_n\|_{\mathcal{G}} \lesssim 2(2 + c_0) \phi_{\max}^{1/2}(2s_\theta) r_0 \sqrt{s_\theta \log(ep/s_\theta) + \log(1 +  L_\theta/r_0) + \log (1/\delta)}. 
	\end{align*}
\end{lemma}

\begin{lemma}\label{lemma:MaxInequalityCovarianceCone}
	Let $\mathcal{T}$ be a compact subset of $(0,1)$. Let $\delta \in (0,1)$ be arbitrary and $\vartheta_k \in [0,\infty]$, $q_k \geq 1$, $J_k(\tau) \subseteq \{1, \ldots, p\}$ for $\tau \in \mathcal{T}$, $s_k = \sup_{\tau \in \mathcal{T}}\mathrm{card}(J_k(\tau))$, $p \geq s_k + 2$, and $k \in \{1,2\}$.  Let $\{(X_i, \xi_i)\}_{i=1}^n$ be a sequence of i.i.d. random vectors. Suppose that Assumption~\ref{assumption:SubGaussianity} holds and $|\xi_i|\leq 1$ a.s. for all $1 \leq i \leq n$.  Define
	\begin{align*}
		\mathcal{G} = \big\{ g : \mathbb{R}^p \times [-1,1] \rightarrow \mathbb{R}: g(X, \xi) = \xi (X'u_1) (X'u_2), \:u_k \in  C^p_{q_k}(J_k(\tau), \vartheta_k) \cap B^p(0, 1), \: k \in \{1, 2\}, \: \tau \in \mathcal{T}\big\}.
	\end{align*} 
	The following holds true:
	\begin{itemize}
		\item[(i)] With probability at least $1-\delta$,
		\begin{align*}
			\|\mathbb{G}_n\|_{\mathcal{G}} &\lesssim (2 + \vartheta_1)(2 + \vartheta_2)\varphi_{\max}^{1/2}(s_1)\varphi_{\max}^{1/2}(s_2) \left( \sqrt{ s_1 \log( ep/s_1) + s_2 \log( ep/s_2) + \log(1/\delta)} \right.\\
			&\quad{}\quad{}\quad{}\quad{}\quad{}\quad{}\quad{}\quad{}\quad{}\quad{}\quad{}\quad{}\quad{} \quad{}\quad{}\quad{}\left.+ n^{-1/2} \big(s_1 \log( ep/s_1) + s_2 \log( ep/s_2) + \log(1/\delta)\big)\right).
		\end{align*}
		\item[(ii)] Statement (i) holds also for the function class $|\mathcal{G}| := \left\{h : \mathbb{R}^{p + 1} \times [-1,1] \rightarrow \mathbb{R}:\:  \exists g \in \mathcal{G} : h = |g| \right\}$.
		\item[(iii)] Statements (i) and (ii) hold with $\vartheta_1 = 1$ also for the function class
		\begin{align*}
			\mathcal{G} &= \big\{ g : \mathbb{R}^p \times [-1, 1] \rightarrow \mathbb{R}: g(X, \xi) = \xi (X'v) (X'u),\\
			&\quad{}\quad{} v \in \mathbb{R}^p, \: \|v\|_0 \leq s, \: \|v\|_2 \leq 1, \:u \in C^p_{q}(J_2(\tau), \vartheta_2) \cap B^p(0, 1), \: \tau \in \mathcal{T}\big\}.
		\end{align*}
	\end{itemize}
	
\end{lemma}

\begin{lemma}\label{lemma:GradientCone} 
	Let $\mathcal{T}$ be a compact subset of $(0,1)$. Let $\delta \in (0,1)$ be arbitrary and $\vartheta \in [0,\infty]$, $q \geq 1$, $J(\tau) \subseteq \{1, \ldots, p\}$ for $\tau \in \mathcal{T}$, $s = \sup_{\tau \in \mathcal{T}}\mathrm{card}(J(\tau))$, and $p \geq s + 2$.  Let $\{(X_i, Y_i, \xi_i)\}_{i=1}^n$ be a sequence of i.i.d. random vectors. Suppose that Assumptions~\ref{assumption:SubGaussianity}--\ref{assumption:SparsityQRVector} hold and $|\xi_i|\leq 1$ a.s. for all $1 \leq i \leq n$. Define 
	\begin{align*}
		\mathcal{G} = \left\{g: \mathbb{R}^{p+1} \times [-1,1] \rightarrow \mathbb{R}: g(X, Y, \xi) = \xi\left(\tau - \mathbf{1}\big\{Y \leq X'\theta_0(\tau) \}\right)X'v, \right.\\
		\left.  v \in  C^p_q(J(\tau), \vartheta) \cap B^p(0, 1), \: \tau \in \mathcal{T}\right\}.
	\end{align*}
	The following holds true:
	\begin{itemize}
		\item[(i)] With probability at least $1-\delta$,
		\begin{align*}
			\|\mathbb{G}_n\|_{\mathcal{G}} \lesssim (2 + \vartheta)\varphi_{\max}^{1/2}(s)\sqrt{s \log (ep/s) + \log(1/\delta)} \sqrt{1 + \pi_{n,1}^2(s \log (ep/s) + \log(1/\delta))},
		\end{align*}
		where $\pi_{n,1}^2(z) = \sqrt{z/n} + z/n$ for $ z \geq  0$.	
		\item[(ii)] Statement (i) holds with $\vartheta = 1$ also for the function class
		\begin{align*}
			\mathcal{G} = \left\{g: \mathbb{R}^{p+1} \times [-1,1] \rightarrow \mathbb{R}: g(X, Y, \xi) = \xi\left(\tau - \mathbf{1}\big\{Y \leq X'\theta_0(\tau) \}\right)X'v, \right.\\
			\left.   v \in \mathbb{R}^p, \: \|v\|_0 \leq s, \: \|v\|_2 \leq 1, \: \tau \in \mathcal{T}\right\}.
		\end{align*}
		\item[(iii)] Statements (i) and (ii) hold for $\xi_1, \ldots, \xi_n$ i.i.d. standard normal random variables with $\pi_{n,1/2}^2(z) = \sqrt{z/n} + z^2/n$ replacing $\pi_{n,1}^2(z)$.
	\end{itemize}
\end{lemma}

The next two lemmata provide bounds that hold uniformly over collections of empirical processes.

\begin{lemma}\label{lemma:LocalizedRankScores} 
	Let $\delta \in (0,1)$ be arbitrary and $\mathcal{T}$ be a compact subset of $(0,1)$. Let $\{(X_i, Y_i,  \xi_i)\}_{i=1}^n$ be a sequence of i.i.d. random vectors. Suppose that Assumptions~\ref{assumption:SubGaussianity}--\ref{assumption:SparsityQRVector} hold and $|\xi_i|\leq 1$ a.s. for all $1 \leq i \leq n$. Define 
	\begin{align*}
		\mathcal{G} = \left\{g: \mathbb{R}^{p+1} \times [-1,1] \rightarrow \mathbb{R}: g(X, Y, \xi) = \xi\left(\mathbf{1}\big\{Y \leq X'\theta \big\} - \mathbf{1}\big\{Y \leq X'\theta_0(\tau) \}\right)X'v,  \:  v, \theta \in \mathbb{R}^p, \right.\\
		\left.  \|v\|_2 \leq 1,\: \|v\|_0 \leq n, \: \|\theta\|_0 \leq n, \: \tau \in \mathcal{T}\right\}.
	\end{align*}
	The following holds true:
	\begin{itemize}
		\item[(i)] With probability at least $1-\delta$,
		\begin{align*}
			\forall g_{v,\theta, \tau} \in \mathcal{G} : \: |\mathbb{G}_n(g_{v, \theta, \tau})| &\lesssim  \varphi_{\max}^{1/2}(\|v\|_0)\sqrt{t_{\|v\|_0, \|\theta\|_0, n, \delta}} \sqrt{1 + \pi_{n,1}^2(t_{\|v\|_0, \|\theta\|_0, n, \delta})}.
		\end{align*}
		where $t_{k,\ell,n, \delta} = k \log (ep/k) + \ell \log (ep/\ell) + \log(n/\delta)$ and $\pi_{n,1}^2(z) = \sqrt{z/n} + z/n$ for $ z \geq  0$;
		\item[(ii)] Let $\mathcal{G}(m) = \left\{g_{v, \theta, \tau} \in \mathcal{G}:  \|v\|_0 \leq m, \|\theta\|_0 \leq m\right\}$. With probability at least $1-\delta$,
		\begin{align*}
			\forall m \leq n \wedge p : \: \|\mathbb{G}_n\|_{\mathcal{G}(m)} &\lesssim \varphi_{\max}^{1/2}(m) \sqrt{ m \log(ep/m) + \log (n/\delta) } \sqrt{1 + \pi_{n,1}^2\big(m \log(ep/m) + \log (n/\delta) \big)},
		\end{align*}
		where $\pi_{n,1}^2(z) = \sqrt{z/n} + z/n$ for $ z \geq  0$;
	\end{itemize}	
\end{lemma}

\begin{lemma}\label{lemma:MaxInequalityCovariance} Let $\delta \in (0,1)$ be arbitrary. Suppose that Assumption~\ref{assumption:SubGaussianity} holds. With probability at least $1-\delta$, 
	\begin{align*}
		&\forall k \leq n : \:\: \sup_{\mathrm{card}(I) \leq k} \sup_{\|u\|_2 \leq 1, \|u\|_0\leq k} \left|\frac{1}{k}\sum_{i \in I} (X_i'u)^2 - \mathbb{E}[(X_i'u)^2]\right|\\
		&\quad{} \lesssim \varphi_{\max}(k) \left(\sqrt{\frac{k \log (epn/k) + \log(1/\delta)}{k}}  + \frac{k \log (epn/k) + \log(1/\delta)}{k}\right).
	\end{align*}
\end{lemma}

\section{The rank-score balanced quantile regression problem}\label{sec:RankScore-QR}
\subsection{Setting}\label{subsec:Setting-RankScore}
We define the rank-score balanced estimator of the CQF at covariate $z \in \mathbb{R}^p$ as
\begin{align}\label{eq:subsec:SettingNotation-Debiased-1}
	\widehat{Q}_{\lambda, \gamma}(\tau;z) : = z'\hat{\theta}_\lambda(\tau) + \frac{1}{\sqrt{n}}\sum_{i=1}^n \widehat{w}_{\gamma, i}(\tau;z) \hat{f}_i^{-1}(\tau)(\tau - 1\{Y_i \leq X_i'\hat{\theta}_\lambda(\tau)\}) \quad{}\quad{} \forall \tau \in \mathcal{T},
\end{align}
where the $\hat{f}_i$'s are estimates of $f_{Y|X}(X_i'\theta_0(\tau)|X_i)$ and the vector $\widehat{w}_\gamma(\tau;z) \in \mathbb{R}^n$ is the solution to the ranks-score balancing program
\begin{align}\label{eq:BalancingWeights-Primal-Appendix}
	\begin{split}
		\min_{w \in \mathbb{R}^n} &\quad{} \sum w_i^2 \hat{f}_i^{-2}(\tau)\\
		\mathrm{s.t.} &\quad{} \|z - n^{-1/2} \mathbf{X}'w\|_\infty \leq \gamma/ n,
	\end{split}
\end{align}
where $\mathbf{X}' = [X_1, \ldots, X_n] \in \mathbb{R}^{p\times n}$.

\subsection{Assumptions}\label{subsec:Assumptions-RankScore}
The following assumptions partially refine the conditions in the main text.

\begin{assumption}[Relative consistency of the density estimator]\label{assumption:RelativeConsistencyDensity}
	The estimates $\{\hat{f}_i(\cdot)\}_{i=1}^n$ of the conditional densities $\{f_{Y|X}(X_i'\theta_0(\cdot) |X_i)\}_{i=1}^n$ are relative consistent in the following sense: There exists $r_f > 0$ such that, with probability at least $1- \eta$,
	\begin{align*}
		\sup_{\tau \in \mathcal{T}} \max_{1 \leq i \leq n}\left|\frac{\hat{f}_i(\tau)}{f_{Y|X}(X_i'\theta_0(\tau)|X_i)} - 1\right| \lesssim r_f.
	\end{align*}
\end{assumption}

\begin{assumption}[Boundedness of the conditional density]\label{assumption:BoundedDensity} The conditional density of $Y$ given $X$ is bounded, i.e. there exists an absolute constant $1 \leq \bar{f}< \infty$ such that
	\begin{align*}
		\sup_{z \in \mathbb{R}} \sup_{x \in \mathbb{R}^p} f_{Y|X}(z| x) \leq \bar{f}.
	\end{align*}
\end{assumption}

\begin{definition}[Exact and approximate solutions to the population dual problem]\label{definition:ExactApproxDualSolution} For $z \in \mathbb{R}^p$ and $\tau \in \mathcal{T}$ denote the exact solution to the population dual by
	\begin{align}\label{eq:PopulationBalancingWeights-DualSolution}
		v_0(\tau;z) := -2 \mathbb{E}\left[ f_{Y|X}^2\big(X'\theta_0(\tau)|X\big)XX' \right]^{-1}z,
	\end{align}
	and the sparse approximation to the solution of the population dual as
	\begin{align}\label{eq:PopulationBalancingWeights-DualApproximation}
		\tilde{v}(\tau;z) := \arg\min_{v \in \mathbb{R}^p} \left\{\|v\|_0 : \|v - v_0(\tau;z)\|_2\leq r_a \|v_0(\tau, z)\|_2\right\},
	\end{align}
	where $0 \leq r_a \leq 1/4$ controls the accuracy of the sparse approximation.
\end{definition}
\begin{remark}
	We do not have to solve this $\ell_0$-minimization problem; we only introduce it for conceptual reasons. In particular, to establish weak convergence of the rank-score balanced quantile regression process we will impose conditions on the tuning parameter $r_a > 0$ and the sparsity of the approximation $\tilde{v}(\tau;z)$. These conditions are rather mild and are be satisfied in a range of situations.
\end{remark}

\begin{assumption}[Sparsity of $\tilde{v}(\tau; z)$]\label{assumption:SparsityDual}
	The approximation to the solution of the population dual $\tilde{v}(\tau;z) \in \mathbb{R}^p$ satisfies
	\begin{align*}
		&T_v(\tau;z) :=\mathrm{support}\big(\tilde{v}(\tau; z)\big) \hspace{20pt}  \mathrm{and} \hspace{20pt} s_v(z) := \sup_{\tau \in \mathcal{T}}\|\tilde{v}(\tau; z)\|_0 < n\wedge p.
	\end{align*}
\end{assumption}

\begin{assumption}[Growth condition for the approximate solution to the population dual problem]\label{assumption:GrowthConditionDual}
	The parameters $\delta \in (0,1)$, $r_f \geq 0$, $1/4 \geq r_a \geq 0$, and $z, s_v(z), p, n \geq 1$ satisfy
	\begin{align*}
		s_v(z) \log\big(ep/s_v(z)\big) =o(n), \quad{}  \log(1/\delta) = O\left(\log\big(ep/s_v(z)\big) \wedge \log n\right),  \quad{} s_v(z) + 2 < p, \quad{} r_f\vee r_a = o(1).
	\end{align*}
\end{assumption}
\begin{assumption}[Positivity of $f_{Y|X}(X'\theta_0(\tau)|X)$]\label{assumption:PositiveDensity}
	Let $\mathcal{T}$ be a compact subset of $(0,1)$. The conditional density function of $Y$ given $X$ is strictly positive uniformly in $\tau \in \mathcal{T}$ and $x \in \mathbb{R}^p$ in the following sense,
	\begin{align*}
		\inf_{\tau \in \mathcal{T}} \inf_{x \in \mathbb{R}^p} f_{Y|X}(x'\theta_0(\tau)|x) > 0.
	\end{align*}
\end{assumption}

\begin{assumption}[Differentiability of $\tau \mapsto Q_Y(\tau; X)$]\label{assumption:DiffCQF}
	Let $\mathcal{T}$ be a compact subset of $(0,1)$. The conditional quantile function of $Y$ given $X$, $\tau \mapsto Q_Y(\tau; X)$, is three times boundedly differentiable on $\mathcal{T}$, i.e. there exists a constant $C_Q \geq 1$ such that $\sup_{\tau \in \mathcal{T}}\sup_{x \in \mathbb{R}^d}\left|\frac{d^3}{d\tau^3}Q_Y(\tau; x)\right| \leq C_Q$.
\end{assumption}

\subsection{The dual problem}\label{subsec:DualProblem}
Consider the following convex optimization problem:
\begin{align}\label{eq:BalancingWeights-Dual-Appendix}
	\min_{v \in \mathbb{R}^p} \:\:\:&\frac{1}{4}\sum_{i =1}^n \hat{f}_i^2(\tau) (X_i'v)^2 +  nz' v + \gamma \| v\|_1,
\end{align}

The next lemma establishes that this program is the dual program to the rank-score balancing program~\eqref{eq:BalancingWeights-Primal-Appendix}.

\begin{lemma}[Dual of the rank-score balancing program]\label{lemma:CharacterizationBalancingEstimator}
	\noindent
	
	\begin{itemize}
		\item[(i)] Programs~\eqref{eq:BalancingWeights-Dual-Appendix} and~\eqref{eq:BalancingWeights-Primal-Appendix} are a primal-dual pair.
		\item[(ii)] Let $\delta \in (0,1)$. Suppose that Assumption~\ref{assumption:SubGaussianity} holds. There exists an absolute constant $c_2 > 1$ such that for all $\gamma > 0$ that satisfy
		\begin{align}\label{eq:BalancingWeights-Dual-00}
			\gamma c_2^{-1} \geq \frac{\bar{f}^2\varphi_{\max}^{1/2}(1)\varphi_{\max}^{1/2}\big(s_v(z)\big)}{\kappa_2(\infty)}\sqrt{\log(p/\delta)}\|z\|_2\sqrt{n},
		\end{align}
		with probability at least $1 - \delta$, 
		\begin{align}\label{eq:BalancingWeights-Dual-0}
			\widehat{w}_{\gamma, i}(\tau; z) = -\frac{\hat{f}_i^2(\tau)}{2\sqrt{n}}X_i'\hat{v}_\gamma(\tau; z), \quad{}\quad{} 1 \leq i \leq n, \quad{} \forall \tau \in \mathcal{T},
		\end{align}
		where $\widehat{w}_\gamma(\tau;z)$ and $\hat{v}_\gamma(\tau;z)$ are the unique solutions to programs ~\eqref{eq:BalancingWeights-Primal-Appendix} and~\eqref{eq:BalancingWeights-Dual-Appendix}, respectively.
	\end{itemize}
\end{lemma}

\begin{theorem}[Consistency]\label{theorem:ConsistencyDual}
	Let $\mathcal{T}$ be a compact subset of $(0,1)$, $\delta, \eta \in (0, 1)$, $c_0 > 1$, and $\gamma > 0$. Set $\bar{c} := (c_0 + 1)/(c_0 -1)$ and
	\begin{align}\label{eq:theorem:ConsistencyDual-r}
		\begin{split}
			r_v &:= C_3 \left(\frac{\|z\|_2}{\kappa_2(\infty)} \vee 1 \right) \left( \sqrt{\frac{s_v(z) \log\big(ep/s_v(z)\big) + s_\theta \log(ep/s_\theta) + \log (n/\delta)}{n}} \vee r_f\vee r_a  \right)\\
			&\quad{}\quad{}\quad{}\quad{}\quad{}+ \left(\frac{\bar{c}\|z\|_2^2}{\kappa_2^2(\infty)} \right) \left(\frac{\mu}{n} \vee \frac{\mu}{\gamma}\right)+ \left(\frac{\bar{c}}{\kappa_2(2\bar{c})}\right) \frac{\gamma \sqrt{s_v}}{n}.
		\end{split}
	\end{align}
	where $C_3 : = \bar{c}^2 \bar{f}^2  L_f L_\theta\big(1 +\varphi_{\max}^{1/2}(2s_\theta)\big)\varphi_{\max}\big(s_v(z)\big)/\kappa_2(2\bar{c})$.
	Suppose that Assumptions~\ref{assumption:SubGaussianity}--\ref{assumption:GrowthCondition} and~\ref{assumption:RelativeConsistencyDensity}--\ref{assumption:GrowthConditionDual} hold and that $\gamma > 0$ and $\mu \geq 0$ satisfy eq.~\eqref{eq:lemma:RestrictedConeProp-Dual-1} and~\eqref{eq:lemma:RestrictedConeProp-Dual-11}, respectively. With probability at least $1- \eta- \delta$,
	\begin{align*}
		\sup_{\tau \in \mathcal{T}} \left\|\hat{v}_\gamma(\tau; z) - \tilde{v}(\tau;z)\right\|_2 \lesssim r_v.
	\end{align*}
\end{theorem}
\begin{remark}
	Note that the rate of consistency of $\hat{v}_\gamma(\tau; z)$ to $\tilde{v}(\tau;z)$ is at most as fast as the rate of consistency $r_f$ of the estimated conditional density and the rate $r_a$ at which the approximation error between $\tilde{v}(\tau; z)$ and $v_0(\tau;z)$ vanishes.
\end{remark}

\begin{corollary}\label{corollary:theorem:ConsistencyDual}
	Let $\mathcal{T}$ be a compact subset of $(0,1)$, $\delta, \eta \in (0, 1)$ and $c_0 > 1$. Set $\bar{c} := (c_0 + 1)/(c_0 -1)$. Suppose that Assumptions~\ref{assumption:SubGaussianity}--\ref{assumption:GrowthCondition} and~\ref{assumption:RelativeConsistencyDensity}--\ref{assumption:GrowthConditionDual} hold. There exists an absolute constant $c_3 > c_0$ such that for all $C_\gamma \geq c_3$ and
	\begin{align}\label{eq:corollary:theorem:ConsistencyDual-00}
		\gamma := C_\gamma \frac{C_3\bar{f} \kappa_2(\bar{c})}{\kappa_2(\infty)}\left(  \sqrt{\log(np/\delta) } + \sqrt{n} r_f\right) \|z\|_2\sqrt{n},
	\end{align}
	with probability at least $1- \eta - \delta$, 
	\begin{align*}
		&\sup_{\tau \in \mathcal{T}} \left\|\hat{v}_\gamma(\tau; z) - \tilde{v}(\tau;z)\right\|_2 \\
		&\lesssim C_3 \left( \frac{C_\gamma \bar{c} \bar{f}\|z\|_2}{\kappa_2(\infty)} \vee \frac{\|z\|_2^2}{C_\gamma\bar{c} \bar{f} \kappa_2(\infty)} \vee \frac{\|z\|_2}{\kappa_2(\infty)} \vee 1\right) \left(\sqrt{\frac{s_v(z) \log(np/\delta) + s_\theta \log(ep/s_\theta)}{n}} \vee  r_f \sqrt{s_v(z)} \vee r_a \vee \frac{r_a^2}{r_f} \right).
	\end{align*}
\end{corollary}
\begin{remark}
	Note that for $C_\gamma \geq c_3$ large enough any $\gamma > 0$ that satisfies eq.~\eqref{eq:corollary:theorem:ConsistencyDual-00} also satisfies inequality~\eqref{eq:BalancingWeights-Dual-00}.
\end{remark}

\subsection{Bahadur-type representation}\label{subsec:BahadurRepresentation}
In this section, we consider a specific estimator for the conditional density. Following~\cite{koenker2005quantile} we observe that $1/f_{Y|X}(X_i'\theta_0(\tau)|X_i) = \frac{d}{d\tau} Q(\tau; X_i)$. Thus, we estimate the conditional densities $f_{Y|X}(X_i'\theta_0(\tau)|X_i)$, $1\leq i \leq n$, by
\begin{align}\label{eq:subsec:BahadurRepresentation-fhat}
	\hat{f}_i(\tau) := \frac{2h}{X_i'\hat{\theta}_\lambda(\tau + h) - X_i'\hat{\theta}_\lambda(\tau - h) }, \quad{} \forall \tau \in \mathcal{T}, \quad{}  1\leq i \leq n,
\end{align}
where $h > 0$ is a bandwidth parameter. This estimator satisfies the following relative consistency result:
\begin{lemma}\label{lemma:RelativeConsistencyDensity}
	Let $\mathcal{T}$ be a compact subset of $(0,1)$, $\delta \in (0,1)$. Let $\{\hat{f}_i(\tau)\}_{i=1}^n$ be as defined in eq.~\eqref{eq:subsec:BahadurRepresentation-fhat}. Set $\bar{c} := (c_0 + 1)/(c_0 -1)$, let $r_\theta > 0$ be as in eq.~\eqref{eq:theorem:Consistency-r}, and let $\lambda > 0$ satisfy eq.~\eqref{eq:lemma:RestrictedConeProp-QRVector-1}and define
	\begin{align}\label{eq:lemmaRelativeConsistencyDensity-rf}
		r_f : = \bar{f} \sqrt{n} h^{-1}r_\theta^2 + \bar{f} C_Q h^2.
	\end{align}
	Suppose that Assumptions~\ref{assumption:SubGaussianity}--\ref{assumption:GrowthCondition},~\ref{assumption:BoundedDensity}--\ref{assumption:DiffCQF}, and Assumption~\ref{assumption:RestrictedIdentifiability} with $(\vartheta, \varrho) = (\bar{c}, r_\theta)$ hold. If $r_f = o(1)$, then, with probability at least $1 - \delta$,
	\begin{align*}
		\sup_{\tau \in \mathcal{T}} \max_{1 \leq i \leq n}\left|\frac{\hat{f}_i(\tau)}{f_{Y|X}(X_i'\theta_0(\tau)|X_i)} - 1\right| \lesssim r_f.
	\end{align*}
\end{lemma}

\begin{theorem}[Bahadur-type representation]\label{theorem:BahadurTypeRep}
	Let $\mathcal{T}$ be a compact subset of $(0,1)$, $\delta \in (0, 1)$, $c_0 > 1$, and $h, \lambda, \gamma > 0$. Let $r_\theta, r_v, r_f > 0$ be as in eq.~\eqref{eq:theorem:Consistency-r},~\eqref{eq:theorem:ConsistencyDual-r} and~\eqref{eq:lemmaRelativeConsistencyDensity-rf}, respectively. Set $\bar{c} := (c_0 + 1)/(c_0 -1)$, $\hat{s}_\lambda := \sup_{\tau \in \mathcal{T}}\|\hat{\theta}_\lambda(\tau)\|_0$, and
	\begin{align}
		\hat{r}_B &: = \sqrt{\frac{ (s_v(z) + s_\theta + \hat{s}_\lambda)\log(L_fL_\theta np/\delta) }{n}}. \label{eq:theorem:BahadurTypeRep-rB}
	\end{align}
	Suppose that Assumptions~\ref{assumption:SubGaussianity}--\ref{assumption:GrowthCondition},~\ref{assumption:BoundedDensity}--\ref{assumption:DiffCQF}, and Assumption~\ref{assumption:RestrictedIdentifiability} with $(\vartheta, \varrho) = (\bar{c}, r_\theta)$ hold. Further, suppose that $\lambda, \gamma, \mu \geq 0$ satisfy eq.~\eqref{eq:lemma:RestrictedConeProp-QRVector-1},~\eqref{eq:BalancingWeights-Dual-00},~\eqref{eq:lemma:RestrictedConeProp-Dual-1}, and~\eqref{eq:lemma:RestrictedConeProp-Dual-11}, respectively. If $\big(s_v(z) + s_\theta\big)^2 \log^2(np/\delta) \log^2(n) = o(n)$ and $r_f \vee r_a = o(1)$, then, for all $\tau \in \mathcal{T}$ and $z \in \mathbb{R}^p$, with probability at least $1 - \delta$,
	\begin{align*}
		\widehat{Q}_{\lambda, \gamma}(\tau;z) - Q(\tau; z) = - \frac{1}{2n}\sum_{i=1}^n f_{Y|X}(X_i'\theta_0(\tau)|X_i)\big(\tau - \mathbf{1}\{Y_i \leq X_i'\theta_0(\tau)\}\big)X_i'v_0(\tau; z) + e_n(\tau; z),
	\end{align*}
	and 
	\begin{align*}
		\sup_{\tau \in \mathcal{T}} |e_n(\tau; z)| & \lesssim C_4 r_\theta r_v + C_4 \left( r_v  + \frac{2\bar{c}\|z\|_2}{\kappa_2(\infty)}  \right) \left(\hat{r}_B \log n + \sqrt{r_\theta} \right) \hat{r}_B  \\
		&\quad{} + C_5 \left(r_v  + \frac{2\bar{c}\|z\|_2}{\kappa_2(\infty)}  \right)(C_Q h^2 + h^{-1}r_\theta) \hat{r}_B,
	\end{align*}
	where $C_4, C_5 \geq 1$ are defined in eq.~\eqref{eq:theorem:BahadurTypeRep-16}.
\end{theorem}

\begin{remark}
	The upper bound on the remainder term $e_n(\tau; z)$ is complicated; however, it has a simple explanation: By the duality result of Lemma~\ref{lemma:CharacterizationBalancingEstimator} the rank-score balanced estimator can be formulated as
	\begin{align*}
		\widehat{Q}_{\lambda, \gamma}(\tau;z) =   z'\hat{\theta}_\lambda(\tau) - \frac{1}{2n}\sum_{i=1}^n\hat{f}_i(\tau)(\tau - 1\{Y_i \leq X_i'\hat{\theta}_\lambda(\tau)\})X_i'\hat{v}_\gamma(\tau; z).
	\end{align*}
	We note that the bias-correction term of the rank-score balanced estimator is a function of $\hat{v}_\gamma(\tau;z)$, $\{\hat{f}_i(\tau)\}_{i=1}^n$, and $\hat{\theta}_\lambda(\tau)$. To establish the Bahadur-type representation we therefore expand the bias-correction term by adding and subtracting the corresponding population versions of these parameters. We then derive individual bounds on the resulting seven terms. With high probability, each of these seven terms can be upper bounded by the supremum of an empirical process indexed by some function class with vanishing $L^2(P)$-diameter (due to the consistency results of Theorems~\ref{theorem:Consistency} and~\ref{theorem:ConsistencyDual}). The bound on the remainder term $e_n(\tau; z)$ then follows from a careful application of empirical process techniques. The specific choice of the estimates $\{\hat{f}_i(\tau)\}_{i=1}^n$ is crucial to the proof. We leave establishing a Bahadur-type representation based on other estimators of the conditional density for future research.
\end{remark}

In Theorem~\ref{theorem:BahadurTypeRep} the upper bound on the remainder term $e_n(\tau; z)$ is a random variable: It depends on the number of non-zero coefficients of the estimated quantile regression vector, i.e. $\hat{s}_\lambda := \sup_{\tau \in \mathcal{T}}\|\hat{\theta}_\lambda(\tau)\|_0$. We know from Theorem~\ref{theorem:EmpiricalSparsity} that for $\lambda > 0$ large enough, $\hat{s}_\lambda \lesssim s_\theta$. This leads to several simplifications which are the content of the following corollary:

\begin{corollary}\label{corollary:theorem:BahadurTypeRep}
	Let $\mathcal{T}$ be a compact subset of $(0,1)$, $\delta \in (0, 1)$, and $c_0 > 1$. Set $\bar{c} := (c_0 + 1)/(c_0 -1)$ and
	\begin{align}\label{eq:corollary:theorem:BahadurTypeRep-rB}
		\bar{r}_B : = \sqrt{\frac{(s_v(z) + s_\theta)\log(np/\delta)}{n}}.
	\end{align}
	Let $\lambda, \gamma > 0$ be as in eq.~\eqref{eq:theorem:EmpiricalSparsity-0} and eq.~\eqref{eq:corollary:theorem:ConsistencyDual-00}, respectively. Let $\{\hat{f}_i(\tau)\}_{i=1}^n$ be as defined in eq.~\eqref{eq:subsec:BahadurRepresentation-fhat}. Suppose that Assumptions~\ref{assumption:SubGaussianity}--\ref{assumption:GrowthCondition},~\ref{assumption:BoundedDensity}--\ref{assumption:DiffCQF}, and Assumption~\ref{assumption:RestrictedIdentifiability} with $(\vartheta, \varrho) = (\bar{c}, \bar{r}_B)$ hold. If $(s_v(z) + s_\theta)^2\log^2( np/\delta)\log^2(n) = o(nh^2)$, $h^2 s_v(z) = o(1)$, and $r_a^2 = O(\sqrt{n} h^{-1} \bar{r}_B^2 + h^2)$, then, for all $\tau \in \mathcal{T}$ and $z \in \mathbb{R}^p$, with probability at least $1 - \delta$,
	\begin{align*}
		\widehat{Q}_{\lambda, \gamma}(\tau;z) - Q(\tau; z) = - \frac{1}{2n}\sum_{i=1}^n f_{Y|X}(X_i'\theta_0(\tau)|X_i)\big(\tau - \mathbf{1}\{Y_i \leq X_i'\theta_0(\tau)\}\big)X_i'v_0(\tau; z) + e_n(\tau; z),
	\end{align*}
	and 
	\begin{align*}
		\sup_{\tau \in \mathcal{T}} |e_n(\tau; z)| &\lesssim C_6 \left( \bar{r}_B^{3/2} (\log n)^{3/4} +  h^2 \bar{r}_B + h^{-1}\bar{r}_B^2\right)
	\end{align*}
	where $C_6 > 0$ is a constant depending on $C_4, C_5, c_0, \bar{f}, L_f, L_\theta, C_Q, \kappa_2(\infty), \varphi_{\max}(p), \|z\|_2$ only.
\end{corollary}

\subsection{Weak convergence}\label{subsec:WeakConvergence-Appendix}
\begin{lemma}[Asymptotic equicontinuity of the leading term in the Bahadur-type representation]\label{lemma:AsymptoticEquicontinuity}
	Let $\mathcal{T}$ be a compact subset of $(0,1)$ and $c_0 > 1$. Set $\bar{c} := (c_0 + 1)/(c_0 -1)$ and $\bar{r}_B : = \sqrt{(s_v(z) + s_\theta)\log(np)/n}$. Suppose that Assumptions~\ref{assumption:SubGaussianity}--\ref{assumption:GrowthCondition},~\ref{assumption:GrowthConditionDual},~\ref{assumption:SparsityDual}, and~\ref{assumption:DiffCQF} hold. Consider
	\begin{align*}
		\mathcal{G} = \{g : \mathbb{R}^{p+1} \rightarrow \mathbb{R}: g(X,Y) = f_{Y|X}(X'\theta_0(\tau)|X)(\tau - 1\{Y \leq X'\theta_0(\tau)\})X'v_0(\tau; z), \:\: \tau \in \mathcal{T} \},
	\end{align*}
	and $\mathcal{G}_\xi = \{ g_\tau - h_{\tau'}: \: g_\tau,h_{\tau'} \in \mathcal{G}, \: |\tau - \tau'| \leq \xi, \: \tau, \tau' \in \mathcal{T}\}$. If $\varphi_{\max}(p) = O(1)$ and $\|z\|_2 \kappa_2^{-1}(\infty) = O(1)$, then the following holds true:
	\begin{itemize}
		\item[(i)] $\mathcal{G}$ is totally bounded with respect to the standard deviation metric;
		\item[(ii)] The process $\mathbb{G}_n(g)$, $g \in \mathcal{G}$ is asymptotic equicontinuous, i.e. $\lim_{\xi \downarrow 0} \lim_{n \rightarrow \infty} \mathbb{P}\left\{\|\mathbb{G}_n\|_{\mathcal{G}_{\xi}} > \varepsilon\right\} = 0$ for all $\varepsilon > 0$.
	\end{itemize}
\end{lemma}

The next result is the main theorem of this section.

\begin{theorem}[Weak convergence of the rank-score balanced estimator]\label{theorem:WeakConvergenceRankScoreBalancedEstimator}
	Let $\mathcal{T}$ be a compact subset of $(0,1)$ and $c_0 > 1$. Set $\bar{c} := (c_0 + 1)/(c_0 -1)$ and $\bar{r}_B : = \sqrt{(s_v(z) + s_\theta)\log(np)/n}$. Let $\lambda, \gamma > 0$ be as in eq.~\eqref{eq:theorem:EmpiricalSparsity-0} and eq.~\eqref{eq:corollary:theorem:ConsistencyDual-00}, respectively. Let $\{\hat{f}_i(\tau)\}_{i=1}^n$ be as defined in eq.~\eqref{eq:subsec:BahadurRepresentation-fhat}. Suppose that Assumptions~\ref{assumption:SubGaussianity}--\ref{assumption:GrowthCondition},~\ref{assumption:BoundedDensity}--\ref{assumption:DiffCQF}, and Assumption~\ref{assumption:RestrictedIdentifiability} with $(\vartheta, \varrho) = (\bar{c}, \bar{r}_B)$ hold. Moreover, suppose that $(s_v(z) + s_\theta)^3\log^3(np)\log^3(n) = o(nh^2)$, $h^2 s_v(z) = o(1)$, $r_a^2 = O(\sqrt{n} h^{-1} \bar{r}_B^2 + h^2), \varphi_{\max}(p) = O(1)$, $\kappa_1(\bar{c}) = O(1)$, and $\|z\|_2 \kappa_2^{-1}(\infty) = O(1)$. If the following limit
	\begin{align*}
		H(\tau_1, \tau_2; z) := \lim_{n\rightarrow\infty} \frac{\tau_1 \wedge \tau_2 - \tau_1 \tau_2}{4} v_0'(\tau_1; z)\mathbb{E}[f_{Y|X}(X'\theta_0(\tau_1)|X)f_{Y|X}(X'\theta_0(\tau_2)|X) XX']v_0(\tau_2; z)
	\end{align*}
	exists for all $\tau_1, \tau_2 \in \mathcal{T}$, then
	\begin{align*}
		\sqrt{n} \left(\widehat{Q}_{\lambda, \gamma}(\cdot;z) - Q(\cdot; z) \right) \leadsto \mathbb{G}(\cdot\:; z) \quad{} \mathrm{in} \quad{} \ell^\infty(\mathcal{T}),
	\end{align*}
	where $\mathbb{G}(\cdot\:;z)$ is a centered Gaussian process with covariance function $(\tau_1, \tau_2) \mapsto H(\tau_1, \tau_2; z)$.
\end{theorem}

\begin{lemma}\label{lemma:ConsistencyCovarianceFunction-2}
	Recall the setup of Corollary~\ref{corollary:theorem:BahadurTypeRep}. With probability at least $1 - \delta$, 
	\begin{align*}
		\sup_{\tau_1, \tau_2 \in \mathcal{T}}\left|\widehat{H}(\tau_1, \tau_2;z)  - H(\tau_1, \tau_2;z) \right| &\lesssim C_7 \left(\frac{\|z\|_2^2}{\kappa_2^2(\infty)} \vee 1\right)\Big(\bar{r}_B + \sqrt{n} \sqrt{s_v(z)} h^{-1} \bar{r}_B^2  + \sqrt{s_v(z)}h^2\Big),
	\end{align*}
	where $C_7 > 0$ is a constant depending on $c_0, \bar{f}, L_f, L_\theta,  C_Q, \kappa_2(\infty), \varphi_{\max}(p)$ only, and
	\begin{align*}
		\widehat{H}(\tau_1, \tau_2;z) = \hat{v}_\gamma'(\tau_1;z) \left(\frac{1}{4}\sum_{i=1}^n \hat{f}_i(\tau_1)\hat{f}_i(\tau_2) X_iX_i'\right)\hat{v}_\gamma(\tau_2;z).
	\end{align*}
\end{lemma}

\subsection{Auxiliary results}\label{subsec:AuxResults-RankScore}

\begin{lemma}\label{lemma:ConeProperties}
	Let $\alpha \geq 0, \beta > 0, \gamma \geq 1$. Let $v \in  C^p_2\left(T, \alpha\right)$ and  $\tilde{v} := t (v_T, 0)' + (0, v_{T^c})'$ with $|t| \geq \beta$. Then, $\tilde{v} \in C^p_2\left(T, \alpha/\beta\right)$. Moreover, if $v \in  C^p_2\left(T, \alpha\right)$, then $v \in  C^p_2\left(T, \alpha\gamma\right)$.
\end{lemma}

\begin{lemma}\label{lemma:InducedRestrictedConeProp-Dual}
	Let $0 \leq r_a \leq 1/4$ and $M \in \mathbb{R}^{p \times r}$, $r \geq 1$. Recall that $T_v(\tau;z) = \mathrm{support}\big(\tilde{v}(\tau; z)\big)$. The following holds true:
	\begin{itemize}
		\item[(i)] $v_0(\tau; z) \in  C^p_2\left(T_v(\tau;z), \frac{r_a}{1- r_a}\right)$;
		\item[(ii)] $\left\|M'\big(v_0(\tau;z) - \tilde{v}(\tau;z)\big) \right\|_2 \leq \sup_{u \in C^p_2\big(T_v(\tau; z), 1\big) \cap B_2^p\big(0, 8r_a\|v_0(\tau)\|_2 \big)} \|M'u\|_2 $.
	\end{itemize}
\end{lemma}

\begin{lemma}[Restricted cone and cross-polytope property]\label{lemma:RestrictedConeProp-Dual}
	For $c_0 > 1$ set $\bar{c} = \frac{c_0 + 1}{c_0 -1}$. Suppose that Assumption~\ref{assumption:SparsityDual} holds,
	\begin{align}\label{eq:lemma:RestrictedConeProp-Dual-1}
		\gamma c_0^{-1} \geq \sup_{\tau \in \mathcal{T}} \left\|\frac{1}{2}\sum_{i=1}^n\hat{f}_i^2(\tau) X_iX_i'v_0(\tau; z) + nz \right\|_\infty,
	\end{align}
	and
	\begin{align}\label{eq:lemma:RestrictedConeProp-Dual-11}
		\mu c_0^{-1} \geq \sup_{\tau \in \mathcal{T}} \sup_{u \in  C^p_2(T_v(\tau; z), 1) \cap  B^p_2(0,r_a) } \left|\frac{1}{2}\sum_{i=1}^n\hat{f}_i^2(\tau) (X_i'u)^2\right|.
	\end{align}
	Then, for all $\tau \in \mathcal{T}$,
	\begin{align}\label{eq:lemma:RestrictedConeProp-Dual-111}
		\bar{c}\sum_{k \in T_v(\tau;z)} \big|\hat{v}_{k}(\tau) - \tilde{v}_k(\tau)\big| + \frac{\bar{c}\|z\|_2^2}{\kappa_2^2(\infty)} \frac{\mu}{\gamma} \geq \sum_{k \in T_v^c(\tau;z)}\big|\hat{v}_{k}(\tau)\big|,
	\end{align}
	and
	\begin{align}\label{eq:lemma:RestrictedConeProp-Dual-1111}
		\hat{v}_\gamma(\tau;z) - \tilde{v}(\tau; z) \in C^p_1\big(T_v(\tau; z), 2 \bar{c}\big) \cup B^p_1\left(0,  \frac{2\bar{c}\|z\|_2^2}{\kappa_2^2(\infty)} \frac{\mu}{\gamma} \right).
	\end{align}
\end{lemma}
\begin{remark}
	If $r_a = 0$, i.e. $\tilde{v}(\tau,z) = v_0(\tau; z)$, we can take $\mu = 0$ and the lemma reduces to the classical restricted cone property~\citep[e.g.][]{bickel2009simultaneous, belloni2011L1penalized, belloni2013least}. Note that inequality~\eqref{eq:lemma:RestrictedConeProp-Dual-111} implies that $\hat{v}_\gamma(\tau;z) - \tilde{v}(\tau;z)$ lies in a so-called star-shaped set. However, since we are unaware of useful metric entropy bounds for star-shaped sets, we will use relation~\eqref{eq:lemma:RestrictedConeProp-Dual-1111}.
\end{remark}

\begin{lemma}[Lipschitz continuity of $\tau \mapsto v_0(\tau; z)$]\label{lemma:LipschitzDual}
	Let $\mathcal{T}$ be a compact subset of $(0,1)$. Suppose that Assumptions~\ref{assumption:LipschitzQRVector},~\ref{assumption:LipschitzDensity},~\ref{assumption:BoundedDensity}, and~\ref{assumption:SparsityDual} hold. There exists an absolute constant $C_v \geq 1$ such that for all $s, \tau \in \mathcal{T}$,
	\begin{align*}
		\left\|v_0(s; z) - v_0(\tau;z)\right\|_2 \leq C_v\bar{f}L_fL_\theta\varphi_{\max}^{1/2}(2s_\theta) \frac{\varphi_{\max}(p)}{\kappa_2(\infty)} \|z\|_2|s - \tau|.
	\end{align*}
\end{lemma}

\begin{lemma}[Maxima of sum of block multi-convex functions]\label{lemma:MaximaSumMulticonvexFunctions}
	Let $f_1, \ldots, f_N : \mathcal{X}_1 \times \cdots \times \mathcal{X}_K \rightarrow \mathbb{R}$ be block multi-convex functions. Then,
	\begin{align*}
		\sup_{x_j \in \mathrm{conv}(\mathcal{X}_j), 1\leq j \leq K} \sum_{i=1}^N f_i(x_1, \ldots, x_K) = \sup_{x_j \in \mathcal{X}_j, 1\leq j \leq K} \sum_{i=1}^N f_i(x_1, \ldots, x_K).
	\end{align*}
	Moreover, the identity remains true if $\sum_{i=1}^Nf_i$ is replaced by $\left|\sum_{i=1}^Nf_i\right|$.
\end{lemma}
\begin{remark}
	This is a generalization of Lemma~\ref{lemma:MaximaBiconvexFunction}.
\end{remark}

\begin{lemma}\label{lemma:GradientConeSquare} 
	Let $\mathcal{T}$ be a compact subset of $(0,1)$. Let $\delta \in (0,1)$ be arbitrary and $\vartheta_k \in [0,\infty]$, $q_k \geq 1$, $J_k(\tau) \subseteq \{1, \ldots, p\}$ for $\tau \in \mathcal{T}$, $s_k = \sup_{\tau \in \mathcal{T}}\mathrm{card}(J_k(\tau))$, $p \geq s_k + 2$, and $k \in \{1,2\}$.  Let $\{(X_i, Y_i, \xi_i)\}_{i=1}^n$ be a sequence of i.i.d. random vectors. Suppose that Assumptions~\ref{assumption:SubGaussianity}--\ref{assumption:SparsityQRVector} hold and the $\xi_i$'s are i.i.d. standard Gaussian random variables. Define 
	\begin{align*}
		\mathcal{G} = \big\{ g : \mathbb{R}^{p+2} \rightarrow \mathbb{R}: g(X, Y, \xi) = \xi \left(\tau - \mathbf{1}\big\{Y \leq X'\theta_0(\tau) \}\right)(X'u_1) (X'u_2),\\
		u_k \in  C^p_{q_k}(J_k(\tau), \vartheta_k) \cap B^p(0, 1), \: k \in \{1, 2\}, \: \tau \in \mathcal{T}\big\}.
	\end{align*} 
	With probability at least $1-\delta$,
	\begin{align*}
		\|\mathbb{G}_n\|_{\mathcal{G}} &\lesssim (2 + \vartheta_1) (2 + \vartheta_2) \varphi_{\max}^{1/2}(s_1)\varphi_{\max}^{1/2}(s_2) \sqrt{s_1 \log (ep/s_1)  + s_2 \log (ep/s_2) + \log(1/\delta)} \\
		&\quad{} \times \sqrt{1 + \pi_{n,1}^2(s_1 \log (ep/s_1) + s_2 \log (ep/s_2) + \log(1/\delta))},
	\end{align*}
	where $\pi_{n,1/3}^2(z) = \sqrt{z/n} + z^3/n$ with $z \geq 0$.
\end{lemma}

\begin{lemma}\label{lemma:LocalizedLossDual} Let $\mathcal{T}$ be a compact subset of $(0,1)$. Let $\delta \in (0,1)$ be arbitrary and $\vartheta_k \in [0,\infty]$, $q_k \geq 1$, $J_k(\tau) \subseteq \{1, \ldots, p\}$ for $\tau \in \mathcal{T}$, $s_k = \sup_{\tau \in \mathcal{T}}\mathrm{card}(J_k(\tau))$, $p \geq s_k + 2$, and $k \in \{1,2\}$. Suppose that Assumptions~\ref{assumption:SubGaussianity}--\ref{assumption:LipschitzDensity}, and~\ref{assumption:BoundedDensity} hold. Define
	\begin{align*}
		\mathcal{G} = \big\{ g : \mathbb{R}^p \rightarrow \mathbb{R}: g(X) = f_{Y|X}^2(X'\theta_0(\tau)|X) (X'u_1) (X'u_2),\\
		\:u_k \in C^p_{q_k}(J_k(\tau), \vartheta_k) \cap B^p(0, 1), \: k \in \{1, 2\}, \: \tau \in \mathcal{T}\big\}.
	\end{align*}
	The following holds true:
	\begin{itemize}
		\item[(i)] With probability at least $1-\delta$,
		\begin{align*}
			\|\mathbb{G}_n\|_{\mathcal{G}} &\lesssim (2 + \vartheta_1)(2+\vartheta_2) \bar{f}^2 \varphi_{\max}^{1/2}(s_1)\varphi_{\max}^{1/2}(s_2) (1 + \varphi_{\max}^{1/2}(2s_\theta)) \psi_n\big(t_{s_\theta, s_1, s_2, n, \delta }\big),
		\end{align*}
		where $t_{s_1, s_2, s_\theta, n, \delta } = s_1 \log(ep/s_1) + s_2 \log(ep/s_2) + s_\theta \log(ep/s_\theta) + \log (nL_f L_\theta/\delta)$ and $\psi_n(z) = \sqrt{z}\big(1 + n^{-1/2} \sqrt{z} + n^{-3/2} z^{3/2}\big)$ for $z \geq 0$.
		\item[(ii)] Statement (i) holds also for the function class $|\mathcal{G}| := \left\{h : \mathbb{R}^{p + 1} \rightarrow \mathbb{R}:\:  \exists g \in \mathcal{G} : h = |g| \right\}$.
		\item[(iii)] Statements (i) and (ii) hold with $\vartheta_1 = 1$ also for the function class
		\begin{align*}
			\mathcal{G} &= \big\{ g : \mathbb{R}^p \rightarrow \mathbb{R}: g(X) = f_{Y|X}^2(X'\theta_0(\tau)|X) (X'v) (X'u),\\
			&\quad{}\quad{} v \in \mathbb{R}^p, \: \|v\|_0 \leq s, \: \|v\|_2 \leq 1, \:u \in C^p_{q}(J_2(\tau), \vartheta_2) \cap B^p(0, 1), \: \tau \in \mathcal{T}\big\}.
		\end{align*}
	\end{itemize}
	
\end{lemma}

\begin{lemma}\label{lemma:GradientCone-2}
	Let $\mathcal{T}$ be a compact subset of $(0,1)$. Let $\delta \in (0,1)$ be arbitrary, $\vartheta \in [0,\infty]$, $q \geq 1$, $J(\tau) \subseteq \{1, \ldots, p\}$ for $\tau \in \mathcal{T}$, $s = \sup_{\tau \in \mathcal{T}}\mathrm{card}(J(\tau))$, and $p \geq s + 2$. Suppose that Assumptions~\ref{assumption:SubGaussianity}--\ref{assumption:LipschitzDensity}, and~\ref{assumption:BoundedDensity} hold. Define 
	\begin{align*}
		\mathcal{G} = \left\{g: \mathbb{R}^{p+1} \rightarrow \mathbb{R}: g(X, Y) = f_{Y|X}(X'\theta_0(\tau)|X)\left(\tau - \mathbf{1}\big\{Y \leq X'\theta_0(\tau) \}\right)X'v, \right.\\
		\left. v \in C^p_q(J(\tau), \vartheta) \cap B^p(0,1), \:\tau \in \mathcal{T}\right\}.
	\end{align*}
	The following holds true:
	\begin{itemize}
		\item[(i)]	With probability at least $1-\delta$,
		\begin{align*}
			\|\mathbb{G}_n\|_{\mathcal{G}} \lesssim (2 + \vartheta) \bar{f}\varphi_{\max}^{1/2}(s)\big(1 + \varphi_{\max}^{1/2}(2s_\theta)\big) \psi_n\big(t_{s, s_\theta, n, \delta }\big),
		\end{align*}
		where $t_{s, s_\theta, n, \delta } = s \log(ep/s) + s_\theta \log(ep/s_\theta) + \log (nL_f L_\theta/\delta)$ and $\psi_n(z) = \sqrt{z}\big(1 + n^{-1/2} \sqrt{z} + n^{-3/2} z^{3/2}\big)$ for $z \geq 0$.
		\item[(ii)] Statement (i) holds with $\vartheta = 1$ also for the function class
		\begin{align*}
			\mathcal{G} = \left\{g: \mathbb{R}^{p+1} \rightarrow \mathbb{R}: g(X, Y) = f_{Y|X}(X'\theta_0(\tau)|X)\left(\tau - \mathbf{1}\big\{Y \leq X'\theta_0(\tau) \}\right)X'v, \right.\\
			\left.  v \in \mathbb{R}^p, \: \|v\|_0 \leq s, \: \|v\|_2 \leq 1, \:\tau \in \mathcal{T}\right\}.
		\end{align*}
	\end{itemize}
	
\end{lemma}

\begin{lemma}\label{lemma:GradientCone-3}
	Let $\mathcal{T}$ be a compact subset of $(0,1)$. Let $\delta \in (0,1)$ be arbitrary and $\vartheta_k \in [0,\infty]$, $q_k \geq 1$, $J_k(\tau) \subseteq \{1, \ldots, p\}$ for $\tau \in \mathcal{T}$, $s_k = \sup_{\tau \in \mathcal{T}}\mathrm{card}(J_k(\tau))$, $p \geq s_k + 2$, and $k \in \{1,2\}$. Suppose that Assumptions~\ref{assumption:SubGaussianity}--\ref{assumption:LipschitzDensity}, and~\ref{assumption:BoundedDensity} hold. Define 
	\begin{align*}
		\mathcal{G} = \left\{g: \mathbb{R}^{p+1} \rightarrow \mathbb{R}: g(X, Y) = f_{Y|X}^2(X'\theta_0(\tau)|X)\left(\tau - \mathbf{1}\big\{Y \leq X'\theta_0(\tau) \}\right) (X'u_1)(X'u_2), \right.\\
		\left. u_k \in C^p_{q_k}(J_k(\tau), \vartheta_k) \cap B^p(0,1), \: k \in \{1, 2\},\:\tau \in \mathcal{T}\right\}.
	\end{align*}
	The following holds true:
	\begin{itemize}
		\item[(i)] 	With probability at least $1-\delta$,
		\begin{align*}
			\|\mathbb{G}_n\|_{\mathcal{G}} \lesssim (2 + \vartheta_1) (2 + \vartheta_2) \bar{f}^2 \varphi_{\max}^{1/2}(s_1)\varphi_{\max}^{1/2}(s_2)\big(1 + \varphi_{\max}^{1/2}(2s_\theta)\big) \psi_n\big(t_{s_1, s_2, s_\theta, n, \delta }\big),
		\end{align*}
		where $t_{s_1, s_2, s_\theta, n, \delta } = s_1 \log(ep/s_1) + s_2\log(ep/s_2) + s_\theta \log(ep/s_\theta) + \log (nL_f L_\theta/\delta)$ and $\psi_n(z) = \sqrt{z}\big(1 + n^{-1/2} \sqrt{z} + n^{-3/2} z^2\big)$ for $z \geq 0$.
		\item[(ii)] Statement (i) holds with $\vartheta_1 = 1$ also for the function class
		\begin{align*}
			\mathcal{G} = \left\{g: \mathbb{R}^{p+1} \rightarrow \mathbb{R}: g(X, Y) = f_{Y|X}^2(X'\theta_0(\tau)|X)\left(\tau - \mathbf{1}\big\{Y \leq X'\theta_0(\tau) \}\right)(X'v)(X'u), \right.\\
			\left.  v \in \mathbb{R}^p, \: \|v\|_0 \leq s, \: \|v\|_2 \leq 1, \: u \in C^p_{q}(J(\tau), \vartheta) \cap B^p(0,1), \: \tau \in \mathcal{T}\right\}.
		\end{align*}
	\end{itemize}
\end{lemma}

\begin{lemma}\label{lemma:LocalizedRankScoresCone} 
	Let $\mathcal{T}$ be a compact subset of $(0,1)$. Let $r_0 > 0$, $\delta \in (0,1)$ be arbitrary, $\vartheta \in [0,\infty]$, $q\geq 1$, $J(\tau) \subseteq \{1, \ldots, p\}$ for $\tau \in \mathcal{T}$, $s = \sup_{\tau \in \mathcal{T}}\mathrm{card}(J(\tau))$, and $p \geq s + 2$. Suppose that Assumptions~\ref{assumption:SubGaussianity}--\ref{assumption:LipschitzDensity} and~\ref{assumption:BoundedDensity} hold. Define 
	\begin{align*}
		\mathcal{G} = \left\{g: \mathbb{R}^{p+1} \rightarrow \mathbb{R}: g(X, Y) = f_{Y|X}(X'\theta_0(\tau)|X)\left(\mathbf{1}\big\{Y \leq X'\theta \big\} - \mathbf{1}\big\{Y \leq X'\theta_0(\tau) \}\right)X'v,  \: \theta \in \mathbb{R}^p, \right.\\
		\left. \|\theta\|_0 \leq n, \: \|\theta - \theta_0(\tau)\|_2 \leq r_0, \: v \in C^p_q(J(\tau), \vartheta) \cap B^p_2(0,1), \:\tau \in \mathcal{T}\right\}.
	\end{align*}
	The following holds true:
	\begin{itemize}
		\item[(i)] With probability at least $1-\delta$,
		\begin{align*}
			\forall g_{v,\theta, \tau} \in \mathcal{G} : \: |\mathbb{G}_n(g_{v, \theta, \tau})| &\lesssim  (2+ \vartheta)\bar{f}^{3/2}\varphi_{\max}^{1/2}(s) \big( 1 +  \varphi_{\max}^{1/2}(2s_\theta)\big)\big(1 + \varphi_{\max}^{1/2}(\|\theta\|_0 + s_\theta)\big)\\
			&\quad{} \times \Big(\upsilon_{r_0, n}\big(\|\theta\|_0 \log (1/r_0) \big) + \upsilon_{r_0, n}(t_{s, \|\theta\|_0, s_\theta, n, \delta}) \Big).
		\end{align*}
		where $t_{s, k, s_\theta, n, \delta} = s\log(ep/s) +  k\log (ep/k) + s_\theta \log (ep/s_\theta) + \log(L_fL_\theta n/\delta)$ and $\upsilon_{r_0, n}(z) = \sqrt{z}\big(\sqrt{r_0} + n^{-1/2}(\log n) \sqrt{z} + n^{-1} (\log n)^{3/2} z\big)$ for $z \geq 0$;
		\item[(ii)] Let $\mathcal{G}(m) = \left\{g_{v, \theta, \tau} \in \mathcal{G}:  \|\theta\|_0 \leq m\right\}$. With probability at least $1-\delta$,
		\begin{align*}
			\forall m \leq n \wedge p : \: \|\mathbb{G}_n\|_{\mathcal{G}(m)} &\lesssim  (2+ \vartheta)\bar{f}^{3/2}\varphi_{\max}^{1/2}(s) \big( 1 +  \varphi_{\max}^{1/2}(2s_\theta)\big)\big(1 + \varphi_{\max}^{1/2}(m + s_\theta)\big)\\
			&\quad{} \times  \Big(\upsilon_{r_0, n}\big(m \log (1/r_0) \big) + \upsilon_{r_0, n}(t_{s, m, s_\theta, n, \delta}) \Big),
		\end{align*}
		where $t_{s, m, s_\theta, n, \delta} = s\log (ep/s) + m \log (ep/m) + s_\theta\log (ep/s_\theta) + \log(L_fL_\theta n/\delta)$ and $\upsilon_{r_0, n}(z) = \sqrt{z}\big(\sqrt{r_0} + n^{-1/2}(\log n) \sqrt{z} + n^{-1} (\log n)^{3/2} z\big)$ for $z \geq 0$.
		\item[(iii)] Statements (i) and (ii) hold also for the function class $|\mathcal{G}| := \left\{h : \mathbb{R}^{p + 1} \rightarrow \mathbb{R}:\:  \exists g \in \mathcal{G} : h = |g| \right\}$.
		\item[(iv)] Statements (i), (ii), and (iii) hold with $\vartheta = 1$ also for the function class
		\begin{align*}
			\mathcal{G} = \left\{g: \mathbb{R}^{p+1} \rightarrow \mathbb{R}: g(X, Y) = f_{Y|X}(X'\theta_0(\tau)|X)\left(\mathbf{1}\big\{Y \leq X'\theta \big\} - \mathbf{1}\big\{Y \leq X'\theta_0(\tau) \}\right)X'v,  \: \theta \in \mathbb{R}^p, \right.\\
			\left. \|\theta\|_0 \leq n, \: \|\theta - \theta_0(\tau)\|_2 \leq r_0, \: \|v\|_0 \leq s, \: \|v\|_2 \leq 1, \:\tau \in \mathcal{T}\right\}.
		\end{align*}
	\end{itemize}	
\end{lemma}

\section{Results from empirical process theory}\label{sec:MaximalInequalities}
\subsection{Maximal and deviation inequalities}
In this section we collect maximal inequalities that we use throughout the proofs. 

\begin{definition}[Exponential Orlicz-norms ($\psi_\alpha$-norms)]
	For $\alpha > 0$ set $\psi_\alpha(x) = \exp(x^\alpha) -1$. We define the exponential Orlicz-norm of a real-valued random variable $\xi$ on $(\Omega, \mathcal{A}, \mathbb{P})$ as
	\begin{align*}
		\|\xi\|_{\psi_\alpha} := \inf\left\{\lambda > 0: \mathbb{E} \psi_\alpha(|\xi|/\lambda) \leq 1\right\},
	\end{align*}
	and the $\psi_\alpha$-norm of a real-valued $Q$-integrable function $f$ on $(S, \mathcal{S})$ as
	\begin{align*}
		\|f\|_{Q, \psi_\alpha} := \inf\left\{\lambda > 0: \int \psi_\alpha(|f|/\lambda) dQ \leq 1\right\}.
	\end{align*}
\end{definition}
\begin{remark}
	For $\alpha \in (0,1]$ the map $t \mapsto \psi_\alpha(t)$ is not convex; hence, $\|\cdot\|_{\psi_\alpha}$ and $\|\cdot\|_{Q,\psi_\alpha}$ are only quasinorms.
\end{remark}

\begin{theorem}[\citeauthor{giessing2020maximal}, \citeyear{giessing2020maximal}]\label{theorem:MaxInequalityBernsteinOrlicz}
	Let $\mathcal{F} \subset L_1(S, \mathcal{S}, P)$ and $\rho$ be a pseudo-metric on $\mathcal{F} \times \mathcal{F}$. For $\delta >0 $ define $\mathcal{F}_\delta = \{f - g : f,g \in \mathcal{F}, \rho(f,g) \leq \delta\}$. Suppose that for $\alpha > 0$ there exists a constant $K > 0$ such that for all $f,g \in \mathcal{F}$,
	\begin{align}\label{eq:theorem:MaxInequalityBernsteinOrlicz-1}
		\|(f - Pf) - (g - Pg)\|_{P, \psi_\alpha} \leq K \rho(f, g).
	\end{align}
	The following holds true:
	\begin{itemize}
		\item[(i)] If $\alpha \in (0,1]$, then for all events $A \in \mathcal{A}$ with $\mathbb{P}\{A\} > 0$,
		\begin{align*}
			\mathbb{E}\left[\|\mathbb{G}_n\|_{\mathcal{F}_\delta}  \mid A\right] \leq C_\alpha K \left[ \int_0^\delta \sqrt{ \log \left(\frac{N(\varepsilon, \mathcal{F}, \rho)}{\mathbb{P}\{A\}}\right)} d\varepsilon + n^{-1/2}\int_0^\delta \left(\log \left(\frac{N(\varepsilon, \mathcal{F}, \rho)}{\mathbb{P}\{A\}}\right) \right)^{1/\alpha} d\varepsilon\right],
		\end{align*}
		where $C_\alpha > 0$ is a constant depending on $\alpha$ only.
		
		\item[(ii)] If $\alpha \in (1,2]$, then for all events $A \in \mathcal{A}$ with $\mathbb{P}\{A\} > 0$,
		\begin{align*}
			\mathbb{E}\left[\|\mathbb{G}_n\|_{\mathcal{F}_\delta}  \mid A\right] \leq C_\alpha K \left[ \int_0^\delta \sqrt{ \log \left(\frac{N(\varepsilon, \mathcal{F}, \rho)}{\mathbb{P}\{A\} }\right)} d\varepsilon + n^{-1/2 + 1/\beta}\int_0^\delta \left(\log \left(\frac{N(\varepsilon, \mathcal{F}, \rho)}{\mathbb{P}\{A\} }\right) \right)^{1/\alpha} d\varepsilon\right],
		\end{align*}
		where $C_\alpha > 0$ is a constant depending on $\alpha$ only and $1/\alpha + 1/\beta = 1$.
	\end{itemize}
\end{theorem}

\begin{remark}\label{remark:theorem:MaxInequalityBernsteinOrlicz-Lipschitz}
	The Lipschitz-type condition~\eqref{eq:theorem:MaxInequalityBernsteinOrlicz-1} on the individual increments can be substituted by a Lipschitz condition on the uncentered individual increments, i.e. cases (i) and (ii) continue to hold true (with a larger constant) if there exist $K > 0$ such that for all $f,g \in \mathcal{F}$, $\|f - g\|_{P, \psi_\alpha} \leq K \rho(f, g)$. We establish this claim as a side result in the proof of Theorem~\ref{theorem:MaxInequalityBernsteinOrlicz}.
\end{remark}
\begin{remark}\label{remark:theorem:MaxInequalityBernsteinOrlicz-IndexSet}
	If $\mathcal{F} = \{f_t : t \in T\}$ and $ \|(f_s - Pf_s) - (f_t - Pf_t)\|_{P, \psi_\alpha} \leq K d(s, t)$ for all $f_s, f_t \in \mathcal{F}$ and a pseudo-metric $d$ on $T \times T$, then we can replace $N(\varepsilon, \mathcal{F}, \rho)$ by $N(\varepsilon, T, d)$.
\end{remark}

\begin{corollary}[\citeauthor{giessing2020maximal}, \citeyear{giessing2020maximal}]\label{corollary:DeviationInequalityBernsteinOrlicz}
	Recall the setup of Theorem~\ref{theorem:MaxInequalityBernsteinOrlicz}. The following holds true:
	\begin{itemize}
		\item[(i)] If $\alpha \in (0,1]$, then, for all $t \geq 0$, with probability at least $1 - e^{-t}$,
		\begin{align*}
			\|\mathbb{G}_n\|_{\mathcal{F}_\delta} &\leq C_\alpha K \left(\int_0^\delta \sqrt{ \log N(\varepsilon, \mathcal{F}, \rho)} d\varepsilon + n^{-1/2} \int_0^\delta \big(\log N(\varepsilon, \mathcal{F}, \rho)\big)^{1/\alpha} d\varepsilon \right)\\
			&\quad{}+   C_\alpha  K \delta \left(\sqrt{t} + n^{-1/2} t^{1/\alpha} \right),
		\end{align*}
		where $C_\alpha > 0$ is a constant depending on $\alpha$ only.
		
		\item[(ii)] If $\alpha \in (1,2]$, then, for all $t \geq 0$, with probability at least $1 - e^{-t}$,
		\begin{align*}
			\|\mathbb{G}_n\|_{\mathcal{F}_\delta} &\leq C_\alpha K \left(\int_0^\delta \sqrt{ \log N(\varepsilon, \mathcal{F}, \rho)} d\varepsilon + n^{-1/2+ 1/\beta} \int_0^\delta \big(\log N(\varepsilon, \mathcal{F}, \rho)\big)^{1/\alpha} d\varepsilon\right)\\
			&\quad{} + C_\alpha K \delta \left(\sqrt{t} + n^{-1/2+1/\beta} t^{1/\alpha} \right),
		\end{align*}
		where $C_\alpha > 0$ is a constant depending on $\alpha$ only and $1/\alpha + 1/\beta = 1$.
	\end{itemize}
\end{corollary}

\begin{lemma}[Theorem 5.2,~\citeauthor{chernozhukov2014gaussian},~\citeyear{chernozhukov2014gaussian}]\label{lemma:MaxInequalityChernozhukov2014-EntropyIntegral}
	Let $\mathcal{F} \subset L_2(S, \mathcal{S}, P)$ with envelope $F \in L_2(S, \mathcal{S}, P)$ and $0 \in \mathcal{F}$. Let $\sigma^2 > 0$ be any positive constant such that $\sup_{f \in \mathcal{F}}Pf^2 \leq \sigma^2 \leq \|F\|_{P,2}^2$. Set $\eta = \sigma/\|F\|_{P, 2}$. Define $M = \max_{1 \leq i \leq n}F(X_i)$. Then,
	\begin{align*}
		\mathbb{E}\|\mathbb{G}_n\|_\mathcal{F} \lesssim J(\eta, \mathcal{F}) \|F\|_{P,2} + \frac{\|M\|_2 J^2(\eta, \mathcal{F})}{\eta^2\sqrt{n}},
	\end{align*}
	where
	\begin{align*}
		J(\eta, \mathcal{F})  = \int_0^\eta \sup_Q \sqrt{\log N\big(\varepsilon\|F\|_{Q,2}, \mathcal{F}, L_2(Q)\big)} d\varepsilon,
	\end{align*}
	and the supremum is taken over all finitely discrete probability measures $Q$.
\end{lemma}

\begin{lemma}[Corollary 5.1,~\citeauthor{chernozhukov2014gaussian},~\citeyear{chernozhukov2014gaussian}]\label{lemma:MaxInequalityChernozhukov2014-VC-Class}
	Let $\mathcal{F} \subset L_2(S, \mathcal{S}, P)$ be a VC type class of functions with envelope $F \in L_2(S, \mathcal{S}, P)$ and $0 \in \mathcal{F}$. Let $\sigma^2 > 0$ be any positive constant such that $\sup_{f \in \mathcal{F}}Pf^2 \leq \sigma^2 \leq \|F\|_{P,2}^2$. Set $\eta = \sigma/\|F\|_{P, 2}$. Define $M = \max_{1 \leq i \leq n}F(X_i)$.
	\begin{align*}
		\mathbb{E}\|\mathbb{G}_n\|_\mathcal{F} \lesssim \sqrt{V \sigma^2  \log \left(\frac{A\|F\|_{P,2}}{\sigma}\right)} + \frac{V\|M\|_2}{\sqrt{n}}\log \left(\frac{A\|F\|_{P,2}}{\sigma}\right).
	\end{align*}
\end{lemma}

\begin{lemma}[Theorem 4,~\citeauthor{adamczak2008tail},~\citeyear{adamczak2008tail}]\label{lemma:MaxInequalityAdamczak2008}
	Let $\mathcal{F} \subset L_2(S, \mathcal{S}, P)$ with envelope $F \in L_2(S, \mathcal{S}, P)$ and $0 \in \mathcal{F}$. Let $\sigma^2 > 0$ be any positive constant such that $\sup_{f \in \mathcal{F}}Pf^2 \leq \sigma^2 \leq \|F\|_{P,2}^2$. Define $M = \max_{1 \leq i \leq n}F(X_i)$. For all $\alpha \in (0,1]$ and $t \geq 0$, with probability at least $1- e^{-t}$, 
	\begin{align*}
		\|\mathbb{G}_n\|_\mathcal{F} \lesssim \mathbb{E}\|\mathbb{G}_n\|_\mathcal{F} + \sigma \sqrt{t} + \|M\|_{\psi_\alpha} n^{-1/2}t^{1/\alpha}.
	\end{align*} 
\end{lemma}

\begin{lemma}[Lemma 7.1,~\citeauthor{kley2016Quantile},~\citeyear{kley2016Quantile}]\label{lemma:MaxInequalityKley2016}
	Let $\{X_t : t \in T \}$ be a separable stochastic process with $\|X_s - X_t\|_{P,\Psi} \leq C d(s,t)$ ($\|\cdot\|_{P, \Psi}$ denotes the $\Psi$-Orlicz with respect to measure $P$) for all $s,t$ satisfying $d(s,t) \geq \bar{\eta}/2 \geq 0$. Denote by $D(\epsilon,T,d)$ the packing number of the metric space $(T, d)$. Then, for any $\delta > 0$, $\eta \geq \bar{\eta}$, there exists a random variable $S_1$ and a constant $K < \infty$ such that
	\begin{align*}
		\sup_{d(s,t) \leq \delta} |X_s - X_t| \leq S_1 + 2 \sup_{d(s,t) \leq \bar{\eta}, t \in \widetilde{T}} |X_s - X_t|,
	\end{align*}
	and
	\begin{align*}
		\|S_1\|_{P,\Psi} \leq K \left[ \int_{\bar{\eta}/2}^\eta \Psi^{-1}\big(D(\epsilon,T,d)\big) d\epsilon + (\delta +2 \bar{\eta}) \Psi^{-1}\big(D^2(\eta,T, d)\big)\right],
	\end{align*}
	where the set $\widetilde{T}$ contains at most $D(\bar{\eta}, T, d)$ points. In particular, by Markov's inequality,
	\begin{align*}
		\mathbb{P}\left\{|S_1| > x\right\} \leq \left( \Psi\left(x\left[8K\left(\int_{\bar{\eta}/2}^\eta \Psi^{-1}\big(D(\epsilon,T, d)\big) d\epsilon + (\delta + 2 \bar{\eta})\Psi^{-1}\big(D^2(\eta,T, d)\big) \right) \right]^{-1} \right)\right)^{-1}.
	\end{align*}
\end{lemma}

\subsection{Auxiliary results}\label{subsec:AuxResults-EP}
In this section we collect technical auxiliary results that we use throughout the proofs.

\begin{corollary}\label{corollary:MaxInequalityBernsteinOrlicz-Gradient}
	Let $\mathcal{G} \subset L_2(S, \mathcal{S}, P)$ a finite collection of functions with $\mathrm{card}(\mathcal{G}) < \infty$. Let $\sigma^2 > 0$ be a positive constant such that $\sup_{g \in \mathcal{G}}Pg^2 \leq \sigma^2$. Let $\rho$ be a semi-metric on $\mathcal{G}^2 \times \mathcal{G}^2$ with $\mathcal{G}^2 = \{g^2: g \in \mathcal{G}\}$ and $\delta > 0$ be a positive constant such that $\sup_{g_1, g_2 \in \mathcal{G}^2} \rho(g_1, g_2) \leq \delta$. Let $\mathcal{H} \subset L_1(S, \mathcal{S}, P)$ be a collection of VC subgraph functions with VC-index $V(\mathcal{H})$ and absolute values bounded by one. Define $\mathcal{F} = \mathcal{G} \mathcal{H} = \{x \mapsto g(x)h(x): g \in \mathcal{G}, \: h \in \mathcal{H}\}$. If there exists a constant $K > 0$ such that for some $\alpha \in (0,1]$ and all $g_1, g_2 \in \mathcal{G}^2$,
	\begin{align}\label{eq:corollary:MaxInequalityBernsteinOrlicz-Gradient-0}
		\|(g_1 - Pg_1) - (g_2- Pg_2)\|_{P, \psi_\alpha} \leq K \rho(g_1, g_2),
	\end{align}
	then, there exist constants $c, c' >0$ (depending on $\alpha$) such that for all $t \geq 0$, with probability at least $1 - ce^{-c't}$,
	\begin{align*}
		\|\mathbb{G}_n\|_{\mathcal{F}} 
		\lesssim \left(\sqrt{V(\mathcal{H}) + \log \mathrm{card}(\mathcal{G})} + \sqrt{t} \right)  \left(\sqrt{\sigma^2 + K\delta \pi_{n,\alpha}^2\big(\log\mathrm{card}(\mathcal{G})\big)} + \sqrt{K \delta\pi_{n,\alpha}^2(t)}\right),
	\end{align*}
	where $\pi_{n,\alpha}^2(z) = \sqrt{z/n} + z^{1/\alpha}/n$ for $ z \geq  0$.
\end{corollary}
\begin{remark}
	A similar result holds for $\alpha \in (1, 2]$ with explicit constants $c = 3e, c' = 1$; however, we do not need such a result in this paper.
\end{remark}
\begin{remark}
	Note that the upper bound does not depend on the envelope of $\mathcal{F}$. The bound is therefore better by at least a $\log n$-factor than if we had used a combination of Lemma~\ref{lemma:MaxInequalityAdamczak2008} and Lemma~\ref{lemma:MaxInequalityChernozhukov2014-VC-Class}. Typically, the quantities $K\delta\pi_{n,\alpha}^2\big(\mathrm{card}(\mathcal{G})\big)$ and $K\delta\pi_{n,\alpha}^2(t)$ will be negligible compared to $\sigma^2$.
\end{remark}
\begin{remark}
	We use this result to derive sharp bounds on the gradient of the check loss (i.e. the loss function of the quantile regression program). To establish the connection between this result and quantile regression note that the (sub)gradient of the check loss $\{\tau - \mathbf{1}\{Y \leq X'\theta_0(\tau)\}X'v$ can be written as the product of two functions, $h(X,Y) = \tau - \mathbf{1}\{Y \leq X'\theta_0(\tau)\} = \tau - \mathbf{1}\{F_{Y|X}(Y|X) \leq \tau\}$ and $g(X) = X'v$. The function $h$ is bounded in absolute value by one and belongs to a class of VC subgraph functions with VC-index at most 3 (the difference of two classes of VC-subgraph functions with VC-index at most 2) and $g$ is indexed by $v \in \mathbb{R}^p$. In our specific applications, we will be able to argue that $v \in \mathcal{M} \subset \mathbb{R}^p$ with $\mathrm{card}(\mathcal{M}) < \infty$. 
\end{remark}

\begin{lemma}[Orlicz-norm of products of random variables]\label{lemma:ProductSubgaussian}
	Let $X_1, \ldots, X_K \in \mathbb{R}$ be random variables with finite $\psi_\alpha$-Orlicz-norm. Then, for $K \geq 1$, $\|\prod_{k=1}^K X_k\|_{\psi_{\alpha/K}} \leq \prod_{k=1}^K \|X_k \|_{\psi_\alpha}$.
\end{lemma}

\begin{lemma}\label{lemma:MomentsToTails}
	Let $\phi, \varphi : \mathbb{R}_+ \rightarrow \mathbb{R}_+$ be increasing functions and $\xi$ be a random variable on $L_1(\Omega, \mathcal{A}, \mathbb{P})$. If for all events $A \in \mathcal{A}$ with $\mathbb{P}\{A\} > 0$, $\mathbb{E}\left[ \xi \mid A \right] \leq \phi\left(\varphi^{-1}\left(1/\mathbb{P}\{A\}\right)\right)$, then for all $u > 0$, $\mathbb{P}\big\{ \xi > \phi(u)\big\} \leq \big(\varphi(u)\big)^{-1}$.
\end{lemma}

\begin{lemma}[Useful generalization of Lemma 1,~\citeauthor{panchenko2003symmetrization},~\citeyear{panchenko2003symmetrization}]\label{lemma:Panchenko2003Handel}
	Let $X, Y$ be random variables such that $\mathbb{E}[F(X)] \leq \mathbb{E}[F(Y)]$ for every convex and increasing function $F$. Further, let $\varphi: \mathbb{R}_+ \rightarrow \mathbb{R}_+$ be a concave and strictly increasing function and $\alpha \in (0,\infty)$. If for some constants $c_1 > 1$, $c_2 > 0$, and for all $t \geq 0$,$\mathbb{P}\left\{Y \geq \varphi\big(t^{1/\alpha}\big) \right\} \leq c_1e^{-c_2 t}$,	then there exist constants $c_3 > 1$, $c_4 > 0$ (depending only on $c_1, c_2, \alpha$) such that, for all $t \geq 0$,$\mathbb{P}\left\{X \geq \varphi\big(t^{1/\alpha}\big)\right\} \leq  c_3 e^{-c_4 t}$.
\end{lemma}
\begin{remark}
	It is possible to find explicit expressions for the constants $c_3, c_4$ in terms of $c_1, c_2, \alpha$. For our purposes the constants are irrelevant, only the nature of the inherited tail behavior is important. However, for completeness, we record that for $\alpha \in [1, \infty)$ the result holds with $c_3 = c_1 e$ and $c_4 = c_2$. The main use of this lemma is to deduce tail bounds for empirical processes based on tail bounds on the corresponding symmetrized empirical process. 
\end{remark}

\begin{lemma}\label{lemma:Composite-GaussianContraction}
	Let $\mathcal{F}, \mathcal{H} \subset L_2(S, \mathcal{S},P)$. Further, $\varphi_i: \mathbb{R} \rightarrow \mathbb{R}$, $i \leq n$, be a contraction with $\varphi_i(0)= 0$ and $F : \mathbb{R}_+ \rightarrow \mathbb{R}_+$ be convex and increasing. Let $\{g_i\}_{i=1}^n$ be a sequence of i.i.d. standard Gaussian random variables independent of $\{X_i\}_{i=1}^n$. Then,
	\begin{align*}
		\mathbb{E} \left[ F\left(\frac{1}{2} \sup_{f \in \mathcal{F}} \sup_{h \in \mathcal{H}} \left|\sum_{i=1}^n g_i \varphi_i\big(f(X_i)\big)h(X_i) \right|\right)\right] \leq	\mathbb{E} \left[F\left(2\sup_{f \in \mathcal{F}} \sup_{h \in \mathcal{H}} \left|\sum_{i=1}^n g_i f(X_i)h(X_i) \right|\right) \right].
	\end{align*}
\end{lemma}
\begin{remark}
	It would be desirable to derive an analogous result with Rademacher instead of standard Gaussian random variables. For bounded function classes such a result can be easily obtained via Theorem 2 in~\cite{maurer2016vectorconcentration}; for unbounded function classes it is unclear whether such a result holds true.
\end{remark}

\section{Proofs of the results in the supplementary materials}\label{sec:Proofs}
\subsection{Proofs of Section~\ref{subsec:Consistency-L1}}
\begin{proof}[\textbf{Proof of Theorem~\ref{theorem:Consistency}}]
	\noindent
	
	\textbf{Ansatz.}
	For $\tau \in \mathcal{T}$ and $r > 0$ (to be specified below) define $B^p(0, r ) := \{v \in \mathbb{R}^p : \|v \|_2 \leq r \}$ and recall that $C^p(J, \vartheta) = \left\{ \theta \in \mathbb{R}^p: \|\theta_{J^c}\|_1 \leq  \vartheta\|\theta_{J}\|_1  \right\}$. Suppose that, with high probability,
	\begin{itemize}
		\item[(a)] $\hat{\theta}(\tau) - \theta_0(\tau) \in C^p\big(T_\theta(\tau), \bar{c}\big)$ for all $\tau \in \mathcal{T}$; and
		\item[(b)] for some absolute constant $c > 0$, the regularized and centered objective function is strictly positive when evaluated at points $(\theta, \tau) \in \mathbb{R}^p \times \mathcal{T}$ satisfying $\theta - \theta_0(\tau) \in C^p(T_\theta(\tau), \bar{c}) \cap \partial B^p(0, r/(2c)) =: K( r/(2c), \tau)$, i.e.	
		\begin{align*}
			\inf_{\tau \in \mathcal{T}}\inf_{ \theta- \theta_0(\tau) \in K(r/(2c), \tau) } \left(\sum_{i=1}^n \rho_\tau(Y_i - X_i'\theta) - \rho_\tau\big(Y_i - X_i'\theta_0(\tau)\big) + \lambda\big(\|\theta\|_1 - \|\theta_0(\tau)\|_1\big)\right) > 0.
		\end{align*}
	\end{itemize}
	
	Since the regularized and centered objective function is convex in $\theta$ and negative at $\widehat{\theta}(\tau)$ for all $\tau \in \mathcal{T}$, it follows that $\sup_{\tau \in \mathcal{T}}\|\widehat{\theta}(\tau) - \theta_0(\tau) \|_2 \lesssim r $. Thus, to establish the claim of the theorem, we only need to prove that statements (a) and (b) hold with high probability.
	
	\textbf{Verification of high probability statements.} By assumption, eq.~\eqref{eq:lemma:RestrictedConeProp-QRVector-1} holds true. Thus, by Lemma~\ref{lemma:RestrictedConeProp-QRVector}, statement (a) holds for all $\tau \in \mathcal{T}$ with probability one. We are left with establishing statement (b). Suppose that $r > 0$ is such that Assumption~\ref{assumption:RestrictedIdentifiability} holds with $(\vartheta, \varrho) = (\bar{c}, r)$. Then, $q_r(\bar{c}) \gtrsim 1$, i.e. there exists an absolute constant $c \geq 1$ such that $c \cdot q_r(\bar{c}) \geq 1$. Set
	\begin{align*}
		\mathcal{G} := \big\{g: \mathbb{R}^{p+1} \rightarrow \mathbb{R} : g(X,Y) = \rho_\tau\big(Y-X'\theta_0(\tau)\big) - \rho_\tau(Y-X'\theta), \:	\theta \in \mathbb{R}^p, \: \tau \in \mathcal{T},\:	\theta - \theta_0(\tau) \in  K(r/(2c), \tau)\big\}.
	\end{align*}
	
	Compute
	\begin{align}\label{eq:theorem:Consistency-1}
		&\inf_{\tau \in \mathcal{T}} \inf_{ \theta- \theta_0(\tau) \in K(r/(2c), \tau) } \left(\frac{1}{n}\sum_{i=1}^n \rho_\tau(Y_i - X_i'\theta) - \rho_\tau\big(Y_i - X_i'\theta_0(\tau)\big) + \frac{\lambda}{n} \big(\|\theta\|_1 - \|\theta_0(\tau)\|_1\big) \right)\nonumber\\
		&\gtrsim - \sup_{g \in \mathcal{G}}\frac{1}{n}\sum_{i=1}^n g(X_i, Y_i) - \sup_{\tau \in \mathcal{T}} \sup_{ \theta- \theta_0(\tau) \in K(r/(2c), \tau) } \frac{\lambda}{n} \big(\|\theta\|_1 - \|\theta_0(\tau)\|_1\big) \nonumber\\
		&\gtrsim - \sup_{g \in \mathcal{G}}\mathbb{E}\left[\frac{1}{n}\sum_{i=1}^n g(X_i, Y_i)\right]  - n^{-1/2}\|\mathbb{G}_n\|_{\mathcal{G}}  - \sup_{\tau \in \mathcal{T}} \sup_{ \theta- \theta_0(\tau) \in K(r/(2c), \tau)} \frac{\lambda}{n} \big(\|\theta\|_1 - \|\theta_0(\tau)\|_1\big).
	\end{align}
	
	We bound the expressions on the far right hand side in above display. By Lemma~\ref{lemma:LocallyQuadraticMinorization}, 
	\begin{align}\label{eq:theorem:Consistency-2}
		-\sup_{g \in \mathcal{G}}\mathbb{E}\left[\frac{1}{n}\sum_{i=1}^n g(X_i, Y_i)\right] = \inf_{g \in \mathcal{G}}\mathbb{E}\left[\frac{1}{n}\sum_{i=1}^n - g(X_i, Y_i)\right] \gtrsim \kappa_1(\bar{c})r^2.
	\end{align}
	
	By Lemma~\ref{lemma:LocalizedLoss}, with probability at least $1 -\delta$,
	\begin{align}\label{eq:theorem:Consistency-3}
		n^{-1/2}\|\mathbb{G}_n\|_{\mathcal{G}} \lesssim 2(2 + \bar{c}) \phi_{\max}^{1/2}(2s_\theta) r \sqrt{\frac{s_\theta \log(ep/s_\theta) + \log (1/\delta) + \log(1 + L_\theta/r)}{n}}.
	\end{align}
	
	By the reverse triangle inequality and Lemma~\ref{lemma:RestrictedConeProp-QRVector}, 
	\begin{align}\label{eq:theorem:Consistency-6}
		&\sup_{\tau \in \mathcal{T}} \sup_{\theta - \theta_0(\tau) \in K(r/(2c), \tau)}\frac{\lambda}{n} \big(\|\theta\|_1 - \|\theta_0(\tau)\|_1\big).\nonumber\\
		&\quad{}\leq \sup_{\tau \in \mathcal{T}} \sup_{\theta - \theta_0(\tau) \in K(r/(2c), \tau)}\frac{\lambda}{n} \left(\sum_{k \in T_\theta} \big(|\theta_k| - |\theta_0(\tau)|\big) + \sum_{k \in T_\theta^c}|\theta_k|\right)\nonumber\\
		&\quad{}\leq \sup_{\tau \in \mathcal{T}} \sup_{\theta - \theta_0(\tau) \in K(r/(2c), \tau)}\frac{\lambda}{n} \left( \sum_{k \in T_\theta}|\theta_k - \theta_0(\tau)| + \sum_{k \in T_\theta^c}|\theta_k| \right)\nonumber\\
		&\quad{}\leq (1 + \bar{c}) \sup_{\tau \in \mathcal{T}} \sup_{\theta - \theta_0(\tau) \in K(r/(2c), \tau)} \frac{\lambda}{n} \sum_{k \in T_\theta} |\theta_k - \theta_0(\tau)|\nonumber\\
		&\quad{}  \lesssim  (1 + \bar{c}) \frac{\lambda}{n} s_\theta^{1/2}r.
	\end{align}
	
	Combine eq.~\eqref{eq:theorem:Consistency-2},~\eqref{eq:theorem:Consistency-3}, and~\eqref{eq:theorem:Consistency-6}, use Assumption~\ref{assumption:GrowthCondition}, and observe that, there exist absolute constants $c_1, c_2, c_3 > 0$ such that with probability at least $1-\delta$, the expression in eq.~\eqref{eq:theorem:Consistency-1} can be lower bounded (up to a multiplicative constant) by
	\begin{align*}
		\kappa_1(\bar{c})r^2 &- c_1 2(2 + \bar{c}) \phi_{\max}^{1/2}(2s_\theta) r\sqrt{\frac{s_\theta \log(ep/s_\theta) + \log(1 + L_\theta/r) + \log (1/\delta)}{n}}\\
		& - c_2(1 +  \bar{c}) \frac{\lambda}{n} s_\theta^{1/2}r\\
		&> 0,
	\end{align*}
	whenever 
	\begin{align*}
		r \geq  c_3\left( C_1 \sqrt{ \frac{s_\theta \log(ep/ s_\theta) + \log n + \log(1/\delta)}{n}} \bigvee \frac{\bar{c}}{\kappa_1(\bar{c})}\frac{\lambda \sqrt{s_\theta}}{n}\right),
	\end{align*}
	where $C_1 := \frac{\bar{c}\phi_{\max}^{1/2}(2s_\theta)L_\theta}{\kappa_1(\bar{c})}$ (recall that $L_\theta \geq 1$) and $\bar{c} > 1$ is given in Lemma~\ref{lemma:RestrictedConeProp-QRVector}. By assumption, $(\bar{c}, r)$ are such that the $(\vartheta, \varrho)$-restricted identifiability Assumption~\ref{assumption:RestrictedIdentifiability} holds. This concludes the proof.
\end{proof}

\begin{proof}[\textbf{Proof of Corollary~\ref{corollary:theorem:Consistency}}]
	The idea is to derive an upper bound on the gradient of the quantile loss function. Any $\lambda > 0$ greater than this upper bound, will satisfy eq.~\eqref{eq:lemma:RestrictedConeProp-QRVector-1}. By Lemma~\ref{lemma:GradientCone} (ii), with probability at least $1- \delta$
	\begin{align*}
		\sup_{\tau \in \mathcal{T}}\left\|\sum_{i=1}^n \big(\tau - \mathbf{1}\{Y_i\leq X_i'\theta_0(\tau)\}\big)X_i \right\|_\infty &= \sup_{\tau \in \mathcal{T}} \sup_{\|v\|_0= 1, \|v\|_2 \leq 1}\left|\sum_{i=1}^n  \big(\tau - \mathbf{1}\{Y_i\leq X_i'\theta_0(\tau)\}\big) X_i'v\right|\\
		&\lesssim \sigma_{\max}\sqrt{n\log(p/\delta)} \sqrt{1 + \pi_{n,1}^2(\log(p/\delta))},
	\end{align*}
	where $\pi_{n,1}^2(z) = \sqrt{z/n} + z/n$ for $ z \geq  0$ and $\sigma_{\max} = \varphi_{\max}^{1/2}(1)$.	Thus, by Assumption~\ref{assumption:GrowthCondition}, there exists a constant $C_1 > 0$ such that with probability at least $1- \delta$,
	\begin{align*}
		\sup_{\tau \in \mathcal{T}}\left\|\sum_{i=1}^n \big(\tau - \mathbf{1}\{Y_i\leq X_i'\theta_0(\tau)\}\big)X_i \right\|_\infty  \leq C_1\sigma_{\max}\sqrt{n\log(p/\delta)}.
	\end{align*}
	Therefore, any $\lambda \geq C_1 c_0\sigma_{\max}\sqrt{n\log(p/\delta)}$ satisfies eq.~\eqref{eq:lemma:RestrictedConeProp-QRVector-1} with probability at least $1 -\delta$. Plug this lower bound on $\lambda > 0$ into the rate $r > 0$ from Theorem~\ref{theorem:Consistency} and simplify the expression using Assumption~\ref{assumption:GrowthCondition}.
\end{proof}

\subsection{Proofs of Section~\ref{subsec:EmpiricalSparsity}}
\begin{proof}[\textbf{Proof of Theorem~\ref{theorem:EmpiricalSparsity}}]
	\noindent
	
	\textbf{Proof of statement (i).} Recall from Lemma 9 in~\cite{belloni2011L1penalized} that $\hat{s} \leq n \wedge p$. From the complementary slackness characterizations (C.3) in~\cite{belloni2011L1penalized} we obtain the following inequality:
	\begin{align*}
		\lambda \hat{s} &= \lambda \sup_{\tau \in \mathcal{T}} \mathrm{sign}\big(\hat{\theta}_\lambda(\tau)\big)'\mathrm{sign}\big(\hat{\theta}_\lambda(\tau)\big)\\ &\leq \sup_{\tau \in \mathcal{T}}\mathrm{sign}\big(\hat{\theta}_\lambda(\tau)\big)'\mathbf{X}'\hat{a}(\tau)\\
		&\leq n \sqrt{\hat{s}} \left(\sup_{\|u\|_2\leq 1, \: \|u\|_0 \leq \hat{s}} \frac{1}{n}\sum_{i=1}^n (X_i'u)^2\right)^{1/2}\\
		&\leq  n \sqrt{\widehat{\varphi}_{\max}(\hat{s}) \hat{s}}.
	\end{align*}
	This implies that for all $\lambda > 0$,
	\begin{align*}
		\lambda \leq n \sqrt{\widehat{\varphi}_{\max}(\hat{s})/\hat{s}}.
	\end{align*}
	Combine above inequality with the lower bound $ \lambda \geq n\sqrt{2} \sqrt{\widehat{\varphi}_{\max}(m)/m}$ to obtain
	\begin{align*}
		\hat{s} \leq \frac{m}{2} \frac{\widehat{\varphi}_{\max}(\hat{s})}{\widehat{\varphi}_{\max}(m)}.
	\end{align*}
	We now proceed by contradiction as in the proof of Lemma 6 in~\cite{belloni2011L1penalized}: Suppose that $\hat{s} > m$. Then there exists $\ell > 1$ such that $\hat{s} = \ell m$. Therefore, by Lemma 13 in~\cite{belloni2011L1penalized},
	\begin{align*}
		\hat{s} \leq \frac{m}{2} \frac{\widehat{\varphi}_{\max}(\ell m)}{\widehat{\varphi}_{\max}(m)} \leq \frac{m}{2} \lceil \ell \rceil < \ell m = \hat{s}.
	\end{align*}
	Thus, $\hat{s} \leq m$. This concludes the proof of statement (i).
	
	\textbf{Proof of statement (ii).}
	The following proof is modeled after the proof of Lemma 7 in~\cite{belloni2011L1penalized} with necessary modifications to accommodate our setup.
	
	First, since $s \mapsto \varphi_{\max}(s) \vee \widehat{\varphi}_{\max}(s) $ is non-decreasing, we have 
	\begin{align*}
		\lambda &= C_\lambda \sqrt{\varphi_{\max}\left(\frac{n}{\log (np/\delta)} \right) \vee \hat{\varphi}_{\max}\left(\frac{n}{\log (np/\delta)}\right) }\sqrt{n \log(np/\delta)}\\
		&\geq C_\lambda \sqrt{\varphi_{\max}(1) \vee \widehat{\varphi}_{\max}(1)} \sqrt{n\log (p/\delta) }.
	\end{align*}
	Thus, arguing as in the proof of Corollary~\ref{corollary:theorem:Consistency}, we have for all $\tau \in \mathcal{T}$
	\begin{align*}
		\hat{\theta}_\lambda(\tau) - \theta_0(\tau) \in C^p(T_\theta(\tau), \bar{c}),
	\end{align*}
	and, with probability at least $1- \delta$,
	\begin{align*}
		\sup_{\tau \in \mathcal{T}} \|\hat{\theta}_\lambda(\tau) - \theta_0(\tau)\|_2 \lesssim \widetilde{C}_1\sqrt{ \frac{s_\theta \log(np/ \delta) }{n}},
	\end{align*}
	where
	\begin{align}\label{eq:theorem:EmpiricalSparsity-00}
		\widetilde{C}_1 : = \frac{\bar{c}\phi_{\max}^{1/2}(2s_\theta)L_\theta}{\kappa_1(\bar{c})} \vee \frac{C_\lambda \varphi_{\max}^{1/2}\big(n/\log(np/\delta)\big)}{\kappa_1(\bar{c})} \vee \frac{C_\lambda \widehat{\varphi}_{\max}^{1/2}\big(n/\log(np/\delta)\big)}{\kappa_1(\bar{c})} \vee 1.
	\end{align}
	
	In the following, to simplify the notation, we write $\widehat{T}_\theta(\tau) = \mathrm{support}\big(\hat{\theta}_\lambda(\tau)\big)$, $\mathbf{X} = [X_1, \ldots, X_n]$, and $\hat{s}(\tau) = \|\hat{\theta}_\lambda(\tau)\|_0$. Also, denote by $\hat{a}(\tau) = (\hat{a}_1(\tau), \ldots, \hat{a}_n(\tau))$, the (vector of) dual optimal rank scores which solve the dual program (C.2) in~\cite{belloni2011L1penalized}. From the complementary slackness characterizations (C.3) and identity (C.4) in~\cite{belloni2011L1penalized} we obtain the following inequality:
	\begin{align}\label{eq:theorem:EmpiricalSparsity-1}
		\lambda \sqrt{\hat{s}} &\leq  \sup_{\tau \in \mathcal{T}} \left\|\sum_{i=1}^n \big(\tau - \mathbf{1}\{Y_i \leq X_i'\hat{\theta}_\lambda(\tau)\}- \hat{a}_i(\tau) \big)X_{i,\widehat{T}_\theta(\tau)} \right\|_2 \nonumber\\
		&\quad{} + \sup_{\tau \in \mathcal{T}} \left\|\sum_{i=1}^n\big( \mathbf{1}\{Y_i \leq X_i'\theta_0(\tau)\} -  \mathbf{1}\{Y_i \leq X_i'\hat{\theta}_\lambda(\tau)\}\big) X_{i,\widehat{T}_\theta(\tau)} \right\|_2 \nonumber\\
		&\quad{} + \sup_{\tau \in \mathcal{T}} \left\|\sum_{i=1}^n\big(\tau - \mathbf{1}\{Y_i \leq X_i'\theta_0(\tau)\}\big)X_{i,\widehat{T}_\theta(\tau)}\right\|_2 \nonumber\\
		&= \mathbf{I} + \mathbf{II} + \mathbf{III}.
	\end{align}
	
	\textbf{Bound on $\mathbf{I}$.} Observe that $\hat{a}_i(\tau) \neq \tau - \mathbf{1}\{Y_i \leq X_i'\hat{\theta}_\lambda(\tau)\}$ only if $Y_i = X_i'\hat{\theta}_\lambda(\tau)$. By Lemma 9 in~\cite{belloni2011L1penalized} the penalized quantile regression fit can only interpolate at most $\hat{s}(\tau) \leq \hat{s}$ points almost surely uniformly over $\tau \in \mathcal{T}$. This implies that $\sum_{i=1}^n \big|\tau - \mathbf{1}\{Y_i \leq X_i'\hat{\theta}_\lambda(\tau)\} - \hat{a}_i(\tau)\big| \leq \hat{s}$. Therefore, by H{\"o}lder's inequality and Lemma~\ref{lemma:MaxInequalityCovariance}, with probability at least $1- \delta$,
	\begin{align}\label{eq:theorem:EmpiricalSparsity-2}
		&\sup_{\tau \in \mathcal{T}} \left\|\sum_{i=1}^n\big(\tau - \mathbf{1}\{Y_i \leq X_i'\hat{\theta}_\lambda(\tau)\} - \hat{a}_i(\tau) \big)  X_{i,\widehat{T}_\theta(\tau)}\right\|_2 \nonumber\\
		&\quad{}\leq \sup_{\mathrm{card}(I) \leq \hat{s}} \sup_{\|u\|_2 \leq 1, \|u\|_0 \leq \hat{s}} \left|\sum_{i \in I} X_i'u\right|\nonumber\\
		&\quad{}\leq \sup_{\mathrm{card}(I) \leq \hat{s}} \sup_{\|u\|_2 \leq 1, \|u\|_0 \leq \hat{s}} \hat{s}\left(\frac{1}{\hat{s}}\sum_{i \in I} (X_i'u)^2 - E[(X_i'u)^2]\right)^{1/2} + \hat{s}\varphi_{\max}^{1/2}(\hat{s}) \nonumber\\
		&\quad{}\lesssim \varphi_{\max}^{1/2}(\hat{s}) \hat{s} \sqrt{\log (epn/\hat{s})} + \varphi_{\max}^{1/2}(\hat{s})\hat{s}^{3/4} \big(\log(1/\delta)\big)^{1/4}  +    \varphi_{\max}^{1/2}(\hat{s})\hat{s}^{1/2} \sqrt{\log(1/\delta)} + \hat{s}\varphi_{\max}^{1/2}(\hat{s}),
	\end{align}
	where we have used that $\phi_{\max}(\hat{s}) \leq \varphi_{\max}(\hat{s})$.
	
	\textbf{Bound on $\mathbf{II}$.} Define	
	\begin{align*}
		\mathcal{G} = \left\{g: \mathbb{R}^{p+1} \rightarrow \mathbb{R}: g(X, Y) = \left(1\big\{Y \leq X'\theta \big\} - 1\big\{Y \leq X'\theta_0(\tau) \}\right)X'v,  \right.\\
		\left.  v, \theta \in \mathbb{R}^p,\: \|v\|_2 \leq 1,\: \|v\|_0 \leq n, \: \|\theta\|_0 \leq n, \: \tau \in \mathcal{T}\right\},
	\end{align*}
	and, for $1 \leq s \leq n \wedge p$ and $r =\widetilde{C}_1 \sqrt{ \frac{s_\theta \log(np/ \delta) }{n}}$ with $\widetilde{C}_1 \geq 1$ defined in eq.~\eqref{eq:theorem:EmpiricalSparsity-00},
	\begin{align*}
		\mathcal{G}(s) = \big\{g_{v, \theta, \tau} \in \mathcal{G}:  \|v\|_0 \leq s, \|\theta\|_0 \leq s\big\} \bigcap \big\{\theta \in \mathbb{R}^p : \: \theta - \theta_0(\tau) \in C^p(T_\theta(\tau), \bar{c}) \cap B^p(0,r),\: \tau \in \mathcal{T}\big\}.
	\end{align*}
	By the triangle inequality and Corollary~\ref{corollary:theorem:Consistency}, with probability at least $1- \delta$,
	\begin{align}\label{eq:theorem:EmpiricalSparsity-4}
		\sup_{\tau \in \mathcal{T}} \left\|\sum_{i=1}^n \big( \mathbf{1}\{Y_i \leq X_i'\theta_0(\tau)\} -  \mathbf{1}\{Y_i \leq X_i'\hat{\theta}_\lambda(\tau)\}\big) X_{i,\widehat{T}_\theta(\tau)} \right\|_2
		\leq \sqrt{n} \|\mathbb{G}_n\|_{\mathcal{G}(\hat{s}) } + n \sup_{g \in \mathcal{G}(\hat{s})}|Pg|.
	\end{align}	
	By Lemma~\ref{lemma:LocalizedRankScores} (ii), with probability at least $1- \delta$,
	\begin{align}
		\sqrt{n}\|\mathbb{G}_n\|_{\mathcal{G}(\hat{s})} &\lesssim  \varphi_{\max}^{1/2}(\hat{s}) \sqrt{n}\sqrt{ \hat{s} \log(ep/\hat{s}) + \log (n/\delta) } \sqrt{1 + \pi_{n,1}^2\big(\hat{s} \log(ep/\hat{s}) + \log (n/\delta) \big)} \nonumber\\
		&\lesssim \varphi_{\max}^{1/2}(\hat{s}) \sqrt{n}\sqrt{ \hat{s} \log(ep/\hat{s}) + \log (n/\delta) },
	\end{align}
	where the second inequality holds since $\hat{s} \leq n/\log(np/\delta)$ by choice of $\lambda > 0$ and statement (i).
	
	By Assumption~\ref{assumption:LipschitzDensity}, for all $g \in \mathcal{G}(\hat{s})$, 
	\begin{align}
		n Pg &= n \mathbb{E}\left[X'v \int_{X'\theta_0(\tau)}^{X'\theta} \big( f_{Y|X}(z|X) -f_{Y|X}(X'\theta_0(\tau)|X)\big)dz\right] + v'\mathbb{E}\left[f_{Y|X}(X'\theta_0(\tau)|X)XX'\right]\big(\theta - \theta_0(\tau)\big) \nonumber\\
		&\leq n \frac{L_f}{2} \mathbb{E}\left[ |X'v| \big(X'(\theta_0(\tau) - \theta)\big)^2\right] + nv'\mathbb{E}\left[f_{Y|X}(X'\theta_0(\tau)|X)XX'\right]\big(\theta - \theta_0(\tau)\big).
	\end{align}
	Since $\tau \mapsto f_{Y|X}(X'\theta_0(\tau)|X)$ is continuous on the compact set $\mathcal{T}$, it is necessarily bounded. Thus, by Lemmas~\ref{lemma:SizeConesDominatedCoordinates} and~\ref{lemma:MaximaBiconvexFunction},
	\begin{align}
		nv'\mathbb{E}\left[f_{Y|X}(X'\theta_0(\tau)|X)XX'\right]\big(\theta - \theta_0(\tau)\big) \lesssim (2 + \bar{c}) n \varphi_{\max}^{1/2}(\hat{s}) \varphi_{\max}^{1/2}(s_\theta) r.
	\end{align}
	By H{\"o}lder's inequality, Assumption~\ref{assumption:SubGaussianity}, and Lemmas~\ref{lemma:SizeConesDominatedCoordinates} and~\ref{lemma:MaximaBiconvexFunction},
	\begin{align}\label{eq:theorem:EmpiricalSparsity-5}
		n \frac{L_f}{2} \mathbb{E}\left[ |X'v| \big(X'(\theta_0(\tau) - \theta)\big)^2\right]  &\leq n \frac{L_f}{2} \big(\mathbb{E}[ |X'v|^3]\big)^{1/3} \Big(\mathbb{E}\big[\big(X'(\theta_0(\tau) - \theta)\big)^3\big]\Big)^{2/3} \nonumber\\
		&\lesssim (2 + \bar{c})^2 n L_f \big(\varphi_{\max}^{1/2}(\hat{s}) + \phi_{\max}^{1/2}(\hat{s})\big)\phi_{\max}(s_\theta) r^2 \nonumber\\
		&\lesssim (2 + \bar{c})^2 n L_f \varphi_{\max}^{1/2}(\hat{s})\phi_{\max}(s_\theta) r^2.
	\end{align}
	
	Combine eq.~\eqref{eq:theorem:EmpiricalSparsity-4}--\eqref{eq:theorem:EmpiricalSparsity-5} and conclude that with probability at least $1- \delta$,
	\begin{align}\label{eq:theorem:EmpiricalSparsity-6}
		\begin{split}
			&\sup_{\tau \in \mathcal{T}} \left\|\sum_{i=1}^n \big( \mathbf{1}\{Y_i \leq X_i'\theta_0(\tau)\} -  \mathbf{1}\{Y_i \leq X_i'\hat{\theta}_\lambda(\tau)\}\big) X_{i,\widehat{T}_\theta(\tau)} \right\|_2\\ &\lesssim  \varphi_{\max}^{1/2}(\hat{s})\sqrt{n} \sqrt{ \hat{s} \log(ep/\hat{s}) + \log (n/\delta) } +(2 + \bar{c})^2 n L_f\varphi_{\max}^{1/2}(\hat{s})\phi_{\max}(s_\theta) r^2 + (2 + \bar{c}) n  \varphi_{\max}^{1/2}(\hat{s})\varphi_{\max}^{1/2}(s_\theta) r.
		\end{split}
	\end{align}
	
	\textbf{Bound on $\mathbf{III}$.} Since $\lambda > 0$ satisfies eq.~\eqref{eq:lemma:RestrictedConeProp-QRVector-1}, H{\"o}lder's inequality yields, with probability at least $1-\delta$,
	\begin{align}\label{eq:theorem:EmpiricalSparsity-8}
		\begin{split}
			&\sup_{\tau \in \mathcal{T}} \left\|\sum_{i=1}^n \big(\tau - \mathbf{1}\{Y_i \leq X_i'\theta_0(\tau)\} \big)X_{i,\widehat{T}_\theta(\tau)} \right\|_2\\
			&\quad{}\leq \sqrt{\hat{s}}\sup_{\tau \in \mathcal{T}} \left\|\sum_{i=1}^n\big(\tau - \mathbf{1}\{Y_i \leq X_i'\theta_0(\tau)\} \big)X_{i,\widehat{T}_\theta(\tau)}  \right\|_\infty \leq  c_0^{-1} \lambda \sqrt{\hat{s}}.
		\end{split}
	\end{align}
	
	\textbf{Conclusion.} Combine eq.~\eqref{eq:theorem:EmpiricalSparsity-1},~\eqref{eq:theorem:EmpiricalSparsity-2},~\eqref{eq:theorem:EmpiricalSparsity-6}, and~\eqref{eq:theorem:EmpiricalSparsity-8} to conclude that there exist absolute constants $c_8, c_9 > 0$ such with probability at least $1 - 3\delta$,
	\begin{align*}
		\lambda \sqrt{\hat{s}} &\leq c_8 \left(\varphi_{\max}^{1/2}(\hat{s}) \hat{s} \sqrt{\log (epn/\hat{s})} + \varphi_{\max}^{1/2}(\hat{s})\hat{s}^{3/4} \big(\log(1/\delta)\big)^{1/4}  +    \varphi_{\max}^{1/2}(\hat{s})\hat{s}^{1/2} \sqrt{\log(1/\delta)} + \hat{s}\varphi_{\max}^{1/2}(\hat{s})\right)\\
		&\quad{} + c_9 \varphi_{\max}^{1/2}(\hat{s}) \sqrt{n\hat{s} \log(ep/\hat{s}) + n\log (n/\delta) }\\
		&\quad{} + c_9(2 + \bar{c})^2 n L_f \varphi_{\max}^{1/2}(\hat{s})\phi_{\max}(s_\theta) r^2\\
		&\quad{} + c_9(2 + \bar{c}) n \varphi_{\max}^{1/2}(\hat{s}) \varphi_{\max}^{1/2}(s_\theta)r + c_0^{-1} \lambda \sqrt{\hat{s}}.
	\end{align*}
	Set $m := n/ \log(np/\delta)$. Recall that $s \mapsto \varphi_{\max}(s)$ is non-decreasing. By statement (i) and since $c_0 > 1$ above inequality can be simplified to
	\begin{align}\label{eq:theorem:EmpiricalSparsity-9}
		\begin{split}
			\lambda \sqrt{\hat{s}} &\leq \frac{c_0 c_8}{c_0 - 1} \sqrt{\varphi_{\max}(m) m}\sqrt{\hat{s}} \left(\sqrt{\log(ep/\hat{s})} +  \left(\frac{\log(1/\delta)}{m}\right)^{1/4}+ \sqrt{\frac{\log(1/\delta)}{m}} + 1\right) \\
			&\quad{} + \frac{c_0 c_9}{c_0 - 1}\sqrt{\varphi_{\max}(m)}\sqrt{\hat{s}} \left(\sqrt{n \log(ep/\hat{s})} + \sqrt{\frac{n\log(n/\delta)}{\hat{s}}}\right)\\
			&\quad{} + \frac{c_0 c_9}{c_0 - 1} \phi_{\max}(s_\theta)(2 + \bar{c})^2L_f \sqrt{\varphi_{\max}(m)} n r^2 \\
			&\quad{} + \frac{c_0 c_9}{c_0 - 1} \varphi_{\max}^{1/2}(s_\theta) (2 + \bar{c})  \sqrt{\varphi_{\max}(m)}n r.
		\end{split}
	\end{align}
	Divide above inequality by $\lambda =  C_\lambda \frac{n}{\sqrt{m}}\sqrt{\varphi_{\max}(m) \vee \widehat{\varphi}_{\max}(m)}$ and conclude that
	\begin{align*}
		\sqrt{\hat{s}} &\lesssim \frac{c_0}{c_0 - 1}  \left(\sqrt{\frac{\log(ep/\hat{s})}{n}} +  \left(\frac{\log(1/\delta)}{n^2m}\right)^{1/4}+ \sqrt{\frac{\log(1/\delta)}{nm}} + n^{-1/2}\right)\sqrt{\frac{m}{n}} \frac{\sqrt{\hat{s}}}{C_\lambda}\\
		&\quad{} + \frac{c_0}{c_0 - 1} \sqrt{\frac{m\log(ep/\hat{s})}{n}}\frac{\sqrt{\hat{s}}}{C_\lambda} +  \frac{c_0}{c_0 - 1} \sqrt{\frac{m\log(n/\delta)}{C_\lambda^2 n}}\\
		&\quad{} + \frac{c_0 }{c_0 - 1} \frac{L_f}{C_\lambda} (2 + \bar{c})^2\phi_{\max}(s_\theta) \sqrt{m} r^2\\
		&\quad{} + \frac{c_0}{c_0 - 1} \frac{1}{C_\lambda}(2 + \bar{c}) \varphi_{\max}^{1/2}(s_\theta) \sqrt{m} r.
	\end{align*}
	
	Since $m \log(p/\delta) = o(n)$, above inequality simplifies to
	\begin{align*}
		\sqrt{\hat{s}} & \lesssim \frac{c_0 }{c_0 - 1}\frac{L_f}{C_\lambda} (2 + \bar{c})^2\phi_{\max}(s_\theta) \sqrt{m} r^2  + \frac{c_0}{c_0 - 1}  \frac{\kappa_1(\bar{c})}{C_\lambda}\frac{\sqrt{m} r}{\sqrt{\varphi_{\max}(m)}}\\
		&\lesssim \frac{c_0 }{c_0 - 1}\frac{\widetilde{C}_1^2}{C_\lambda} L_f (2 + \bar{c})^2\phi_{\max}(s_\theta) \sqrt{\frac{s_\theta \log(np/\delta)}{n}} \sqrt{s_\theta} + \frac{c_0 }{c_0 - 1}\frac{\widetilde{C}_1}{C_\lambda}  (2 + \bar{c}) \varphi_{\max}^{1/2}(s_\theta) \sqrt{s_\theta}\\
		&\lesssim \frac{c_0 }{c_0 - 1}\frac{\widetilde{C}_1}{C_\lambda} \left( L_f (2 + \bar{c})^2\phi_{\max}(s_\theta) + (2 + \bar{c}) \varphi_{\max}^{1/2}(s_\theta)\right)\sqrt{s_\theta}.
	\end{align*}
	This concludes the proof of the second statement.
\end{proof}

\subsection{Proofs of Section~\ref{subsec:AuxResults-L1}}
\begin{proof}[\textbf{Proof of Lemma~\ref{lemma:RestrictedConeProp-QRVector}}]
	The proof strategy is standard~\citep[e.g.][]{bickel2009simultaneous, belloni2011L1penalized, belloni2013least}. By optimality of $\hat{\theta}_\lambda(\tau)$ and the premise, for all $\tau \in \mathcal{T}$,
	\begin{align*}
		0 &\geq \sum_{i=1}^n \rho_\tau\big(Y_i - X_i'\hat{\theta}_\lambda(\tau)\big) - \sum_{i=1}^n \rho_\tau\big(Y_i - X_i'\theta_0(\tau)\big) + \lambda \|\hat{\theta}_\lambda(\tau)\|_1 - \lambda \|\theta_0(\tau)\|_1\\
		&\geq -\sum_{i=1}^n \big(\tau-1\big\{Y_i \leq X_i'\theta_0(\tau)\big\}\big)X_i'\big(\hat{\theta}_\lambda(\tau) - \theta_0(\tau) \big) + \lambda \|\hat{\theta}_\lambda(\tau)\|_1 - \lambda \|\theta_0(\tau)\|_1\\
		&\geq \lambda \left( - c_0^{-1} \|\hat{\theta}_\lambda(\tau) - \theta_0(\tau)\|_1 + \|\hat{\theta}_\lambda(\tau)\|_1 - \|\theta_0(\tau)\|_1 \right).
	\end{align*}
	Thus,
	\begin{align}\label{eq:lemma:RestrictedConeProp-QRVector-2}
		\left(1 + c_0^{-1}\right)  \sum_{k=1}^p \big|\hat{\theta}_{\lambda,k}(\tau) - \theta_{0,k}(\tau)\big|\geq  \sum_{k=1}^p \big|\hat{\theta}_{\lambda,k}(\tau) - \theta_{0,k}(\tau)\big| + \sum_{k=1}^p \big|\hat{\theta}_{\lambda,k}(\tau)\big| -\sum_{k=1}^p\big|\theta_{0,k}(\tau)\big|.
	\end{align}
	By Assumption~\ref{assumption:SparsityQRVector} and the reverse triangle inequality,
	\begin{align}\label{eq:lemma:RestrictedConeProp-QRVector-3}
		\sum_{k=1}^p \big|\hat{\theta}_{\lambda,k}(\tau) - \theta_{0,k}(\tau)\big| + \sum_{k=1}^p \big|\hat{\theta}_{\lambda,k}(\tau)\big| -\sum_{k=1}^p\big|\theta_{0,k}(\tau)\big|\geq 2 \sum_{k \in T_\theta^c(\tau)}\big|\hat{\theta}_{\lambda,k}(\tau)\big|.
	\end{align}
	Combine eq.~\eqref{eq:lemma:RestrictedConeProp-QRVector-2} and~\eqref{eq:lemma:RestrictedConeProp-QRVector-3} to conclude that
	\begin{align}\label{eq:lemma:RestrictedConeProp-QRVector-4}
		\frac{c_0+1}{c_0-1}\sum_{k \in T_\theta(\tau)} \big|\hat{\theta}_{\lambda,k}(\tau) - \theta_{0,k}\big| \geq \sum_{k \in T_\theta^c(\tau)}\big|\hat{\theta}_{\lambda,k}(\tau)\big|.
	\end{align}
	This concludes the proof.
\end{proof}
\begin{proof}[\textbf{Proof of Lemma~\ref{lemma:NewKnightsIdentity}}]
	Obviously, $\phi_{\tau,x, y}(0) = 0$ and $|\phi_{\tau,x,y}(a) - \phi_{\tau,x,y}(b)| \leq |a-b|$ for all $a, b \in \mathbb{R}$. Moreover, by a change of variables,
	\begin{align*}
		\phi_{\tau, x,y}\big(x'\theta - x'\theta_0(\tau)\big) &= \int_0^{x'\theta - x'\theta_0(\tau)} 1\big\{y \leq x'\theta_0(s) + u + \big(x'\theta_0(\tau) - x'\theta_0(s)\big) \big\} du\\
		&= \int_{x'\theta_0(\tau) - x'\theta_0(s)}^{x'\theta - x'\theta_0(s)} 1\{y \leq x'\theta_0(s) + u\} du\\
		&= \phi_{s,x,y}\big(x'\theta - x'\theta_0(s)\big) - \phi_{s,x,y}\big(x'\theta_0(\tau) - x'\theta_0(s)\big).
	\end{align*}
	Lastly, re-arrange Knight's identity and obtain
	\begin{align*}
		\rho_\tau(y - x'\theta) - \rho_\tau\big(y - x'\theta_0(\tau)\big) = - \tau \big(x'\theta - x'\theta_0(\tau)\big) + \phi_{\tau, x, y}\big(x'\theta - x'\theta_0(\tau)\big).
	\end{align*}
\end{proof}
\begin{proof}[\textbf{Proof of Lemma~\ref{lemma:LocallyQuadraticMinorization}}] 
	
	\textbf{Proof of Case (i).}
	Define
	\begin{align*}
		L(\theta \mid X) : = \mathbb{E}\left[ \rho_\tau(Y-X'\theta) \mid X\right] = (\tau - 1) \int_{-\infty}^{X'\theta} (z - X'\theta) d F_{Y| X}(z) + \tau \int_{X'\theta}^\infty (z - X'\theta)dF_{Y|X}(z).
	\end{align*}
	Hence, by Leibniz' rule for differentiating parameter integrals,
	\begin{align}\label{eq:lemma:QuadraticMinorization-1}
		\frac{d}{d\theta}L(\theta\mid X) = (\tau - 1)\int_{-\infty}^{X'\theta} - X' d F_{Y| X}(z) + \tau \int_{X'\theta}^\infty -X' dF_{Y|X}(z) =\left(F_{Y|X}(X'\theta|X) - \tau \right)X',
	\end{align}
	and, by differentiating eq.~\eqref{eq:lemma:QuadraticMinorization-1},
	\begin{align}\label{eq:lemma:QuadraticMinorization-2}
		\frac{d^2}{d\theta d\theta'}L(\theta\mid X) =f_{Y|X}(X'\theta|X) XX'.
	\end{align}
	By optimality of $\theta_0(\tau)$ we have $\frac{d}{d\theta}L(\theta_0(\tau)\mid X)= 0$. Hence, eq.~\eqref{eq:lemma:QuadraticMinorization-1}, eq.~\eqref{eq:lemma:QuadraticMinorization-2}, and a second order Taylor approximation around $\theta_0(\tau)$ yield
	\begin{align}\label{eq:lemma:QuadraticMinorization-3}
		\mathbb{E}\left[ \rho_\tau(Y-X'\theta) - \rho_\tau(Y-X'\theta_0(\tau)) \mid X\right] = \frac{1}{2}\big(\theta - \theta_0(\tau)\big)'f_{Y|X}(X'\xi(\tau)|X)XX'\big(\theta - \theta_0(\tau)\big),
	\end{align}
	where $\xi(\tau) = \lambda \big(\theta - \theta_0(\tau)\big) + \theta_0(\tau)$ for some $\lambda \in [0,1]$. Whence, by Assumption~\ref{assumption:RestrictedIdentifiability}, 
	\begin{align*}
		&\mathbb{E}\left[ \rho_\tau(Y-X'\theta) - \rho_\tau(Y-X'\theta_0(\tau))\right]\\
		&\quad{}\geq \frac{1}{2}\big(\theta - \theta_0(\tau)\big)'\mathbb{E}\big[f_{Y|X}(X'\theta_0(\tau)|X)XX'\big]\big(\theta - \theta_0(\tau)\big)\\
		&\quad{}\quad{} \times\big(\theta - \theta_0(\tau)\big)' \mathbb{E}[f_{Y|X}(X'\xi(\tau)|X)XX']\big(\theta - \theta_0(\tau)\big) \Big/\big(\theta - \theta_0(\tau)\big)'\mathbb{E}\big[f_{Y|X}(X'\theta_0(\tau)|X)XX'\big]\big(\theta - \theta_0(\tau)\big)\\
		&\quad{} \geq \frac{\kappa_1(c_0)}{2} r_0^2 \inf_{\tau \in \mathcal{T}} \inf_{\|\zeta\|_2 \leq r_0} \inf_{u \in C^p_1(T_\theta(\tau), c_0) \cap \partial B^p(0,1)} \frac{\mathbb{E}[f_{Y|X}(X'\theta_0(\tau) + X'\zeta)|X)(X'u)^2]}{\mathbb{E}[f_{Y|X}(X'\theta_0(\tau)|X)(X'u)^2]}\\
		&\quad{} \gtrsim \frac{\kappa_1(c_0)}{2} r_0^2.
	\end{align*}
	
	This proves the first statement.
	
	\textbf{Proof of Case (ii).} By Assumption~\ref{assumption:LipschitzDensity} the right hand side of eq.~\eqref{eq:lemma:QuadraticMinorization-3} can be lower bounded by
	\begin{align*}
		&\frac{1}{2}\big(\theta - \theta_0(\tau)\big)'f_{Y|X}(X'\theta_0(\tau)|X)XX'\big(\theta - \theta_0(\tau)\big) - \frac{L_f}{2} \big|X'\big(\theta - \theta_0(\tau)\big)\big|^3.
	\end{align*}
	By Assumption~\ref{assumption:RestrictedPrimeIdentifiability} there exists an absolute constant  $c > 1$ such that $q_1(c_0) (r_0/c) \leq 1/2$. Whence, for all $(\theta, \tau) \in \mathbb{R}^p \times \mathcal{T}$ such that $\theta - \theta_0(\tau) \in C^p1(J, c_0) \cap \partial B^p(0, r_0/c)$, we have
	\begin{align*}
		&\mathbb{E}\left[ \rho_\tau(Y-X'\theta) - \rho_\tau(Y-X'\theta_0(\tau))\right]\\
		&\quad{}\geq \frac{1}{2}\big(\theta - \theta_0(\tau)\big)'\mathbb{E}\big[f_{Y|X}(X'\theta_0(\tau)|X)XX'\big]\big(\theta - \theta_0(\tau)\big)\\
		&\quad{}\quad{} \times \left[1-  \sup_{\tau \in \mathcal{T}}\sup_{u \in C^p_1(T_\theta(\tau), c_0) \cap \partial B^p(0,1)} \frac{L_f \mathbb{E}[|X'u|^3]\|\theta - \theta_0(\tau)\|_2}{\mathbb{E}[f_{Y|X}(X'\theta_0(\tau)|X)(X'u)^2]} \right]\\
		&\quad{} \geq  \frac{\kappa_1(c_0)}{2} \frac{r_0^2}{c^2} \left[1-  \sup_{\tau \in \mathcal{T}}\sup_{u \in C^p_1(T_\theta(\tau), c_0) \cap \partial B^p(0,1)} \frac{L_f \mathbb{E}[|X'u|^3] (r_0/c)}{\mathbb{E}[f_{Y|X}(X'\theta_0(\tau)|X)(X'u)^2]} \right]\\
		&\quad{} \gtrsim \frac{\kappa_1(c_0)}{4} r_0^2.
	\end{align*}
	
	This proves the second statement.
	
	\textbf{Proof of Case (iii).} Let $(\theta, \tau) \in \mathbb{R}^p \times \mathcal{T}$ be such that $\theta - \theta_0(\tau) \in C^p_1(J, c_0) \cap \partial B^p(0, r_0)$. Then, by Assumption~\ref{assumption:Restricted2PrimeIdentifiability} there exists an absolute constant $\underline{f} > 0$ such that the right hand side of eq.~\eqref{eq:lemma:QuadraticMinorization-3} can be lower bounded by
	\begin{align*}
		(\underline{f}/2)\Big(X'\big(\theta - \theta_0(\tau)\big)\Big)^2\mathbf{1}\left\{ |X'(\theta - \theta_0(\tau))| \leq2 q_0(c_0)r_0\right\},
	\end{align*}
	and, therefore,
	\begin{align*}
		&\mathbb{E}\left[ \rho_\tau(Y-X'\theta) - \rho_\tau(Y-X'\theta_0(\tau))\right]\\
		&\quad{}\geq (\underline{f}/2) \mathbb{E}\left[\Big(X'\big(\theta - \theta_0(\tau)\big)\Big)^2\right] -  (\underline{f}/2) \mathbb{E}\left[\Big(X'\big(\theta - \theta_0(\tau)\big)\Big)^2\mathbf{1}\left\{ \Big|X'\big(\theta - \theta_0(\tau)\big)\Big| > 2 q_0(c_0)r_0\right\}\right]\\
		&\quad{} \geq (\underline{f}/2)  \mathbb{E}\left[\Big(X'\big(\theta - \theta_0(\tau)\big)\Big)^2\right]  -  (\underline{f}/2) \mathbb{E}\left[\big|X'\big(\theta - \theta_0(\tau)\big)\big|^3\right] / (2 q_0(c_0)r_0)\\
		&\quad{} \overset{(a)}{\geq} (\underline{f}/4)  \mathbb{E}\left[\Big(X'\big(\theta - \theta_0(\tau)\big)\Big)^2\right]\\
		&\quad{} \gtrsim (\underline{f}/4) \kappa_0(c_0) r_0^2,
	\end{align*}
	where (a) holds because for all $(\theta, \tau) \in \mathbb{R}^p \times \mathcal{T}$ with $\theta - \theta_0(\tau) \in C^p_1(J, c_0) \cap \partial B^p(0, r_0)$,
	\begin{align*}
		q_0(c_0)r_0 \geq \mathbb{E}\left[\big|X'\big(\theta - \theta_0(\tau)\big)\big|^3\right]\Big/ \mathbb{E}\left[\Big(X'\big(\theta - \theta_0(\tau)\big)\Big)^2\right].
	\end{align*}
	This proves the third statement.
\end{proof}

\begin{proof}[\textbf{Proof of Lemma~\ref{lemma:SizeConesDominatedCoordinates}}]
	By Lemma 7.1 (ii) in~\cite{koltchinskii2011oracle} the set $\mathcal{M} \subset B^p(0,1)$ is such that $C^p_1(J, \vartheta) \cap B^p(0,1) \subset 2(2 + \vartheta) \mathrm{conv}(\mathcal{M})$, $\|u\|_0 \leq s$ for all $u \in \mathcal{M}$, and $\mathrm{card}(\mathcal{M}) \leq 5^s \sum_{k=0}^s {p \choose k}$. By Proposition 3.6.4 in~\cite{gine2015mathematical} we have $\sum_{k=0}^s {p \choose k} \leq \frac{3}{2}\frac{p^s}{s!}  \leq \frac{3}{2}\left(\frac{ep}{s}\right)^s$ for $p \geq s + 2$.	A trivial modification of Koltchinskii's arguments yields the claim of the lemma for all $q \geq 1$.
\end{proof}

\begin{proof}[\textbf{Proof of Lemma~\ref{lemma:MaximaBiconvexFunction}}]
	For each $x \in \mathrm{conv}(\mathcal{X})$ there exist $n \in \mathbb{N}$ and $\lambda_1, \ldots, \lambda_n \geq 0$, $\sum_{i=1}^n \lambda_i = 1$, such that $x = \sum_{i=1}^n \lambda_i x_i$ for some $x_i \in \mathcal{X}$. Similarly, for each $y \in \mathrm{conv}(\mathcal{Y})$ there exist $m \in \mathbb{N}$ and $\mu_1, \ldots, \mu_m \geq 0$, $\sum_{j=1}^m \mu_j = 1$, such that $y = \sum_{j=1}^m \mu_i y_i$ for some $y_i \in \mathcal{Y}$. Thus, by biconvexity of $f$ and two applications of Jensen's inequality, for all $x \in \mathrm{conv}(\mathcal{X})$ and for all $y \in \mathrm{conv}(\mathcal{Y})$,
	\begin{align*}
		f(x,y) \leq \sum_{i=1}^n \sum_{j=1}^m \lambda_i \mu_j f(x_i, y_j) \leq \sum_{i=1}^n \sum_{j=1}^m \lambda_i \mu_j \sup_{x \in \mathcal{X}} \sup_{y \in \mathcal{Y}} f(x,y) = \sup_{x \in \mathcal{X}} \sup_{y \in \mathcal{Y}} f(x,y). 
	\end{align*}
	Thus, $\sup_{x \in \mathrm{conv}(\mathcal{X})} \sup_{y \in \mathrm{conv}(\mathcal{Y})} f(x,y) \leq \sup_{x \in \mathcal{X}} \sup_{y \in \mathcal{Y}} f(x,y)$. The reverse inequality holds trivially true since $\mathcal{X} \subseteq \mathrm{conv}(\mathcal{X})$ and $\mathcal{Y} \subseteq \mathrm{conv}(\mathcal{Y})$. The same arguments hold if $f$ is replaced by $|f|$. This concludes the proof.
\end{proof}
\begin{proof}[\textbf{Proof of Lemma~\ref{lemma:LocalizedLoss}}] 
	
	We plan to apply the maximal inequality from Theorem~\ref{theorem:MaxInequalityBernsteinOrlicz}. However, instead of applying Theorem~\ref{theorem:MaxInequalityBernsteinOrlicz} to the original process, we apply it to various auxiliary processes. We then use Lemma~\ref{lemma:Panchenko2003Handel} to assemble a bound for the original empirical process from the bounds of those auxiliary processes. The rationale behind this complicated approach is that a direct application of Theorem~\ref{theorem:MaxInequalityBernsteinOrlicz} leads to an upper bound involving the metric entropy integral over $C^p(\cup_{\tau \in \mathcal{T}}T_\theta(\tau), c_0) \cap B^p(0, r_0)$, for which we lack the tools to obtain tight estimates. In contrast, the bounds on the auxiliary processes involve metric entropy integrals of finite sets only which are easy to estimate.
	
	Denote the underlying probability space by $(\Omega, \mathcal{A}, \mathbb{P})$. To simplify notation, write $K(r_0, \tau) = C^p(T_\theta(\tau), c_0) \cap B^p(0, r_0)$. Let $\eta  \in (0,1)$ be arbitrary and $\mathcal{T}_\eta$ be an $\eta$-net with cardinality $\mathrm{card}(T_\eta) \leq 1 + 1/\eta$. We have the following decomposition:
	\begin{align}\label{eq:lemma:LocalizedLoss-1}
		\|\mathbb{G}_n\|_{\mathcal{G}} &\overset{(a)}{\leq} \sup_{\tau \in \mathcal{T}}\sup_{u \in K(r_0, \tau)} \left|\frac{1}{\sqrt{n}} \sum_{i=1}^n X_i'u - \mathbb{E}[X_i'u] \right| \nonumber\\
		& + \sup_{\tau \in \mathcal{T}}\sup_{\theta - \theta_0(\tau) \in K(r_0, \tau)} \left|\frac{1}{\sqrt{n}}\sum_{i=1}^n \phi_{\tau, X_i, Y_i}\big(X_i'\theta - X_i'\theta_0(\tau)\big) - \mathbb{E}\big[\phi_{\tau, X_i, Y_i}\big(X_i'\theta - X_i'\theta_0(\tau)\big)\big] \right| \nonumber\\
		&\overset{(b)}{\leq}\sup_{\tau \in \mathcal{T}}\sup_{u \in K(r_0, \tau)} \left|\frac{1}{\sqrt{n}} \sum_{i=1}^n X_i'u - \mathbb{E}[X_i'u] \right| \nonumber\\
		&+ \sup_{s \in \mathcal{T}_\eta} \sup_{\tau: |\tau -s| \leq \eta }\: \sup_{\theta - \theta_0(\tau) \in K(r_0, \tau)} \left|\frac{1}{\sqrt{n}}\sum_{i=1}^n \phi_{s,X_i,Y_i}\big(X_i'\theta - X_i'\theta_0(s)\big) - \mathbb{E}\big[\phi_{s,X_i,Y_i}\big(X_i'\theta - X_i'\theta_0(s)\big)\big] \right| \nonumber\\
		&+ \sup_{s \in \mathcal{T}_\eta} \sup_{\tau: |\tau -s| \leq \eta } \left| \frac{1}{\sqrt{n}}\sum_{i=1}^n \phi_{s,X_i,Y_i}\big(X_i'\theta_0(\tau) - X_i'\theta_0(s)\big)- \mathbb{E}\big[\phi_{s,X_i,Y_i}\big(X_i'\theta_0(\tau) - X_i'\theta_0(s)\big)\big]\right| \nonumber\\
		& = \mathbf{I} +  \mathbf{II} +  \mathbf{III},
	\end{align}
	where $(a)$ follows from Lemma~\ref{lemma:NewKnightsIdentity} $(iii)$ and $(b)$ follows from Lemma~\ref{lemma:NewKnightsIdentity} $(ii)$.
	
	\textbf{Bound on $\mathbf{I}$.} By Lemma~\ref{lemma:SizeConesDominatedCoordinates} and Assumption~\ref{assumption:SparsityQRVector} there exists $\mathcal{M} \subset B^p(0,1)$ with cardinality $\mathrm{card}(\mathcal{M}) \leq \frac{3}{2}\left(\frac{5ep}{s_\theta}\right)^{s_\theta}$, $\|u\|_0 \leq s_\theta$ for all $u \in \mathcal{M}$, and $K(1, \tau) \subset 2(2 + c_0) \mathrm{conv}(\mathcal{M})$ for all $\tau \in \mathcal{T}$. Hence, for $A \in \mathcal{A}$ arbitrary,
	\begin{align*}
		\mathbb{E}\left[\sup_{\tau \in \mathcal{T}}\sup_{u \in K(r_0, \tau)}\left|\frac{1}{\sqrt{n}} \sum_{i=1}^n X_i'u - \mathbb{E}[X_i'u] \right|\mid A\right] &\leq  2(2 + c_0) r_0 \: \mathbb{E}\left[\sup_{u \in \mathrm{conv}(\mathcal{M})}\left|\frac{1}{\sqrt{n}} \sum_{i=1}^n X_i'u - \mathbb{E}[X_i'u] \right| \mid A\right] .
	\end{align*}
	Note that $u \mapsto \frac{1}{n}\sum_{i=1}^n (X_i - \mathbb{E}[X_i])'u$ is linear. Hence, by Lemma~\ref{lemma:MaximaBiconvexFunction} the term on the right hand side in above display is equal to
	\begin{align}\label{eq:lemma:LocalizedLoss-2}
		2(2 + c_0) r_0 \: \mathbb{E}\left[\sup_{u \in \mathcal{M}}\left|\frac{1}{\sqrt{n}} \sum_{i=1}^n X_i'u - \mathbb{E}[X_i'u] \right|\mid A\right].
	\end{align}
	Now, note that for all $u_1,u_2 \in \mathcal{M}$,
	\begin{align*}
		\big\| (X - \mathbb{E}[X])'u_1 - (X - \mathbb{E}[X])'u_2 \big\|_{\psi_2} \lesssim \phi_{\max}^{1/2}(s_\theta)\|u_1 - u_2\|_2.
	\end{align*}
	Thus, by Theorem~\ref{theorem:MaxInequalityBernsteinOrlicz} and Remark~\ref{remark:theorem:MaxInequalityBernsteinOrlicz-IndexSet}, eq.~\eqref{eq:lemma:LocalizedLoss-2} can be upper bounded (up to a multiplicative constant) by
	\begin{align*}
		2(2 + c_0) r_0 \phi_{\max}^{1/2}(s_\theta) \sqrt{s_\theta \log(ep/s_\theta) + \log(1/ \mathbb{P}\{A\})}.
	\end{align*}
	Hence, by Lemma~\ref{lemma:MomentsToTails}, for all $t \geq 0$, with probability at least $1 - e^{-t}$,
	\begin{align}\label{eq:lemma:LocalizedLoss-3}
		\mathbf{I} \lesssim 2(2 + c_0) r_0 \phi_{\max}^{1/2}(s_\theta) \left(\sqrt{s_\theta \log(ep/s_\theta)} + \sqrt{t} \right).
	\end{align}
	\textbf{Bound on $\mathbf{II}$.}  Let $\{\varepsilon_i\}_{i=1}^n$ be a sequence of i.i.d. Rademacher random variables independent of $\{(Y_i, X_i)\}_{i=1}^n$. For $A \in \mathcal{A}$ arbitrary consider
	\begin{align}
		&\mathbb{E}\left[\sup_{\tau: |\tau -s| \leq \eta }\: \sup_{\theta - \theta_0(\tau) \in K(r_0, \tau)} \left|\frac{8}{\sqrt{n}}\sum_{i=1}^n 
		\varepsilon_i\big(X_i - \mathbb{E}[X_i]\big)'\big(\theta -\theta_0(s)\big)\right| \mid A\right] \nonumber\\
		&\leq \mathbb{E}\left[\sup_{\tau \in \mathcal{T} }\: \sup_{u \in K(r_0, \tau)} \left|\frac{8}{\sqrt{n}}\sum_{i=1}^n \varepsilon_i\big(X_i - \mathbb{E}[X_i]\big)'u\right| \mid A\right] \label{eq:lemma:LocalizedLoss-6}\\
		& +  \mathbb{E}\left[\sup_{\tau: |\tau -s| \leq \eta } \left|\frac{8}{\sqrt{n}}\sum_{i=1}^n \varepsilon_i\big(X_i - \mathbb{E}[X_i]\big)'\big(\theta_0(\tau) -\theta_0(s)\big)\right| \mid A\right].\label{eq:lemma:LocalizedLoss-7}
	\end{align}
	Note that the symmetrized summands $\varepsilon_i\big(X_i - \mathbb{E}[X_i]\big)'u$ have zero mean, are sub-Gaussian, and have the same second moments as $(X_i -  \mathbb{E}[X_i]\big)'u$. Hence, $\|\varepsilon_i\big(X_i - \mathbb{E}[X_i]\big)'u\|_{P, \psi_2} \lesssim \phi_{\max}^{1/2}(\|u\|_0)\|u\|_2$. Proceeding as in Step 1, we can therefore upper bound the term in eq.~\eqref{eq:lemma:LocalizedLoss-6} (up to a multiplicative constant) by
	\begin{align*}
		16(2 + c_0) r_0 \phi_{\max}^{1/2}(s_\theta) \sqrt{s_\theta \log(ep/s_\theta) + \log(1/ \mathbb{P}\{A\})}.
	\end{align*}
	Note that $\sup_{\tau, s \in \mathcal{T}}\|\theta(\tau) - \theta(s)\|_0 \leq 2s_\theta$. Hence, by Assumption~\ref{assumption:LipschitzQRVector}, for all $\tau, s \in \mathcal{T}$,
	\begin{align*}
		\big\| \varepsilon\big(X - \mathbb{E}[X]\big)'\theta(\tau) - \varepsilon\big(X - \mathbb{E}[X]\big)'\theta(s)\big\|_{\psi_2} \lesssim \phi_{\max}^{1/2}(2s_\theta) \|\theta(\tau) - \theta(s)\|_2 \lesssim \phi_{\max}^{1/2}(2s_\theta)  L_\theta|\tau - s|.
	\end{align*}
	Therefore, by Theorem~\ref{theorem:MaxInequalityBernsteinOrlicz}, Remark~\ref{remark:theorem:MaxInequalityBernsteinOrlicz-Lipschitz}, and Remark~\ref{remark:theorem:MaxInequalityBernsteinOrlicz-IndexSet}, eq.~\eqref{eq:lemma:LocalizedLoss-7} can be upper bounded (up to a multiplicative constant) by
	\begin{align*}
		2(2 + c_0) \eta \phi_{\max}^{1/2}(2s_\theta) L_\theta \sqrt{2s_\theta \log(ep/2s_\theta) + \log(1/ \mathbb{P}\{A\})}.
	\end{align*}
	Thus, by Lemma~\ref{lemma:MomentsToTails}, for all $t \geq 0$, with probability at least $1 - 2e^{-t}$,
	\begin{align}\label{eq:lemma:LocalizedLoss-4}
		\begin{split}
			&\sup_{\tau: |\tau -s| \leq \eta }\: \sup_{\theta - \theta_0(\tau) \in K(r_0, \tau)} \left|\frac{8}{\sqrt{n}}\sum_{i=1}^n \varepsilon_i\big(X_i - \mathbb{E}[X_i]\big)'\big(\theta -\theta_0(s)\big)\right| \\
			&\lesssim 16(2 + c_0) r_0 \phi_{\max}^{1/2}(s_\theta) \left(\sqrt{s_\theta \log(ep/s_\theta)} + \sqrt{t} \right) + 16(2 + c_0) \eta \phi_{\max}^{1/2}(2s_\theta)  L_\theta\left(\sqrt{2s_\theta \log(ep/2s_\theta)} + \sqrt{t} \right)
		\end{split}
	\end{align}
	
	We now turn the bound on this symmetrized process into a bound on $\mathbf{II}$. For any increasing and convex function $F$ we have
	\begin{align}
		&\mathbb{E}\left[F\left(\sup_{\tau: |\tau -s| \leq \eta }\: \sup_{\theta - \theta_0(\tau) \in K(r_0, \tau)} \left|\frac{1}{\sqrt{n}}\sum_{i=1}^n \phi_{s,X_i,Y_i}\big(X_i'\theta - X_i'\theta_0(s)\big) - \mathbb{E}\big[\phi_{s,X_i,Y_i}\big(X_i'\theta - X_i'\theta_0(s)\big)\big] \right|\right)\right] \nonumber\\
		&\overset{(a)}{\leq} \mathbb{E}\left[F\left(2\sup_{\tau: |\tau -s| \leq \eta }\: \sup_{\theta - \theta_0(\tau) \in K(r_0, \tau)} \left|\frac{1}{\sqrt{n}}\sum_{i=1}^n \varepsilon_i\phi_{s,X_i,Y_i}\big(X_i'\theta - X_i'\theta_0(s)\big)\right|\right)\right]\nonumber\\
		&\overset{(b)}{\leq}  \mathbb{E}\left[F\left(4\sup_{\tau: |\tau -s| \leq \eta }\: \sup_{\theta - \theta_0(\tau) \in K(r_0, \tau)} \left|\frac{1}{\sqrt{n}}\sum_{i=1}^n \varepsilon_i\big(X_i'\theta - X_i'\theta_0(s)\big)\right| \right)\right]\nonumber\\
		&\overset{(c)}{\leq}\mathbb{E}\left[F\left(8\sup_{\tau: |\tau -s| \leq \eta }\: \sup_{\theta - \theta_0(\tau) \in K(r_0, \tau)} \left|\frac{1}{\sqrt{n}}\sum_{i=1}^n \varepsilon_i\big(X_i - \mathbb{E}[X_i]\big)'\big(\theta -\theta_0(s)\big)\right| \right)\right]	\label{eq:lemma:LocalizedLoss-5}
	\end{align}
	where $(a)$ holds by Lemma 2.3.6 in~\cite{vaartwellner1996weak}, $(b)$ holds by Theorem 4.12 in~\cite{ledoux1996probability} and since by Lemma~\ref{lemma:NewKnightsIdentity} (i) $\phi_{s,X,Y}$ is a contraction, and $(c)$ holds again by Lemma 2.3.6 in~\cite{vaartwellner1996weak}.	
	
	By Lemma~\ref{lemma:Panchenko2003Handel}, eq.~\eqref{eq:lemma:LocalizedLoss-4} and~\eqref{eq:lemma:LocalizedLoss-5}, and the union bound over $\eta \in \mathcal{T}_\eta$, for all $t \geq 0$, with probability at least $1 - 2 e^{1-t}$,
	\begin{align}\label{eq:lemma:LocalizedLoss-8}
		\begin{split}
			\mathbf{II} &\lesssim 16(2 + c_0) r_0 \phi_{\max}^{1/2}(s_\theta) \left(\sqrt{s_\theta \log(ep/s_\theta)} + \sqrt{ \log(1 + 1/\eta) + t} \right)\\
			& + 16(2 + c_0) \eta \phi_{\max}^{1/2}(2s_\theta) L_\theta \left(\sqrt{2s_\theta \log(ep/2s_\theta)} + \sqrt{ \log(1 + 1/\eta) + t} \right).
		\end{split}
	\end{align}
	
	\textbf{Bound on $\mathbf{III}$.} We obtain a bound for this term using similar arguments as in Step 2. We skip the redundant details and simply note that, for all $t \geq 0$, with probability at least $1 -e^{1-t}$,
	\begin{align}\label{eq:lemma:LocalizedLoss-9}
		\mathbf{III} \lesssim 16(2 + c_0) \eta \phi_{\max}^{1/2}(2s_\theta)  L_\theta\left(\sqrt{2s_\theta \log(ep/2s_\theta)} + \sqrt{ \log(1 + 1/\eta) + t} \right).
	\end{align}
	
	\textbf{Conclusion.} Since $\eta \in (0,1)$ is arbitrary, we can choose $\eta \asymp r_0/ L_\theta$. Combine the bounds in eq.~\eqref{eq:lemma:LocalizedLoss-1},~\eqref{eq:lemma:LocalizedLoss-3},~\eqref{eq:lemma:LocalizedLoss-8} and~\eqref{eq:lemma:LocalizedLoss-9}, adjust the constants, and conclude that with probability at least $1 - \delta$,
	\begin{align*}
		\|\mathbb{G}_n\|_{\mathcal{G}} \lesssim 2(2 + c_0)\phi_{\max}^{1/2}(2s_\theta) r_0\sqrt{s_\theta \log(ep/s_\theta) + \log(1 +  L_\theta/r_0) + \log (1/\delta)}.
	\end{align*}
\end{proof}
\begin{proof}[\textbf{Proof of Lemma~\ref{lemma:MaxInequalityCovarianceCone}}] 
	To simplify notation, we write $K_k(\tau) = C^p_{q_k}(J_k(\tau), \vartheta_k) \cap B^p(0,1)$ for $k \in \{1, 2\}$.
	
	\textbf{Proof of Case (i).} By Lemma~\ref{lemma:SizeConesDominatedCoordinates} there exist $\mathcal{M}_1, \mathcal{M}_2 \subset B^p(0,1)$ such that
	\begin{align*}
		&\mathrm{card}(\mathcal{M}_k) \leq \frac{3}{2}\left(\frac{5ep}{s_k}\right)^{s_k}, \quad{} \forall u \in \mathcal{M}_k: \: \|u\|_0 \leq s_k, \quad{} \forall \tau \in \mathcal{T}, \: k \in \{1, 2\} : \: K_k(\tau) \subset 2(2 + \vartheta_k) \mathrm{conv}(\mathcal{M}_k).
	\end{align*}
	For these $\mathcal{M}_1, \mathcal{M}_2$ we define
	\begin{align*}
		\mathcal{G}_{\mathcal{M}_1, \mathcal{M}_2} = \big\{g(X, \xi) = \xi (X'u_1)(X'u_2):  \: u_k \in \mathcal{M}_k,  \: k \in \{1, 2\}\big\}.
	\end{align*}
	By Assumption~\ref{assumption:SubGaussianity} and Lemma~\ref{lemma:ProductSubgaussian},  for all $v_1, u_1 \in \mathcal{M}_1$ and $v_2, u_2 \in \mathcal{M}_2$,
	\begin{align*}
		&\left\| \big(\xi(X'u_1)(X'u_2) - \mathbb{E}[\xi(X'u_1)(X'u_2)]\big) - \big(\xi(X'v_1)(X'v_2) - \mathbb{E}[\xi(X'v_1)(X'v_2)]\big) \right\|_{P, \psi_1} \nonumber\\
		&\quad{} \lesssim \left\|\xi(X'u_1)(X'u_2) - \xi(X'v_1)(X'v_2)\right\|_{P, \psi_1} \nonumber\\
		&\quad{} \lesssim \sup_{w_1, w_2}\left\| (X'w_1)(X'w_2) \right\|_{P, \psi_1} \left( \|u_1 - v_1\|_2 + \|u_2 - v_2\|_2 \right) \nonumber\\
		&\quad{} \lesssim \varphi_{\max}^{1/2}(s_1)\varphi_{\max}^{1/2}(s_2)\left( \|u_1 - v_1\|_2 + \|u_2 - v_2\|_2 \right),
	\end{align*}
	where the supremum in the third line is taken over all $w_1, w_2$ such that $\|w_k\|_2 \leq 1$ and $\|w_k\|_0 \leq s_k$ for $k \in \{1, 2\}$. Thus, by Corollary~\ref{corollary:DeviationInequalityBernsteinOrlicz}, with probability at least $1 -\delta$, 
	\begin{align*}
		\|\mathbb{G}_n\|_{\mathcal{G}_{\mathcal{M}_1, \mathcal{M}_2}} \lesssim \varphi_{\max}^{1/2}(s_1)\varphi_{\max}^{1/2}(s_2) \left( \sqrt{t_{s_1, s_2, \delta}} + n^{-1/2} t_{s_1, s_2, \delta}\right),
	\end{align*}
	where $t_{s_1, s_2, \delta} = s_1 \log( ep/s_1) + s_2 \log( ep/s_2) + \log(1/\delta)$. Hence, by Lemma~\ref{lemma:MaximaBiconvexFunction} and construction of $\mathcal{M}$ we have, with probability at least $1- \delta$,
	\begin{align*}
		\|\mathbb{G}_n\|_\mathcal{G} \lesssim (2 + \vartheta_1)(2 + \vartheta_2) \varphi_{\max}^{1/2}(s_1)\varphi_{\max}^{1/2}(s_2) \left( \sqrt{t_{s_1, s_2, \delta}} + n^{-1/2} t_{s_1, s_2, \delta}\right).
	\end{align*}
	
	\textbf{Proof of Case (ii).} Let $\mathcal{M}_k$ be as in the proof of case (i) and define $\widetilde{\mathcal{M}}_k = \{u \in \mathbb{R}^p: u \in \mathcal{M}_k \mathrm{\:or\:} -u \in \mathcal{M}_k\}$. Then, $\mathrm{card}(\widetilde{\mathcal{M}}_k) \leq 2 \mathrm{card}(\mathcal{M}_k)$, $\|u\|_0 \leq s_k$ for all $u \in\widetilde{\mathcal{M}}_k$ and $K(\tau) \subset 2(2+\vartheta_k) \mathrm{conv}(\widetilde{\mathcal{M}}_k)$. The claim follows now by the same arguments used to proof case (i).
	
	\textbf{Proof of Case (iii).} The claim is a simple consequence from the fact that the proofs of cases (i) and (ii) rely on an $\varepsilon$-net approximation of $s$-sparse sets. In fact, the proofs of these cases establish case (iii) and then use Lemma~\ref{lemma:SizeConesDominatedCoordinates} to deduce the case of $v \in C^p_{q_k}(J_k(\tau), \vartheta_k) \cap B^p_2(0,1)$. This completes the proof.
	
\end{proof}
\begin{proof}[\textbf{Proof of Lemma~\ref{lemma:GradientCone}}]
	
	To simplify notation, we write $K(\tau) = C^p_q(J(\tau), \vartheta) \cap B^p(0,1)$.
	
	\textbf{Proof of Case (i).} By Lemma~\ref{lemma:SizeConesDominatedCoordinates} there exist $\mathcal{M} \subset B^p(0,1)$ such that
	\begin{align*}
		&\mathrm{card}(\mathcal{M}) \leq \frac{3}{2}\left(\frac{5ep}{s}\right)^s, \quad{} \forall u \in \mathcal{M}: \: \|u\|_0 \leq s, \quad{} \forall \tau \in \mathcal{T} : \: K(\tau) \subset 2(2 + \vartheta) \mathrm{conv}(\mathcal{M}).
	\end{align*}
	For this $\mathcal{M}$ we define
	\begin{align*}
		\mathcal{G}_\mathcal{M} &= \left\{g(X, Y, \xi) = \xi \left(\tau - \mathbf{1}\big\{Y \leq X'\theta_0(\tau) \}\right)X'v:  \: v \in \mathcal{M}, \: \tau \in \mathcal{T}\right\}.
	\end{align*}
	We observe the following: First, $\mathcal{G}_\mathcal{M} =\{hj: h \in \mathcal{H}, \: j \in \mathcal{J}_\mathcal{M}\} $, where $\mathcal{H} = \big\{h(X,Y) =  \tau - \mathbf{1}\{F_{Y|X}(Y|X) \leq \tau \} : \: \tau \in \mathcal{T} \big\}$ and $\mathcal{J}_\mathcal{M} = \{j(X, \xi) = \xi X'v :  v \in \mathcal{M}\}$. The set $\mathcal{H}$ is the difference of two VC-subgraph classes with VC-indices at most 2, respectively~\citep[][Lemma 2.6.15 and Example 2.6.1]{vaartwellner1996weak}. Thus, $\mathcal{H}$ is VC-subgraph class with VC-index at most $3$~\citep[][Lemma 2.6.18]{vaartwellner1996weak}. The function class $\mathcal{J}_\mathcal{M}$ is finite with $\mathrm{card}(\mathcal{J}_\mathcal{M}) = \mathrm{card}(\mathcal{M})$. By Assumption~\ref{assumption:SubGaussianity} and Lemma~\ref{lemma:ProductSubgaussian}, for $u_1, u_2, v_1, v_2 \in \mathcal{M}$ arbitrary,
	\begin{align*}
		&\left\| \xi^2(u_1'XX'u_2 - \mathbb{E}[\xi^2u_1'XX'u_2]) - (\xi^2v_1'XX'v_2 - \mathbb{E}[\xi^2v_1'XX'v_2]) \right\|_{P, \psi_1} \nonumber\\
		&\quad{} \lesssim \left\|\xi^2 u_1'XX'u_2 - \xi^2v_1'XX'v_2\right\|_{P, \psi_1} \nonumber\\
		&\quad{} \lesssim \sup_{w_1, w_2}\left\| (X'w_1)(X'w_2) \right\|_{P, \psi_1} \left( \|u_1 - v_1\|_2 + \|u_2 - v_2\|_2 \right) \nonumber\\
		&\quad{} \lesssim \varphi_{\max}(|S|)\left( \|u_1 - v_1\|_2 + \|u_2 - v_2\|_2 \right),
	\end{align*}
	where the supremum in the third line is taken over all $w_1, w_2$ such that $\|w_1\|_2, \|w_2\|_2 \leq 1$ and $\|w_1\|_0, \|w_2\|_0 \leq |S|$. Therefore, for $j_v, j_u \in \mathcal{J}_\mathcal{M}$,
	\begin{align*}
		\| (j_v^2 - Pj_v^2) - (j_u^2 - Pj_u^2) \|_{P, \psi_1} \lesssim \varphi_{\max}(|S|)\|v - u\|_2.
	\end{align*}
	Thus, by Corollary~\ref{corollary:MaxInequalityBernsteinOrlicz-Gradient}, with probability at least $1- \delta$,
	\begin{align*}
		\|\mathbb{G}_n\|_{\mathcal{G}_\mathcal{M}}  \lesssim  \varphi_{\max}^{1/2}(s) \sqrt{t_{s, \delta}} \sqrt{1 + \pi_{n,1}^2(t_{s, \delta})}
	\end{align*}
	where $ t_{s, \delta} = s \log(ep/s) + \log(1/\delta)$ and $\pi_{n,1}^2(z) = \sqrt{z/n} + z/n$ for $ z \geq  0$. Hence, by Lemma~\ref{lemma:MaximaBiconvexFunction} and construction of $\mathcal{M}$ we have, with probability at least $1- \delta$,
	\begin{align*}
		\|\mathbb{G}_n\|_\mathcal{G} \lesssim 2 (2 + \vartheta) \varphi_{\max}^{1/2}(s) \sqrt{t_{s, \delta}} \sqrt{1 + \pi_{n,1}^2(t_{s, \delta})}.
	\end{align*}
	
	\textbf{Proof of Case (ii).} The claim is a simple consequence from the fact that the proof of case (i) uses an $\varepsilon$-net approximation of $s$-sparse sets. In fact, the proof of case (i) establish first case (ii) and then invokes Lemma~\ref{lemma:SizeConesDominatedCoordinates} to deduce the case of $v \in C^p_q(J(\tau), \vartheta) \cap B^p_2(0,1)$. This completes the proof.
	
	\textbf{Proof of Case (iii).} Repeat the proofs of cases (i) and (ii) but use the $\psi_{1/2}$-Orlicz norm to bound the increments and apply Corollary~\ref{corollary:MaxInequalityBernsteinOrlicz-Gradient} with $\alpha = 1/2$.
\end{proof}
\begin{proof}[\textbf{Proof of Lemma~\ref{lemma:LocalizedRankScores}}]
	\noindent
	
	\textbf{Proof of Case (i).}	Let $S \subset \{1, \ldots, p\}$ be arbitrary.  By Lemma~\ref{lemma:SizeConesDominatedCoordinates} there exist $\mathcal{M} \subset B^p(0,1)$ such that
	\begin{align*}
		&\mathrm{card}(\mathcal{M}) \leq \frac{3}{2}\left(\frac{5ep}{|S|}\right)^{|S|}, \quad{} \forall u \in \mathcal{M}: \: \|u\|_0 \leq |S|, \quad{} B^p(0,1) \subset 4 \mathrm{conv}(\mathcal{M}).
	\end{align*}
	In the following it is understood that $\mathcal{M}$ is a function of $S$ and we will not make this dependence explicit. For $T \subset \{1, \ldots, p\}$ and the pair $(S, \mathcal{M})$ we define
	\begin{align*}
		\mathcal{G}_{S,T, \mathcal{M}} &= \left\{g(X, Y, \xi) = \xi \left(\mathbf{1}\big\{Y \leq X'\theta \big\} - \mathbf{1}\big\{Y \leq X'\theta_0(\tau) \}\right)X'v:  \theta \in \mathbb{R}^p, \: \mathrm{supp}(\theta) = T, \: v \in \mathcal{M}, \: \tau \in \mathcal{T}\right\}.
	\end{align*}
	We make the following observations: First, each $g \in \mathcal{G}_{S, T, \mathcal{M}}$ is parameterized (uniquely) by the triple $(v, \theta, \tau)$. Hence, we will write $g = g_{v, \theta, \tau}$ whenever we need to highlight the dependence on the parameters. Second, $\mathcal{G}_{S, T, \mathcal{M}} =\{hj: h \in \mathcal{H}_T, \: j \in \mathcal{J}_{S, \mathcal{M}}\} $, where $\mathcal{H}_T = \big\{h(X,Y) =  \mathbf{1}\{Y \leq X'\theta \} - \mathbf{1}\{F_{Y|X}(Y|X) \leq \tau \} : \theta \in \mathbb{R}^p,\: \mathrm{supp}(\theta) = T, \: \tau \in \mathcal{T} \big\}$ and $\mathcal{J}_{S,\mathcal{M}} = \{j(X, \xi) = \xi X'v :  v \in \mathcal{M}\}$. In particular, for every $g= g_{v, \theta, \tau} \in \mathcal{G}_{S,T, \mathcal{M}}$ there exist unique $h_{\theta, \tau}  \in \mathcal{H}_T$ and $j_v \in \mathcal{J}_{S, \mathcal{M}}$ such that $g_{v, \theta, \tau} = h_{\theta, \tau}j_v$. The set $\mathcal{H}_T$ is the difference of two VC-subgraph classes with VC-indices at most $|T| + 3$~\citep[][Lemma 2.6.15]{vaartwellner1996weak} and 2~\citep[][Theorem 4.10 (a)]{dudley2014uniform}, respectively. Thus, $\mathcal{H}_T$ is VC-subgraph class with VC-index at most $|T| + 4$~\citep[][Lemma 2.6.18]{vaartwellner1996weak}. The function class $\mathcal{J}_{S,\mathcal{M}}$ is finite with $\mathrm{card}(\mathcal{J}_{S, \mathcal{M}}) = \mathrm{card}(\mathcal{M})$. By Assumption~\ref{assumption:SubGaussianity} and Lemma~\ref{lemma:ProductSubgaussian}, for $u_1, u_2, v_1, v_2 \in \mathcal{M}$ arbitrary,
	\begin{align*}
		&\left\| \xi^2(u_1'XX'u_2 - \mathbb{E}[\xi^2u_1'XX'u_2]) - (\xi^2v_1'XX'v_2 - \mathbb{E}[\xi^2v_1'XX'v_2]) \right\|_{P, \psi_1} \nonumber\\
		&\quad{} \lesssim \left\|\xi^2 u_1'XX'u_2 - \xi^2v_1'XX'v_2\right\|_{P, \psi_1} \nonumber\\
		&\quad{} \lesssim \sup_{w_1, w_2}\left\| (X'w_1)(X'w_2) \right\|_{P, \psi_1} \left( \|u_1 - v_1\|_2 + \|u_2 - v_2\|_2 \right) \nonumber\\
		&\quad{} \lesssim \varphi_{\max}(|S|)\left( \|u_1 - v_1\|_2 + \|u_2 - v_2\|_2 \right),
	\end{align*}
	where the supremum in the third line is taken over all $w_1, w_2$ such that $\|w_1\|_2, \|w_2\|_2 \leq 1$ and $\|w_1\|_0, \|w_2\|_0 \leq |S|$. Therefore, for $j_v, j_u \in \mathcal{J}_{S,\mathcal{M}}$,
	\begin{align*}
		\| (j_v^2 - Pj_v^2) - (j_u^2 - Pj_u^2) \|_{P, \psi_1} \lesssim \varphi_{\max}(|S|)\|v - u\|_2.
	\end{align*}	
	Thus, by Corollary~\ref{corollary:MaxInequalityBernsteinOrlicz-Gradient} there exists an absolute constant $C_1 > 0$ (independent of $S, T, n, p, \mathcal{M}$) such that for all $t > 0$,
	\begin{align*}
		\mathbb{P}\left\{ \|\mathbb{G}_n\|_{\mathcal{G}_{S, T, \mathcal{M}}}  > C_1 \varphi_{\max}^{1/2}(|S|) \Pi_{n,p}(|S|, |T|, t) \right\}	\leq 3e^{1-t},
	\end{align*}
	where
	\begin{align*}
		\Pi_{n,p}(|S|, |T|, t) = \left(\sqrt{|T| + |S|\log(ep/|S|)} + \sqrt{t} \right) \left(\sqrt{1 + \pi_{n,1}^2\big(|S|\log(ep/|S|)\big)} +  \sqrt{\pi_{n,1}^2(t)}\right),
	\end{align*}
	and $\pi_{n,1}^2(z) = \sqrt{z/n} + z/n$ for $ z \geq  0$.
	
	Note that $\mathrm{card}\big(\{ S\subseteq \{1, \ldots, p\} : |S| \leq k \}\big) \leq \left(ep/k\right)^k$ and $\mathrm{card}\big(\{ T \subseteq \{1, \ldots, p\} : |T| \leq \ell \}\big) = \sum_{i=1}^\ell  { p \choose i} \leq \left(ep/\ell\right)^\ell$. Therefore, setting $t_{k,\ell,n, \delta} = k \log (ep/k) + \ell \log (ep/\ell) + 2\log n + \log(3e/\delta)$ and applying the union bound over above tail probabilities gives
	\begin{align}\label{eq:lemma:LocalizedRankScores-1}
		\mathbb{P}\left\{\sup_{1\leq k, \ell \leq n}\sup_{\mathrm{card}(S) \leq k} \sup_{\mathrm{card}(T) \leq\ell}\frac{  \|\mathbb{G}_n\|_{\mathcal{G}_{S, T, \mathcal{M}}}}{\varphi_{\max}^{1/2}(k) \Pi_{n,p}(k, \ell, t_{k, \ell, n, \delta}) }  > C_1 \right\}\leq 3e\sum_{k=1}^n\sum_{\ell=1}^n \left(\frac{ep}{k}\right)^k \left(\frac{ep}{\ell}\right)^\ell e^{-t_{k,\ell, n, \delta}} \leq \delta.
	\end{align}
	
	Next, for $S, T\subset \{1, \ldots, p\}$ arbitrary, define
	\begin{align*}
		\mathcal{G}_{S,T} = \left\{g(X, Y, \xi) = \xi\left(\mathbf{1}\big\{Y \leq X'\theta \big\} - \mathbf{1}\big\{Y \leq X'\theta_0(\tau) \}\right)X'v : \right.\\
		\left. \theta, v \in \mathbb{R}^p, \: \mathrm{supp}(\theta) = T, \: \mathrm{supp}(v) = S, \: \|v\|_2 \leq 1, \: \tau \in \mathcal{T}\right\}.
	\end{align*}
	By Lemma~\ref{lemma:MaximaBiconvexFunction} and construction of $\mathcal{M}$ we have
	\begin{align}\label{eq:lemma:LocalizedRankScores-2}
		\|\mathbb{G}_n\|_{\mathcal{G}_{S,T}} \leq 4 \|\mathbb{G}_n\|_{\mathcal{G}_{S, T, \mathcal{M}}},
	\end{align}
	and
	\begin{align}\label{eq:lemma:LocalizedRankScores-3}
		\mathcal{G} = \bigcup_{S , T\subset \{1, \ldots p\}, \:\mathrm{card}(S), \mathrm{card}(T) \leq n} \mathcal{G}_{S,T}.
	\end{align}	
	Hence, eq.~\eqref{eq:lemma:LocalizedRankScores-1}--\eqref{eq:lemma:LocalizedRankScores-3} imply	
	\begin{align*}
		&\mathbb{P}\left\{\exists g_{v, \theta, \tau} \in \mathcal{G} :\frac{\mathbb{G}_n(g_{v, \theta, \tau})}{\varphi_{\max}^{1/2}(\|v\|_0) \Pi_{n,p}(\|v\|_0, \|\theta\|_0, t_{\|v\|_0, \|\theta\|_0, n, \delta}) }  > 4C_1 \right\} \leq \delta.
	\end{align*}
	Lastly, there exists an absolute constant $C_2 > 0$ such that
	\begin{align*}
		\Pi_{n,p}(\|v\|_0, \|\theta\|_0, t_{\|v\|_0, \|\theta\|_0, n, \delta}) \leq C_2\sqrt{t_{\|v\|_0, \|\theta\|_0, n,\delta}} \sqrt{1 + \pi_{n,1}^2(t_{\|v\|_0, \|\theta\|_0, n, \delta})}.
	\end{align*}
	This completes the proof of case (i).
	
	\textbf{Proof of Case (ii).} Observe that $s \mapsto s \log (ep/s)$ and $s \mapsto \varphi_{\max}(s)$ are monotone increasing on $[1, p]$. Thus, the bound of case (i) for $g_{v, \theta, \tau} \in \mathcal{G}$ with $\|v\|_0 = \|\theta\|_0 = m$ holds also for all $g_{v', \theta', \tau'} \in \mathcal{G}$ with $\|v'\|_0, \|\theta'\|_0 \leq m$. To conclude, adjust some constants.
\end{proof}
\begin{proof}[\textbf{Proof of Lemma~\ref{lemma:MaxInequalityCovariance}}]
	Fix $k \leq n \wedge p$, fix $I \subseteq \{1, \ldots, n\}$ with $\mathrm{card}(I)=k$,  and fix the support set of $u \in \mathbb{R}^p$ with $\|u\|_0 = \ell$. By Lemma~\ref{lemma:MaxInequalityCovarianceCone} with $\vartheta = 0$, there exists an absolute constant $C > 0$ such that with probability at least $1- 2e^{-t}$,
	\begin{align*}
		\sup_{\|u\|_2 \leq 1} \left|\frac{1}{k}\sum_{i \in I} (X_i'u)^2 - \mathbb{E}[(X_i'u)^2]\right| \leq C \varphi_{\max}(\ell) \left(\sqrt{\frac{\ell + t}{k}}+ \frac{\ell + t}{k}\right).
	\end{align*}
	
	Note that $\mathrm{card}\big(\{ I \subseteq \{1, \ldots, n\} : \mathrm{card}(I) \leq k \}\big) = \sum_{i=1}^k  { n \choose i} \leq \left(\frac{en}{k}\right)^k$ and $\mathrm{card}\big(\{ u \in \{0,1\}^p: \|u\|_0 \leq k  \}\big) = \sum_{i=1}^k { p \choose i } \leq \left(\frac{ep}{k}\right)^k$. Therefore, setting $t_{k,\ell, n} = k \log (ep/k) + k\log(en/k) + \log( 2n/\delta)$ and applying the union bound gives
	\begin{align*}
		&\mathbb{P} \left\{\exists k \leq n : \frac{\sup_{\mathrm{card}(I) \leq k} \sup_{\ell \leq k}\sup_{\|u\|_2 \leq 1, \|u\|_0 = \ell} \left|\frac{1}{k}\sum_{i \in I} (X_i'u)^2 - \mathbb{E}[(X_i'u)^2]\right|}{ \varphi_{\max}(\ell) \left(\sqrt{\frac{\ell +t_{k,\ell, n}}{k}}+ \frac{\ell + t_{k,\ell, n}}{k}\right)} \geq C\right\}\\
		&\quad{} \leq 2\sum_{k=1}^n\left(\frac{ep}{k}\right)^k\left(\frac{en}{k}\right)^k e^{-t_{k,\ell, n}} \leq \delta.
	\end{align*}
	Upper bound $\ell$ by $k$, simplify the expression, and conclude.
\end{proof}

\subsection{Proofs of Section~\ref{subsec:DualProblem}}
\begin{proof}[\textbf{Proof of Lemma~\ref{lemma:CharacterizationBalancingEstimator}}]
	\noindent
	
	\textbf{Proof of statement (i).} To simplify the discussion we introduce the following matrices and vectors
	\begin{align*}
		\widehat{\Psi}(\tau) &= 2\: \mathrm{diag}\left(\hat{f}_1^{-2}(\tau), \: \ldots, \: \hat{f}_n^{-2}(\tau)\right)\:\:\in \mathbb{R}^{n \times n},\\
		A &= n^{-1/2}\left[- \mathbf{X},\: \mathbf{X} \right]' \:\:\in \mathbb{R}^{2p \times n},\\
		b &= \left[\frac{\gamma}{n}\mathbf{1}_p'  - z' ,\: \frac{\gamma}{n}\mathbf{1}_p' + z'\right]' \:\: \in \mathbb{R}^{2p},
	\end{align*}
	and rewrite the convex optimization problem~\eqref{eq:BalancingWeights-Primal-Appendix} in standard matrix form as
	\begin{align*}
		\min_{w \in \mathbb{R}^n}\:\:\:& \frac{1}{2}w'\widehat{\Psi}(\tau)w \\
		\mathrm{s.t.\:\:\:}& A w \preceq b.
	\end{align*}
	Recall that the dual of above optimization problem is given by
	\begin{align*}
		\max_{\lambda \in \mathbb{R}^{2p}}\:\:\:& -\frac{1}{2} \lambda'A \widehat{\Psi}^{-1}(\tau)A'\lambda - \lambda'b \\
		\mathrm{s.t.\:\:\:}& \lambda \succeq 0.
	\end{align*}
	By assumption the primal problem is feasible. Thus, strong duality holds and the solution to the primal, $\widehat{w}(\tau; z) \in \mathbb{R}^n$, and the solution to the dual, $\hat{\lambda}(\tau) \in \mathbb{R}^{2p}$, satisfy
	\begin{align}\label{eq:CharacterizationBalancingEstimator-1}
		\widehat{w}(\tau; z) = - \widehat{\Psi}^{-1}(\tau)A' \hat{\lambda}(\tau).
	\end{align}
	Since $\hat{\lambda}(\tau) \in \mathbb{R}^{2p}$ is just the vector of optimal Lagrange multipliers associated with the original problem formulation in~\eqref{eq:BalancingWeights-Primal-Appendix}, it also satisfies the complementary slackness conditions
	\begin{align*}
		\hat{\lambda}_k(\tau) \geq 0 \hspace{20pt} \mathrm{and} \hspace{20pt} \big[A\widehat{w}(\tau)\big]_k < b_k \implies \hat{\lambda}_k(\tau) = 0, \hspace{20pt} k=1, \ldots, 2p.
	\end{align*}
	Write $\hat{\lambda}(\tau) = \left(\overline{\mu}, \: \underline{\mu} \right)$, where $\overline{\mu}, \underline{\mu} \in \mathbb{R}^p_+$ are the optimal Lagrange multipliers associate with the box constraints of COP~\eqref{eq:BalancingWeights-Primal-Appendix}. The complementary slackness conditions imply that
	\begin{align*}
		\overline{\mu}_k \underline{\mu}_k = 0, \quad{} k=1, \ldots, p.
	\end{align*}
	Set $\hat{v}(\tau; z) := \left(-\overline{\mu} + \underline{\mu}\right) \in \mathbb{R}^p$. Above identity implies that
	\begin{align*}
		|\hat{v}_k(\tau; z)| = \overline{\mu}_k+ \underline{\mu}_k, \quad{} k=1, \ldots, p,
	\end{align*}
	and hence
	\begin{align}\label{eq:CharacterizationRankBalancedEstimator-3}
		-\frac{1}{2} \hat{\lambda}'(\tau)A \widehat{\Psi}^{-1}(\tau)A'\hat{\lambda}(\tau) - \hat{\lambda}'(\tau)b = -\frac{1}{2n}\hat{v}'(\tau; z) \mathbf{X}'\widehat{\Psi}^{-1}(\tau) \mathbf{X}\hat{v}(\tau; z)   - z' \hat{v}(\tau; z) - \frac{\gamma}{n}\|\hat{v}(\tau; z)\|_1.
	\end{align}
	Since every $v \in \mathbb{R}^p$ can be written as $v = \left(-\overline{\mu} + \underline{\mu}\right)$ for some $\overline{\mu}, \underline{\mu} \in \mathbb{R}^p_+$, identity~\eqref{eq:CharacterizationRankBalancedEstimator-3} implies that $\hat{v}(\tau;z)$ is the solution to the unconstrained COP
	\begin{align*}
		\min_{v \in \mathbb{R}^p} \frac{1}{2n}v'\mathbf{X}'\widehat{\Psi}^{-1}(\tau) \mathbf{X}v + z'v + \frac{\gamma}{n}\|v\|_1.
	\end{align*}
	Lastly, note that 
	\begin{align*}
		\frac{1}{2n}v'\mathbf{X}'\widehat{\Psi}^{-1}(\tau)\mathbf{X}v = \frac{1}{4n}\sum_{i=1}^n \hat{f}_i^2(\tau)\big( X_i'v\big)^2.
	\end{align*}
	This concludes the proof of the first statement.
	
	\textbf{Proof of statement (ii).} Note that the primal problem~\ref{eq:BalancingWeights-Primal-Appendix} is convex. To establish the claim it is therefore sufficient to verify that Slater's condition holds with probability at least $1 - \delta$. To this end, define
	\begin{align*}
		w^*(z) = n^{-1/2}\mathbf{X} \mathbb{E}[XX']^{-1} z.
	\end{align*}
	Note that $n^{-1/2} \mathbb{E}[\mathbf{X}'w^*(z)] = z$. Hence, by Assumption~\ref{assumption:SubGaussianity} the $z_1 - n^{-1/2}e_1'\mathbf{X}'w^*(z), \ldots, z_p - n^{-1/2}e_p'\mathbf{X}'w^*(z)$ are centered (non-identical and dependent) sub-exponential random variables. (Here, $e_k$ denotes the $k$th standard unit vector in $\mathbb{R}^p$.) Moreover, $z_k - n^{-1/2}e_k'\mathbf{X}'w^*(z) = -\frac{1}{n}\sum_{i=1}^n (X_{ik}X_i'  \mathbb{E}[XX']^{-1} z - z_k)$ for all $1 \leq k \leq p$. By Lemma~\ref{lemma:ProductSubgaussian} and Assumption~\ref{assumption:SubGaussianity}, for all $1 \leq i \leq n$
	\begin{align*}
		&\max_{1 \leq k \leq p} \|X_{ik}X_i' \mathbb{E}[XX']^{-1} z - z_k\|_{\psi_1}\\
		&\quad{}\lesssim \max_{1 \leq k \leq p} \left\|\ X_{ik}X_i'  \mathbb{E}[XX']^{-1} z \right\|_{\psi_1} \\
		&\quad{}\lesssim \max_{1 \leq k \leq p}\max_{1 \leq i \leq n}\|X_{ik}\|_{\psi_2} \|X_i' \mathbb{E}[XX']^{-1} z\|_{\psi_2}\\
		&\quad{}\lesssim \frac{\bar{f}^2\varphi_{\max}^{1/2}(1)\varphi_{\max}^{1/2}(s_v)}{\kappa_2(\infty)}\|z\|_2.
	\end{align*}
	Hence, the union bound followed by Bernstein's inequality implies that for all $t \geq 0$, there exists an absolute constant $C > 0$ such that
	\begin{align*}
		\mathbb{P}\left\{\big\|z - n^{-1/2}\mathbf{X}'w^*(z)\big\|_\infty > t\right\} &\leq p \max_{1 \leq k \leq p} \mathbb{P}\left\{\left|\frac{1}{n}\sum_{i=1}^n (X_{ik}X_i'\mathbb{E}[XX']^{-1} z - z_k)\right| > t\right\} \\
		&\leq 2p \exp\left(-C \min\left\{\frac{t\kappa_2(\infty)}{\bar{f}^2\varphi_{\max}^{1/2}(1)\varphi_{\max}^{1/2}(s_v)\|z\|_2}, \: \frac{t^2\kappa_2(\infty)}{\bar{f}^2\varphi_{\max}(1)\varphi_{\max}^{1/2}(s_v)\|z\|_2^2} \right\} n\right).
	\end{align*}
	Set $t^* = \max\{s, s^2\}$ with $s = \frac{\bar{f}^2\varphi_{\max}^{1/2}(1)\varphi_{\max}^{1/2}(s_v)}{\kappa_2(\infty)}\|z\|_2 \left(\frac{\log 2p + \log (C/\delta)}{n}\right)^{1/2}$ and conclude that with probability at least $1-\delta$ the constraint set is non-empty whenever $\gamma > t^*$, i.e. Slater's condition holds. Lastly, observe that by eq.~\eqref{eq:CharacterizationBalancingEstimator-1},
	\begin{align*}
		\widehat{w}(\tau; z) = - n^{-1/2}\widehat{\Psi}^{-1}(\tau)\mathbf{X}\hat{v}(\tau; z).
	\end{align*}
	This concludes the proof of the second statement.
\end{proof}
\begin{proof}[\textbf{Proof of Theorem~\ref{theorem:ConsistencyDual}}]
	
	\noindent
	
	\textbf{Ansatz.}
	Let $\tau \in \mathcal{T}$ and $r > \frac{2\bar{c}\|z\|_2^2}{\kappa^2_2(\infty)}\frac{\mu}{\gamma} \geq 0$ (to be specified below). Recall that  $B^p_q(0, r ) := \{u \in \mathbb{R}^p : \|u\|_q \leq r \}$ and that $C^p_q(J, \vartheta) = \left\{ u \in \mathbb{R}^p: \|u_{J^c}\|_q \leq  \vartheta\|u_{J}\|_q \right\}$. To simplify notation, we write $\tilde{v}(\tau)$, $\hat{v}(\tau)$,  $T_v(\tau)$, and $s_v$ for $\tilde{v}(\tau; z)$, $\hat{v}(\tau;z)$, $T_v(\tau;z)$, and $s_v(z)$, respectively.
	Define
	\begin{align*}
		K(\tau) : = \left\{ v \in \mathbb{R}^p :
		\begin{matrix}
			v - \tilde{v}(\tau) \in C^p_1\big(T_v(\tau), 2 \bar{c}\big) \cup B^p_1\left(0,  \frac{2\bar{c}\|z\|_2^2}{\kappa_2^2(\infty)}\frac{\mu}{\gamma} \right),\\
			\bar{c}\sum_{k \in T_v(\tau)} \big|v_{k}(\tau) - \tilde{v}_k(\tau)\big| + \frac{\bar{c}\|z\|_2^2}{\kappa_2^2(\infty)} \frac{\mu}{\gamma}  \geq \sum_{k \in T_v^c(\tau)}\big|v_{k}(\tau) \big|
		\end{matrix}
		\right\}.
	\end{align*}
	Suppose that, with high probability,
	\begin{itemize}
		\item[(a)] $\hat{v}(\tau) \in  K(\tau)$ for all $\tau \in \mathcal{T}$; and
		\item[(b)] the centered dual objective function is strictly positive when evaluated at points $(v, \tau) \in \mathbb{R}^p \times \mathcal{T}$ satisfying $v \in K(r, \tau) : = K(\tau) \cap B^p_2(\tilde{v}(\tau), r)$, i.e.	
		\begin{align*}
			\inf_{\tau \in \mathcal{T}}\inf_{ v \in K(r, \tau) }  \left(\frac{1}{4}\sum_{i=1}^n \left(\hat{f}_i^2(\tau) (X_i'v)^2 - \hat{f}_i^2(\tau) \big(X_i'\tilde{v}(\tau)\big)^2\right) + n z'\big(v- \tilde{v}(\tau)\big) + \gamma\big(\|v\|_1 - \|\tilde{v}(\tau)\|_1\big)\right) > 0.
		\end{align*}
	\end{itemize}
	Since the dual objective function is convex in $v$ and negative at $\hat{v}(\tau)$ for all $\tau \in \mathcal{T}$, it then follows that $\sup_{\tau \in \mathcal{T}}\|\hat{v}_\gamma(\tau) - \tilde{v}(\tau) \|_2 \lesssim r$. Thus, to establish the claim of the theorem, we only need to prove  that statements (a) and (b) hold with high probability.
	
	\textbf{Verification of high probability statements.} By assumption, eq.~\eqref{eq:lemma:RestrictedConeProp-Dual-1} and~\eqref{eq:lemma:RestrictedConeProp-Dual-11} hold true. Thus, by Lemma~\ref{lemma:RestrictedConeProp-Dual}, statement (a) holds for all $\tau \in \mathcal{T}$ with probability one. We now establish statement (b). To this end, we define
	\begin{align*}
		\mathcal{G}_1 &= \{ g : \mathbb{R}^p \rightarrow \mathbb{R}: g(X) = f_{Y|X}^2(X'\theta_0(\tau)|X) v_0'(\tau)XX'\big(v - \tilde{v}(\tau)\big), \:v \in K(r, \tau),\: \tau \in \mathcal{T}\},\\
		\mathcal{G}_2 &= \{ g : \mathbb{R}^p \rightarrow \mathbb{R}: g(X) = f_{Y|X}^2(X'\theta_0(\tau)|X) \big(X'(\tilde{v}(\tau) - v)\big)^2,\:v \in K(r, \tau), \: \tau \in \mathcal{T}\},\\
		\mathcal{G}_3 &= \{ g : \mathbb{R}^p \rightarrow \mathbb{R}: g(X) = f_{Y|X}^2(X'\theta_0(\tau)|X) \big(\tilde{v}(\tau) - v_0(\tau)\big)' XX'\big(v - \tilde{v}(\tau)\big),\:v \in K(r, \tau), \: \tau \in \mathcal{T}\}.
	\end{align*}	
	We begin with two preliminary observations.
	
	First, note that by Assumptions~\ref{assumption:RelativeConsistencyDensity} and~\ref{assumption:BoundedDensity}, with probability at least $1- \eta$,
	\begin{align}\label{eq:theorem:ConsistencyDual-0}
		\begin{split}
			\left|\hat{f}_i^2(\tau) - f_i^2(\tau)\right| &= \left|\big(\hat{f}_i(\tau) - f_i(\tau)\big)\big( \hat{f}_i(\tau) + f_i(\tau)\big)\right| =  \big(\hat{f}_i(\tau) - f_i(\tau)\big)^2 + 2 \big|\hat{f}_i(\tau) - f_i(\tau)\big| f_i(\tau)\\
			&\leq \left|\frac{\hat{f}_i(\tau)}{f_i(\tau)}-1\right|^2 f_i^2(\tau) + 2 \left|\frac{\hat{f}_i(\tau)}{f_i(\tau)} - 1\right|f_i^2(\tau) \lesssim r_f f_i^2(\tau).
		\end{split}
	\end{align}	
	Therefore, with probability at least $1- \eta$, for all $\tau \in \mathcal{T}$,
	\begin{align*}
		&\frac{1}{4n}\sum_{i=1}^n \hat{f}_i^2(\tau)(X_i'v)^2 - \hat{f}_i^2(\tau) \big(X_i'\tilde{v}(\tau)\big)^2 + z'\big(v- \tilde{v}(\tau)\big)\\
		&\quad{}=\frac{1}{2n} \sum_{i=1}^n f_i^2(\tau) \tilde{v}'(\tau)X_iX_i'\big( v- \tilde{v}(\tau)\big) + \frac{1}{4n} \sum_{i=1}^n f_i^2(\tau) \big( v-\tilde{v}(\tau)\big)'X_iX_i'\big( v- \tilde{v}(\tau)\big) + z'\big(v- \tilde{v}(\tau)\big)\nonumber\\
		&\quad{}\quad{} + \frac{1}{2n} \sum_{i=1}^n \big(\hat{f}_i^2(\tau)- f_i^2(\tau)\big)\tilde{v}'(\tau)X_iX_i'\big( v- \tilde{v}(\tau)\big) + \frac{1}{4n} \sum_{i=1}^n \big(\hat{f}_i^2(\tau)- f_i^2(\tau)\big)\big( v- \tilde{v}(\tau)\big)'X_iX_i'\big( v- \tilde{v}(\tau)\big)\nonumber\\
		&\quad{}=\frac{1}{2n} \sum_{i=1}^n f_i^2(\tau) v_0(\tau)X_iX_i'\big( v- \tilde{v}(\tau)\big) + z'\big(v- \tilde{v}(\tau)\big)\nonumber\\
		&\quad{}\quad{} + \frac{1}{2n} \sum_{i=1}^n f_i^2(\tau) \big(\tilde{v}(\tau)- v_0(\tau)\big)X_iX_i'\big( v- \tilde{v}(\tau)\big) + \frac{1}{4n} \sum_{i=1}^n f_i^2(\tau) \big( v-\tilde{v}(\tau)\big)'X_iX_i'\big( v- \tilde{v}(\tau)\big) \nonumber\\
		&\quad{}\quad{} + \frac{1}{2n} \sum_{i=1}^n \big(\hat{f}_i^2(\tau)- f_i^2(\tau)\big)v_0'(\tau)X_iX_i'\big( v- \tilde{v}(\tau)\big) + \frac{1}{2n} \sum_{i=1}^n \big(\hat{f}_i^2(\tau)- f_i^2(\tau)\big)\big(\tilde{v}(\tau) - v_0(\tau)\big)'X_iX_i'\big( v- \tilde{v}(\tau)\big) \nonumber\\
		&\quad{}\quad{} + \frac{1}{4n} \sum_{i=1}^n \big(\hat{f}_i^2(\tau)- f_i^2(\tau)\big)\big( v- \tilde{v}(\tau)\big)'X_iX_i'\big( v- \tilde{v}(\tau)\big)\nonumber\\
		&\quad{}\gtrsim \mathbb{E}\left[\frac{1}{4n} \sum_{i=1}^n f_i^2(\tau) \big( v- \tilde{v}(\tau)\big)'X_iX_i'\big( v- \tilde{v}(\tau)\big)\right] + \mathbb{E}\left[\frac{1}{n} \sum_{i=1}^n f_i^2(\tau) \big(\tilde{v}(\tau)- v_0(\tau)\big)'X_iX_i'\big( v- \tilde{v}(\tau)\big)\right]\\
		&\quad{}\quad{} - n^{-1/2}\|\mathbb{G}_n\|_{\mathcal{G}_1} - n^{-1/2} \|\mathbb{G}_n\|_{\mathcal{G}_2} -  n^{-1/2} \|\mathbb{G}_n\|_{\mathcal{G}_3} - r_f \sup_{g \in \mathcal{G}_1} \|g\|_{P_n,1} - r_f \sup_{g \in \mathcal{G}_2} \|g\|_{P_n,1} - r_f \sup_{g \in \mathcal{G}_3} \|g\|_{P_n,1}.
	\end{align*} 
	
	Second, observe that
	\begin{align}\label{eq:theorem:ConsistencyDual-00}
		K(r, \tau) =\left\{ v \in \mathbb{R}^p :
		\begin{matrix}
			v - \tilde{v}(\tau) \in C^p_1\big(T_v(\tau), 2 \bar{c}\big),\\
			\bar{c}\sum_{k \in T_v(\tau)} \big|v_{k}(\tau) - \tilde{v}_k(\tau)\big| + \frac{\bar{c}\|z\|_2^2}{\kappa_2^2(\infty)}\frac{\mu}{\gamma}\geq \sum_{k \in T_v^c(\tau)}\big|v_{k}(\tau) \big|
		\end{matrix}
		\right\}\cap \partial B^p_2(0, r),
	\end{align}
	since $r > \frac{2\bar{c}\|z\|_2^2}{\kappa^2_2(\infty)}\frac{\mu}{\gamma}$ and $B^p_1\left(0,  \frac{2\bar{c}\|z\|_2^2}{\kappa_2^2(\infty)}\frac{\mu}{\gamma}\right) \cap \partial B^p_2(0, r) \neq \emptyset$ only if the radius of the $\ell_1$-ball is greater or equal to the radius of the $\ell_2$-sphere.
	
	Now, with probability at least $1- \eta$, 
	\begin{align}\label{eq:theorem:ConsistencyDual-1}
		&\inf_{\tau \in \mathcal{T}} \inf_{ v \in K(r, \tau)} \left(\frac{1}{4n}\sum_{i=1}^n \left(\hat{f}_i^2(\tau) (X_i'v)^2 - \hat{f}_i^2(\tau) \big(X_i'\tilde{v}(\tau)\big)^2\right) + z'\big(v- \tilde{v}(\tau)\big) + \frac{\gamma}{n}\big(\|v\|_1 - \|\tilde{v}(\tau)\|_1\big)\right) \nonumber\\
		&\gtrsim \inf_{\tau \in \mathcal{T}} \inf_{ v \in K(r, \tau)} \mathbb{E}\left[\frac{1}{n}\sum_{i=1}^n f_i^2(\tau) \big(v - \tilde{v}(\tau)\big)'X_iX_i'\big(v - \tilde{v}(\tau)\big)\right] \nonumber\\
		&\quad{} - \sup_{\tau \in \mathcal{T}} \sup_{ v \in K(r, \tau)} \left|\mathbb{E}\left[\frac{1}{4n} \sum_{i=1}^n f_i^2(\tau) \big(\tilde{v}(\tau)- v_0(\tau)\big)'X_iX_i'\big( v- \tilde{v}(\tau)\big)\right] \right| \nonumber\\
		&\quad{} -n^{-1/2}\|\mathbb{G}_n\|_{\mathcal{G}_1} - n^{-1/2} \|\mathbb{G}_n\|_{\mathcal{G}_2} -  n^{-1/2} \|\mathbb{G}_n\|_{\mathcal{G}_3} - r_f \sup_{g \in \mathcal{G}_1} \|g\|_{P_n,1} - r_f \sup_{g \in \mathcal{G}_2} \|g\|_{P_n,1} - r_f \sup_{g \in \mathcal{G}_3} \|g\|_{P_n,1} \nonumber\\
		&\quad{} -\sup_{\tau \in \mathcal{T}} \sup_{ v \in K(r, \tau)} \frac{\gamma}{n}\big(\|v\|_1 - \|\tilde{v}(\tau)\|_1\big) \nonumber\\
		&= \mathbf{I} - \mathbf{II} - \mathbf{III} - \mathbf{IV} - \mathbf{V} - \mathbf{VI} - \mathbf{VII} - \mathbf{VIII} - \mathbf{IX}.
	\end{align}
	In the following, we bound the expressions on the far right hand side in eq.~\eqref{eq:theorem:ConsistencyDual-1}.
	
	\textbf{Bound on $\mathbf{I}$.} Thus, by eq.~\eqref{eq:theorem:ConsistencyDual-00},
	\begin{align}\label{eq:theorem:ConsistencyDual-2}
		\mathbf{I} \gtrsim \inf_{\tau \in \mathcal{T}} \inf_{ u \in C^p_1(T_v(\tau), 2\bar{c}) \cap \partial B^p_2(0 ,1)} \mathbb{E}\left[\frac{1}{n}\sum_{i=1}^n f_i^2(\tau) (X_i'u)^2 \right]r^2 \geq \kappa_2(2\bar{c}) r^2.
	\end{align}
	
	\textbf{Bound on $\mathbf{II}$.} By Lemma~\ref{lemma:InducedRestrictedConeProp-Dual}, eq.~\eqref{eq:theorem:ConsistencyDual-00}, and two applications of Cauchy-Schwarz inequality,
	\begin{align}
		&\sup_{\tau \in \mathcal{T}} \sup_{ v \in K(r, \tau)}\mathbb{E}\left[\frac{1}{n} \sum_{i=1}^n f_i^2(\tau) \big(\tilde{v}(\tau)- v_0(\tau)\big)'X_iX_i'\big( v- \tilde{v}(\tau)\big)\right] \nonumber\\
		&\leq \sup_{\tau \in \mathcal{T}} \sup_{ v \in K(r, \tau)} \left(\mathbb{E}\left[\frac{1}{n} \sum_{i=1}^n f_i^2(\tau) \Big(\big(\tilde{v}(\tau)- v_0(\tau)\big)'X_i\Big)^2 \right]\right)^{1/2}  \left(\mathbb{E}\left[\frac{1}{n} \sum_{i=1}^n f_i^2(\tau) \Big(X_i'\big( v- \tilde{v}(\tau)\big)\Big)^2\right]\right)^{1/2} \nonumber\\
		&\leq \frac{\|z\|_2}{\kappa_2(\infty)}r_a r \times \sup_{\tau \in \mathcal{T}} \sup_{u \in C^p_2(T_v(\tau), 1) \cap \partial B^P_2(0,1) } \left(\mathbb{E}\left[\frac{1}{n} \sum_{i=1}^n f_i^2(\tau) (X_i'u)^2\right]\right)^{1/2}  \nonumber\\
		&\quad{}\quad{}\quad{}\quad{}\quad{}\quad{} \times \sup_{ u  \in C^p_1(T_v(\tau), 2\bar{c}) \cap \partial B^P_2(0,1) }\left(\mathbb{E}\left[\frac{1}{n} \sum_{i=1}^n f_i^2(\tau) (X_i'u)^2\right]\right)^{1/2} \nonumber.
	\end{align} 
	By Lemmas~\ref{lemma:SizeConesDominatedCoordinates},~\ref{lemma:MaximaBiconvexFunction}, and~\ref{lemma:MaxInequalityCovarianceCone},
	\begin{align*}
		\sup_{\tau \in \mathcal{T}} \sup_{u \in C^p_2(T_v(\tau), 1) \cap \partial B^P_2(0,1) } \left(\mathbb{E}\left[\frac{1}{n} \sum_{i=1}^n f_i^2(\tau) (X_i'u)^2\right]\right)^{1/2} \leq 6\bar{f} \varphi_{\max}^{1/2}(s_v),
	\end{align*}
	and
	\begin{align*}
		\sup_{\tau \in \mathcal{T}} \sup_{ u  \in C^p_1(T_v(\tau), 2\bar{c}) \cap \partial B^P_2(0,1) }\left(\mathbb{E}\left[\frac{1}{n} \sum_{i=1}^n f_i^2(\tau) (X_i'u)^2\right]\right)^{1/2} \leq 2(2+2\bar{c})\bar{f} \varphi_{\max}^{1/2}(s_v),
	\end{align*}
	and, therefore,
	\begin{align}\label{eq:theorem:ConsistencyDual-3}
		\mathbf{II} \lesssim (2+2\bar{c}) \bar{f}^2\frac{\varphi_{\max}(s_v)}{\kappa_2(\infty)} \|z\|_2r_a r.
	\end{align}
	
	\textbf{Bound on $\mathbf{III}$.} Define the following function class:
	\begin{align*}
		\mathcal{G}_4 &= \Big\{ g : \mathbb{R}^p \rightarrow \mathbb{R}: g(X) = f_{Y|X}^2(X'\theta_0(\tau)|X) (X'u)(X'w),\\
		&\quad{}\quad{} \: u \in \mathbb{R}^p, \: \|u\|_0 \leq s_v, \: \|u\|_2 \leq 1,\: w \in C^p_1(T_v(\tau), 2\bar{c}) \cap \partial B^p_2(0 ,1), \: \tau \in \mathcal{T}\Big\}.
	\end{align*}
	By the triangle inequality,
	\begin{align*}
		n^{-1/2}\|\mathbb{G}_n\|_{\mathcal{G}_1} \leq n^{-1/2}\|\mathbb{G}_n\|_{\mathcal{G}_3} + r \frac{2\|z\|_2}{\kappa_2(\infty)} n^{-1/2} \|\mathbb{G}_n\|_{\mathcal{G}_4}.
	\end{align*}
	Obviously, upper bounding the supremum over the set $\{v_0(\tau) : \tau \in \mathcal{T}\}$ by the supremum over the set of all $s$-sparse vectors is wasteful. However, in the present case, doing so does not change the asymptotic rate. (We derive a tighter bound on a similar quantity in the proof of Corollary~\ref{corollary:theorem:ConsistencyDual}, where the tighter bound makes a difference.)
	
	By eq.~\eqref{eq:theorem:ConsistencyDual-00}, Lemma~\ref{lemma:LocalizedLossDual}, and Assumptions~\ref{assumption:GrowthCondition} and~\ref{assumption:GrowthConditionDual}, with probability at least $1-\delta$, 
	\begin{align*}
		n^{-1/2}\|\mathbb{G}_n\|_{\mathcal{G}_3}  &\lesssim (2 +2\bar{c})\bar{f}^2(1 + \varphi_{\max}^{1/2}(2s_\theta)) \frac{\varphi_{\max}(s_v)}{\kappa_2(\infty)} \|z\|_2 \\
		&\quad{}\times\sqrt{\frac{s_v \log(ep/s_v) + s_\theta \log(ep/s_\theta) + \log (n L_f L_\theta )+  \log (1/\delta)}{n}} \:r_a r.
	\end{align*}
	and
	\begin{align*}
		n^{-1/2}\|\mathbb{G}_n\|_{\mathcal{G}_4} &\lesssim (2 +2\bar{c}) \bar{f}^2(1 + \varphi_{\max}^{1/2}(2s_\theta))\frac{\varphi_{\max}(s_v)}{\kappa_2(\infty)} \|z\|_2\\
		&\quad{}\times\sqrt{\frac{s_v \log(ep/s_v) + s_\theta \log(ep/s_\theta) + \log (n L_f L_\theta )+  \log (1/\delta)}{n}} \: r,
	\end{align*}
	Combine the preceding two upper bounds and the fact that $0 \leq r_a < 1/2$ to conclude that, with probability at least $1-\delta$, 
	\begin{align}\label{eq:theorem:ConsistencyDual-4}
		\begin{split}
			n^{-1/2}\|\mathbb{G}_n\|_{\mathcal{G}_1}  &\lesssim (2 +2\bar{c})\bar{f}^2(1 + \varphi_{\max}^{1/2}(2s_\theta)) \frac{\varphi_{\max}(s_v)}{\kappa_2(\infty)} \|z\|_2 \\
			&\quad{}\times\sqrt{\frac{s_v \log(ep/s_v) + s_\theta \log(ep/s_\theta) + \log (n L_f L_\theta )+  \log (1/\delta)}{n}} \:r.
		\end{split}
	\end{align}

	\textbf{Bound on $\mathbf{IV}$.} By eq.~\eqref{eq:theorem:ConsistencyDual-00}, Lemma~\ref{lemma:LocalizedLossDual}, and Assumptions~\ref{assumption:GrowthCondition} and~\ref{assumption:GrowthConditionDual}, with probability at least $1-\delta$, 
	\begin{align}\label{eq:theorem:ConsistencyDual-5}
		\begin{split}
			n^{-1/2}\|\mathbb{G}_n\|_{\mathcal{G}_2} &\lesssim (2 + 2\bar{c})^2 \bar{f}^2(1 + \varphi_{\max}^{1/2}(2s_\theta)) \varphi_{\max}(s_v)\\
			&\quad{} \times \sqrt{\frac{s_v \log(ep/s_v) + s_\theta \log(ep/s_\theta) + \log(n L_f L_\theta)+  \log (1/\delta)}{n}} \:r^2.
		\end{split}
	\end{align}
	
	\textbf{Bound on $\mathbf{V}$.} By eq.~\eqref{eq:theorem:ConsistencyDual-00}, Lemma~\ref{lemma:LocalizedLossDual}, and Assumptions~\ref{assumption:GrowthCondition} and~\ref{assumption:GrowthConditionDual}, with probability at least $1-\delta$, 
	\begin{align}\label{eq:theorem:ConsistencyDual-6}
		\begin{split}
			n^{-1/2}\|\mathbb{G}_n\|_{\mathcal{G}_3} &\lesssim (2 +2\bar{c})\bar{f}^2(1 + \varphi_{\max}^{1/2}(2s_\theta)) \frac{\varphi_{\max}(s_v)}{\kappa_2(\infty)} \|z\|_2 \\
			&\quad{}\times\sqrt{\frac{s_v \log(ep/s_v) + s_\theta \log(ep/s_\theta) + \log (n L_f L_\theta )+  \log (1/\delta)}{n}} \:r_a r.
		\end{split}
	\end{align}
	
	\textbf{Bound on $\mathbf{VI}$.} 
	We introduce the following additional function classes:
	\begin{align*}
		\mathcal{G}_5 &= \{ g : \mathbb{R}^p \rightarrow \mathbb{R}: g(X) = f_{Y|X}^2(X'\theta_0(\tau)|X) (X'u)^2, \: u 
		\in C^p_1(T_v(\tau),2\bar{c}) \cap B^p_2(0, 1) \: \tau \in \mathcal{T}\},\\
		\mathcal{G}_6 &= \{ g : \mathbb{R}^p \rightarrow \mathbb{R}: g(X) = f_{Y|X}^2(X'\theta_0(\tau)|X) (X'u)^2, \: u 
		\in C^p_2(T_v(\tau),1) \cap B^p_2(0, 1) \: \tau \in \mathcal{T}\},\\
		\mathcal{G}_7 &= \{ g : \mathbb{R}^p \rightarrow \mathbb{R}: g(X) = f_{Y|X}^2(X'\theta_0(\tau)|X) (X'u)^2, \:u \in \mathbb{R}^p, \: \|u\|_2 \leq 1, \: \|u\|_0\leq s_v, \: \tau \in \mathcal{T}\}.
	\end{align*}
	Then, by the triangle inequality, the bound on term $\mathbf{II}$, and the fact that $0 \leq r_a < 1/2$, 
	\begin{align*}
		r_f \sup_{g \in \mathcal{G}_1} \|g\|_{P_n,1} & \lesssim r_f \sup_{g \in \mathcal{G}_3} \|g\|_{P_n,1} + 	r_f r \frac{\|z\|_2}{\kappa_2(\infty)} \sup_{g \in \mathcal{G}_4} \|g\|_{P_n,1}\\	
		&\lesssim r_f n^{-1/2}  \|\mathbb{G}_n\|_{|\mathcal{G}_3|} + r_f r \frac{\|z\|_2}{\kappa_2(\infty)} n^{-1/2} \|\mathbb{G}_n\|_{|\mathcal{G}_4|} + (2+2\bar{c}) \bar{f}^2\frac{\varphi_{\max}(s_v)}{\kappa_2(\infty)} \|z\|_2 r_f r\\
		&\lesssim r_f r_a r \frac{\|z\|_2}{\kappa_2(\infty)} n^{-1/2} \|\mathbb{G}_n\|_{\mathcal{G}_5} + r_f r \frac{\|z\|_2}{\kappa_2(\infty)} n^{-1/2} \Big(\|\mathbb{G}_n\|_{\mathcal{G}_6} + \|\mathbb{G}_n\|_{\mathcal{G}_7} \Big)  \\
		&\quad{} + (2+2\bar{c}) \bar{f}^2\frac{\varphi_{\max}(s_v)}{\kappa_2(\infty)} \|z\|_2 r_f r.
	\end{align*}
	Thus, by Lemma~\ref{lemma:LocalizedLossDual} and Assumptions~\ref{assumption:GrowthCondition} and~\ref{assumption:GrowthConditionDual}, with probability at least $1-\delta$,
	\begin{align}\label{eq:theorem:ConsistencyDual-7}
		\begin{split}
			r_f \sup_{g \in \mathcal{G}_1} \|g\|_{P_n,1} & \lesssim (2+2\bar{c}) \bar{f}^2\frac{\varphi_{\max}(s_v)}{\kappa_2(\infty)} \|z\|_2 r_f r + (2 +2\bar{c})\bar{f}^2(1 + \varphi_{\max}^{1/2}(2s_\theta)) \frac{\varphi_{\max}(s_v)}{\kappa_2(\infty)} \|z\|_2 \\
			&\quad{}\times\sqrt{\frac{s_v \log(ep/s_v) + s_\theta \log(ep/s_\theta) + \log (n L_f L_\theta )+  \log (1/\delta)}{n}} \:r_f r. 
		\end{split}
	\end{align}
	
	\textbf{Bound on $\mathbf{VII}$.}
	By Lemma~\ref{lemma:LocalizedLossDual} and Assumptions~\ref{assumption:GrowthCondition} and~\ref{assumption:GrowthConditionDual}, with probability at least $1-\delta$,
	\begin{align}\label{eq:theorem:ConsistencyDual-8}
		r_f \sup_{g \in \mathcal{G}_2} \|g\|_{P_n,1} &\lesssim r_f n^{-1/2}\|\mathbb{G}_n\|_{|\mathcal{G}_2|} +  (2 + 2\bar{c})^2 \bar{f}^2 \varphi_{\max}(s_v) r_f r^2 \nonumber\\
		&\lesssim r_f n^{-1/2}\|\mathbb{G}_n\|_{\mathcal{G}_2} +  (2 +2\bar{c})^2 \bar{f}^2 \varphi_{\max}(s_v) r_f r^2 \nonumber\\
		\begin{split}
			&\lesssim (2 +2\bar{c})^2 \bar{f}^2 \varphi_{\max}(s_v) r_f r^2 + (2 + 2\bar{c})^2 \bar{f}^2(1 + \varphi_{\max}^{1/2}(2s_\theta)) \varphi_{\max}(s_v)\\
			&\quad{} \times \sqrt{\frac{s_v \log(ep/s_v) + s_\theta \log(ep/s_\theta) + \log(n L_f L_\theta)+  \log (1/\delta)}{n}} \: r_f r^2.
		\end{split}
	\end{align}
	
	\textbf{Bound on $\mathbf{VIII}$.} By Lemma~\ref{lemma:LocalizedLossDual} and Assumptions~\ref{assumption:GrowthCondition} and~\ref{assumption:GrowthConditionDual}, with probability at least $1-\delta$,
	\begin{align}\label{eq:theorem:ConsistencyDual-9}
		r_f \sup_{g \in \mathcal{G}_3} \|g\|_{P_n,1} &\lesssim  r_f n^{-1/2}  \|\mathbb{G}_n\|_{|\mathcal{G}_3|} + (2+2\bar{c}) \bar{f}^2\frac{\varphi_{\max}(s_v)}{\kappa_2(\infty)} \|z\|_2 r_fr_a  r \nonumber\\
		&\lesssim r_f r_a r \frac{\|z\|_2}{\kappa_2(\infty)} n^{-1/2} \Big(\|\mathbb{G}_n\|_{\mathcal{G}_5} + \|\mathbb{G}_n\|_{\mathcal{G}_6} \Big) + (2+2\bar{c}) \bar{f}^2\frac{\varphi_{\max}(s_v)}{\kappa_2(\infty)} \|z\|_2 r_fr_a  r  \nonumber \\
		\begin{split}
			&\lesssim (2+2\bar{c}) \bar{f}^2\frac{\varphi_{\max}(s_v)}{\kappa_2(\infty)} \|z\|_2 r_f r_a r + (2 +2\bar{c})\bar{f}^2(1 + \varphi_{\max}^{1/2}(2s_\theta)) \frac{\varphi_{\max}(s_v)}{\kappa_2(\infty)} \|z\|_2 \\
			&\quad{}\times\sqrt{\frac{s_v \log(ep/s_v) + s_\theta \log(ep/s_\theta) + \log (n L_f L_\theta )+  \log (1/\delta)}{n}} \:r_f r_a r. 
		\end{split}
	\end{align}
	
	\textbf{Bound on $\mathbf{IX}$.} By the reverse triangle inequality and  Lemma~\ref{lemma:RestrictedConeProp-Dual}, 
	\begin{align}\label{eq:theorem:ConsistencyDual-10}
		&\sup_{\tau \in \mathcal{T}} \sup_{ v \in K(r, \tau)} \frac{\gamma}{n}\big(\|v\|_1 - \|\tilde{v}(\tau)\|_1\big) \nonumber\\
		&\quad{} \leq \sup_{\tau \in \mathcal{T}} \sup_{ v \in K(r, \tau)} \frac{\gamma}{n}\left( \sum_{k \in T_v(\tau)} \big(|v_k| - |\tilde{v}_k(\tau)|\big)  +  \sum_{k \in T_v^c(\tau)} |v_k| \right) \nonumber\\
		&\quad{} \leq \sup_{\tau \in \mathcal{T}} \sup_{ v \in K(r, \tau)} \frac{\gamma}{n}\left( \sum_{k \in T_v(\tau)} |v_k - \tilde{v}_k(\tau)|  +  \sum_{k \in T_v^c(\tau)} |v_k| \right) \nonumber\\
		&\quad{}  \leq \frac{\bar{c}\|z\|_2^2}{\kappa_2^2(\infty)}\frac{\gamma}{n}\frac{\mu}{\gamma}+ (1 + \bar{c}) \sup_{\tau \in \mathcal{T}} \sup_{ v \in K(r, \tau)} \frac{\gamma}{n} \sum_{k \in T_v(\tau)} |v_k - \tilde{v}_k(\tau)| \nonumber\\
		&\quad{}  \lesssim \frac{\bar{c}\|z\|_2^2}{\kappa_2^2(\infty)}\frac{\mu}{n}+ (1 + \bar{c})\frac{\gamma}{n} s_v^{1/2} r.
	\end{align}

	\textbf{Conclusion.} Combine Assumptions~\ref{assumption:GrowthCondition} and~\ref{assumption:GrowthConditionDual}, eq.~\eqref{eq:theorem:ConsistencyDual-2}--\eqref{eq:theorem:ConsistencyDual-10}, and observe that there exist absolute constants $c_2, \ldots, c_7 > 0$ such that, with probability at least $1 - \eta - \delta$, the expression in eq.~\eqref{eq:theorem:ConsistencyDual-1} can be lower bounded (up to a multiplicative constant) by
	\begin{align*}
		&\kappa_2(2\bar{c}) r^2\\
		&\quad{} - c_2 (2+2\bar{c}) \bar{f}^2\frac{\varphi_{\max}(s_v)}{\kappa_2(\infty)} \|z\|_2 \big(r_a + r_f\big) r\\
		&\quad{} - c_3 (2 +2\bar{c})\bar{f}^2(1 + \varphi_{\max}^{1/2}(2s_\theta)) \frac{\varphi_{\max}(s_v)}{\kappa_2(\infty)} \|z\|_2 \\
		&\quad{}\quad{}\times\sqrt{\frac{s_v \log(ep/s_v) + s_\theta \log(ep/s_\theta) + \log (n L_f L_\theta )+  \log (1/\delta)}{n}}\: r\\
		&\quad{} - c_4 (2 + 2\bar{c})^2 \bar{f}^2(1 + \varphi_{\max}^{1/2}(2s_\theta)) \varphi_{\max}(s_v)\\
		&\quad{} \quad{} \times \sqrt{\frac{s_v \log(ep/s_v) + s_\theta \log(ep/s_\theta) + \log(n L_f L_\theta)+  \log (1/\delta)}{n}} \: r^2\\
		&\quad{} - c_5 (2 +2\bar{c})^2 \bar{f}^2 \varphi_{\max}(s_v) r_f r^2\\
		&\quad{} - c_6\frac{\bar{c}\|z\|_2^2}{\kappa_2^2(\infty)}\frac{\mu}{n}- c_6 (1 + \bar{c}) \frac{\gamma}{n} s_v^{1/2} r\\
		&\quad{} > 0,
	\end{align*}
	whenever
	\begin{align*}
		r &\geq  c_7 C\left(\frac{\|z\|_2}{\kappa_2(\infty)} \vee 1\right)\left( \sqrt{\frac{s_v \log(ep/s_v) + s_\theta \log(ep/s_\theta) + \log (n L_f L_\theta )+  \log (1/\delta)}{n}} \vee r_f\vee r_a \right) \\
		&\quad{}\bigvee c_7 \left(\frac{\bar{c}\|z\|_2^2}{\kappa_2^2(\infty)} \right) \left(\frac{\mu}{n} \vee \frac{\mu}{\gamma}\right) \bigvee c_7 \left(\frac{\bar{c}}{\kappa_2(2\bar{c})}\right) \frac{\gamma \sqrt{s_v}}{n}.
	\end{align*}
	where $C=  (1 + \varphi_{\max}^{1/2}(2s_\theta))\bar{c}^2 \bar{f}^2 \varphi_{\max}(s_v)/\kappa_2(2\bar{c})$. (Note that $c_7 > 0$ can be chosen such that $r >  \frac{2\bar{c}\|z\|_2^2}{\kappa_2^2(\infty)}\frac{\mu}{\gamma}$.) To conclude, adjust constants.

\end{proof}
\begin{proof}[\textbf{Proof of Corollary~\ref{corollary:theorem:ConsistencyDual}}]
	The proof consists of two parts. First, we derive an upper bound on the gradient of the objective function. Then, we derive an upper bound on the empirical covariance matrix. Any $\gamma, \mu > 0$ grater than or equal to these upper bounds will satisfy eq.~\eqref{eq:lemma:RestrictedConeProp-Dual-1} and~\eqref{eq:lemma:RestrictedConeProp-Dual-11}. To simplify notation, we write $v_0(\tau)$, $\hat{v}(\tau)$,  $T_v(\tau)$, and $s_v$ for $v_0(\tau; z)$, $\hat{v}(\tau;z)$, $T_v(\tau;z)$, and $s_v(z)$, respectively. Let $\zeta  \in (0,1)$ be arbitrary and $\mathcal{T}_\zeta$ be an $\zeta$-net with cardinality $\mathrm{card}(T_\zeta) \leq 1 + 1/\zeta$.
	
	\textbf{First part.}
	
	By the definition of $v_0(\tau)$ and repeated applications of the triangle inequality,
	\begin{align}\label{eq:corollary:theorem:ConsistencyDual-0}
		&\sup_{\tau \in \mathcal{T}}\left\|\frac{1}{2}\sum_{i=1}^n\hat{f}_i^2(\tau) X_iX_i'v_0(\tau) + nz \right\|_\infty \nonumber\\
		&\quad{}\leq \sup_{\tau \in \mathcal{T}} \left\|\frac{1}{2}\sum_{i=1}^n \big(f_i^2(\tau) X_iX_i' - \mathbb{E}[f_i^2(\tau) X_iX_i'] \big)v_0(\tau)\right\|_\infty + \sup_{\tau \in \mathcal{T}}\left\|\frac{1}{2}\sum_{i=1}^n \big(\hat{f}_i^2(\tau) - f_i^2(\tau)\big)X_iX_i'v_0(\tau)\right\|_\infty\nonumber\\
		&\quad{} \leq \sup_{s \in T_\zeta} \sup_{s' \in \mathcal{T}} \left\|\frac{1}{2}\sum_{i=1}^n \big(f_i^2(s') X_iX_i' - \mathbb{E}[f_i^2(s') X_iX_i'] \big)v_0(s)\right\|_\infty\nonumber\\
		&\quad{}\quad{} +  \sup_{s \in T_\zeta}\sup_{\tau : |\tau - s| \leq \zeta}\sup_{s' \in \mathcal{T}} \left\|\frac{1}{2}\sum_{i=1}^n \big(f_i^2(s') X_iX_i' - \mathbb{E}[f_i^2(s') X_iX_i'] \big)\big(v_0(\tau) - v_0(s)\big)\right\|_\infty\nonumber\\
		&\quad{}\quad{} + \sup_{s \in T_\zeta}\sup_{s' \in \mathcal{T}}  \left\|\frac{1}{2}\sum_{i=1}^n \big(\hat{f}_i^2(s') - f_i^2(s')\big)X_iX_i'v_0(s)\right\|_\infty\nonumber\\
		&\quad{}\quad{} +  \sup_{s \in T_\zeta}\sup_{\tau : |\tau - s| \leq \zeta} \sup_{s' \in \mathcal{T}}\left\|\frac{1}{2}\sum_{i=1}^n \big(\hat{f}_i^2(s') - f_i^2(s')\big)X_iX_i'\big(v_0(\tau) - v_0(s)\big)\right\|_\infty\nonumber\\
		&\quad{} = \mathbf{I} +  \mathbf{II} +  \mathbf{III} +  \mathbf{IV}.
	\end{align}
	We now bound the four terms on the far right hand side in above display.
	
	\textbf{Bound on $\mathbf{I}$.} By Lemma~\ref{lemma:LocalizedLossDual}, Assumptions~\ref{assumption:GrowthCondition} and~\ref{assumption:GrowthConditionDual}, and the union bound over $s \in \mathcal{T}_\zeta$, with probability at least $1-\delta$, 
	\begin{align}\label{eq:corollary:theorem:ConsistencyDual-1}
		\mathbf{I} \lesssim \bar{c}\bar{f}^2\varphi_{\max}^{1/2}(s_v)\varphi_{\max}^{1/2}(1)\big(1 + \varphi_{\max}^{1/2}(2s_\theta)\big) \frac{\|z\|_2}{\kappa_2(\infty)} \sqrt{n} \sqrt{\log p + \log (nL_f L_\theta) + \log(1 + 1/\zeta)+ \log(1/\delta)}.
	\end{align}
	
	\textbf{Bound on $\mathbf{II}$.} By Lemmas~\ref{lemma:LipschitzDual} and~\ref{lemma:LocalizedLossDual}, Assumptions~\ref{assumption:GrowthCondition} and~\ref{assumption:GrowthConditionDual}, and the union bound over $s \in \mathcal{T}_\zeta$, with probability at least $1-\delta$, 
	\begin{align}\label{eq:corollary:theorem:ConsistencyDual-2}
		\begin{split}
			\mathbf{II} &\lesssim \bar{c} \bar{f}^3 L_fL_\theta\varphi_{\max}^{1/2}(2s_v)\varphi_{\max}^{1/2}(1)\big(1 + \varphi_{\max}^{1/2}(2s_\theta)\big)\frac{\|z\|_2}{\kappa_2(\infty)}\varphi_{\max}^{1/2}(2s_\theta)\varphi_{\max}(p)\\
			&\quad{}\times \sqrt{n} \sqrt{s_v \log(ep/s_v) + \log p + \log(nL_f L_\theta) + \log(1 + 1/\zeta)+ \log(1/\delta)} \: \zeta.
		\end{split}
	\end{align}

	\textbf{Bound on $\mathbf{III}$.} We introduce the following two function classes:
	\begin{align*}
		\mathcal{G}_1&= \{ g : \mathbb{R}^p \rightarrow \mathbb{R}: g(X) = f_{Y|X}^2(X'\theta_0(s')|X) X_k^2, \: 1 \leq k \leq p, \: s' \in \mathcal{T}\},\\
		\mathcal{G}_2 &= \{ g : \mathbb{R}^p \rightarrow \mathbb{R}: g(X) = f_{Y|X}^2(X'\theta_0(s')|X) (X'v)^2, \: v \in \big\{v_0(s) \kappa_2(\infty)/\|z\|_2: s\in T_\zeta \big\}, \: s' \in \mathcal{T}\}.
	\end{align*}	
	
	Recall that by eq.~\eqref{eq:theorem:ConsistencyDual-0}, with probability at least $1- \eta$, we have $|\hat{f}_i^2(\tau) - f_i^2(\tau)| \lesssim r_f f_i^2(\tau).$ Thus, with probability at least $1- \eta$, 
	\begin{align*}
		\mathbf{III} &\lesssim r_f \sup_{s \in T_\zeta}\sup_{s' \in \mathcal{T}} \max_{1 \leq k \leq p}\left( \frac{1}{2}\sum_{i=1}^n f_i^2(s')\big)|X_{ik}| |X_i'v_0(s)| \right)\\
		&\lesssim r_f \sqrt{n}\frac{ \|z\|_2}{\kappa_2(\infty)} \big(\|\mathbb{G}_n\|_{\mathcal{G}_1} + \|\mathbb{G}_n\|_{\mathcal{G}_2} \big) + r_f n \: \bar{f}^2\varphi_{\max}(s_v)\frac{ \|z\|_2}{\kappa_2(\infty)}.
	\end{align*}
	Therefore, by Lemma~\ref{lemma:LocalizedLossDual}, Assumptions~\ref{assumption:GrowthCondition} and~\ref{assumption:GrowthConditionDual}, and the union bound over $s \in \mathcal{T}_\zeta$, we have, with probability at least $1- \eta - \delta$,
	\begin{align}\label{eq:corollary:theorem:ConsistencyDual-3}
		\begin{split}
			\mathbf{III} &\lesssim \bar{c}^2\bar{f}^2\varphi_{\max}(s_v)\big(1 + \varphi_{\max}^{1/2}(2s_\theta)\big) \frac{\|z\|_2}{\kappa_2(\infty)} \sqrt{n} \sqrt{\log p + \log (nL_f L_\theta ) + \log(1 + 1/\zeta)+ \log(1/\delta)} \:r_f\\
			&\quad{} + \bar{c}^2\bar{f}^2\varphi_{\max}(s_v) \frac{ \|z\|_2}{\kappa_2(\infty)}\: n r_f.
		\end{split}
	\end{align}
	
	\textbf{Bound on $\mathbf{IV}$.} Analogous to the bounds of $\mathbf{II}$ and $\mathbf{III}$, we have, with probability at least $1 -\eta - \delta$,
	\begin{align}\label{eq:corollary:theorem:ConsistencyDual-4}
		\begin{split}
			\mathbf{IV} &\lesssim \bar{c}^2 \bar{f}^3 L_fL_\theta\varphi_{\max}(2s_v)\big(1 + \varphi_{\max}^{1/2}(2s_\theta)\big)\frac{\|z\|_2}{\kappa_2(\infty)}\varphi_{\max}^{1/2}(2s_\theta)\varphi_{\max}(p)\\
			&\quad{}\times \sqrt{n} \sqrt{s_v \log(ep/s_v) + \log p + \log (nL_f L_\theta) + \log(1 + 1/\zeta)+ \log(1/\delta)} \: r_f\: \zeta\\
			&\quad{} +\bar{c}^2 \bar{f}^3 L_fL_\theta\varphi_{\max}(2s_v)\frac{\|z\|_2}{\kappa_2(\infty)}\varphi_{\max}^{1/2}(2s_\theta)\varphi_{\max}(p)\: n r_f\: \zeta.
		\end{split}
	\end{align}
	
	Since $\zeta \in (0,1)$ is arbitrary, we can choose $\zeta \asymp 1/\big(L_f L_\theta \varphi_{\max}^{1/2}(2s_\theta)\varphi_{\max}(p)\sqrt{p}\big) $. Combine the bounds in eq.~\eqref{eq:corollary:theorem:ConsistencyDual-0}--\eqref{eq:corollary:theorem:ConsistencyDual-4}, adjust some constants, and conclude that with probability at least $1 - \eta - \delta$,
	\begin{align*}
		&\sup_{\tau \in \mathcal{T}}\left\|\frac{1}{2}\sum_{i=1}^n\hat{f}_i^2(\tau) X_iX_i'v_0(\tau) + nz \right\|_\infty\\
		&\lesssim \bar{c}^2\bar{f}^3\varphi_{\max}(2s_v)\big(1 + \varphi_{\max}^{1/2}(2s_\theta)\big)\frac{\|z\|_2 }{\kappa_2(\infty)}  \left(  \sqrt{\log(np/\delta) + \log\big(L_fL_\theta\varphi_{\max}(2s_\theta)\varphi_{\max}(p)\big) } + \sqrt{n} r_f\right) \sqrt{n}.
	\end{align*}
	We further upper bound (and simplify) the term on the right hand side in above display.  Recall from Theorem~\ref{theorem:ConsistencyDual} $C_3 = \bar{c}^2 \bar{f}^2  L_f L_\theta \big(1 + \varphi_{\max}^{1/2}(2s_\theta)\big)\varphi_{\max}(s_v)/\kappa_2(\bar{c})$. Then,
	\begin{align*}
		&\bar{c}^2\bar{f}^3\varphi_{\max}(2s_v)\big(1 + \varphi_{\max}^{1/2}(2s_\theta)\big)\frac{\|z\|_2 }{\kappa_2(\infty)}  \left(  \sqrt{\log(np/\delta) + \log\big(L_fL_\theta\varphi_{\max}(2s_\theta)\varphi_{\max}(p)\big) } + \sqrt{n} r_f\right) \sqrt{n}\\
		&\overset{(a)}{\lesssim} \bar{c}^2\bar{f}^3L_fL_\theta\varphi_{\max}(2s_v)\big(1 + \varphi_{\max}(2s_\theta)\big)\frac{\|z\|_2 }{\kappa_2(\infty)}  \left(  \sqrt{\log(np/\delta)} + \sqrt{n} r_f\right) \sqrt{n}\\
		&\lesssim C_3 \bar{f} \frac{\varphi_{\max}(2s_v)}{\varphi_{\max}(s_v)} \frac{\kappa_2(\bar{c})\|z\|_2}{\kappa_2(\infty)}\left(  \sqrt{\log(np/\delta)} + \sqrt{n} r_f\right) \sqrt{n}\\
		&\overset{(b)}{\lesssim} C_3 \bar{f} \frac{\kappa_2(\bar{c})\|z\|_2}{\kappa_2(\infty)}\left(  \sqrt{\log(np/\delta)} + \sqrt{n} r_f\right) \sqrt{n},
	\end{align*}
	where (a) and (b) follow from Lemma 13 in~\cite{belloni2011L1penalized} (more specifically, (a) holds since $\log\big(\varphi_{\max}(2s_\theta)\varphi_{\max}(p)\big) \lesssim 2\log\big(\varphi_{\max}(2s_\theta)\big) + \log p$ by Lemma 13 in~\cite{belloni2011L1penalized} and (b) holds since $\varphi_{\max}(2s_v)/\varphi_{\max}(s_v) \leq 2$ and $\varphi_{\max}(1)/\varphi_{\max}(s_v) \leq 1$.)
	
	Therefore, there exists an absolute constant $C_\gamma> 0$ such that for all 
	\begin{align}\label{eq:corollary:theorem:ConsistencyDual-5}
		\gamma \geq  C_\gamma c_0C_3\bar{f} \frac{\kappa_2(\bar{c})\|z\|_2}{\kappa_2(\infty)}\left(  \sqrt{\log(np/\delta) } + \sqrt{n} r_f\right) \sqrt{n},
	\end{align}
	eq.~\eqref{eq:lemma:RestrictedConeProp-Dual-1} holds with probability at least $1 - \eta - \delta$.
	
	\textbf{Second part.}
	Next, by the triangle inequality,
	\begin{align*}
		&\sup_{\tau \in \mathcal{T}} \sup_{u \in  C^p_2(T_v(\tau; z), 1) \cap  B^p_2(0,r_a) } \left|\frac{1}{2}\sum_{i=1}^n\hat{f}_i^2(\tau) (X_i'u)^2\right|\\
		&\quad{} \leq \sup_{\tau \in \mathcal{T}} \sup_{u \in  C^p_2(T_v(\tau; z), 1) \cap  B^p_2(0,r_a) } \left| \frac{1}{2} \sum_{i=1}^n \big(\hat{f}_i^2(\tau) - f_i^2(\tau)\big) (X_i'u)^2\right|\\
		&\quad{}\quad{} + \sup_{\tau \in \mathcal{T}} \sup_{u \in  C^p_2(T_v(\tau; z), 1) \cap  B^p_2(0,r_a) } \left| \frac{1}{2} \sum_{i=1}^n f_i^2(\tau) (X_i'u)^2 - \mathbb{E}\big[f_i^2(\tau)(X_i'u)^2\big] \right|\\
		&\quad{}\quad{} + \sup_{\tau \in \mathcal{T}} \sup_{u \in  C^p_2(T_v(\tau; z), 1) \cap  B^p_2(0,r_a) } \mathbb{E}\left[ \frac{1}{2} \sum_{i=1}^n f_i^2(\tau) (X_i'u)^2 \right].
	\end{align*}
	Recall eq.~\ref{eq:theorem:ConsistencyDual-0} and upper bound the right hand side in above inequality (up to a multiplicative constant), with probability at least $1- \eta$, by
	\begin{align*}
		&(1 + r_f)\sup_{\tau \in \mathcal{T}} \sup_{u \in  C^p_2(T_v(\tau; z), 1) \cap  B^p_2(0,r_a) } \left| \frac{1}{2} \sum_{i=1}^n f_i^2(\tau) (X_i'u)^2 - \mathbb{E}\big[f_i^2(\tau)(X_i'u)^2\big] \right|\\
		&\quad{} + (1 + r_f)\sup_{\tau \in \mathcal{T}} \sup_{u \in  C^p_2(T_v(\tau; z), 1) \cap  B^p_2(0,r_a) } \mathbb{E}\left[ \frac{1}{2} \sum_{i=1}^n f_i^2(\tau) (X_i'u)^2 \right].
	\end{align*}
	By Lemma~\ref{lemma:LocalizedLossDual}, Assumptions~\ref{assumption:GrowthCondition} and~\ref{assumption:GrowthConditionDual}, we upper bound these two terms (up to a multiplicative constant), with probability at least $1- \eta - \delta$, by
	\begin{align*}
		\bar{f}^2 \varphi_{\max}(s_v)  (1 + \varphi_{\max}^{1/2}(2s_\theta)) \sqrt{s_v\log(ep/s_v) + \log(1/\delta)} \sqrt{n} r_a^2 + \bar{f}^2\varphi_{\max}(s_v) n r_a^2.
	\end{align*}	
	Therefore, there exists an absolute constant $C_\mu> 0$ such that for all 
	\begin{align}\label{eq:corollary:theorem:ConsistencyDual-6}
		\mu \geq C_\mu c_0 \left(\bar{f}^2 \varphi_{\max}(s_v)  (1 + \varphi_{\max}^{1/2}(2s_\theta)) \sqrt{s_v\log(ep/s_v) + \log(1/\delta)} \sqrt{n} r_a^2 + \bar{f}^2\varphi_{\max}(s_v) n r_a^2\right),
	\end{align}
	eq.~\eqref{eq:lemma:RestrictedConeProp-Dual-11} holds with probability at least $1 - \eta - \delta$.

	\textbf{Conclusion.} Plug the lower bounds in eq.~\eqref{eq:corollary:theorem:ConsistencyDual-5} and~\eqref{eq:corollary:theorem:ConsistencyDual-6} into the rate $r > 0$ from Theorem~\ref{theorem:ConsistencyDual} and simplify the expression using Assumptions~\ref{assumption:GrowthCondition} and~\ref{assumption:GrowthConditionDual} to conclude that with probability at least $1- \eta - \delta$, 
	\begin{align*}
		&\sup_{\tau \in \mathcal{T}} \left\|\hat{v}_\gamma(\tau) - \tilde{v}(\tau)\right\|_2 \\
		&\quad{} \lesssim C_3 \left( \frac{C_\gamma \bar{c} \bar{f}\|z\|_2}{\kappa_2(\infty)} \vee \frac{C_\mu \|z\|_2^2}{C_\gamma \bar{c} \bar{f} \kappa_2(\infty)} \vee \frac{\|z\|_2}{\kappa_2(\infty)} \vee 1\right) \left(\sqrt{\frac{s_v \log(np/\delta) + s_\theta \log(ep/s_\theta)}{n}} \vee  r_f \sqrt{s_v} \vee r_a  \vee \frac{r_a^2}{r_f} \right),
	\end{align*}
	where $C_3=  (1 + \varphi_{\max}^{1/2}(2s_\theta))\bar{c}^2 \bar{f}^2 \varphi_{\max}(s_v)/\kappa_2(2\bar{c})$. 	
\end{proof}

\subsection{Proofs of Section~\ref{subsec:BahadurRepresentation}}
\begin{proof}[\textbf{Proof of Lemma~\ref{lemma:RelativeConsistencyDensity}}]
	
	To simplify notation we write $f_i(\tau)$ for $f_{Y|X}(X_i'\theta_0(\tau)|X_i)$. We have
	\begin{align*}
		\left|\hat{f}_i(\tau) - f_i(\tau)\right| &= \left|1/\hat{f}_i(\tau) - 1/f_i(\tau)\right|\hat{f}_i(\tau)f_i(\tau) \\
		&\leq \left|1/\hat{f}_i(\tau) - 1/f_i(\tau)\right| \left| \hat{f}_i(\tau) - f_i(\tau)\right| f_i(\tau) + \left|1/\hat{f}_i(\tau) - 1/f_i(\tau)\right|f_i^2(\tau).
	\end{align*}
	If $\sup_{\tau \in \mathcal{T}}\max_{1 \leq i \leq n}|1/\hat{f}_i(\tau) - 1/f_i(\tau)|f_i(\tau) < 1/2$, this can be rearranged to yield
	\begin{align}\label{eq:lemma:RelativeConsistencyDensity-1}
		\sup_{\tau \in \mathcal{T}}\max_{1 \leq i \leq n}\left|\frac{\hat{f}_i(\tau)}{f_i(\tau)} -1 \right| \leq  2\bar{f}\sup_{\tau \in \mathcal{T}}\max_{1 \leq i \leq n}\left|\frac{1}{\hat{f}_i(\tau)} - \frac{1}{f_i(\tau)}\right|.
	\end{align}
	Thus, the claim of the lemma follows if $ \sup_{\tau \in \mathcal{T}}\max_{1 \leq i \leq n} |1/\hat{f}_i(\tau) - 1/f_i(\tau)| \rightarrow 0$ at the prescribed rate.
	
	Let $\tau \in \mathcal{T}$ and $h > 0$ be arbitrary. Two third-order Taylor expansions of the CQF give
	\begin{align*}
		Q_Y(\tau + h; X) &= Q_Y(\tau; X) + Q_Y'(\tau; X)h + Q_Y''(\tau; X)h^2/2 + Q_Y'''(\zeta_+; X) h^3/6,\\
		Q_Y(\tau - h; X) &= Q_Y(\tau; X) - Q_Y'(\tau; X)h + Q_Y''(\tau; X)h^2/2 -  Q_Y'''(\zeta_-; X) h^3/6,
	\end{align*}
	where $\zeta_+ \in (\tau, \tau + h)$ and $\zeta_- \in (\tau - h, \tau)$. Combine both expansions and conclude that
	\begin{align}\label{eq:lemma:RelativeConsistencyDensity-2}
		Q_Y'(\tau; X) = \frac{Q_Y(\tau + h; X) - Q_Y(\tau - h; X)}{2h} + \left(Q_Y'''(\zeta_+; X) -  Q_Y'''(\zeta_-; X)\right)h^2/12.
	\end{align}
	Recall the identity $Q_Y'(\tau; X)  = 1/f_{Y|X}(X'\theta_0(\tau)|X)$ and invoke Assumption~\ref{assumption:DiffCQF} to arrive at
	\begin{align}\label{eq:lemma:RelativeConsistencyDensity-3}
		\begin{split}
			&\sup_{\tau \in \mathcal{T}}\max_{1 \leq i \leq n}\left|\frac{1}{\hat{f}_i(\tau)} - \frac{1}{f_i(\tau)}\right|\\
			&\leq \sup_{\tau \in \mathcal{T}} \max_{1 \leq i \leq n}\left|\frac{X_i'\hat{\theta}_\lambda(\tau + h) - X_i'\hat{\theta}_\lambda(\tau - h)  - Q(\tau + h; X_i) + Q(\tau - h; X_i)}{2h}\right| + C_Q h^2.
		\end{split}
	\end{align}
	By Assumption~\ref{assumption:SubGaussianity}, Lemma~\ref{lemma:SizeConesDominatedCoordinates} and~\ref{lemma:MaximaBiconvexFunction},
	\begin{align*}
		\big\|\max_{1 \leq i \leq n} \sup_{\tau \in \mathcal{T}}\sup_{u \in C^p(T_\theta(\tau), \bar{c}) \cap B^p(0,1)} X_i'u \big\|_{\psi_2} \lesssim \sqrt{ \log n + s_\theta \log (ep/s_\theta) } \varphi_{\max}^{1/2}(s_\theta),
	\end{align*}
	and by Theorem~\ref{theorem:Consistency}, with probability at least $1 - \delta$,
	\begin{align*}
		\sup_{\tau \in \mathcal{T}}\|\hat{\theta}_\lambda(\tau) - \theta_0(\tau)\|_2 \lesssim \left(\frac{\bar{c}\phi_{\max}^{1/2}(2s_\theta)L_\theta}{\kappa_1(\bar{c})} \vee 1\right)\sqrt{ \frac{s_\theta \log(ep/ s_\theta) + \log n + \log(1/\delta)}{n}} \bigvee \frac{\bar{c}}{\kappa_1(\bar{c})}\frac{\lambda \sqrt{s_\theta}}{n} =: r_\theta.
	\end{align*}
	Hence, with probability at least $1-\delta$,
	\begin{align*}
		\sup_{\tau \in \mathcal{T}}\max_{1 \leq i \leq n}\left|\frac{1}{\hat{f}_i(\tau)} - \frac{1}{f_i(\tau)}\right| \lesssim  \sqrt{n} h^{-1}r_\theta^2 + C_Q h^2.
	\end{align*}
	By assumption the right hand side in above display vanishes as $n, p \rightarrow \infty$. Combine this with eq.~\eqref{eq:lemma:RelativeConsistencyDensity-1} to conclude the proof. 
\end{proof}
\begin{proof}[\textbf{Proof of Theorem~\ref{theorem:BahadurTypeRep}}]
	To simplify notation, we write $f_i(\tau)$, $\tilde{v}(\tau)$, $v_0(\tau)$, and $T_v(\tau)$ instead of $f_{Y|X}(X_i'\theta_0(\tau)|X_i)$, $\tilde{v}(\tau;z)$, $v_0(\tau;z)$, and $T_v(\tau; z)$, respectively. Note that $\hat{a}\hat{b}\hat{c} - abc = (\hat{a} - a)(\hat{b} - b)(\hat{c} - c) + (\hat{a} - a)(\hat{b} - b)c+ (\hat{a} - a)b(\hat{c} -c) + a(\hat{b} -b)(\hat{c} - c)+ (\hat{a}-a)bc + a(\hat{b} - b) c + ab(\hat{c} - c)$ for arbitrary $\hat{a}, \hat{b}, \hat{c}, a, b, c \in \mathbb{R}$. Hence, the error term can be expanded in the following way:
	\begin{align}\label{eq:theorem:BahadurTypeRep-1}
		&\sup_{\tau \in \mathcal{T}} |e_n(\tau; z)| \nonumber\\
		&\quad{}=\sup_{\tau \in \mathcal{T}} \left|z'\hat{\theta}_\lambda(\tau) - z'\theta_0(\tau) - \frac{1}{2n}\sum_{i=1}^n \hat{f}_i(\tau)\big(\tau - \mathbf{1}\{Y_i \leq X_i'\hat{\theta}_\lambda(\tau)\}\big)X_i'\hat{v}_\gamma(\tau) \right.\nonumber\\
		&\quad{}\quad{}\left. + \frac{1}{2n}\sum_{i=1}^n f_i(\tau)\big(\tau - \mathbf{1}\{Y_i \leq X_i'\theta_0(\tau)\}\big)X_i'v_0(\tau) \right|\nonumber\\
		&\quad{}\leq \sup_{\tau \in \mathcal{T}} \left|\frac{1}{2n}\sum_{i=1}^n f_i(\tau)\big(\mathbf{1}\{Y_i \leq X_i'\hat{\theta}_\lambda(\tau)\}- \mathbf{1}\{Y_i \leq X_i'\theta_0(\tau)\}\big)X_i'v_0(\tau) + z'\big(\hat{\theta}_\lambda(\tau) - \theta_0(\tau)\big)\right|\nonumber\\
		&\quad{}\quad{} + \sup_{\tau \in \mathcal{T}} \left|\frac{1}{2n}\sum_{i=1}^n \big(\hat{f}_i(\tau) - f_i(\tau)\big)\big(\mathbf{1}\{Y_i \leq X_i'\theta_0(\tau)\} - \mathbf{1}\{Y_i \leq X_i'\hat{\theta}_\lambda(\tau)\}\big)X_i'\big(\hat{v}_\gamma(\tau) -v_0(\tau)\big) \right| \nonumber\\
		&\quad{} \quad{} + \sup_{\tau \in \mathcal{T}} \left|\frac{1}{2n}\sum_{i=1}^n f_i(\tau)\big(\mathbf{1}\{Y_i \leq X_i'\theta_0(\tau)\}  - \mathbf{1}\{Y_i \leq X_i'\hat{\theta}_\lambda(\tau)\}\big)X_i'\big(\hat{v}_\gamma(\tau) -v_0(\tau)\big) \right| \nonumber\\
		&\quad{}\quad{} + \sup_{\tau \in \mathcal{T}}\left|\frac{1}{2n}\sum_{i=1}^n \big(\hat{f}_i(\tau) - f_i(\tau)\big)\big(\mathbf{1}\{Y_i \leq X_i'\theta_0(\tau)\}  - \mathbf{1}\{Y_i \leq X_i'\hat{\theta}_\lambda(\tau)\}\big)X_i'v_0(\tau)\right|\nonumber\\
		&\quad{}\quad{} + \sup_{\tau \in \mathcal{T}} \left|\frac{1}{2n}\sum_{i=1}^n \big(\hat{f}_i(\tau) - f_i(\tau)\big)\big(\tau - \mathbf{1}\{Y_i \leq X_i'\theta_0(\tau)\}\big)X_i'\big(\hat{v}_\gamma(\tau) -v_0(\tau)\big) \right| \nonumber\\
		&\quad{} \quad{} + \sup_{\tau \in \mathcal{T}} \left|\frac{1}{2n}\sum_{i=1}^n f_i(\tau)\big(\tau - \mathbf{1}\{Y_i \leq X_i'\theta_0(\tau)\}\big)X_i'\big(\hat{v}_\gamma(\tau) -v_0(\tau)\big) \right| \nonumber\\
		&\quad{}\quad{} + \sup_{\tau \in \mathcal{T}}\left|\frac{1}{2n}\sum_{i=1}^n \big(\hat{f}_i(\tau) - f_i(\tau)\big)\big(\tau - \mathbf{1}\{Y_i \leq X_i'\theta_0(\tau)\}\big)X_i'v_0(\tau)\right|\nonumber\\
		&\quad{} = \mathbf{I} + \mathbf{II} + \mathbf{III} + \mathbf{IV} + \mathbf{V} + \mathbf{VI} + \mathbf{VII}.
	\end{align}	
	
	Unless $v_0(\tau)$ is sparse it is impossible to directly bound the difference between $\hat{v}_\gamma(\tau) - v_0(\tau)$ and to control the supremum of $a'v_0(\tau)$ uniformly in $\tau \in \mathcal{T}$ and arbitrary $a \in \mathbb{R}^p$ tightly enough to allow for $p \gg n$. To get around this problem, we will add and subtract the sparse approximation $\tilde{v}(\tau)$ in each of the terms $\mathbf{I}$ -- $\mathbf{VII}$. By construction of $\tilde{v}(\tau)$, we can then bound the differences $\hat{v}_\gamma(\tau) - \tilde{v}(\tau)$ and $\tilde{v}(\tau) - v_0(\tau)$ and control the supremum of $a'\tilde{v}(\tau)$ uniformly in $\tau \in \mathcal{T}$ and arbitrary $a \in \mathbb{R}^p$. In particular, we have, by Lemmas~\ref{lemma:InducedRestrictedConeProp-Dual} (i) and~\ref{lemma:RestrictedConeProp-Dual},
	\begin{align}
		v_0(\tau) - \tilde{v}(\tau) &\in \left\{ u \in \mathbb{R}^p: (u_{T_v(\tau)}, u_{T_v^c(\tau)}) = (0, w_{T_v^c(\tau)}), \: w \in C^p_2(T_v(\tau), 1)\right\}, \label{eq:theorem:BahadurTypeRep-1-1}\\
		\hat{v}_\gamma(\tau) - \tilde{v}(\tau) &\in C_1^p(T_v(\tau), 2\bar{c}) \cup B^p_1\left(0, \frac{2\bar{c}\|z\|_2^2}{\kappa_2^2(\infty)} \frac{\mu}{\gamma}\right),\label{eq:theorem:BahadurTypeRep-1-2}
	\end{align}	
	and, by Definition~\ref{definition:ExactApproxDualSolution},
	\begin{align}
		&\sup_{\tau \in \mathcal{T}}\|v_0(\tau) - \tilde{v}(\tau)\|_2 \lesssim \frac{\|z\|_2}{\kappa_2(\infty)} r_a,\label{eq:theorem:BahadurTypeRep-1-3}\\
		&\tilde{v}(\tau) \in C^p_1(T_v(\tau), 1) \cap B^p_2\left(0, \frac{2\bar{c}\|z\|_2}{\kappa_2(\infty)}\right).\label{eq:theorem:BahadurTypeRep-1-4}
	\end{align}

	We are now ready to bound the seven terms on the far right hand side in above display~\eqref{eq:theorem:BahadurTypeRep-1}.
	
	\textbf{Bound on $\mathbf{I}$.} Define
	\begin{align*}
		\mathcal{G}_1 &= \{g : \mathbb{R}^{p+1} \rightarrow \mathbb{R} : g(X,Y) = f_{Y|X}(X'\theta_0(\tau)|X)\big(1\{Y \leq X'\theta_0(\tau)\} - 1\{Y \leq X'\theta\}\big)X'v, \:\:\theta \in \mathbb{R}^p,\\
		&\quad{}\quad{}\|\theta\|_0 \leq n, \: \|\theta - \theta_0(\tau)\|_2 \leq r_\theta, \: v \in C^p_1(T_v(\tau), 1) \cap B^p_2(0,1),\: \tau \in \mathcal{T} \},\\
		\mathcal{G}_2 &= \{g : \mathbb{R}^{p+1} \rightarrow \mathbb{R} : g(X,Y) = f_{Y|X}(X'\theta_0(\tau)|X)\big(1\{Y \leq X'\theta_0(\tau)\} - 1\{Y \leq X'\theta\}\big)X'v, \:\:\theta \in \mathbb{R}^p,\\
		&\quad{}\quad{}\|\theta\|_0 \leq n, \: \|\theta - \theta_0(\tau)\|_2 \leq r_\theta, \: v \in U^p \cap B^p_2(0, 1), \: \tau \in \mathcal{T}\},	
	\end{align*}
	where  $U^p:= \left\{ u \in \mathbb{R}^p: (u_{T_v(\tau)}, u_{T_v^c(\tau)}) = (0, w_{T_v^c(\tau)}), \: w \in C^p_2(T_v(\tau), 1)\right\}$. Every $g \in \mathcal{G}_q$ is uniquely determined by a triplet $(v, \theta, \tau)$; hence, we may write $g = g_{v, \theta, \tau}$ Let $\hat{s}_\lambda := \sup_{\tau \in \mathcal{T}}\|\hat{\theta}_\lambda(\tau)\|_2$ and define $\mathcal{G}_q(\hat{s}_\lambda) = \{g_{v, \theta, \tau} \in \mathcal{G}_q : \|\theta\|_0 \leq \hat{s}_\lambda\}$, $q \in \{1, 2\}$. Thus, by the geometric constraints~\eqref{eq:theorem:BahadurTypeRep-1-1},~\eqref{eq:theorem:BahadurTypeRep-1-3}, and~\eqref{eq:theorem:BahadurTypeRep-1-4}, and Theorem~\ref{theorem:Consistency}, with probability at least $1- \delta$,
	\begin{align*} 
		\mathbf{I} &\lesssim  \frac{1}{\sqrt{n}}\frac{2\bar{c}\|z\|_2}{\kappa_2(\infty)}\|\mathbb{G}_n\|_{\mathcal{G}_1(\hat{s}_\lambda)} + \frac{r_a}{\sqrt{n}}\frac{\|z\|_2}{\kappa_2(\infty)}\|\mathbb{G}_n\|_{\mathcal{G}_2(\hat{s}_\lambda)}  \\
		&\quad{} +  \sup_{\tau \in \mathcal{T}}\sup_{\theta} \left| \frac{1}{2}\mathbb{E}\big[ X'v_0(\tau) f_{Y|X}(X'\theta_0(\tau)|X)\big(1\{Y \leq X'\theta\} - 1\{Y \leq X'\theta_0(\tau)\}\big)\big] + z'\theta - z'\theta_0(\tau) \right|\\
		& \equiv \mathbf{I}_1 + \mathbf{I}_2 + \mathbf{I}_3,
	\end{align*}
	where the supremum in $\theta$ is taken over $\|\theta\|_0 \leq \hat{s}_\lambda$ and $\|\theta - \theta_0(\tau)\|_2 \leq r_\theta$.
	
	By Lemma~\ref{lemma:LocalizedRankScoresCone} (ii), with probability at least $1- \delta$,
	\begin{align}\label{eq:theorem:BahadurTypeRep-3}
		\begin{split}
			\mathbf{I}_1  &\lesssim  \bar{f}^{3/2}\varphi_{\max}^{1/2}(s_v) \big( 1 +  \varphi_{\max}^{1/2}(2s_\theta)\big)\big(1 + \varphi_{\max}^{1/2}(\hat{s}_\lambda + s_\theta)\big)\\
			&\quad{}\times\frac{1}{\sqrt{n}}\frac{2\bar{c}\|z\|_2}{\kappa_2(\infty)}\big(\upsilon_{r_\theta, n}\big(\hat{s}_\lambda \log (1/r_\theta) \big) + \upsilon_{r_\theta, n}(t_{s_v, \hat{s}_\lambda, s_\theta, n, \delta}) \big),
		\end{split}
	\end{align}
	and, by Lemma~\ref{lemma:LocalizedRankScoresCone} (ii) combined with Lemma~\ref{lemma:InducedRestrictedConeProp-Dual} (ii), with probability at least $1- \delta$,
	\begin{align}\label{eq:theorem:BahadurTypeRep-3-2}
		\begin{split}
			\mathbf{I}_2&\lesssim  \bar{f}^{3/2}\varphi_{\max}^{1/2}(s_v) \big( 1 +  \varphi_{\max}^{1/2}(2s_\theta)\big)\big(1 + \varphi_{\max}^{1/2}(\hat{s}_\lambda + s_\theta)\big)\\
			&\quad{}\times\frac{r_a}{\sqrt{n}}\frac{\|z\|_2}{\kappa_2(\infty)}\big(\upsilon_{r_\theta, n}\big(\hat{s}_\lambda \log (1/r_\theta) \big) + \upsilon_{r_\theta, n}(t_{s_v, \hat{s}_\lambda, s_\theta, n, \delta}) \big),
		\end{split}
	\end{align}
	where $t_{s_v, \hat{s}_\lambda, s_\theta, n, \delta} = s_v\log (ep/s_v) + \hat{s}_\lambda\log (ep/\hat{s}_\lambda) + s_\theta\log (ep/s_\theta) + \log(L_fL_\theta n/\delta)$ and $\upsilon_{r_\theta, n}(z) = \sqrt{z}\big(\sqrt{r_\theta} + n^{-1/2}(\log n) \sqrt{z} + n^{-1} (\log n)^{3/2} z\big)$ for $z \geq 0$.
	
	By a first-order Taylor approximation of $F_{Y|X}$ in $X'\theta_0(\tau)$  (with Peano's remainder term) and by the definition of $v_0(\tau)$,
	\begin{align}\label{eq:theorem:BahadurTypeRep-4}
		\mathbf{I}_3 &\leq\sup_{\tau \in \mathcal{T}}\sup_{\theta} \left| \frac{1}{2}\mathbb{E}\big[ X'v_0(\tau) f_{Y|X}^2(X'\theta_0(\tau)|X)X'\big(\theta - \theta_0(\tau)\big) + z'\theta - z'\theta_0(\tau) \right| \nonumber\\
		&\quad{} + \sup_{\tau \in \mathcal{T}}\sup_{\theta}\left|\frac{1}{2}\mathbb{E}\big[ X'v_0(\tau) f_{Y|X}(X'\theta_0(\tau)|X) L_f\big(X'(\theta - \theta_0(\tau))\big)^2\big]  \right| \nonumber\\
		& = \frac{L_f}{2}\sup_{\tau \in \mathcal{T}}\sup_{\theta}\left|\mathbb{E}\big[ X'v_0(\tau) f_{Y|X}(X'\theta_0(\tau)|X) \big(X'(\theta - \theta_0(\tau))\big)^2\big]  \right| \nonumber\\
		&\lesssim \bar{f} L_f \varphi_{\max}^{1/2}(s_v)\varphi_{\max}^{1/2}(\hat{s}_\lambda + s_\theta) \frac{\|z\|_2}{\kappa_2(\infty)} r_\theta^2.
	\end{align}
	Combine eq.~\eqref{eq:theorem:BahadurTypeRep-3}--\eqref{eq:theorem:BahadurTypeRep-4}, use that $(s_v + s_\theta)^2 \log^2(np/\delta) \log^2(n) = o(n)$, and conclude that, with probability at least $1 - \delta$,
	\begin{align}\label{eq:theorem:BahadurTypeRep-5}
		\begin{split}
			\mathbf{I} 
			&\lesssim  L_f \bar{f}^{3/2}\varphi_{\max}^{1/2}(s_v) \big( 1 +  \varphi_{\max}^{1/2}(2s_\theta)\big)\big(1 + \varphi_{\max}^{1/2}(\hat{s}_\lambda + s_\theta)\big)\\
			& \quad{}\times \left( \frac{2\bar{c}\|z\|_2}{\kappa_2(\infty)} + r_a\frac{\|z\|_2}{\kappa_2(\infty)}\right)  \left(\hat{r}_B^2 (\log n) + \sqrt{r_\theta} \hat{r}_B + r_\theta^2\right).
		\end{split}
	\end{align}
	
	\textbf{Bound on $\mathbf{II}$.} Recall the function classes $\mathcal{G}_1$ and $\mathcal{G}_2$. Define	
	\begin{align*}
		\mathcal{G}_3 &= \{g : \mathbb{R}^{p+1} \rightarrow \mathbb{R} : g(X,Y) = f_{Y|X}(X'\theta_0(\tau)|X)\big(1\{Y \leq X'\theta_0(\tau)\} - 1\{Y \leq X'\theta\}\big)(X'v), \:\:\theta \in \mathbb{R}^p,\\
		&\quad{}\quad{}\|\theta\|_0 \leq n, \: \|\theta - \theta_0(\tau)\|_2 \leq r_\theta, \: v \in \mathbb{R}^p, \: \|v\|_0 =1,\:  \|v\|_1=1, \: \tau \in \mathcal{T} \},
	\end{align*}
	Every $g \in \mathcal{G}_3$ is uniquely determined by atriplet $(v, \theta, \tau)$; hence, we may write $g = g_{v, \theta, \tau}$. Define $\mathcal{G}_3(\hat{s}_\lambda) = \{g_{v, \theta, \tau} \in \mathcal{G}_q : \|\theta\|_0 \leq \hat{s}_\lambda\}$.
	
	Using the geometric constraints~\eqref{eq:theorem:BahadurTypeRep-1-1},~\eqref{eq:theorem:BahadurTypeRep-1-2}, and~\eqref{eq:theorem:BahadurTypeRep-1-3}, Theorems~\ref{theorem:Consistency} and~\ref{theorem:ConsistencyDual}, and Lemma~\ref{lemma:RelativeConsistencyDensity} with $r_f = o(1)$ followed by Lemmas~\ref{lemma:SizeConesDominatedCoordinates} and~\ref{lemma:MaximaBiconvexFunction}, we have, with probability at least $1- \delta$,
	\begin{align*}
		\mathbf{II} &\lesssim r_fr_v\|P_n\|_{|\mathcal{G}_1(\hat{s}_\lambda)|}  + r_f \frac{2\bar{c}\|z\|_2^2}{\kappa_2^2(\infty)}\frac{\mu}{\gamma}\|P_n\|_{|\mathcal{G}_3(\hat{s}_\lambda)|} + r_fr_a\frac{\|z\|_2}{\kappa_2(\infty)}\|P_n\|_{|\mathcal{G}_2(\hat{s}_\lambda)|}\equiv \mathbf{II}_1 + \mathbf{II}_2 + \mathbf{II}_3.
	\end{align*}
	
	\textbf{Bound on $\mathbf{II}_1$.}  Since $(s_v + s_\theta)^2 \log^2(np/\delta) \log^2(n) = o(n)$ we have, by Lemma~\ref{lemma:LocalizedRankScoresCone} (iii), with probability at least $1-\delta$,
	\begin{align}\label{eq:theorem:BahadurTypeRep-6}
		\mathbf{II}_1 &\leq \frac{r_fr_v}{\sqrt{n}}\|\mathbb{G}_n\|_{|\mathcal{G}_1(\hat{s}_\lambda)|} + r_f r_v \sup_{\tau} \sup_{\theta} \sup_{v} \frac{1}{2} \mathbb{E}\left[f_{Y|X}(X'\theta_0(\tau)|X)\big|\mathbf{1}\{Y \leq X'\theta_0(\tau)\} - \mathbf{1}\{Y \leq X'\theta\}\big| |X'v|\right] \nonumber\\
		&\leq \frac{r_fr_v}{\sqrt{n}} \|\mathbb{G}_n\|_{|\mathcal{G}_1(\hat{s}_\lambda)|}  +  L_f \bar{f} \varphi_{\max}^{1/2}(s_v)\varphi_{\max}^{1/2}(s_\theta + \hat{s}_\lambda) r_f r_v r_\theta \nonumber\\
		&\lesssim L_f \bar{f}^{3/2} \varphi_{\max}^{1/2}(s_v) \big( 1 +  \varphi_{\max}^{1/2}(2s_\theta)\big)\big(1 + \varphi_{\max}^{1/2}(\hat{s}_\lambda + s_\theta)\big)\nonumber\\
		&\quad{}\times \frac{r_f r_v}{\sqrt{n}}\big(\upsilon_{r_\theta, n}\big(\hat{s}_\lambda \log (1/r_\theta) \big) + \upsilon_{r_\theta, n}(t_{s_v, \hat{s}_\lambda, s_\theta, n, \delta}) + \sqrt{n}r_\theta \big)\nonumber\\
		&\lesssim L_f\bar{f}^{3/2} \varphi_{\max}^{1/2}(s_v) \big( 1 +  \varphi_{\max}^{1/2}(2s_\theta)\big)\big(1 + \varphi_{\max}^{1/2}(\hat{s}_\lambda + s_\theta)\big) \times r_f r_v\left(\hat{r}_B^2 (\log n) + \sqrt{r_\theta} \hat{r}_B + r_\theta\right),
	\end{align}
	where the supremum in the first line is taken over $\tau \in \mathcal{T}$, $v \in C^p_1(T_v(\tau), 1) \cap B^p_2(0,1)$, and $\theta \in \mathbb{R}^p$ such that $\|\theta\|_0 \leq \hat{s}_\lambda$ and $\|\theta - \theta_0(\tau)\|_2 \leq r_\theta$ and the expected value can be bounded by the term in the second line $|\mathbf{1}_A - \mathbf{1}_B| = \mathbf{1}_{A\setminus B} + \mathbf{1}_{B\setminus A}$ for arbitrary sets $A, B$ (see also proof of Lemma~\ref{lemma:LocalizedRankScoresCone}).
	
	\textbf{Bound on $\mathbf{II}_2$.} As above, since $s_\theta^2 \log^2(np/\delta) \log^2(n) = o(n)$ we have, by Lemma~\ref{lemma:LocalizedRankScoresCone} (iv), with probability at least $1-\delta$,
	\begin{align}\label{eq:theorem:BahadurTypeRep-7}
		\mathbf{II}_2 &\leq \frac{r_f}{\sqrt{n}} \frac{2\bar{c}\|z\|_2^2}{\kappa_2^2(\infty)}\frac{\mu}{\gamma}\|\mathbb{G}_n\|_{|\mathcal{G}_3(\hat{s}_\lambda)|}\nonumber\\
		& \quad{} + r_f \frac{2\bar{c}\|z\|_2^2}{\kappa_2^2(\infty)}\frac{\mu}{\gamma} \sup_{\tau} \sup_{\theta} \sup_{v} \frac{1}{2} \mathbb{E}\left[f_{Y|X}(X'\theta_0(\tau)|X)\big|\mathbf{1}\{Y \leq X'\theta_0(\tau)\} - \mathbf{1}\{Y \leq X'\theta\}\big| |X'v|\right] \nonumber\\
		&\leq \frac{r_f}{\sqrt{n}} \frac{2\bar{c}\|z\|_2^2}{\kappa_2^2(\infty)}\frac{\mu}{\gamma} \|\mathbb{G}_n\|_{|\mathcal{G}_3(\hat{s}_\lambda)|}  +  L_f \bar{f} \varphi_{\max}^{1/2}(1)\varphi_{\max}^{1/2}(s_\theta + \hat{s}_\lambda) r_f r_\theta \frac{2\bar{c}\|z\|_2^2}{\kappa_2^2(\infty)}\frac{\mu}{\gamma} \nonumber\\
		&\lesssim L_f \bar{f}^{3/2} \varphi_{\max}^{1/2}(1) \big( 1 +  \varphi_{\max}^{1/2}(2s_\theta)\big)\big(1 + \varphi_{\max}^{1/2}(\hat{s}_\lambda + s_\theta)\big)\nonumber\\
		&\quad{}\times \frac{r_f}{\sqrt{n}} \frac{2\bar{c}\|z\|_2^2}{\kappa_2^2(\infty)}\frac{\mu}{\gamma} \big(\upsilon_{r_\theta, n}\big(\hat{s}_\lambda \log (1/r_\theta) \big) + \upsilon_{r_\theta, n}(t_{1, \hat{s}_\lambda, s_\theta, n, \delta}) + \sqrt{n}r_\theta \big)\nonumber\\
		&\lesssim L_f\bar{f}^{3/2} \varphi_{\max}^{1/2}(1) \big( 1 +  \varphi_{\max}^{1/2}(2s_\theta)\big)\big(1 + \varphi_{\max}^{1/2}(\hat{s}_\lambda + s_\theta)\big) \times r_f \frac{2\bar{c}\|z\|_2^2}{\kappa_2^2(\infty)}\frac{\mu}{\gamma}\left(\hat{r}_B^2 (\log n) + \sqrt{r_\theta} \hat{r}_B + r_\theta\right),
	\end{align}
	where the supremum in the first line is taken over $\tau \in \mathcal{T}$, $v \in B^p_1(0,1)$, and $\theta \in \mathbb{R}^p$ such that $\|\theta\|_0 \leq \hat{s}_\lambda$ and $\|\theta - \theta_0(\tau)\|_2 \leq r_\theta$. (This bound is loose, but a tighter bound  on $\mathbf{II}_2$ does not result in a tighter over all bound because the upper bound on $\mathbf{II}_1$ is of the same order.)
	
	\textbf{Bound on $\mathbf{II}_3$.} Again, since $s_\theta^2 \log^2(np/\delta) \log^2(n) = o(n)$ we have, by Lemma~\ref{lemma:LocalizedRankScoresCone} (iii) combined with Lemma~\ref{lemma:InducedRestrictedConeProp-Dual}, with probability at least $1-\delta$,
	\begin{align}\label{eq:theorem:BahadurTypeRep-8}
		\mathbf{II}_3 &\leq \frac{r_fr_a}{\sqrt{n}}\frac{\|z\|_2}{\kappa_2(\infty)}\|\mathbb{G}_n\|_{|\mathcal{G}_2(\hat{s}_\lambda)|}\nonumber\\
		& \quad{} + r_fr_a\frac{\|z\|_2}{\kappa_2(\infty)} \sup_{\tau} \sup_{\theta} \sup_{v} \frac{1}{2} \mathbb{E}\left[f_{Y|X}(X'\theta_0(\tau)|X)\big|\mathbf{1}\{Y \leq X'\theta_0(\tau)\} - \mathbf{1}\{Y \leq X'\theta\}\big| |X'v|\right] \nonumber\\
		&\leq \frac{r_fr_a}{\sqrt{n}}\frac{\|z\|_2}{\kappa_2(\infty)}\|\mathbb{G}_n\|_{|\mathcal{G}_2(\hat{s}_\lambda)|}  +  L_f \bar{f} \varphi_{\max}^{1/2}(1)\varphi_{\max}^{1/2}(s_\theta + \hat{s}_\lambda) r_fr_ar_\theta \frac{\|z\|_2}{\kappa_2(\infty)} \nonumber\\
		&\lesssim L_f \bar{f}^{3/2} \varphi_{\max}^{1/2}(s_v) \big( 1 +  \varphi_{\max}^{1/2}(2s_\theta)\big)\big(1 + \varphi_{\max}^{1/2}(\hat{s}_\lambda + s_\theta)\big)\nonumber\\
		&\quad{}\times \frac{r_fr_a}{\sqrt{n}}\frac{\|z\|_2}{\kappa_2(\infty)}\big(\upsilon_{r_\theta, n}\big(\hat{s}_\lambda \log (1/r_\theta) \big) + \upsilon_{r_\theta, n}(t_{s_v, \hat{s}_\lambda, s_\theta, n, \delta}) + \sqrt{n}r_\theta \big)\nonumber\\
		&\lesssim L_f\bar{f}^{3/2} \varphi_{\max}^{1/2}(s_v) \big( 1 +  \varphi_{\max}^{1/2}(2s_\theta)\big)\big(1 + \varphi_{\max}^{1/2}(\hat{s}_\lambda + s_\theta)\big) \times r_fr_a \frac{\|z\|_2}{\kappa_2(\infty)}\left(\hat{r}_B^2 (\log n) + \sqrt{r_\theta} \hat{r}_B + r_\theta\right),
	\end{align}
	where the supremum in the first line is taken over $\tau \in \mathcal{T}$, $v \in C^p_2(T_v(\tau),1) \cap B^p_2(0,1)$ (by Lemma~\ref{lemma:InducedRestrictedConeProp-Dual} (ii)), and $\theta \in \mathbb{R}^p$ such that $\|\theta\|_0 \leq \hat{s}_\lambda$ and $\|\theta - \theta_0(\tau)\|_2 \leq r_\theta$.
	
	Combine the bounds~\eqref{eq:theorem:BahadurTypeRep-6}--\eqref{eq:theorem:BahadurTypeRep-8} to conclude that with probability at least $1-\delta$,
	\begin{align}\label{eq:theorem:BahadurTypeRep-9}
		\begin{split}
			\mathbf{II} &\lesssim L_f\bar{f}^{3/2} \varphi_{\max}^{1/2}(s_v) \big( 1 +  \varphi_{\max}^{1/2}(2s_\theta)\big)\big(1 + \varphi_{\max}^{1/2}(\hat{s}_\lambda + s_\theta)\big)\\
			&\quad{} \times\left( r_f r_v + r_f \frac{2\bar{c}\|z\|_2^2}{\kappa_2^2(\infty)}\frac{\mu}{\gamma} + r_fr_a \frac{\|z\|_2}{\kappa_2(\infty)}\right)\left(\hat{r}_B^2 (\log n) + \sqrt{r_\theta} \hat{r}_B + r_\theta\right).
		\end{split}
	\end{align}
	
	\textbf{Bound on $\mathbf{III}$.} Recall the definition of the function classes $\mathcal{G}_1, \mathcal{G}_2, \mathcal{G}_3$. Then, by the geometric constraints~\eqref{eq:theorem:BahadurTypeRep-1-1},~\eqref{eq:theorem:BahadurTypeRep-1-2}, and~\eqref{eq:theorem:BahadurTypeRep-1-3}, Theorems~\ref{theorem:Consistency} and~\ref{theorem:ConsistencyDual}, and Lemma~\ref{lemma:RelativeConsistencyDensity} with $r_f = o(1)$ followed by Lemmas~\ref{lemma:SizeConesDominatedCoordinates} and~\ref{lemma:MaximaBiconvexFunction}, we have, with probability at least $1- \delta$,
	\begin{align*}
		\mathbf{III} &\lesssim r_v\|P_n\|_{\mathcal{G}_1(\hat{s}_\lambda)}  + \frac{2\bar{c}\|z\|_2^2}{\kappa_2^2(\infty)}\frac{\mu}{\gamma}\|P_n\|_{\mathcal{G}_3(\hat{s}_\lambda)} + r_a\frac{\|z\|_2}{\kappa_2(\infty)}\|P_n\|_{\mathcal{G}_2(\hat{s}_\lambda)}\equiv\mathbf{III}_1 + \mathbf{III}_2 + \mathbf{III}_3.
	\end{align*}
	We can bound $ \mathbf{III}_1,  \mathbf{III}_2,  \mathbf{III}_3$ using the same arguments as those used to bound $\mathbf{II}_1,  \mathbf{II}_2,  \mathbf{II}_3$. The only difference is that we apply Lemma~\ref{lemma:LocalizedRankScoresCone} (iv) instead of (iii). Thus, with probability at least $1- \delta$,
	\begin{align}\label{eq:theorem:BahadurTypeRep-10}
		\begin{split}
			\mathbf{III} &\lesssim L_f\bar{f}^{3/2} \varphi_{\max}^{1/2}(s_v) \big( 1 +  \varphi_{\max}^{1/2}(2s_\theta)\big)\big(1 + \varphi_{\max}^{1/2}(\hat{s}_\lambda + s_\theta)\big)\\
			&\quad{} \times\left( r_v + \frac{2\bar{c}\|z\|_2^2}{\kappa_2^2(\infty)}\frac{\mu}{\gamma} + r_a \frac{\|z\|_2}{\kappa_2(\infty)}\right)\left(\hat{r}_B^2 (\log n) + \sqrt{r_\theta} \hat{r}_B + r_\theta\right).
		\end{split}
	\end{align}
	
	\textbf{Bound on $\mathbf{IV}$.} Recall the definition of the function classes $\mathcal{G}_1, \mathcal{G}_2$. By the geometric constraints~\eqref{eq:theorem:BahadurTypeRep-1-1},~\eqref{eq:theorem:BahadurTypeRep-1-3}, and~\eqref{eq:theorem:BahadurTypeRep-1-4}, Theorem~\ref{theorem:Consistency}, and Lemma~\ref{lemma:RelativeConsistencyDensity} with $r_f = o(1)$, we have, with probability at least $1- \delta$,
	\begin{align*}
		\mathbf{IV} &\lesssim  r_f\frac{2\bar{c}\|z\|_2}{\kappa_2(\infty)}\|P_n\|_{|\mathcal{G}_1(\hat{s}_\lambda)|} + r_fr_a\frac{\|z\|_2}{\kappa_2(\infty)}\|P_n\|_{|\mathcal{G}_2(\hat{s}_\lambda)|} \equiv \mathbf{IV}_1 + \mathbf{IV}_2.
	\end{align*}
	To bound $\mathbf{IV}_1, \mathbf{IV}_2$ we use Lemma~\ref{lemma:LocalizedRankScoresCone} (iii): Recall that $s_\theta^2 \log^2(np/\delta) \log^2(n) = o(n)$. Then, with probability at least $1- \delta$,
	\begin{align*}
		\mathbf{IV}_1 &\leq \frac{r_f}{\sqrt{n}}\frac{2\bar{c}\|z\|_2}{\kappa_2(\infty)}\|\mathbb{G}_n\|_{|\mathcal{G}_1(\hat{s}_\lambda)|}\\
		&\quad{} + r_f\frac{2\bar{c}\|z\|_2}{\kappa_2(\infty)}  \sup_{\tau} \sup_{\theta} \sup_{v} \frac{1}{2} \mathbb{E}\left[f_{Y|X}(X'\theta_0(\tau)|X)\big|\mathbf{1}\{Y \leq X'\theta_0(\tau)\} - \mathbf{1}\{Y \leq X'\theta\}\big| |X'v|\right]\\
		&\lesssim  L_f\bar{f}^{3/2} \varphi_{\max}^{1/2}(s_v) \big( 1 +  \varphi_{\max}^{1/2}(2s_\theta)\big)\big(1 + \varphi_{\max}^{1/2}(\hat{s}_\lambda + s_\theta)\big) \times r_f \frac{2\bar{c}\|z\|_2}{\kappa_2(\infty)}\left(\hat{r}_B^2 (\log n) + \sqrt{r_\theta} \hat{r}_B + r_\theta\right),
	\end{align*}
	and, with probability at least $1- \delta$,
	\begin{align*}
		\mathbf{IV}_2 &\leq \frac{r_fr_a}{\sqrt{n}}\frac{2\bar{c}\|z\|_2}{\kappa_2(\infty)}\|\mathbb{G}_n\|_{|\mathcal{G}_2(\hat{s}_\lambda)|}\\
		&\quad{} + r_fr_a\frac{2\bar{c}\|z\|_2}{\kappa_2(\infty)}  \sup_{\tau} \sup_{\theta} \sup_{v} \frac{1}{2} \mathbb{E}\left[f_{Y|X}(X'\theta_0(\tau)|X)\big|\mathbf{1}\{Y \leq X'\theta_0(\tau)\} - \mathbf{1}\{Y \leq X'\theta\}\big| |X'v|\right]\\
		&\lesssim  L_f\bar{f}^{3/2} \varphi_{\max}^{1/2}(s_v) \big( 1 +  \varphi_{\max}^{1/2}(2s_\theta)\big)\big(1 + \varphi_{\max}^{1/2}(\hat{s}_\lambda + s_\theta)\big) \times r_fr_a \frac{2\bar{c}\|z\|_2}{\kappa_2(\infty)}\left(\hat{r}_B^2 (\log n) + \sqrt{r_\theta} \hat{r}_B + r_\theta\right).
	\end{align*}
	Hence, with probability at least $1- \delta$,
	\begin{align}\label{eq:theorem:BahadurTypeRep-11} 
		\begin{split}
			\mathbf{IV} &\lesssim L_f\bar{f}^{3/2} \varphi_{\max}^{1/2}(s_v) \big( 1 +  \varphi_{\max}^{1/2}(2s_\theta)\big)\big(1 + \varphi_{\max}^{1/2}(\hat{s}_\lambda + s_\theta)\big)\\
			&\quad{} \times\left( r_f \frac{2\bar{c}\|z\|_2}{\kappa_2(\infty)} + r_fr_a \frac{\|z\|_2}{\kappa_2(\infty)}\right)\left(\hat{r}_B^2 (\log n) + \sqrt{r_\theta} \hat{r}_B + r_\theta\right).
		\end{split}
	\end{align}
	
	\textbf{Bound on $\mathbf{V}$.} To derive a tight bound on this term we leverage the special structure of the estimates $\{\hat{f}_i(\tau)\}_{i=1}^n$ and properties of quantiles and associated densities: Note that 
	\begin{align*}
		\frac{\hat{f}_i(\tau)}{f_i(\tau)} - 1 = \left(\frac{1}{f_i(\tau)}-\frac{1}{\hat{f}_i(\tau)}\right) \left(\frac{\hat{f}_i(\tau)}{f_i(\tau)} - 1 \right) f_i(\tau) + \left(\frac{1}{f_i(\tau)}-\frac{1}{\hat{f}_i(\tau)}\right)f_i(\tau).
	\end{align*}
	Further, recall that
	\begin{align*}
		Q_Y'(\tau; X)  = \frac{1}{f_{Y|X}(X'\theta_0(\tau)|X)},
	\end{align*}
	and that by eq.~\eqref{eq:lemma:RelativeConsistencyDensity-2} from the proof of Lemma~\ref{lemma:RelativeConsistencyDensity}, for $h > 0$ arbitrary,
	\begin{align*}
		Q_Y'(\tau; X) = \frac{Q_Y(\tau + h; X) - Q_Y(\tau - h; X)}{2h} + \left(Q_Y'''(\zeta_+; X) -  Q_Y'''(\zeta_-; X)\right)h^2.
	\end{align*}
	Combine above identities to conclude that
	\begin{align}\label{eq:theorem:BahadurTypeRep-12}
		\begin{split}
			\hat{f}_i(\tau) - f_i(\tau) &=  f_i^2(\tau)h^2 \big(Q_Y'''(\zeta_+; X) -  Q_Y'''(\zeta_-; X)\big)\left( \left(\frac{\hat{f}_i(\tau)}{f_i(\tau)} - 1\right)+1\right)\\
			&\quad{} + f_i^2(\tau)\left( \frac{X_i'\big(\theta_0(\tau + h) - \hat{\theta}_\lambda(\tau + h)\big)}{2h}\right) 
			\left(\left(\frac{\hat{f}_i(\tau)}{f_i(\tau)} - 1\right) +  1\right) \\
			&\quad{}+ f_i^2(\tau)\left( \frac{X_i'\big(\theta_0(\tau - h) - \hat{\theta}_\lambda(\tau - h)\big)}{2h}\right) 
			\left(\left(\frac{\hat{f}_i(\tau)}{f_i(\tau)} - 1\right) +  1\right).
		\end{split}
	\end{align}
	The remaining steps are similar to those that we have already developed. Define	
	\begin{align*}
		\mathcal{G}_4 &= \{g : \mathbb{R}^{p+1} \rightarrow \mathbb{R} : g(X,Y) = (X'u)(X'v),\:v \in C^p_1(T_v(\tau), 1) \cap B^p_2(0,1), \: u \in C^p_1(T_\theta(\tau), \bar{c}) \cap B^p_2(0,1), \: \tau \in \mathcal{T} \},\\
		\mathcal{G}_5 &= \{g : \mathbb{R}^{p+1} \rightarrow \mathbb{R} : g(X,Y) = (X'u)(X'v),\: v \in U^p \cap B^p_2(0, 1), \: u \in C^p_1(T_\theta(\tau), \bar{c}) \cap B^p_2(0,1), \: \tau \in \mathcal{T}\},	\\
		\mathcal{G}_6 &= \{g : \mathbb{R}^{p+1} \rightarrow \mathbb{R} : g(X,Y) = (X'u)(X'v),\: v \in \mathbb{R}^p, \: \|v\|_0=1, \: \|v\|_1 \leq 1,\: u \in C^p_1(T_\theta(\tau), \bar{c}) \cap B^p_2(0,1), \:  \tau \in \mathcal{T} \},\\
		\mathcal{G}_7 &= \{g : \mathbb{R}^{p+1} \rightarrow \mathbb{R} : g(X,Y) = f_{Y|X}^2(X'\theta_0(\tau)|X)\big(\tau - \mathbf{1}\{Y_i \leq X_i'\theta_0(\tau)\}\big)(X'u)(X'v),\\
		&\quad{}\quad{}v \in C^p_1(T_v(\tau), 1) \cap B^p_2(0,1), \: u \in C^p_1(T_\theta(\tau), \bar{c}) \cap B^p_2(0,1), \: \tau \in \mathcal{T} \},\\
		\mathcal{G}_8 &= \{g : \mathbb{R}^{p+1} \rightarrow \mathbb{R} : g(X,Y) = f_{Y|X}^2(X'\theta_0(\tau)|X)\big(\tau - \mathbf{1}\{Y_i \leq X_i'\theta_0(\tau)\}\big)(X'u)(X'v),\\
		&\quad{}\quad{}v \in U^p \cap B^p_2(0, 1), \: u \in C^p_1(T_\theta(\tau), \bar{c}) \cap B^p_2(0,1), \: \tau \in \mathcal{T}\},	\\
		\mathcal{G}_9 &= \{g : \mathbb{R}^{p+1} \rightarrow \mathbb{R} : g(X,Y) = f_{Y|X}^2(X'\theta_0(\tau)|X)\big(\tau - \mathbf{1}\{Y_i \leq X_i'\theta_0(\tau)\}\big)(X'u)(X'v),\\
		&\quad{}\quad{}v \in \mathbb{R}^p, \: \|v\|_0=1, \: \|v\|_1 \leq 1,\:u \in C^p_1(T_\theta(\tau), \bar{c}) \cap B^p_2(0,1), \:  \tau \in \mathcal{T} \},
	\end{align*}
	where  $U^p:= \left\{ u \in \mathbb{R}^p: (u_{T_v(\tau)}, u_{T_v^c(\tau)}) = (0, w_{T_v^c(\tau)}), \: w \in C^p_2(T_v(\tau), 1)\right\}$. 
	
	Using the geometric constraints~\eqref{eq:theorem:BahadurTypeRep-1-1},~\eqref{eq:theorem:BahadurTypeRep-1-2}, and~\eqref{eq:theorem:BahadurTypeRep-1-3}, the expansion~\eqref{eq:theorem:BahadurTypeRep-12}, Theorems~\ref{theorem:Consistency} and~\ref{theorem:ConsistencyDual}, and Lemma~\ref{lemma:RelativeConsistencyDensity} with $r_f = o(1)$ followed by Lemmas~\ref{lemma:SizeConesDominatedCoordinates} and~\ref{lemma:MaximaBiconvexFunction}, we have, with probability at least $1- \delta$,
	\begin{align*}
		\mathbf{V} &\lesssim \bar{f}^2(C_Q h^2 + h^{-1}r_\theta) \left(r_fr_v\|P_n\|_{|\mathcal{G}_4|}  + r_f\frac{2\bar{c}\|z\|_2^2}{\kappa_2^2(\infty)}\frac{\mu}{\gamma}\|P_n\|_{|\mathcal{G}_6|} + r_f r_a\frac{\|z\|_2}{\kappa_2(\infty)}\|P_n\|_{|\mathcal{G}_5|} \right)\\
		&\quad{} + (C_Q h^2 + h^{-1}r_\theta) \left(r_v\|P_n\|_{\mathcal{G}_7}  + \frac{2\bar{c}\|z\|_2^2}{\kappa_2^2(\infty)}\frac{\mu}{\gamma}\|P_n\|_{\mathcal{G}_9} + r_a\frac{\|z\|_2}{\kappa_2(\infty)}\|P_n\|_{\mathcal{G}_8} \right)\\
		&\equiv  (C_Q h^2 + h^{-1}r_\theta) \left(\mathbf{V}_1 + \mathbf{V}_2 + \mathbf{V}_3 +\mathbf{V}_4 + \mathbf{V}_5 + \mathbf{V}_6 \right).
	\end{align*}
	Recall that $s_\theta^2 \log^2(np/\delta) \log^2(n) = o(n)$. Whence, by Lemma~\ref{lemma:MaxInequalityCovarianceCone} (ii), with probability at least $1- \delta$,
	\begin{align*}
		\mathbf{V}_1 &\leq \bar{f}^2	\frac{r_fr_v}{\sqrt{n}} \|\mathbb{G}_n\|_{|\mathcal{G}_4|} +  \bar{f}^2r_fr_v \sup_v \sup_u \mathbb{E}\left[ |X'u||X'v|\right]\\
		& \lesssim (2 + 2\bar{c}) \bar{f}^2\varphi_{\max}^{1/2}(s_\theta)\varphi_{\max}^{1/2}(s_v) \times \left(r_fr_v \hat{r}_B +r_f r_v\right),
	\end{align*} 
	by Lemma~\ref{lemma:MaxInequalityCovarianceCone} (ii) and Lemma~\ref{lemma:InducedRestrictedConeProp-Dual}, with probability at least $1- \delta$,
	\begin{align*}
		\mathbf{V}_2 &\leq	\frac{r_f }{\sqrt{n}}\frac{\bar{f}^2 2\bar{c}\|z\|_2^2}{\kappa_2^2(\infty)}\frac{\mu}{\gamma}\|\mathbb{G}_n\|_{|\mathcal{G}_6|} + 	r_f\frac{\bar{f}^2 2\bar{c}\|z\|_2^2}{\kappa_2^2(\infty)}\frac{\mu}{\gamma} \sup_v \sup_u \mathbb{E}\left[ |X'u||X'v|\right]\\
		&\lesssim  (2 + 2\bar{c}) \bar{f}^2\varphi_{\max}^{1/2}(s_\theta)\varphi_{\max}^{1/2}(s_v) \times \left(r_f\frac{2\bar{c}\|z\|_2^2}{\kappa_2^2(\infty)}\frac{\mu}{\gamma}\hat{r}_B +r_f\frac{2\bar{c}\|z\|_2^2}{\kappa_2^2(\infty)}\frac{\mu}{\gamma}\right),
	\end{align*} 
	and by Lemma~\ref{lemma:MaxInequalityCovarianceCone} (iii), with probability at least $1- \delta$,
	\begin{align*}
		\mathbf{V}_3 &\leq	\frac{r_fr_a}{\sqrt{n}} \frac{\bar{f}^2\|z\|_2}{\kappa_2(\infty)} \|\mathbb{G}_n\|_{|\mathcal{G}_5|} +   r_fr_a\frac{\bar{f}^2\|z\|_2}{\kappa_2(\infty)} \sup_v \sup_u \mathbb{E}\left[ |X'u||X'v|\right]\\
		& \lesssim (2 + 2\bar{c})\bar{f}^2\varphi_{\max}^{1/2}(s_\theta)\varphi_{\max}^{1/2}(s_v) \times \left( r_fr_a\frac{\|z\|_2}{\kappa_2(\infty)} \hat{r}_B +  r_fr_a\frac{\|z\|_2}{\kappa_2(\infty)}\right).
	\end{align*} 
	By Lemma~\ref{lemma:GradientCone-3} (i), with probability at least $1- \delta$,
	\begin{align*}
		\mathbf{V}_4 = \frac{r_v}{\sqrt{n}}\|\mathbb{G}_n\|_{\mathcal{G}_7} \lesssim  (2 + 2\bar{c}) \bar{f}^2 \varphi_{\max}^{1/2}(s_\theta)\varphi_{\max}^{1/2}(s_v)\big(1 + \varphi_{\max}^{1/2}(2s_\theta)\big) \times r_v \hat{r}_B,
	\end{align*}
	by Lemma~\ref{lemma:GradientCone-3} (i), with probability at least $1- \delta$,
	\begin{align*}
		\mathbf{V}_5 = \frac{2\bar{c}}{\sqrt{n}}\frac{\|z\|_2^2}{\kappa_2^2(\infty)}\frac{\mu}{\gamma}\|\mathbb{G}_n\|_{\mathcal{G}_9} \lesssim  (2 + 2\bar{c}) \bar{f}^2 \varphi_{\max}^{1/2}(s_\theta)\varphi_{\max}^{1/2}(s_v)\big(1 + \varphi_{\max}^{1/2}(2s_\theta)\big) \times \frac{2\bar{c}\|z\|_2^2}{\kappa_2^2(\infty)}\frac{\mu}{\gamma}\hat{r}_B,
	\end{align*}
	and by Lemma~\ref{lemma:GradientCone-3} (ii), with probability at least $1- \delta$,
	\begin{align*}
		\mathbf{V}_6 =  \frac{ r_a}{\sqrt{n}} \frac{\|z\|_2}{\kappa_2(\infty)} \|\mathbb{G}_n\|_{\mathcal{G}_8} \lesssim  (2 + 2\bar{c}) \bar{f}^2 \varphi_{\max}^{1/2}(s_\theta)\varphi_{\max}^{1/2}(s_v)\big(1 + \varphi_{\max}^{1/2}(2s_\theta)\big) \times r_a\frac{\bar{f}^2\|z\|_2}{\kappa_2(\infty)}\hat{r}_B.
	\end{align*}
	
	Combine these bounds with $r_f = o(1)$ and conclude that with probability at least $1- \delta$,
	\begin{align}\label{eq:theorem:BahadurTypeRep-13}
		\begin{split}
			\mathbf{V} &\lesssim (2 + 2\bar{c})\bar{f}^2 \varphi_{\max}^{1/2}(s_\theta)\varphi_{\max}^{1/2}(s_v)\big(1 + \varphi_{\max}^{1/2}(2s_\theta)\big)\\
			&\quad{} \times (C_Q h^2 + h^{-1}r_\theta)  \left(r_v + \frac{2\bar{c}\|z\|_2^2}{\kappa_2^2(\infty)}\frac{\mu}{\gamma} + r_a\frac{\|z\|_2}{\kappa_2(\infty)}\right) \hat{r}_B.
		\end{split}
	\end{align}
	
	\textbf{Bound on $\mathbf{VI}$.}  Define
	\begin{align*}
		\mathcal{G}_{10} &= \left\{g: \mathbb{R}^{p+1} \rightarrow \mathbb{R}: g(X, Y) = f_{Y|X}(X'\theta_0(\tau)|X)\left(\tau - \mathbf{1}\big\{Y \leq X'\theta_0(\tau) \}\right)X'v, \right.\\
		&\quad{}\quad{}\left. v \in C^p_1(T_v(\tau), 1) \cap B^p_2(0,1), \:\tau \in \mathcal{T}\right\},\\
		\mathcal{G}_{11} &= \left\{g: \mathbb{R}^{p+1} \rightarrow \mathbb{R}: g(X, Y) = f_{Y|X}(X'\theta_0(\tau)|X)\left(\tau - \mathbf{1}\big\{Y \leq X'\theta_0(\tau) \}\right)X'v, \right.\\
		&\quad{}\quad{}\left. v \in U^p \cap B^p_2(0,1), \:\tau \in \mathcal{T}\right\},\\
		\mathcal{G}_{12} &= \left\{g: \mathbb{R}^{p+1} \rightarrow \mathbb{R}: g(X, Y) = f_{Y|X}(X'\theta_0(\tau)|X)\left(\tau - \mathbf{1}\big\{Y \leq X'\theta_0(\tau) \}\right)X'v,\right.\\
		&\quad{}\quad{}\left. \: v \in  \mathbb{R}^p, \: \|v\|_0 = 1, \: \|v\|_1 \leq 1, \:\tau \in \mathcal{T}\right\},
	\end{align*}
	where  $U^p:= \left\{ u \in \mathbb{R}^p: (u_{T_v(\tau)}, u_{T_v^c(\tau)}) = (0, w_{T_v^c(\tau)}), \: w \in C^p_2(T_v(\tau), 1)\right\}$. 
	
	Using the geometric constraints~\eqref{eq:theorem:BahadurTypeRep-1-1},~\eqref{eq:theorem:BahadurTypeRep-1-2}, and~\eqref{eq:theorem:BahadurTypeRep-1-3}, Theorems~\ref{theorem:Consistency} and~\ref{theorem:ConsistencyDual} followed by Lemmas~\ref{lemma:SizeConesDominatedCoordinates} and~\ref{lemma:MaximaBiconvexFunction}, we have, with probability at least $1- \delta$,
	\begin{align*}
		\mathbf{VI} \lesssim r_v\|P_n\|_{\mathcal{G}_{10}}  + 	\frac{2\bar{c}\|z\|_2^2}{\kappa_2^2(\infty)}\frac{\mu}{\gamma}\|P_n\|_{\mathcal{G}_{12}} + r_a\frac{\|z\|_2}{\kappa_2(\infty)}\|P_n\|_{\mathcal{G}_{11}} \equiv \mathbf{VI}_1 + \mathbf{VI}_2 + \mathbf{VI}_3.
	\end{align*}
	Recall that $s_\theta^2 \log^2(np/\delta) \log^2(n) = o(n)$. Whence, by Lemma~\ref{lemma:GradientCone-2} (i), with probability at least $1- \delta$,
	\begin{align*}
		\mathbf{VI}_1 = \frac{r_v}{\sqrt{n}} \|\mathbb{G}_n\|_{\mathcal{G}_{10}} \lesssim \bar{f}\varphi_{\max}^{1/2}(s_v)\big(1 + \varphi_{\max}^{1/2}(2s_\theta)\big) \times r_v \hat{r}_B, 
	\end{align*} 
	by Lemma~\ref{lemma:GradientCone-2} (i) and Lemma~\ref{lemma:InducedRestrictedConeProp-Dual}, with probability at least $1- \delta$,
	\begin{align*}
		\mathbf{VI}_2 =	\frac{2\bar{c}}{\sqrt{n}}\frac{\|z\|_2^2}{\kappa_2^2(\infty)}\frac{\mu}{\gamma}\|\mathbb{G}_n\|_{\mathcal{G}_{12}} \lesssim  \bar{f}\varphi_{\max}^{1/2}(s_v)\big(1 + \varphi_{\max}^{1/2}(2s_\theta)\big) \times \frac{2\bar{c}\|z\|_2^2}{\kappa_2^2(\infty)}\frac{\mu}{\gamma}\hat{r}_B,
	\end{align*} 
	by Lemma~\ref{lemma:GradientCone-2} (ii), with probability at least $1- \delta$,
	\begin{align*}
		\mathbf{VI}_3 =	\frac{r_a}{\sqrt{n}} \frac{\|z\|_2}{\kappa_2(\infty)} \|\mathbb{G}_n\|_{\mathcal{G}_{11}}  \lesssim \bar{f}\varphi_{\max}^{1/2}(s_v)\big(1 + \varphi_{\max}^{1/2}(2s_\theta)\big) \times r_a\frac{\|z\|_2}{\kappa_2(\infty)} \hat{r}_B.
	\end{align*} 
	Combine these bounds to conclude that with probability at least $1- \delta$,
	\begin{align}\label{eq:theorem:BahadurTypeRep-14}
		\mathbf{VI} &\lesssim  \bar{f}\varphi_{\max}^{1/2}(s_v)\big(1 + \varphi_{\max}^{1/2}(2s_\theta)\big) \times \left(r_v + \frac{2\bar{c}\|z\|_2^2}{\kappa_2^2(\infty)}\frac{\mu}{\gamma} + r_a\frac{\|z\|_2}{\kappa_2(\infty)}\right) \hat{r}_B.
	\end{align}

	\textbf{Bound on $\mathbf{VII}.$} Recall the function classes $\mathcal{G}_4, \mathcal{G}_5, \mathcal{G}_7$, and $\mathcal{G}_8$. By the geometric constraints~\eqref{eq:theorem:BahadurTypeRep-1-1},~\eqref{eq:theorem:BahadurTypeRep-1-3}, and~\eqref{eq:theorem:BahadurTypeRep-1-4}, the expansion~\eqref{eq:theorem:BahadurTypeRep-12}, Theorems~\ref{theorem:Consistency} and~\ref{theorem:ConsistencyDual}, and Lemma~\ref{lemma:RelativeConsistencyDensity} with $r_f = o(1)$, we have, with probability at least $1- \delta$,
	\begin{align*}
		\mathbf{VII} &\lesssim \bar{f}^2(C_Q h^2 + h^{-1}r_\theta) \left(r_f\frac{2\bar{c}\|z\|_2}{\kappa_2(\infty)}\|P_n\|_{|\mathcal{G}_4 |} + r_fr_a\frac{2\bar{c}\|z\|_2}{\kappa_2(\infty)}\|P_n\|_{|\mathcal{G}_5 |}  \right)\\
		&\quad{} +  (C_Q h^2 + h^{-1}r_\theta) \left(\frac{2\bar{c}\|z\|_2}{\kappa_2(\infty)}\|P_n\|_{\mathcal{G}_7} +  r_a\frac{2\bar{c}\|z\|_2}{\kappa_2(\infty)}\|P_n\|_{\mathcal{G}_8}  \right)\\
		&\equiv (1 + r_f)(C_Q h^2 + h^{-1}r_\theta) \left(\mathbf{VII}_1 + \mathbf{VII}_2+ \mathbf{VII}_3 + \mathbf{VII}_4 \right).
	\end{align*}
	Since $s_\theta^2 \log^2(np/\delta) \log^2(n) = o(n)$ we have by Lemma~\ref{lemma:MaxInequalityCovarianceCone} (ii), with probability at least $1- \delta$,
	\begin{align*}
		\mathbf{VII}_1 &\leq \frac{r_f}{\sqrt{n}}\frac{\bar{f}^22\bar{c}\|z\|_2}{\kappa_2(\infty)}\|\mathbb{G}_n\|_{|\mathcal{G}_4|} + r_f\frac{\bar{f}^2 2\bar{c}\|z\|_2}{\kappa_2(\infty)} \sup_v \sup_u \mathbb{E}\left[ |X'u||X'v|\right]\\
		& \lesssim (2 + 2\bar{c})\bar{f}^2\varphi_{\max}^{1/2}(s_\theta)\varphi_{\max}^{1/2}(s_v) \times \left( r_f\frac{2\bar{c}\|z\|_2}{\kappa_2(\infty)} \hat{r}_B + 	r_f\frac{2\bar{c}\|z\|_2}{\kappa_2(\infty)}\right),
	\end{align*} 
	and by Lemma~\ref{lemma:MaxInequalityCovarianceCone} (ii), with probability at least $1- \delta$,
	\begin{align*}
		\mathbf{VII}_2 &\leq \frac{r_fr_a}{\sqrt{n}}\frac{\bar{f}^2 2\bar{c}\|z\|_2}{\kappa_2(\infty)}\|\mathbb{G}_n\|_{|\mathcal{G}_5|} + r_fr_a\frac{\bar{f}^2 2\bar{c}\|z\|_2}{\kappa_2(\infty)} \sup_v \sup_u \mathbb{E}\left[ |X'u||X'v|\right]\\
		& \lesssim (2 + 2\bar{c})\bar{f}^2\varphi_{\max}^{1/2}(s_\theta)\varphi_{\max}^{1/2}(s_v) \times \left(r_fr_a\frac{2\bar{c}\|z\|_2}{\kappa_2(\infty)} \hat{r}_B + 	r_fr_a\frac{2\bar{c}\|z\|_2}{\kappa_2(\infty)}\right).
	\end{align*} 
	By Lemma~\ref{lemma:GradientCone-3} (i), with probability at least $1- \delta$,
	\begin{align*}
		\mathbf{VII}_3 = \frac{2\bar{c}}{\sqrt{n}}\frac{\|z\|_2}{\kappa_2(\infty)}\|\mathbb{G}_n\|_{\mathcal{G}_7} \lesssim (2 + 2\bar{c})\bar{f}^2\varphi_{\max}^{1/2}(s_\theta)\varphi_{\max}^{1/2}(s_v) \big(1 + \varphi_{\max}^{1/2}(2s_\theta)\big) \times \frac{2\bar{c}\|z\|_2}{\kappa_2(\infty)} \hat{r}_B,
	\end{align*} 
	and Lemma~\ref{lemma:GradientCone-3} (i), with probability at least $1- \delta$,
	\begin{align*}
		\mathbf{VII}_4 = \frac{r_a}{\sqrt{n}}\frac{2\bar{c}\|z\|_2}{\kappa_2(\infty)}\|\mathbb{G}_n\|_{\mathcal{G}_8} \lesssim (2 + 2\bar{c})\bar{f}^2\varphi_{\max}^{1/2}(s_\theta)\varphi_{\max}^{1/2}(s_v)\big(1 + \varphi_{\max}^{1/2}(2s_\theta)\big)  \times r_a \frac{2\bar{c}\|z\|_2}{\kappa_2(\infty)} \hat{r}_B.
	\end{align*} 
	Since $r_f \vee r_a = o(1)$, we conclude that, with probability at least $1 -\delta$,
	\begin{align}\label{eq:theorem:BahadurTypeRep-15}
		\mathbf{VII} \lesssim (2 + 2\bar{c})\bar{f}^2 \varphi_{\max}^{1/2}(s_\theta)\varphi_{\max}^{1/2}(s_v)\big(1 + \varphi_{\max}^{1/2}(2s_\theta)\big)\times (C_Q h^2 + h^{-1}r_\theta)  \left(r_f\frac{2\bar{c}\|z\|_2}{\kappa_2(\infty)}  + \frac{2\bar{c}\|z\|_2}{\kappa_2(\infty)}\hat{r}_B \right).
	\end{align}
	
	\textbf{Conclusion.} Since $r_f \vee r_a = o(1)$ and by definition of $r_v$, we have
	\begin{align*}
		\mathbf{I} + \mathbf{II} + \mathbf{III} + \mathbf{IV} &\lesssim L_f \bar{f}^{3/2}\varphi_{\max}^{1/2}(s_v) \big( 1 +  \varphi_{\max}^{1/2}(2s_\theta)\big)\big(1 + \varphi_{\max}^{1/2}(\hat{s}_\lambda + s_\theta)\big)\\
		&\quad{}\quad{} \times \left(\hat{r}_B^2 (\log n) + \sqrt{r_\theta} \hat{r}_B \right) \left( \frac{2\bar{c}\|z\|_2}{\kappa_2(\infty)} + \frac{2\bar{c}\|z\|_2^2}{\kappa_2^2(\infty)}\frac{\mu}{\gamma} \right)\\
		&\quad{} + L_f \bar{f}^{3/2}\varphi_{\max}^{1/2}(s_v) \big( 1 +  \varphi_{\max}^{1/2}(2s_\theta)\big)\big(1 + \varphi_{\max}^{1/2}(\hat{s}_\lambda + s_\theta)\big)\\
		&\quad{}\quad{} \times r_\theta \left(r_v + r_f \frac{2\bar{c}\|z\|_2}{\kappa_2(\infty)} + r_a  \frac{\|z\|_2}{\kappa_2(\infty)}  + r_\theta \frac{2\bar{c}\|z\|_2}{\kappa_2(\infty)}+\frac{2\bar{c}\|z\|_2^2}{\kappa_2^2(\infty)}\frac{\mu}{\gamma} \right)\\
		&\lesssim L_f \bar{f}^{3/2}\varphi_{\max}^{1/2}(s_v) \big( 1 +  \varphi_{\max}^{1/2}(2s_\theta)\big)\big(1 + \varphi_{\max}^{1/2}(\hat{s}_\lambda + s_\theta)\big)\\
		&\quad{}\quad{} \times \left[ \left(\hat{r}_B^2 (\log n) + \sqrt{r_\theta} \hat{r}_B + r_\theta \right)r_v +  \left(\hat{r}_B^2 (\log n) + \sqrt{r_\theta} \hat{r}_B \right)\frac{2\bar{c}\|z\|_2}{\kappa_2(\infty)} \right],
	\end{align*}
	and
	\begin{align*}
		\mathbf{V} + \mathbf{VII} &\lesssim (2 + 2\bar{c})\bar{f}^2 \varphi_{\max}^{1/2}(s_\theta)\varphi_{\max}^{1/2}(s_v)\big(1 + \varphi_{\max}^{1/2}(2s_\theta)\big)\\
		&\quad \times (C_Q h^2 + h^{-1}r_\theta) \left( r_f \frac{2\bar{c}\|z\|_2}{\kappa_2(\infty)} + \frac{2\bar{c}\|z\|_2}{\kappa_2(\infty)}\hat{r}_B + \frac{2\bar{c}\|z\|_2^2}{\kappa_2^2(\infty)} \frac{\mu}{\gamma} \hat{r}_B \right)\\
		&\lesssim (2 + 2\bar{c})\bar{f}^2 \varphi_{\max}^{1/2}(s_\theta)\varphi_{\max}^{1/2}(s_v)\big(1 + \varphi_{\max}^{1/2}(2s_\theta)\big) \times (C_Q h^2 + h^{-1}r_\theta) \left[ r_v\hat{r}_B  + \hat{r}_B \frac{2\bar{c}\|z\|_2}{\kappa_2(\infty)}  \right],
	\end{align*}
	and
	\begin{align*}
		\mathbf{VI} \lesssim (2 + 2\bar{c})\bar{f}^2 \varphi_{\max}^{1/2}(s_\theta)\varphi_{\max}^{1/2}(s_v)\big(1 + \varphi_{\max}^{1/2}(2s_\theta)\big) \times (C_Q h^2 + h^{-1}r_\theta) r_v\hat{r}_B .
	\end{align*}
	Combine these bounds with $s_\theta^2 \log^2(np/\delta) \log^2(n) = o(n)$ and conclude that
	\begin{align*}
		\sup_{\tau \in \mathcal{T}} |e_n(\tau; z)| &\lesssim C_4 \left(\hat{r}_B (\log n) + \sqrt{r_\theta} \right) \left( r_v \hat{r}_B + \hat{r}_B \frac{2\bar{c}\|z\|_2}{\kappa_2(\infty)}  \right) + C_4 r_\theta r_v \\
		&\quad{} + C_5 (C_Q h^2 + h^{-1}r_\theta) \left(r_v \hat{r}_B  + \hat{r}_B \frac{2\bar{c}\|z\|_2}{\kappa_2(\infty)}  \right),
	\end{align*}
	where 
	\begin{align}\label{eq:theorem:BahadurTypeRep-16}
		\begin{split}
			C_4 &:= L_f \bar{f}^{3/2}\varphi_{\max}^{1/2}(s_v) \big( 1 +  \varphi_{\max}^{1/2}(2s_\theta)\big)\big(1 + \varphi_{\max}^{1/2}(\hat{s}_\lambda + s_\theta)\big), \\
			C_5 &:= (2 + 2\bar{c})\bar{f}^2 \varphi_{\max}^{1/2}(s_\theta)\varphi_{\max}^{1/2}(s_v)\big(1 + \varphi_{\max}^{1/2}(2s_\theta)\big).
		\end{split}
	\end{align}
\end{proof}
\begin{proof}[\textbf{Proof of Corollary~\ref{corollary:theorem:BahadurTypeRep}}]
	The claim follows by combining Theorem~\ref{theorem:BahadurTypeRep} with Theorem~\ref{theorem:EmpiricalSparsity} (ii) and Corollary~\ref{corollary:theorem:ConsistencyDual}. The rate on the remainder term can be simplified by exploiting that, by assumption, $ (s_v + s_\theta)^2 \log^2( np/\delta)\log^2(n) = o(nh^2)$ and $h^2 s_v = o(1)$.
\end{proof}

\subsection{Proofs of Section~\ref{subsec:WeakConvergence-Appendix}}
\begin{proof}[\textbf{Proof of Lemma~\ref{lemma:AsymptoticEquicontinuity}}]
	To simplify notation, we write $f_i(\tau)$,  $v_0(\tau)$, and $T_v(\tau)$ instead of $f_{Y|X}(X_i'\theta_0(\tau)|X_i)$, $v_0(\tau;z)$, and $T_v(\tau; z)$, respectively.
	
	\textbf{Proof of (i).}  Let $q \geq 1$ be arbitrary.
	For all $g_\tau, g_{\tau'} \in \mathcal{G}$
	\begin{align}\label{eq:lemma:AsymptoticEquicontinuity-1}
		&\left\|(g_\tau - Pg_\tau) - (g_{\tau'} - Pg_{\tau'})\right\|_{P,q}\nonumber\\
		&\quad{}= \left\|g_\tau - g_{\tau'}\right\|_{P, q}\nonumber\\
		&\quad{}\leq \left\| \big(f_{Y|X}(X'\theta_0(\tau)|X) - f_{Y|X}(X'\theta_0(\tau')|X)\big)\big(\tau - \mathbf{1}\{Y \leq X'\theta_0(\tau)\}\big) X'v_0(\tau)\right\|_{P, q} \nonumber\\
		&\quad{}\quad{}+ \left\| f_{Y|X}(X'\theta_0(\tau)|X)\big(\tau - \mathbf{1}\{Y \leq X'\theta_0(\tau)\}\big) X'\big(v_0(\tau) - v_0(\tau')\big)\right\|_{P, q}\nonumber\\
		&\quad{}\quad{}+\left\| f_{Y|X}(X'\theta_0(\tau')|X) X'v_0(\tau) \right\|_{P, q} |\tau - \tau'|\nonumber\\
		&\quad{}\quad{}+\left\| f_{Y|X}(X'\theta_0(\tau')|X) \big(\mathbf{1}\{Y \leq X'\theta_0(\tau)\} - \mathbf{1}\{Y \leq X'\theta_0(\tau')\}\big) X'v_0(\tau) \right\|_{P,q} \nonumber\\
		& = \mathbf{I} + \mathbf{II} + \mathbf{III} + \mathbf{IV}.
	\end{align}
	\textbf{Bound on $\mathbf{I}$.} By Assumption~\ref{assumption:LipschitzDensity},
	\begin{align*}
		\mathbf{I} &\leq \left\| \big(f_{Y|X}(X'\theta_0(\tau)|X) - f_{Y|X}(X'\theta_0(\tau')|X)\big) X'v_0(\tau)\right\|_{P, q}\\
		&\leq L_f \left\| \big(\theta_0(\tau) - \theta_0(\tau')\big) XX'v_0(\tau)\right\|_{P, q}\\
		&\leq L_f \|v_0(\tau)\|_2 \| \theta_0(\tau) - \theta_0(\tau')\|_2 \sup_{u ,w} \left(\mathbb{E}\left[ |X'u|^q |X'w|^q\right]\right)^{1/q},
	\end{align*}
	where the supremum over is taken over all $u, w$ satisfying $\|u\|_2= \|w\|_2=1$, $\|u\|_0 \leq s_v$ and $\|w\|_0 \leq 2s_\theta$. By Assumption~\ref{assumption:SubGaussianity} we can upper bound the expected values in the last displayed line as follows:
	\begin{align*}
		&\sup_{u ,w} \left(\mathbb{E}\left[  |X'u|^q |X'w|^q \right]\right)^{1/q} \leq \sup_{u ,w} \left(\mathbb{E}\left[ (X'u)^{2q}\right]\right)^{1/(2q)} \left(\mathbb{E}\left[ (X'w)^{2q} \right]\right)^{1/(2q)}\\
		&\quad{} \leq \sup_{u,w} \left(\| (X - \mathbb{E}[X])'u\|_{2q} + |\mathbb{E}[X'u]|\right)\left( \| (X - \mathbb{E}[X])'w\|_{2q} + |\mathbb{E}[X'w]|\right)\\
		&\quad{} \lesssim \left(\phi_{\max}^{1/2}(s_v) + \varphi_{\max}^{1/2}(s_v)\right) \left( \phi_{\max}^{1/2}(2s_\theta) + \varphi_{\max}^{1/2}(2s_\theta)\right)\\
		&\quad{} \lesssim \varphi_{\max}^{1/2}(s_v)\varphi_{\max}^{1/2}(2s_\theta).
	\end{align*}
	Also, by Assumption~\ref{assumption:LipschitzQRVector},
	\begin{align*}
		\|v_0(\tau)\|_2 \leq \frac{\|z\|_2}{\kappa_2(\infty)} \quad{} \quad{} \mathrm{and} \quad{}\quad{} \|\theta_0(\tau) - \theta_0(\tau')\|_2 \leq L_\theta |\tau - \tau'|.
	\end{align*}
	Hence,
	\begin{align}\label{eq:lemma:AsymptoticEquicontinuity-2}
		\mathbf{I} \leq L_f L_\theta  \varphi_{\max}^{1/2}(2s_\theta)  \varphi_{\max}^{1/2}(s_v)\frac{\|z\|_2}{\kappa_2(\infty)}|\tau - \tau'|.
	\end{align}
	
	\textbf{Bound on $\mathbf{II}$.} By Assumptions~\ref{assumption:SubGaussianity} and~\ref{assumption:BoundedDensity} and Lemma~\ref{lemma:LipschitzDual},
	\begin{align}\label{eq:lemma:AsymptoticEquicontinuity-3}
		\mathbf{II} &\leq  \bar{f} \|v_0(\tau) - v_0(\tau')\|_2 \sup_{u}\left\| X'u\right\|_{P,q} \nonumber\\
		&\lesssim \bar{f} \|v_0(\tau) - v_0(\tau')\|_2 \varphi_{\max}^{1/2}(2s_v) \nonumber\\
		&\lesssim  C_v\bar{f}^2L_fL_\theta\varphi_{\max}^{1/2}(2s_v)\varphi_{\max}^{1/2}(2s_\theta) \varphi_{\max}(p) \frac{  \|z\|_2}{\kappa_2(\infty)} |\tau - \tau'|,
	\end{align}
	where the supremum over is taken over all $u$ satisfying $\|u\|_2 =1$, $\|u\|_0 \leq 2s_v$.
	
	\textbf{Bound on $\mathbf{III}$.} By By Assumptions~\ref{assumption:SubGaussianity} and~\ref{assumption:BoundedDensity},
	\begin{align}\label{eq:lemma:AsymptoticEquicontinuity-4}
		\mathbf{III} \leq  \bar{f}  \frac{  \|z\|_2}{\kappa_2(\infty)}   \sup_{u}\left\| X'u\right\|_{P,q} |\tau - \tau'|  \leq \bar{f}\varphi_{\max}^{1/2}(s_v)  \frac{  \|z\|_2}{\kappa_2(\infty)}  |\tau - \tau'|,
	\end{align}
	where the supremum over is taken over all $u$ satisfying $\|u\|_2 =1$, $\|u\|_0 \leq s_v$.
	
	\textbf{Bound on $\mathbf{IV}$.} We compute
	\begin{align*}
		\mathbf{IV} &\leq \left(\mathbb{E}\left[ f_{Y|X}^q(X'\theta_0(\tau')|X) \big|X'v_0(\tau)\big|^q \big|\mathbf{1}\{Y \leq X'\theta_0(\tau)\} - \mathbf{1}\{Y \leq X'\theta_0(\tau')\}\big|\right]\right)^{1/q}\\
		&\leq \left(\mathbb{E}\left[ f_{Y|X}^q(X'\theta_0(\tau')|X) \big|X'v_0(\tau)\big|^q\mathbf{1}\{ X'\theta_0(\tau')< Y \leq X'\theta_0(\tau)\}\right]\right)^{1/q}\\
		&\quad{} + \left(\mathbb{E}\left[ f_{Y|X}^q(X'\theta_0(\tau')|X) \big|X'v_0(\tau)\big|^q\mathbf{1}\{ X'\theta_0(\tau)< Y \leq X'\theta_0(\tau')\}\right]\right)^{1/q}\\
		& \leq 2 \left(\mathbb{E}\left[ f_{Y|X}^q(X'\theta_0(\tau')|X) \big|X'v_0(\tau)\big|^q\big|F_{Y|X}(X'\theta_0(\tau')|X) -F_{Y|X}(X'\theta_0(\tau)|X)\big|\right]\right)^{1/q}\\
		& \leq 2 \bar{f}^{1 + 1/q}\|v_0(\tau)\|_2 \|\theta_0(\tau) - \theta_0(\tau')\|_2^{1/q} \sup_{u ,w} \left(\mathbb{E}\left[ |X'u|^q |X'w|\right]\right)^{1/q},
	\end{align*}
	where the supremum over is taken over all $u, w$ satisfying $\|u\|_2= \|w\|_2=1$, $\|u\|_0 \leq s_v$ and $\|w\|_0 \leq 2s_\theta$. By Assumption~\ref{assumption:SubGaussianity} we can upper bound the expected values in the last displayed line as follows:
	\begin{align*}
		&\sup_{u ,w} \left(\mathbb{E}\left[ |X'u|^q |X'w|\right]\right)^{1/q} \leq \sup_{u ,w} \left(\mathbb{E}\left[ |X'u|^{2q}\right]\right)^{1/(2q)} \left(\mathbb{E}\left[ (X'w)^2 \right]\right)^{1/(2q)}\\
		&\quad{} \leq \sup_{u,w} \left(\| (X - \mathbb{E}[X])'u\|_{2q} + |\mathbb{E}[X'u]|\right)\left( \| (X - \mathbb{E}[X])'w\|_2 + |\mathbb{E}[X'w]|\right)^{1/2}\\
		&\quad{} \lesssim \left(\phi_{\max}^{1/2}(s_v) + \varphi_{\max}^{1/2}(s_v)\right) \left( \phi_{\max}^{1/2}(2s_\theta) + \varphi_{\max}^{1/2}(2s_\theta)\right)^{1/2}\\
		&\quad{} \lesssim \varphi_{\max}^{1/2}(s_v)\varphi_{\max}^{1/4}(2s_\theta).
	\end{align*}
	Hence, by Assumption~\ref{assumption:LipschitzQRVector},
	\begin{align}\label{eq:lemma:AsymptoticEquicontinuity-5}
		\mathbf{IV} \leq \bar{f}^{1+ 1/q}L_\theta^{1/q} \varphi_{\max}^{1/2}(s_v)\varphi_{\max}^{1/4}(2s_\theta) \frac{\|z\|_2}{\kappa_2(\infty)}|\tau - \tau'|^{1/q}.
	\end{align}
	
	\textbf{Conclusion.} Combining eq.~\eqref{eq:lemma:AsymptoticEquicontinuity-1} and the upper bounds~\eqref{eq:lemma:AsymptoticEquicontinuity-2}--\eqref{eq:lemma:AsymptoticEquicontinuity-5}, we obtain
	\begin{align}\label{eq:lemma:AsymptoticEquicontinuity-6}
		&\left\|(g_\tau - Pg_\tau) - (g_{\tau'} - Pg_{\tau'})\right\|_{P,q}\nonumber\\
		&\quad{}\lesssim L_fL_\theta\varphi_{\max}^{1/2}(2s_\theta)\varphi_{\max}^{1/2}(s_v)\frac{\|z\|_2}{\kappa_2(\infty)}|\tau - \tau'|\nonumber\\
		&\quad{}\quad{}+\bar{f}^2L_fL_\theta \varphi_{\max}^{1/2}(s_v)\varphi_{\max}^{1/2}(2s_\theta) \varphi_{\max}(p) \frac{ \|z\|_2}{\kappa_2(\infty)} |\tau- \tau'|\nonumber\\
		&\quad{}\quad{} + \bar{f} \varphi_{\max}^{1/2}(s_v) \frac{\|z\|_2}{\kappa_2(\infty)}|\tau - \tau'| + \bar{f}^2L_\theta \varphi_{\max}^{1/2}(s_v)\varphi_{\max}^{1/4}(2s_\theta) \frac{\|z\|_2}{\kappa_2(\infty)}|\tau - \tau'|^{1/q}\nonumber\\
		&\quad{} \equiv K \rho_q(\tau, \tau'),
	\end{align}
	where $\rho_q(\tau, \tau') = |\tau - \tau'|^{1/q}$ (because $\tau, \tau' \in \mathcal{T} \subseteq (0,1)$). Total boundedness of $\mathcal{G}$ in the standard deviation metric now follows from eq.~\eqref{eq:lemma:AsymptoticEquicontinuity-6} since $\mathcal{T}$ is totally bounded with respect to $\rho_2$. 
	
	\textbf{Proof of (ii).} We adapt the approach of the proof of Lemma A.3 in~\cite{chao2017quantile} to our setting. Let $\bar{q} \in \mathbb{N}$ and $q = (q_1, \ldots, q_n)' \in \mathbb{R}^n$. By eq.~\eqref{eq:lemma:AsymptoticEquicontinuity-6}, we have, for all $g_\tau, g_{\tau'} \in \mathcal{G}$,
	\begin{align*}
		\mathbb{E}\left[ \left|\mathbb{G}_n(g_\tau- g_{\tau'})\right|^{2\bar{q}} \right] &\leq  n^{-\bar{q}}\mathbb{E}\left[ \left(\sum_{i=1}^n g_\tau(X_i, Y_i) - g_{\tau'}(X_i, Y_i)\right)^{2\bar{q}} \right]\\
		& =  n^{-\bar{q}}\sum_{\|q\|_1 = 2\bar{q}} { 2\bar{q} \choose q_1, \ldots, q_n} \prod_{i=1}^n\mathbb{E}\left[ \big(g_\tau(X_i, Y_i) - g_{\tau'}(X_i, Y_i)\big)^{q_i}\right]\\
		&=  n^{-\bar{q}}\sum_{k=1}^{2\bar{q}} \sum_{\substack{\|q\|_1 = 2\bar{q}, \: q_i \geq 2,\\ \|q\|_0 =k}} { 2\bar{q} \choose q_1, \ldots, q_n} \prod_{i=1}^n\mathbb{E}\left[ \big(g_\tau(X_i, Y_i) - g_{\tau'}(X_i, Y_i)\big)^{q_i}\right]\\
		&\lesssim  n^{-\bar{q}}\sum_{k=1}^{\bar{q}} { n \choose k} (2\bar{q})! K^{2\bar{q}} |\tau - \tau'|^k\\
		&\lesssim  (2\bar{q})! K^{2\bar{q}} \sum_{k=1}^{\bar{q}} \left(\frac{e}{k}\right)^k \frac{|\tau - \tau'|^k}{n^{\bar{q}-k}}.
	\end{align*}
	Hence, for all $\tau, \tau' \in \mathcal{T}$ satisfying $|\tau - \tau'| \geq 1/n$, we have
	\begin{align}\label{eq:lemma:AsymptoticEquicontinuity-7}
		\left\|\mathbb{G}_n(g_\tau- g_{\tau'})\right\|_{P, 2\bar{q}} & \lesssim  K \rho_2(\tau, \tau').
	\end{align}
	Note that the packing number $D(\epsilon, T, \rho_2)$ the $\epsilon$-packing number of $T$ with respect to $\rho_2$ satisfies
	\begin{align}\label{eq:lemma:AsymptoticEquicontinuity-8}
		D(\epsilon, T, \rho_2) \lesssim 1/\epsilon^2.
	\end{align}
	Thus, by Lemma~\ref{lemma:MaxInequalityKley2016} with $\Psi(x) = x^{12}$, $\eta = \xi^{1/2}$, $\bar{\eta} = n^{-1/2}$ (WLOG we can assume that $n$ is so large that $n^{-1} < \xi$), and eq.~\eqref{eq:lemma:AsymptoticEquicontinuity-7} and~\eqref{eq:lemma:AsymptoticEquicontinuity-8} there exists a random variable $S_n(\xi)$ such that
	\begin{align}\label{eq:lemma:AsymptoticEquicontinuity-9}
		\sup_{|\tau - \tau'| \leq \xi} |\mathbb{G}_n(g_\tau- g_{\tau'})| = \sup_{\rho_2(\tau, \tau') \leq \xi^{1/2}} |\mathbb{G}_n(g_\tau- g_{\tau'})| \leq S_n(\xi) + 2 \sup_{\substack{\rho_2(\tau, \tau') \leq n^{-1/2}\\ \tau' \in \widetilde{T}, \tau \in T}} |\mathbb{G}_n(g_\tau- g_{\tau'})|,
	\end{align}
	where the set $\widetilde{T}$ contains at most $D(n^{-1/2}, T, \rho_2) \lesssim n$ points, and
	\begin{align*}
		\|S_n(\xi)\|_{P,\Psi} &\lesssim  \int_{n^{-1/2}/2}^{\xi^{1/2}} \Psi^{-1}\big(D(\epsilon,T,\rho_2)\big) d\epsilon + (\xi^{1/2} +2 n^{-1/2}) \Psi^{-1}\big(D^2(\xi,T, \rho_2)\big) \\
		&\lesssim  \int_{n^{-1/2}/2}^{\xi^{1/2}} \epsilon^{-1/6} d\epsilon + (\xi^{1/2} +2 n^{-1/2}) \xi^{-1/3}\\
		&\lesssim \xi^{5/12} - n^{-5/12} + \xi^{1/6}  + n^{-1/2} \xi^{-1/3}.
	\end{align*}
	Hence, for $\varepsilon > 0$ arbitrary, using Markov's inequality we bound the first term on the far right hand side of eq.~\eqref{eq:lemma:AsymptoticEquicontinuity-9} by
	\begin{align}\label{eq:lemma:AsymptoticEquicontinuity-10}
		\mathbb{P}\left\{|S_n(\xi)| > \varepsilon \right\} \leq \varepsilon^{-12} \left(\xi^{5/12} - n^{-5/12} + \xi^{1/6}  + n^{-1/2} \xi^{-1/3} \right)^{12} \rightarrow 0 \:\: \mathrm{as}\:\: n \rightarrow \infty \:\:\mathrm{followed \: by} \:\: \xi \rightarrow 0.
	\end{align}
	Bounding the second term on the far right hand side of eq.~\eqref{eq:lemma:AsymptoticEquicontinuity-9} is slightly more involved. We have
	\begin{align}\label{eq:lemma:AsymptoticEquicontinuity-11}
		\sup_{\substack{\rho_2(\tau, \tau') \leq n^{-1/2}\\ \tau' \in \widetilde{T}, \tau \in T}} |\mathbb{G}_n(g_\tau- g_{\tau'})| &\leq \sup_{\substack{\rho_2(\tau, \tau') \leq n^{-1/2}\\ \tau', \tau \in T}} |\mathbb{G}_n(g_\tau- g_{\tau'})| \nonumber\\
		&\leq \frac{\bar{c}\|z\|_2}{\kappa_2(\infty)} \|\mathbb{G}_n\|_{\mathcal{G}_1} +\frac{\bar{c}\|z\|_2}{\kappa_2(\infty)} \|\mathbb{G}_n\|_{\mathcal{G}_2} + \|\mathbb{G}_n\|_{\mathcal{G}_3 }\nonumber\\
		&= \mathbf{I} +  \mathbf{II} +  \mathbf{III},
	\end{align}
	where
	\begin{align*}
		\mathcal{G}_1 &= \left\{g: \mathbb{R}^{p+1} \rightarrow \mathbb{R}: g(X, Y) = f_{Y|X}(X'\theta_0(\tau)|X)\left(\tau - \mathbf{1}\big\{Y \leq X'\theta_0(\tau) \big\} - \tau' + \mathbf{1}\big\{Y \leq X'\theta_0(\tau') \}\right)X'v,  \right.\\
		&\quad{}\quad{}\left.|\tau - \tau'| \leq 1/n, \: v \in C^p_1(T_v(\tau), 2\bar{c}) \cap B^p_2(0,1), \:\tau, \tau' \in \mathcal{T}\right\},\\
		\mathcal{G}_2 &= \left\{g: \mathbb{R}^{p+1} \rightarrow \mathbb{R}: g(X, Y) = \big(f_{Y|X}(X'\theta_0(\tau)|X)-f_{Y|X}(X'\theta_0(\tau')|X)\big)\left(\tau' - \mathbf{1}\big\{Y \leq X'\theta_0(\tau') \big\}\right)X'v, \right.\\
		&\quad{}\quad{}\left.|\tau - \tau'| \leq 1/n, \: v \in C^p_1(T_v(\tau), 2\bar{c}) \cap B^p_2(0,1), \:\tau, \tau' \in \mathcal{T}\right\},\\
		\mathcal{G}_3 &= \left\{g: \mathbb{R}^{p+1} \rightarrow \mathbb{R}: g(X, Y) = f_{Y|X}(X'\theta_0(\tau')|X)\left(\tau' - \mathbf{1}\big\{Y \leq X'\theta_0(\tau') \big\}\right)X'\big(v_0(\tau; z) - v_0(\tau'; z)\big), \right.\\
		&\quad{}\quad{}\left.|\tau - \tau'| \leq 1/n, \:\tau, \tau' \in \mathcal{T}\right\}.
	\end{align*}
	
	\textbf{Bound on $\mathbf{I}$.} By Lemma~\ref{lemma:LocalizedRankScoresCone} with $r_0 = L_f/n$ and $s = s_v$ we have, with probability at least $1 - \delta$,
	\begin{align}\label{eq:lemma:AsymptoticEquicontinuity-12}
		\mathbf{I}  &\lesssim \bar{f}^{3/2}\varphi_{\max}^{1/2}(s_v) \big( 1 +  \varphi_{\max}^{1/2}(2s_\theta)\big)^2 \frac{\bar{c}\|z\|_2}{\kappa_2(\infty)}\Big(\upsilon_{L_f/n, n}\big(s_\theta \log (n/L_f) \big) + \upsilon_{L_f/n, n}(t_{s_v, s_\theta, s_\theta, n, \delta}) \Big) \nonumber\\
		\begin{split}
			&\lesssim \bar{f}^{3/2}\varphi_{\max}^{1/2}(s_v) \big( 1 +  \varphi_{\max}^{1/2}(2s_\theta)\big)^2 \frac{\bar{c}\|z\|_2}{\kappa_2(\infty)} \\
			&\quad{} \quad{} \times \left( \frac{s_\theta \log(n/L_f) \log(n)}{\sqrt{n}} + \frac{s_\theta \log(ep/s_\theta) \log(n)}{\sqrt{n}}  + \frac{ \log(L_fL\theta n/\delta) \log(n)}{\sqrt{n}}  \right),
		\end{split}
	\end{align}
	where $t_{s, k, s_\theta, n, \delta} = s\log(ep/s) +  k\log (ep/k) + s_\theta \log (ep/s_\theta) + \log(L_fL_\theta n/\delta)$ and $\upsilon_{r_0, n}(z) = \sqrt{z}\big(\sqrt{r_0} + n^{-1/2}(\log n) \sqrt{z} + n^{-1} (\log n)^{3/2} z\big)$ for $z \geq 0$ and the second, simplified inequality follows from the premises of the lemma.
	
	\textbf{Bound on $\mathbf{II}$.} Let $\{g_i\}_{i=1}^n$ be a sequence of i.i.d. standard normal random variables independent of $\{(X_i, Y_i)\}_{i=1}^n$. Define $K(\tau'') =  C^p_1(T_v(\tau''), 2\bar{c}) \cap B^p_2(0,1)$ and
	\begin{align*}
		\mathcal{F} = \left\{f: \mathbb{R}^{p+1} \times \mathbb{R} \rightarrow \mathbb{R}: f(X, Y, g) = g\left(\tau - \mathbf{1}\big\{Y \leq X'\theta_0(\tau) \}\right)(X'v)(X'u), \right.\\
		\left. v \in K(\tau),\: \|u\|_0 \leq 2s_\theta, \: \|u\|_2 = 1, \: \tau \in \mathcal{T}\right\}.
	\end{align*}
	By Lemma~\ref{lemma:GradientConeSquare}, with probability at least $1-\delta$,
	\begin{align}\label{eq:lemma:AsymptoticEquicontinuity-13}
		\begin{split}
			\|\mathbb{G}_n\|_{\mathcal{F}} &\lesssim \bar{c}\varphi_{\max}^{1/2}(s_1)\varphi_{\max}^{1/2}(s_2) \sqrt{s_1 \log (ep/s_1)  + s_2 \log (ep/s_2) + \log(1/\delta)} \\
			&\quad{} \times \sqrt{1 + \pi_{n,1/3}^2(s_1 \log (ep/s_1) + s_2 \log (ep/s_2) + \log(1/\delta))},
		\end{split}
	\end{align}
	where $\pi_{n,1/3}^2(z) = \sqrt{z/n} + z^3/n$ with $z \geq 0$. We now turn the bound on this process into a bound on $\mathbf{II}$. For any increasing and convex function $F$ we have,
	\begin{align}\label{eq:lemma:AsymptoticEquicontinuity-14}
		&\mathbb{E}\left[F\left(\|\mathbb{G}_n\|_{\mathcal{G}_2} \right)\right] \nonumber\\
		&\overset{(a)}{\leq} \mathbb{E}\left[F\left(\sqrt{2\pi} \sup_{\tau, \tau'' \in \mathcal{T}} \sup_{\tau' : |\tau' - \tau| \leq 1/n} \sup_{v \in K(\tau'')} \left|\frac{1}{\sqrt{n}} \sum_{i=1}^n  g_i\big(f_i(\tau) - f_i(\tau')\big)\big(\tau'' - \mathbf{1}\{Y_i \leq X_i'\theta_0(\tau'')\}\big)X_i'v \right|\right)\right] \nonumber\\
		&\overset{(b)}{\leq} \mathbb{E}\left[F\left(4\sqrt{2 \pi} L_f  \sup_{\tau, \tau''} \sup_{\tau' : |\tau' - \tau| \leq 1/n}\sup_{v \in K(\tau'')}  \left|\frac{1}{\sqrt{n}} \sum_{i=1}^n g_i\big(X_i'\theta_0(\tau) - X_i'\theta_0(\tau')\big)\big(\tau'' - \mathbf{1}\{Y_i \leq X_i'\theta_0(\tau'')\}\big)X_i'v\right|\right)\right]\nonumber\\
		&\overset{(c)}{\leq} \mathbb{E}\left[F\left(4\sqrt{2 \pi} L_f L_\theta/n  \sup_{\tau'' \in \mathcal{T}}\sup_{\|u\|_0 \leq 2s_\theta, \: \|u\|_2 = 1} \sup_{v \in K(\tau'')}  \left|\frac{1}{\sqrt{n}} \sum_{i=1}^n g_i\big(\tau'' - \mathbf{1}\{Y_i \leq X_i'\theta_0(\tau'')\}\big)(X_i'v)(X_i'u)\right|\right)\right]\nonumber\\
		&= \mathbb{E}\left[F\left(4\sqrt{2 \pi} L_f L_\theta/n \|\mathbb{G}_n\|_\mathcal{F}\right)\right],
	\end{align}
	where $(a)$ holds by Lemma 6.3 followed by Lemma 4.5 in~\cite{ledoux1996probability}, $(b)$ holds by Lemma~\ref{lemma:Composite-GaussianContraction} (note that by Assumption~\ref{assumption:LipschitzDensity} the map $z \mapsto f_{Y|X}(z - X'\theta_0(\tau)|X) - f_{Y|X}(X'\theta_0(\tau)|X)$ is Lipschitz-continuous and vanishes at 0), and $(c)$ holds by Assumption~\ref{assumption:LipschitzQRVector}. 
	Thus, by Lemma~\ref{lemma:Panchenko2003Handel} and eq.~\eqref{eq:lemma:AsymptoticEquicontinuity-13} and~\eqref{eq:lemma:AsymptoticEquicontinuity-14}, with probability at least $1- \delta$, 
	\begin{align}\label{eq:lemma:AsymptoticEquicontinuity-15}
		\begin{split}
			\mathbf{II} &\lesssim  L_f L_\theta \varphi_{\max}^{1/2}(s_1)\varphi_{\max}^{1/2}(s_2) \frac{\bar{c}^2\|z\|_2}{\kappa_2(\infty)} \sqrt{ \frac{s_1 \log (ep/s_1)  + s_2 \log (ep/s_2) + \log(1/\delta)}{n}} \\
			&\quad{} \times n^{-1/2} \times \sqrt{1 + \pi_{n,1/3}^2(s_1 \log (ep/s_1) + s_2 \log (ep/s_2) + \log(1/\delta))},
		\end{split}
	\end{align}
	where $\pi_{n,1/3}^2(z) = \sqrt{z/n} + z^3/n$ with $z \geq 0$. 
	
	\textbf{Bound on $\mathbf{III}$.} Define
	\begin{align*}
		\mathcal{H} = \left\{  h : \mathbb{R}^{p+1} \rightarrow \mathbb{R} : h(X,Y) = f_{X | Y}(X'\theta_0(\tau)|X) (\tau - \mathbf{1}\{Y \leq X'\theta_0(\tau)\})X'v, \right.\\
		\left. v \in C^p_1(T_v(\tau), 2\bar{c}) \cap B^p_2(0,1), \:\tau \in \mathcal{T}\right\}.
	\end{align*}
	By Lemma~\ref{lemma:LipschitzDual} followed by Lemma~\ref{lemma:GradientCone-2}, with probability at least $1- \delta$,
	\begin{align}\label{eq:lemma:AsymptoticEquicontinuity-16}
		\mathbf{III} &\lesssim C_v\bar{f}L_fL_\theta\varphi_{\max}^{1/2}(2s_\theta) \frac{\varphi_{\max}(p)}{\kappa_2(\infty)} \frac{\|z\|_2}{n} \|\mathbb{G}_n\|_{\mathcal{H}} \nonumber\\
		&\lesssim C_v\bar{c}\bar{f}^2 L_fL_\theta\varphi_{\max}^{1/2}(2s_\theta)\varphi_{\max}^{1/2}(s_v)\big(1 + \varphi_{\max}^{1/2}(2s_\theta)\big) \frac{\varphi_{\max}(p)}{\kappa_2(\infty)} \frac{\|z\|_2}{n} \psi_n\big(t_{s_v, s_\theta, n, \delta }\big) \nonumber\\
		\begin{split}
			&\lesssim C_v\bar{c}\bar{f}^2 L_fL_\theta\varphi_{\max}^{1/2}(2s_\theta)\varphi_{\max}^{1/2}(s_v)\big(1 + \varphi_{\max}^{1/2}(2s_\theta)\big) \frac{\varphi_{\max}(p)}{\kappa_2(\infty)} \frac{\|z\|_2}{\sqrt{n}}\\
			&\quad{} \times \sqrt{\frac{s_v \log(ep/s_v) + s_\theta \log(ep/s_\theta) + \log (nL_f L_\theta/\delta}{n}},
		\end{split}
	\end{align}
	where $t_{s_v, s_\theta, n, \delta } = s_v \log(ep/s_v) + s_\theta \log(ep/s_\theta) + \log (nL_f L_\theta/\delta)$ and $\psi_n(z) = \sqrt{z}\big(1 + n^{-1/2} \sqrt{z} + n^{-3/2} z^{3/2}\big)$ for $z \geq 0$, and the last, simplified inequality follows from the premises of the lemma.
	
	Combine eq.~\eqref{eq:lemma:AsymptoticEquicontinuity-11}--\eqref{eq:lemma:AsymptoticEquicontinuity-16} and set $\delta = 1/n$. Then, for $\varepsilon > 0$ arbitrary and under the premises of the lemma, the second term in eq.~\eqref{eq:lemma:AsymptoticEquicontinuity-9} satisfies
	\begin{align}\label{eq:lemma:AsymptoticEquicontinuity-17}
		\mathbb{P}\left\{ \sup_{\substack{\rho_2(\tau, \tau') \leq n^{-1/2}\\ \tau' \in \widetilde{T}, \tau \in T}} |\mathbb{G}_n(g_\tau- g_{\tau'})|  > \varepsilon \right\} \rightarrow 0 \quad{}\quad{} \mathrm{as} \quad{}\quad{} n \rightarrow \infty,
	\end{align}
	Hence, eq.~\eqref{eq:lemma:AsymptoticEquicontinuity-10} and eq.~\eqref{eq:lemma:AsymptoticEquicontinuity-17} imply that, for $\varepsilon > 0$ arbitrary,
	\begin{align*}
		\lim_{\xi \downarrow 0} \lim_{n \rightarrow \infty} \mathbb{P}\left\{\|\mathbb{G}_n\|_{\mathcal{G}_\xi} > \varepsilon \right\} = 0.
	\end{align*}
	This completes the proof.
\end{proof}

\begin{proof}[\textbf{Proof of Theorem~\ref{theorem:WeakConvergenceRankScoreBalancedEstimator}}]
	By Theorem 1.5.7 and Addendum 1.5.8 in~\cite{vaartwellner1996weak} it suffices to establish asymptotic equicontinuity of the leading term of the Bahadur-type representation from Theorem~\ref{theorem:BahadurTypeRep}, total boundedness of $\mathcal{G}$ with respect to the standard deviation metric, and finite dimensional convergence to a Gaussian random vector. Asymptotic equicontinuity and total boundedness are established in Lemma~\ref{lemma:AsymptoticEquicontinuity}; hence we are left to show the finite dimensional convergence.
	
	Finite dimensional convergence to a Gaussian random vector follows the strong moment conditions (Assumption~\ref{assumption:SubGaussianity}), the assumption that the limit covariance function exists, and the Lindeberg Feller CLT combined with the Cram{\'e}r-Wold theorem. We omit this standard argument; for details see proof of Theorem 2.1 on p. 3292 in~\cite{chao2017quantile}. We are left to show that the asymptotic mean equals zero and that the (finite dimensional) covariance is as stated in the theorem. First, for all $\tau \in \mathcal{T}$,
	\begin{align}\label{eq:theorem:WeakConvergenceDebiasedEstimator-1}
		\begin{split}
			&\mathbb{E}\left[-\frac{1}{2\sqrt{n}}\sum_{i=1}^n f_{Y|X}(X_i'\theta_0(\tau)|X_i)(\tau - 1\{Y_i \leq X_i'\theta_0(\tau)\})X_i'v_0(\tau; z)\right]\\
			&\quad{}= \mathbb{E}\left[\frac{1}{2\sqrt{n}}\sum_{i=1}^n f_{Y|X}(X_i'\theta_0(\tau)|X_i)\big(\tau - F_{Y|X}(X_i'\theta_0(\tau)|X_i)\big)X_i'v_0(\tau; z)\right]\\
			&\quad{} = 0.
		\end{split}
	\end{align}
	Second, for arbitrary $\tau_1, \ldots, \tau_K \in \mathcal{T}$, the asymptotic covariance matrix is
	\begin{align*}
		&\lim_{n\rightarrow\infty}\left(\mathbb{E}\left[\frac{1}{2\sqrt{n}}\sum_{i=1}^n f_{Y|X}(X_i'\theta_0(\tau_j)|X_i)(\tau_j - 1\{Y_i \leq X_i'\theta_0(\tau_j)\})\big(X_i'v_0(\tau_j; z)\big) \right. \right.\\
		&\left.\left. \quad{}\quad{} \times \frac{1}{2\sqrt{n}}\sum_{i=1}^n f_{Y|X}(X_i'\theta_0(\tau_k)|X_i)^2(\tau_k - 1\{Y_i \leq X_i'\theta_0(\tau_k)\})\big(X_i'v_0(\tau_k; z)\big)\right] \right)_{j,k=1}^K\\
		&\quad{} = \lim_{n\rightarrow\infty}\Big((\tau_j \wedge \tau_k - \tau_j \tau_k)/4 v_0'(\tau_j; z)\mathbb{E}[f_{Y|X}(X'\theta_0(\tau_j)|X)f_{Y|X}(X'\theta_0(\tau_k)|X) XX']v_0'(\tau_k; z) \Big)_{j,k=1}^K
	\end{align*}
	where we have used the definition of $v_0(\tau;z)$, Assumption~\ref{assumption:LinearCQF}, and eq.~\eqref{eq:theorem:WeakConvergenceDebiasedEstimator-1}.
\end{proof}
\begin{proof}[\textbf{Proof of Lemma~\ref{lemma:ConsistencyCovarianceFunction-2}}] We only prove the special case $\tau_1 = \tau_2 = \tau$. The general case follows by the obvious generalization of the arguments; see the comment at the end of this proof.
	
	For notational convenience, we write $v_0(\tau)$ for $v_0(\tau;z)$, $\hat{v}(\tau)$ for $\hat{v}_\gamma(\tau; z)$, $f(\tau)$ for $f_{Y|X}(X'\theta_0(\tau)|X)$, and $f_i(\tau)$ for $f_{Y|X}(X_i'\theta_0(\tau)|X_i)$.
	Define
	\begin{align*}
		D(\tau) := \frac{1}{4}\mathbb{E}\left[f^2(\tau)XX'\right], \quad{} \quad{}	\widetilde{D}(\tau) := \frac{1}{4n}\sum_{i=1}^n f_i^2(\tau) X_iX_i',\quad{}\quad{} \mathrm{and} \quad{}\quad{} \widehat{D}(\tau) := \frac{1}{4n}\sum_{i=1}^n \hat{f}_i^2(\tau) X_iX_i',
	\end{align*}
	and write
	\begin{align}\label{eq:lemma:ConsistencyCovarianceFunction-2-0}
		&\left|\widehat{H}(\tau, \tau; z)-H(\tau, \tau; z) \right| =\left|\hat{v}'(\tau)\widehat{D}(\tau )\hat{v}(\tau) -v_0'(\tau)D(\tau)v_0(\tau) \right| \nonumber\\
		&\leq \left| \big(\hat{v}(\tau) -v_0(\tau) \big)'\left(\widetilde{D}(\tau)-D(\tau)\right) \big(\hat{v}(\tau) -v_0(\tau) \big)\right| + \left| \big(\hat{v}(\tau) -v_0(\tau) \big)'\left(\widehat{D}(\tau)-\widetilde{D}(\tau)\right) \big(\hat{v}(\tau) -v_0(\tau) \big)\right| \nonumber\\
		&\quad{} + 2\left|v_0'(\tau)\left(\widetilde{D}(\tau)-D(\tau)\right)  \big(\hat{v}(\tau) -v_0(\tau) \big)\right| + 2\left|v_0'(\tau)\left(\widehat{D}(\tau)-\widetilde{D}(\tau)\right)  \big(\hat{v}(\tau) -v_0(\tau) \big)\right| \nonumber\\
		&\quad{} + \left|v_0'(\tau)\left(\widetilde{D}(\tau)-D(\tau)\right) v_0(\tau)\right| + \left|v_0'(\tau)\left(\widehat{D}(\tau)-\widetilde{D}(\tau)\right) v_0(\tau)\right|\nonumber\\
		&\quad{} +  \left| \big(\hat{v}(\tau) -v_0(\tau) \big)'D(\tau)\big(\hat{v}(\tau) - v_0(\tau) \big)\right|\nonumber\\
		&\quad{} +  2\left| v_0'(\tau)D(\tau)\big(\hat{v}(\tau) -v_0(\tau) \big)\right| \nonumber\\
		&= \mathbf{I} + \mathbf{II} + \mathbf{III} + \mathbf{IV} + \mathbf{V} + \mathbf{VI} +  \mathbf{VII}  +  \mathbf{VIII} .
	\end{align}	
	Recall that by Lemmas~\ref{lemma:InducedRestrictedConeProp-Dual} (i) and~\ref{lemma:RestrictedConeProp-Dual},
	\begin{align}
		v_0(\tau) - \tilde{v}(\tau) &\in \left\{ u \in \mathbb{R}^p: (u_{T_v(\tau)}, u_{T_v^c(\tau)}) = (0, w_{T_v^c(\tau)}), \: w \in C^p_2(T_v(\tau), 1)\right\}, \label{eq:lemma:ConsistencyCovarianceFunction-2-1}\\
		\hat{v}_\gamma(\tau) - \tilde{v}(\tau) &\in C_1^p(T_v(\tau), 2\bar{c}) \cup B^p_1\left(0, \frac{2\bar{c}\|z\|_2^2}{\kappa_2^2(\infty)} \frac{\mu}{\gamma}\right),\label{eq:lemma:ConsistencyCovarianceFunction-2-2}
	\end{align}	
	and, by Definition~\ref{definition:ExactApproxDualSolution},
	\begin{align}
		&\sup_{\tau \in \mathcal{T}}\|v_0(\tau) - \tilde{v}(\tau)\|_2 \lesssim \frac{\|z\|_2}{\kappa_2(\infty)} r_a,\label{eq:lemma:ConsistencyCovarianceFunction-2-3}\\
		&\tilde{v}(\tau) \in C^p_1(T_v(\tau), 1) \cap B^p_2\left(0, \frac{2\bar{c}\|z\|_2}{\kappa_2(\infty)}\right).\label{eq:lemma:ConsistencyCovarianceFunction-2-4}
	\end{align}
	We now bound the eight terms in eq.~\eqref{eq:lemma:ConsistencyCovarianceFunction-2-0}.
	
	\textbf{Bound on $\mathbf{I}$.} Define
	\begin{align*}
		\mathcal{G}_1 &= \Big\{ g : \mathbb{R}^p \rightarrow \mathbb{R}: g(X) = f_{Y|X}^2(X'\theta_0(\tau)|X) (X'u)^2, \: u \in C^P_1(T_v(\tau, 2\bar{c})\cap B^p_2(0,1),\: \tau \in \mathcal{T}\Big\}, \\
		\mathcal{G}_2 &= \Big\{ g : \mathbb{R}^p \rightarrow \mathbb{R}: g(X) = f_{Y|X}^2(X'\theta_0(\tau)|X) (X'u)^2, \: u \in U^p \cap B^p_2(0 ,1),\: \tau \in \mathcal{T}\Big\},\\
		\mathcal{G}_3 &= \Big\{ g : \mathbb{R}^p \rightarrow \mathbb{R}: g(X) = f_{Y|X}^2(X'\theta_0(\tau)|X) (X'u)^2, \: u \in \mathbb{R}^p, \: \|u\|_0 \leq 1,\: \|u\|_2 \leq 1,\: \tau \in \mathcal{T}\Big\}, \\
		\mathcal{G}_4 &= \Big\{ g : \mathbb{R}^p \rightarrow \mathbb{R}: g(X) = f_{Y|X}^2(X'\theta_0(\tau)|X) (X'u)(X'w), \: u \in C^P_1(T_v(\tau, 2\bar{c})\cap B^p_2(0,1),\\
		&\quad{}\quad{} w\in U^p \cap  B^p_2(0 ,1), \: \tau \in \mathcal{T}\Big\},\\
		\mathcal{G}_5 &= \Big\{ g : \mathbb{R}^p \rightarrow \mathbb{R}: g(X) = f_{Y|X}^2(X'\theta_0(\tau)|X) (X'u)(X'w), \:  u \in \mathbb{R}^p, \: \|u\|_0 \leq 1,\: \|u\|_2 \leq 1,\\
		&\quad{}\quad{} w\in U^p \cap  B^p_2(0 ,1), \: \tau \in \mathcal{T}\Big\},
	\end{align*}
	where  $U^p:= \left\{ u \in \mathbb{R}^p: (u_{T_v(\tau)}, u_{T_v^c(\tau)}) = (0, w_{T_v^c(\tau)}), \: w \in C^p_2(T_v(\tau), 1)\right\}$. 
	
	By the geometric constraints~\eqref{eq:lemma:ConsistencyCovarianceFunction-2-1}--\eqref{eq:lemma:ConsistencyCovarianceFunction-2-3}, Lemmas~\ref{lemma:InducedRestrictedConeProp-Dual} (ii) and~\eqref{lemma:SizeConesDominatedCoordinates}, and Theorem~\ref{theorem:ConsistencyDual}, with probability at least $1-\delta$,
	\begin{align*}
		\mathbf{I} &\lesssim \frac{r_v^2}{\sqrt{n}} \|\mathbb{G}_n\|_{\mathcal{G}_1}  + \frac{r_a^2}{\sqrt{n}}\frac{\|z\|_2^2}{\kappa_2^2(\infty)} \|\mathbb{G}_n\|_{\mathcal{G}_2} + \left(\frac{2\bar{c}\|z\|_2^2}{\kappa_2^2(\infty)} \frac{\mu}{\gamma} \right)^2\frac{1}{\sqrt{n}} \|\mathbb{G}_n\|_{\mathcal{G}_3} \\
		&\quad{}+ \left(\frac{2\bar{c}\|z\|_2^2}{\kappa_2^2(\infty)} \frac{\mu}{\gamma} \right)\frac{r_a}{\sqrt{n}} \frac{\|z\|_2}{\kappa_2(\infty)} \|\mathbb{G}_n\|_{\mathcal{G}_4}+ \frac{r_vr_a}{\sqrt{n}} \frac{\|z\|_2}{\kappa_2(\infty)} \|\mathbb{G}_n\|_{\mathcal{G}_5}.
	\end{align*}
	Whence, by Lemma~\ref{lemma:LocalizedLossDual} (i) and (iii), and Assumptions~\ref{assumption:GrowthCondition} and~\ref{assumption:GrowthConditionDual} and $s_\theta^2 \log^2(np/\delta) \log^2(n) = o(n)$, with probability at least $1-\delta$, 
	\begin{align}\label{eq:lemma:ConsistencyCovarianceFunction-2-5}
		\begin{split}
			\mathbf{I} &\lesssim (2 +2\bar{c})^2 \bar{f}^2 \varphi_{\max}(s_v)(1 + \varphi_{\max}^{1/2}(2s_\theta))  \\
			&\quad{} \times \left( r_v^2 + \frac{r_a^2 \|z\|_2^2}{\kappa_2^2(\infty)}  + \left(\frac{2\bar{c}\|z\|_2^2}{\kappa_2^2(\infty)} \frac{\mu}{\gamma} \right)^2 + \left(\frac{2\bar{c}\|z\|_2^2}{\kappa_2^2(\infty)} \frac{\mu}{\gamma} \right)\frac{r_a\|z\|_2}{\kappa_2(\infty)}  + \frac{r_v r_a \|z\|_2}{\kappa_2(\infty)} \right)  \\  
			& \quad{} \times \sqrt{\frac{s_v \log(ep/s_v) + s_\theta \log(ep/s_\theta) + \log (n L_f L_\theta )+  \log (1/\delta)}{n}}.
		\end{split}
	\end{align}
	
	\textbf{Bound on $\mathbf{II}$.} Recall the definition of function classes $\mathcal{G}_1, \mathcal{G}_2, \mathcal{G}_3, \mathcal{G}_4$ and define
	\begin{align*}
		\mathcal{G}_6 &= \Big\{ g : \mathbb{R}^p \rightarrow \mathbb{R}: g(X) = f_{Y|X}^2(X'\theta_0(\tau)|X) |X'u||X'w|, \: u \in C^P_1(T_v(\tau, 2\bar{c})\cap B^p_2(0,1),\\
		&\quad{}\quad{} w\in U^p \cap  B^p_2(0 ,1), \: \tau \in \mathcal{T}\Big\},\\
		\mathcal{G}_7 &= \Big\{ g : \mathbb{R}^p \rightarrow \mathbb{R}: g(X) = f_{Y|X}^2(X'\theta_0(\tau)|X) |X'u||X'w|, \:  u \in \mathbb{R}^p, \: \|u\|_0 \leq 1,\: \|u\|_2 \leq 1,\\
		&\quad{}\quad{} w\in U^p \cap  B^p_2(0 ,1), \: \tau \in \mathcal{T}\Big\},
	\end{align*}
	where  $U^p:= \left\{ u \in \mathbb{R}^p: (u_{T_v(\tau)}, u_{T_v^c(\tau)}) = (0, w_{T_v^c(\tau)}), \: w \in C^p_2(T_v(\tau), 1)\right\}$. 	
	
	By the geometric constraints~\eqref{eq:lemma:ConsistencyCovarianceFunction-2-1}--\eqref{eq:lemma:ConsistencyCovarianceFunction-2-3}, Lemmas~\ref{lemma:InducedRestrictedConeProp-Dual} (ii) and~\eqref{lemma:SizeConesDominatedCoordinates}, Theorem~\ref{theorem:ConsistencyDual}, and Lemma~\ref{lemma:RelativeConsistencyDensity}, we have, with probability at least $1- \delta$,
	\begin{align*}
		\mathbf{II} &\lesssim r_fr_v^2 \|\mathbb{P}_n\|_{\mathcal{G}_1}  + r_fr_a^2\frac{\|z\|_2^2}{\kappa_2^2(\infty)} \|\mathbb{P}_n\|_{\mathcal{G}_2} + r_f\left(\frac{2\bar{c}\|z\|_2^2}{\kappa_2^2(\infty)} \frac{\mu}{\gamma} \right)^2\|\mathbb{P}_n\|_{\mathcal{G}_3} \\
		&\quad{}+r_f r_a \left(\frac{2\bar{c}\|z\|_2^2}{\kappa_2^2(\infty)} \frac{\mu}{\gamma} \right)\frac{\|z\|_2}{\kappa_2(\infty)} \|\mathbb{P}_n\|_{\mathcal{G}_6}+ r_fr_vr_a\frac{\|z\|_2}{\kappa_2(\infty)} \|\mathbb{P}_n\|_{\mathcal{G}_7}\\
		&\lesssim r_f\left(\frac{r_v^2}{\sqrt{n}} \|\mathbb{G}_n\|_{\mathcal{G}_1}  + \frac{r_a^2}{\sqrt{n}}\frac{\|z\|_2^2}{\kappa_2^2(\infty)} \|\mathbb{G}_n\|_{\mathcal{G}_2} + \left(\frac{2\bar{c}\|z\|_2^2}{\kappa_2^2(\infty)} \frac{\mu}{\gamma} \right)^2\frac{1}{\sqrt{n}} \|\mathbb{G}_n\|_{\mathcal{G}_3}  \right.\\
		& \left.\quad{}\quad{}+ \left(\frac{2\bar{c}\|z\|_2^2}{\kappa_2^2(\infty)} \frac{\mu}{\gamma} \right)\frac{r_a}{\sqrt{n}} \frac{\|z\|_2}{\kappa_2(\infty)} \|\mathbb{G}_n\|_{\mathcal{G}_6}+ \frac{r_vr_a}{\sqrt{n}} \frac{\|z\|_2}{\kappa_2(\infty)} \|\mathbb{G}_n\|_{\mathcal{G}_7}\right)\\
		&\quad{} + r_f \left( r_v^2 \sup_{g \in \mathcal{G}_1} \mathbb{E}[g(X)]  + r_a^2\frac{\|z\|_2^2}{\kappa_2^2(\infty)}\sup_{g \in \mathcal{G}_2} \mathbb{E}[g(X)] + \left(\frac{2\bar{c}\|z\|_2^2}{\kappa_2^2(\infty)} \frac{\mu}{\gamma} \right)^2\sup_{g \in \mathcal{G}_3} \mathbb{E}[g(X)]   \right.\\
		&\left. \quad{}\quad{}  + r_a \left(\frac{2\bar{c}\|z\|_2^2}{\kappa_2^2(\infty)} \frac{\mu}{\gamma} \right)\frac{\|z\|_2}{\kappa_2(\infty)}   \sup_{g \in \mathcal{G}_6 }\mathbb{E}[g(X)]   + r_vr_a\frac{\|z\|_2}{\kappa_2(\infty)} \sup_{g \in \mathcal{G}_7 }\mathbb{E}[g(X)] \right) 
	\end{align*}
	Whence, by Lemma~\ref{lemma:LocalizedLossDual} (i), (ii), and (iii), and Assumptions~\ref{assumption:GrowthCondition} and~\ref{assumption:GrowthConditionDual}, $s_\theta^2 \log^2(np/\delta) \log^2(n) = o(n)$, and $r_f = o(1)$, with probability at least $1-\delta$, 
	\begin{align}\label{eq:lemma:ConsistencyCovarianceFunction-2-6}
		\begin{split}
			\mathbf{II} &\lesssim (2 +2\bar{c})^2 \bar{f}^2 \varphi_{\max}(s_v)(1 + \varphi_{\max}^{1/2}(2s_\theta))  \\
			&\quad{} \times r_f \left( r_v^2 + \frac{r_a^2 \|z\|_2^2}{\kappa_2^2(\infty)}  + \left(\frac{2\bar{c}\|z\|_2^2}{\kappa_2^2(\infty)} \frac{\mu}{\gamma} \right)^2 + \left(\frac{2\bar{c}\|z\|_2^2}{\kappa_2^2(\infty)} \frac{\mu}{\gamma} \right)\frac{r_a\|z\|_2}{\kappa_2(\infty)}  + \frac{r_v r_a \|z\|_2}{\kappa_2(\infty)} \right) .
		\end{split}
	\end{align}
	
	\textbf{Bounds on $\mathbf{III}$ and $\mathbf{IV}$.} These two terms can be upper bounded by the sum of the upper bounds on terms $\mathbf{I} + \mathbf{V}$ and $\mathbf{II} +  \mathbf{VI}$.

	\textbf{Bound on $\mathbf{V}$.} Recall the definition of function classes $\mathcal{G}_1, \mathcal{G}_2, \mathcal{G}_4$. By the geometric constraints~\eqref{eq:lemma:ConsistencyCovarianceFunction-2-1},~\eqref{eq:lemma:ConsistencyCovarianceFunction-2-3} and~\eqref{eq:lemma:ConsistencyCovarianceFunction-2-4}, Lemmas~\ref{lemma:InducedRestrictedConeProp-Dual} (ii) and~\eqref{lemma:SizeConesDominatedCoordinates},
	\begin{align*}
		\mathbf{V} &\leq \left(\frac{2\|z\|_2}{\kappa_2(\infty)}\right)^2 \|\mathbb{G}_n\|_{\mathcal{G}_1} +  r_a^2\frac{\|z\|_2^2}{\kappa_2^2(\infty)} \|\mathbb{G}_n\|_{\mathcal{G}_2} +r_a^2\frac{4\|z\|_2^2}{\kappa_2^2(\infty)} \|\mathbb{G}_n\|_{\mathcal{G}_4}.
	\end{align*}
	Hence, by Lemma~\ref{lemma:LocalizedLossDual}, and Assumptions~\ref{assumption:GrowthCondition} and~\ref{assumption:GrowthConditionDual} and $s_\theta^2 \log^2(np/\delta) \log^2(n) = o(n)$, with probability at least $1-\delta$, 
	\begin{align}\label{eq:lemma:ConsistencyCovarianceFunction-2-7}
		\begin{split}
			\mathbf{V} &\lesssim (2 +2\bar{c})^2 \bar{f}^2 \varphi_{\max}(s_v)(1 + \varphi_{\max}^{1/2}(2s_\theta))  \left( \frac{\|z\|_2^2}{\kappa_2^2(\infty)} +  \frac{r_a^2\|z\|_2^2}{\kappa_2^2(\infty)} \right)  \\  
			& \quad{} \times \sqrt{\frac{s_v \log(ep/s_v) + s_\theta \log(ep/s_\theta) + \log (n L_f L_\theta )+  \log (1/\delta)}{n}}.
		\end{split}
	\end{align}
	
	\textbf{Bound on $\mathbf{VI}$.} Recall the definition of function classes $\mathcal{G}_1, \mathcal{G}_2, \mathcal{G}_6$. By the geometric constraints~\eqref{eq:lemma:ConsistencyCovarianceFunction-2-1},~\eqref{eq:lemma:ConsistencyCovarianceFunction-2-3} and~\eqref{eq:lemma:ConsistencyCovarianceFunction-2-4}, Lemmas~\ref{lemma:InducedRestrictedConeProp-Dual} (ii) and~\eqref{lemma:SizeConesDominatedCoordinates}, Theorem~\ref{theorem:ConsistencyDual}, and Lemma~\ref{lemma:RelativeConsistencyDensity}, we have, with probability at least $1- \delta$,
	\begin{align*}
		\mathbf{VI} &\lesssim r_f\left(\frac{2\|z\|_2}{\kappa_2(\infty)}\right)^2 \|\mathbb{P}_n\|_{\mathcal{G}_1} +  r_f r_a^2 \frac{\|z\|_2^2}{\kappa_2^2(\infty)} \|\mathbb{P}_n\|_{\mathcal{G}_2} +r_f r_a^2 \frac{4\|z\|_2^2}{\kappa_2^2(\infty)} \|\mathbb{P}_n\|_{\mathcal{G}_4}\\
		&\lesssim  r_f \left(\frac{1}{\sqrt{n}} \left(\frac{2\|z\|_2}{\kappa_2(\infty)}\right)^2 \|\mathbb{G}_n\|_{\mathcal{G}_1} +  \frac{r_a^2}{\sqrt{n}}\frac{\|z\|_2^2}{\kappa_2^2(\infty)} \|\mathbb{G}_n\|_{\mathcal{G}_2} +\frac{r_a^2}{\sqrt{n}}\frac{4\|z\|_2^2}{\kappa_2^2(\infty)} \|\mathbb{G}_n\|_{\mathcal{G}_6}  \right.\\
		&\left. \quad{} \quad{} + \left(\frac{2\|z\|_2}{\kappa_2(\infty)}\right)^2 \sup_{g \in \mathcal{G}_1}\mathbb{E}[g(X)]  + r_a^2 \frac{\|z\|_2^2}{\kappa_2^2(\infty)}  \sup_{g \in \mathcal{G}_2}\mathbb{E}[g(X)]  + r_a^2 \frac{4\|z\|_2^2}{\kappa_2^2(\infty)} \sup_{g \in \mathcal{G}_6}\mathbb{E}[g(X)] \right)
	\end{align*}
	Whence, by Lemma~\ref{lemma:LocalizedLossDual} (i), (ii), and (iii), and Assumptions~\ref{assumption:GrowthCondition} and~\ref{assumption:GrowthConditionDual}, $s_\theta^2 \log^2(np/\delta) \log^2(n) = o(n)$, and $r_f = o(1)$, with probability at least $1-\delta$, 
	\begin{align}\label{eq:lemma:ConsistencyCovarianceFunction-2-8}
		\begin{split}
			\mathbf{VI} &\lesssim (2 +2\bar{c})^2 \bar{f}^2 \varphi_{\max}(s_v)(1 + \varphi_{\max}^{1/2}(2s_\theta))  \times r_f \left( \frac{\|z\|_2^2}{\kappa_2^2(\infty)} +  \frac{r_a^2\|z\|_2^2}{\kappa_2^2(\infty)} \right)  .
		\end{split}
	\end{align}
	
	\textbf{Bound on $\mathbf{VII}$.} By the geometric constraints~\eqref{eq:lemma:ConsistencyCovarianceFunction-2-1}--\eqref{eq:lemma:ConsistencyCovarianceFunction-2-3} and Lemmas~\ref{lemma:InducedRestrictedConeProp-Dual} (ii) and~\eqref{lemma:SizeConesDominatedCoordinates},
	\begin{align*}
		\mathbf{VII} &\leq \frac{1}{4}  (2 + 2\bar{c})\bar{f}^2 \varphi_{\max}(s_\theta) \big\| \hat{v}(\tau) - \tilde{v}(\tau) \big\|_2^2\\
		&\quad{} + \frac{1}{4}  \bar{f}^2 \varphi_{\max}(1) \left(\frac{2\bar{c}\|z\|_2^2}{\kappa_2^2(\infty)} \frac{\mu}{\gamma} \right)^2\\
		&\quad{} + \frac{1}{4}  (2 + 2\bar{c})\bar{f}^2 \varphi_{\max}(s_\theta) \big\| \tilde{v}(\tau) - v_0(\tau) \big\|_2^2\\
		&\quad{} + \frac{1}{2}  (2 + 2\bar{c})\bar{f}^2 \varphi_{\max}(s_\theta) \big\| \hat{v}(\tau) - \tilde{v}(\tau) \big\|_2\big\| \tilde{v}(\tau) - v_0(\tau) \big\|_2\\
		&\quad{} + \frac{1}{2}  (2 + 2\bar{c})\bar{f}^2 \varphi^{1/2}_{\max}(s_\theta)\varphi^{1/2}_{\max}(1) \left(\frac{2\bar{c}\|z\|_2^2}{\kappa_2^2(\infty)} \frac{\mu}{\gamma} \right)\big\| \tilde{v}(\tau) - v_0(\tau) \big\|_2.
	\end{align*}
	Thus, by Theorem~\ref{theorem:ConsistencyDual}, with probability at least $1- \delta$,
	\begin{align}\label{eq:lemma:ConsistencyCovarianceFunction-2-9}
		\mathbf{VII} &\lesssim (2 + 2\bar{c})\bar{f}^2 \varphi_{\max}(s_\theta) \times \left( r_v^2 + \left(\frac{2\bar{c}\|z\|_2^2}{\kappa_2^2(\infty)} \frac{\mu}{\gamma} \right)^2 + \frac{r_a^2\|z\|_2^2}{\kappa_2^2(\infty)}  + \frac{r_vr_a \|z\|_2}{\kappa_2(\infty)}  + \left(\frac{2\bar{c}\|z\|_2^2}{\kappa_2^2(\infty)} \frac{\mu}{\gamma} \right) \frac{r_a\|z\|_2}{\kappa_2(\infty)} \right).
	\end{align}
	
	\textbf{Bound on $\mathbf{VIII}$.} By the geometric constraints~\eqref{eq:lemma:ConsistencyCovarianceFunction-2-1}--\eqref{eq:lemma:ConsistencyCovarianceFunction-2-4} and Lemmas~\ref{lemma:InducedRestrictedConeProp-Dual} (ii) and~\eqref{lemma:SizeConesDominatedCoordinates},
	\begin{align*}
		\mathbf{VIII} &\leq \frac{1}{2}  (2 + 2\bar{c})\bar{f}^2 \varphi_{\max}(s_\theta) \big\| \hat{v}(\tau) - \tilde{v}(\tau) \big\|_2\big\| \tilde{v}(\tau) - v_0(\tau) \big\|_2\\
		&\quad{} + \frac{1}{2}  (2 + 2\bar{c})\bar{f}^2 \varphi^{1/2}_{\max}(s_\theta)\varphi^{1/2}_{\max}(1) \left(\frac{2\bar{c}\|z\|_2^2}{\kappa_2^2(\infty)} \frac{\mu}{\gamma} \right)\big\| \tilde{v}(\tau) - v_0(\tau) \big\|_2\\
		&\quad{} + \frac{1}{4}  (2 + 2\bar{c})\bar{f}^2 \varphi_{\max}(s_\theta) \big\| \tilde{v}(\tau) - v_0(\tau) \big\|_2^2\\
		&\quad{} +\frac{1}{4}  (2 + 2\bar{c})\bar{f}^2 \varphi_{\max}(s_\theta) \big\| \tilde{v}(\tau) \big\|_2\big\| \tilde{v}(\tau) - v_0(\tau) \big\|_2\\
		&\quad{} + \frac{1}{4}  (2 + 2\bar{c})\bar{f}^2 \varphi_{\max}(s_\theta) \big\| \hat{v}(\tau) - \tilde{v}(\tau) \big\|_2\big\| \tilde{v}(\tau)\big\|_2\\
		&\quad{} + \frac{1}{4}  (2 + 2\bar{c})\bar{f}^2 \varphi^{1/2}_{\max}(s_\theta)\varphi^{1/2}_{\max}(1) \left(\frac{2\bar{c}\|z\|_2^2}{\kappa_2^2(\infty)} \frac{\mu}{\gamma} \right)\big\| \tilde{v}(\tau)\big\|_2.
	\end{align*}
	Hence, by Theorem~\ref{theorem:ConsistencyDual}, with probability at least $1- \delta$,
	\begin{align}\label{eq:lemma:ConsistencyCovarianceFunction-2-10}
		\begin{split}
			\mathbf{VIII} & \lesssim (2 + 2\bar{c})\bar{f}^2 \varphi_{\max}(s_\theta)\\
			&\quad{} \times \left( \frac{r_a^2\|z\|_2^2}{\kappa_2^2(\infty)}  +  \left(\frac{2\bar{c}\|z\|_2^2}{\kappa_2^2(\infty)} \frac{\mu}{\gamma} \right) \frac{r_a\|z\|_2}{\kappa_2(\infty)} +  \frac{r_a^2\|z\|_2^2}{\kappa_2^2(\infty)}  + \frac{r_a\|z\|_2^2}{\kappa_2^2(\infty)}+  \frac{r_v\|z\|_2^2}{\kappa_2^2(\infty)} + \left(\frac{2\bar{c}\|z\|_2^2}{\kappa_2^2(\infty)} \frac{\mu}{\gamma} \right ) \frac{\|z\|_2}{\kappa_2(\infty)}  \right).
		\end{split}
	\end{align}
	
	\textbf{Conclusion.} Combine eq.~\eqref{eq:lemma:ConsistencyCovarianceFunction-2-5}--\eqref{eq:lemma:ConsistencyCovarianceFunction-2-10}, $s_\theta^2 \log^2(np/\delta) \log^2(n) = o(n)$, and $r_f \vee r_a = o(1)$ to conclude that, with probability at least $1-\delta$,
	\begin{align*}
		\sup_{\tau \in \mathcal{T}}\left|\widehat{H}(\tau, \tau; z)-H(\tau, \tau; z) \right| &\lesssim (2 + 2\bar{c})  \bar{f}^2  \varphi_{\max}(s_v)(1 + \varphi_{\max}^{1/2}(2s_\theta)) \times r_v.
	\end{align*}
	Under the assumptions of Corollary~\ref{corollary:theorem:BahadurTypeRep} this upper bound can be further simplified as in the statement of the lemma. This completes the proof.
	
	To establish the claim for $\tau_1 \neq \tau_2$, we replace $f_{Y|X}^2(X'\theta_0(\tau)|X)$ by $f_{Y|X}(X'\theta_0(\tau_1)|X)f_{Y|X}(X'\theta_0(\tau_2)|X)$ in the definitions of the function classes $\mathcal{G}_1, \ldots, \mathcal{G}_7$. We then need a generalization of Lemma~\ref{lemma:LocalizedLossDual} which handles these function classes. By inspecting the proof of  Lemma~\ref{lemma:LocalizedLossDual} it is obvious that such a generalization holds true. We omit a formal, lengthy, and uneventful proof.
\end{proof}

\subsection{Proofs of Section~\ref{subsec:AuxResults-RankScore}}
\begin{proof}[\textbf{Proof of Lemma~\ref{lemma:ConeProperties}}]	
	Only the first statement needs a proof. We compute
	\begin{align*}
		\|\tilde{v}_{T^c}\|_2 = \|v_{T^c}\|_2 \leq \alpha \|v_T\|_2 \leq  \alpha \frac{|t|}{\beta}\|v_T\|_2 = \frac{\alpha}{\beta}\|\tilde{v}_{T^c}\|_2.
	\end{align*}
\end{proof}

\begin{proof}[\textbf{Proof of Lemma~\ref{lemma:InducedRestrictedConeProp-Dual}}]	
	For notational convenience, we write $v_0(\tau)$, $\tilde{v}(\tau)$, and $T_v(\tau)$ for $v_0(\tau;z)$, $\tilde{v}(\tau;z)$, and $T_v(\tau;z)$, respectively.
	
	To prove part (i), compute
	\begin{align*}
		\|v_0(\tau) - \tilde{v}(\tau)\|_2 &\leq r_a \|v_0(\tau)\|_2
		\overset{(a)}{\leq} \sum_{k=1}^\infty r_a^k \|\tilde{v}(\tau)\|_2
		= \frac{r_a}{1- r_a}\|\tilde{v}(\tau)\|_2,
	\end{align*}
	where (a) holds by iterating $\|v_0(\tau)\|_2 \leq \|v_0(\tau) - \tilde{v}(\tau)\|_2  + \|\tilde{v}(\tau)\|_2 \leq r_a\|v_0(\tau)\|_2 + \|\tilde{v}(\tau)\|_2$.
	To prove part (ii), write
	\begin{align}\label{eq:8}
		\big(\tilde{v}(\tau)-v_0(\tau)\big)'MM'\big(\tilde{v}(\tau)-v_0(\tau)\big) &= \|\tilde{v}(\tau)'M\|_2^2 + \|v_0(\tau)'M\|_2^2 - 2 \big(M'\tilde{v}(\tau)\big)' \big(M'v_0(\tau)\big),
	\end{align}
	and upper bound the right-hand side in above identity as follows:
	\begin{align}\label{eq:9}
		&\|\tilde{v}(\tau)'M\|_2^2 + \|v_0(\tau)'M\|_2^2 - 2 \big(M'\tilde{v}(\tau)\big)' \big(M'v_0(\tau)\big) \nonumber\\
		&= (1 -r_a^2) \left[  \left(\frac{1}{1 + r_a}\right)^2 \|\tilde{v}(\tau)'M\|_2^2 +  \left(\frac{1}{1 - r_a}\right)^2 \|v_0(\tau)'M\|_2^2 - \frac{2}{1-r_a^2} \big(M'\tilde{v}(\tau)\big)' \big(M'v_0(\tau)\big)\right] \nonumber\\
		&\quad + (1-r_a^2) \left[\frac{1}{1-r_a^2} - \left(\frac{1}{1 + r_a}\right)^2\right] \|\tilde{v}(\tau)'M\|_2^2 + (1-r_a^2) \left[\frac{1}{1-r_a^2} - \left(\frac{1}{1 - r_a}\right)^2\right] \|v_0(\tau)'M\|_2^2  \nonumber\\
		\begin{split}
			&\leq (1 -r_a^2) \left(\frac{1}{1 + r_a}\tilde{v}(\tau) - \frac{1}{1-r_a} v_0(\tau)\right)'MM' \left(\frac{1}{1 + r_a}\tilde{v}(\tau) - \frac{1}{1-r_a} v_0(\tau)\right)\\
			&\quad{} + \frac{2r_a}{1+r_a}  \|\tilde{v}(\tau)'M\|_2^2.
		\end{split}
	\end{align}
	Next, by construction of $\tilde{v}(\tau)$,
	\begin{align}
		\frac{1}{1 + r_a}\tilde{v}(\tau) - \frac{1}{1-r_a} v_0(\tau) &= \left( \frac{1}{1 + r_a} - \frac{1}{1-r_a} \right) \tilde{v}(\tau) - \frac{1}{1-r_a}\big( v_0(\tau) -\tilde{v}(\tau)\big) \nonumber\\
		&=\frac{-2r_a}{1-r_a^2} \tilde{v}(\tau) - \frac{1}{1-r_a}\big( v_0(\tau) -\tilde{v}(\tau)\big)\label{eq:4}\\          
		&= \frac{1}{1-r_a} \left[ \frac{-2r_a}{1 + r_a} \big(v_{0,T}(\tau), 0_{T^c}\big) + \big(0_T, v_{0,T^c}(\tau)\big) \right] \label{eq:5}.
	\end{align}
	From eq.~\eqref{eq:4} we infer that
	\begin{align}\label{eq:6}
		\left\| \frac{1}{1 + r_a}\tilde{v}(\tau) - \frac{1}{1-r_a} v_0(\tau)\right\|_2 \leq \frac{2r_a}{1-r_a^2}\|\tilde{v}(\tau)\|_2 + \frac{r_a}{1-r_a}\|v_0(\tau)\|_2 \leq r_a\frac{3 + r_a}{1-r_a^2}\|v_0(\tau)\|_2,
	\end{align}
	and from eq.~\eqref{eq:5} we learn that
	\begin{align}\label{eq:7}
		\frac{1}{1 + r_a}\tilde{v}(\tau) - \frac{1}{1-r_a} v_0(\tau) \:\: \in \:\: C_2^p(T_v(\tau), 1)
	\end{align}
	because $\frac{2r_a}{1 + r_a} \geq \frac{r_a}{1 - r_a}$ for all $0 \leq r_a \leq 1/3$ and Lemma~\ref{lemma:ConeProperties} applied with $\alpha = \beta = \frac{r_a}{1 - r_a}$. 
	
	Combining~\eqref{eq:6} and~\eqref{eq:7} we upper bound~\eqref{eq:9} by
	\begin{align*}
		(1 - r_a^2) \sup_{u \in C_2^p(T,1) \cap B_2^p(0, \alpha\|v_0(\tau)\|_2)} u'MM'u +  \sup_{u \in C_2^p(T,1) \cap B_2^p(0, \alpha \|v_0(\tau)\|_2)} u'MM'u,
	\end{align*}
	where $\alpha = r_a\frac{3 + r_a}{1 - r_a^2}$. To conclude the proof of part (ii) note that for $0 \leq r_a \leq 1/3$ we can upper bound $(2 -r_a^2)\alpha < 8 r_a$.
\end{proof}

\begin{proof}[\textbf{Proof of Lemma~\ref{lemma:InducedRestrictedConeProp-Dual}}]
	For notational convenience, we write $v_0(\tau)$, $\tilde{v}(\tau)$, and $T_v(\tau)$ for $v_0(\tau;z)$, $\tilde{v}(\tau;z)$, and $T_v(\tau;z)$, respectively.
	By construction of $\tilde{v}(\tau)$,
	\begin{align*}
		\left(\sum_{k \in T^c_v(\tau)} |v_{0,k}(\tau)|^2\right)^{1/2} = \|\tilde{v}(\tau) - v_0(\tau)\|_2 \leq \frac{1}{2} \left(\sum_{k \in T^c_v(\tau)} |v_{0,k}(\tau)|^2 \right)^{1/2} +  \frac{1}{2} \left(\sum_{k \in T_v(\tau)} |v_{0,k}(\tau)|^2 \right)^{1/2},
	\end{align*}
	and, hence,
	\begin{align*}
		\left(\sum_{k \in T^c_v(\tau)} |v_{0,k}(\tau)|^2\right)^{1/2}  \leq \left(\sum_{k \in T_v(\tau)} |v_{0,k}(\tau)|^2 \right)^{1/2},
	\end{align*}
	i.e. $v_0(\tau) \in C^p_2(T_v(\tau), 1)$. Therefore,
	\begin{align*}
		\tilde{v}(\tau)- v_0(\tau) \in U^p(\tau) :=\{u \in \mathbb{R}^p: (u_{T_v(\tau)}, u_{T_v^c(\tau)}) = (0, v_{T_v^c(\tau)}), \: v \in C^p_2(T_v(\tau),1)\}.
	\end{align*}
	Thus, since $MM'$ is positive semi-definite, 
	\begin{align*}
		\big(\tilde{v}(\tau)-v_0(\tau)\big)'MM'\big(\tilde{v}(\tau)-v_0(\tau)\big) \leq \sup_{u \in U^p(\tau)} u'MM'u \leq \sup_{u \in C^p_2(T_v(\tau), 1)} u'MM'u.
	\end{align*}
	This concludes the proof.
\end{proof}

\begin{proof}[\textbf{Proof of Lemma~\ref{lemma:RestrictedConeProp-Dual}}]
	
	For notational convenience, we write $v_0(\tau)$, $\tilde{v}(\tau)$,  $\hat{v}(\tau)$,  $f(\tau)$, $f_i(\tau)$, and $T_v(\tau)$  for $v_0(\tau;z)$, $\tilde{v}(\tau;z)$, $\hat{v}_\gamma(\tau; z)$, $f_{Y|X}(X'\theta_0(\tau)|X)$, $f_{Y|X}(X_i'\theta_0(\tau)|X_i)$, and $T_v(\tau;z)$, respectively.
	
	By optimality of $\hat{v}(\tau)$ and Lemma~\ref{lemma:InducedRestrictedConeProp-Dual}, for all $\tau \in \mathcal{T}$,
	\begin{align*}
		0 &\geq \frac{1}{4} \sum_{i=1}^n\hat{f}_i^2(\tau) \big(X_i'\hat{v}(\tau)\big)^2 + n z'\hat{v}(\tau) + \gamma \|\hat{v}(\tau)\|_1 - \left(\frac{1}{4} \sum_{i=1}^n\hat{f}_i^2(\tau) \big(X_i'\tilde{v}(\tau)\big)^2 + n z'\tilde{v}(\tau) + \gamma \|\tilde{v}(\tau)\|_1 \right)\\
		& = \frac{1}{4} \sum_{i=1}^n\hat{f}_i^2(\tau) \left( \big(X_i'\hat{v}(\tau)\big)^2 - \big(X_i'v_0(\tau)\big)^2 \right) + \frac{1}{4} \sum_{i=1}^n\hat{f}_i^2(\tau) \left( \big(X_i'v_0(\tau)\big)^2 - \big(X_i'\tilde{v}(\tau)\big)^2 \right) \\
		&\quad{} + n z'\big(\hat{v}(\tau) - \tilde{v}(\tau)\big)+ \gamma \|\hat{v}(\tau)\|_1 - \gamma \|\tilde{v}(\tau)\|_1 \\
		& \geq \left(\frac{1}{2} \sum_{i=1}^n\hat{f}_i^2(\tau) X_iX_i'v_0(\tau) \right)'\big(\hat{v}(\tau) - v_0(\tau)\big)  + \left(\frac{1}{2} \sum_{i=1}^n\hat{f}_i^2(\tau) X_iX_i'\tilde{v}(\tau) \right)' \big(v_0(\tau) - \tilde{v}(\tau)\big)\\
		&\quad{}   + n z'\big(\hat{v}(\tau) - \tilde{v}(\tau)\big)+ \gamma \|\hat{v}(\tau)\|_1 - \gamma \|\tilde{v}(\tau)\|_1 \\
		& =\left(\frac{1}{2} \sum_{i=1}^n\hat{f}_i^2(\tau) X_iX_i'v_0(\tau) + n z \right)'\big(\hat{v}(\tau) - \tilde{v}(\tau)\big)  + \gamma \|\hat{v}(\tau)\|_1 - \gamma \|\tilde{v}(\tau)\|_1\\
		&\quad{}  - \big(v_0(\tau) - \tilde{v}(\tau)\big)'\left(\frac{1}{2} \sum_{i=1}^n\hat{f}_i^2(\tau) X_iX_i' \right) \big(v_0(\tau) - \tilde{v}(\tau)\big)\\		
		&\geq \gamma \left( -c_0^{-1}\|\hat{v}(\tau) - \tilde{v}(\tau)\|_1 + \|\hat{v}(\tau)\|_1 - \|\tilde{v}(\tau)\|_1  - \gamma^{-1}\mu c_0^{-1}\|v_0(\tau)\|_2^2\right).
	\end{align*}
	Thus,
	\begin{align}\label{eq:lemma:RestrictedConeProp-Dual-2}
		\left(1 + c_0^{-1}\right)  \sum_{k=1}^p \big|\hat{v}_{k}(\tau) - \tilde{v}_k(\tau)\big| + c_0^{-1}\frac{\mu}{\gamma} \|v_0(\tau)\|_2^2 \geq  \sum_{k=1}^p \big|\hat{v}_{k}(\tau) - \tilde{v}_k(\tau)\big| + \sum_{k=1}^p \big|\hat{v}_{k}(\tau)\big| -\sum_{k=1}^p\big|\tilde{v}_k(\tau)\big|.
	\end{align}
	By Assumption~\ref{assumption:SparsityDual} and the reverse triangle inequality,
	\begin{align}\label{eq:lemma:RestrictedConeProp-Dual-3}
		\sum_{k=1}^p \big|\hat{v}_{k}(\tau) -\tilde{v}_k(\tau)\big| + \sum_{k=1}^p \big|\hat{v}_{k}(\tau)\big| -\sum_{k=1}^p\big|\tilde{v}_k(\tau)\big|\geq 2 \sum_{k \in T_v^c(\tau)}\big|\hat{v}_{k}(\tau)\big|.
	\end{align}
	Combine eq.~\eqref{eq:lemma:RestrictedConeProp-Dual-2} and~\eqref{eq:lemma:RestrictedConeProp-Dual-3} and note that $\|v_0(\tau)\|_2 \leq \|z\|_2/ \kappa_2(\infty)$ to conclude that
	\begin{align*}
		\frac{c_0+1}{c_0-1}\sum_{k \in T_v(\tau)} \big|\hat{v}_{k}(\tau) - \tilde{v}_k(\tau)\big| + \frac{1}{c_0-1}\frac{\mu}{\gamma}\frac{\|z\|_2^2}{\kappa_2^2(\infty)} \geq \sum_{k \in T_v^c(\tau)}\big|\hat{v}_{k}(\tau)\big|.
	\end{align*}
	This completes the proof of the first statement of the lemma.
	
	Next, if $\frac{1}{c_0-1} \frac{\mu}{\gamma}\frac{\|z\|_2^2}{\kappa_2^2(\infty)} \leq \frac{c_0+1}{c_0-1}\sum_{k \in T_v(\tau)} \big|\hat{v}_{k}(\tau) - \tilde{v}_k(\tau)\big| $, then
	\begin{align*}
		2\frac{c_0+1}{c_0-1}\sum_{k \in T_v(\tau)} \big|\hat{v}_{k}(\tau) - \tilde{v}_k(\tau)\big| \geq \sum_{k \in T_v^c(\tau)}\big|\hat{v}_{k}(\tau)\big|,
	\end{align*}
	whereas if $\frac{1}{c_0-1} \frac{\mu}{\gamma}\frac{\|z\|_2^2}{\kappa_2^2(\infty)}   > \frac{c_0+1}{c_0-1}\sum_{k \in T_v(\tau)} \big|\hat{v}_{k}(\tau) - \tilde{v}_k(\tau)\big|$, then
	\begin{align*}
		2\frac{c_0+1}{c_0-1} \frac{\mu}{\gamma}\frac{\|z\|_2^2}{\kappa_2^2(\infty)}  \geq \|\hat{v}(\tau) - v_0(\tau)\|_1.
	\end{align*}
	Thus, we conclude that
	\begin{align*}
		\hat{v}(\tau) - \tilde{v}(\tau) \in C^p_1(T_v(\tau), 2 \bar{c}) \cup B^p_1\left(0, \frac{2\bar{c} \|z\|_2^2}{\kappa_2^2(\infty)} \frac{\mu}{\gamma} \right).
	\end{align*}
	This completes the proof of the second statement of the lemma.
\end{proof}
\begin{proof}[\textbf{Proof of Lemma~\ref{lemma:LipschitzDual}}]
	Let $s, \tau \in \mathcal{T}$ be arbitrary. For notational convenience, we introduce $D_2(\tau) := \mathbb{E}\left[f_{Y|X}^2(X'\theta_0(\tau)|X)XX'\right]$ and write $s_v$ for $s_v(z)$.
	\begin{align*}
		&\left\|v_0(s; z) - v_0(\tau;z)\right\|_2\\
		&\quad{}  \lesssim \|D_2^{-1}(s)\|_{op}\sup_{\|u\|_2 \leq 1, \|u\|_0 \leq s_v} \big\|\big(D_2(\tau) - D_2(s)\big)u\|_2 \|D_2^{-1}(\tau)z\|_2\\
		&\quad{} \lesssim \frac{2\|z\|_2}{\kappa_2(\infty)}\sup_{\|u\|_2 \leq 1, \|u\|_0 \leq s_v} \sup_{\|v\|_2 \leq 1} \left|\mathbb{E}\left[ v'\big(f_{Y|X}^2(X'\theta_0(s)|X)-f_{Y|X}^2(X'\theta_0(\tau)|X)\big)XX'u\right]\right|\\ 
		&\quad{}  \lesssim \frac{2\bar{f}L_fL_\theta\|z\|_2}{\kappa_2(\infty)}\sup_{\|u\|_2 \leq 1, \|u\|_0 \leq s_v} \sup_{\|v\|_2\leq 1} \sup_{ \|w\|_2 \leq 1,\: \|w\|_0 \leq 2s_\theta} \mathbb{E}|X'w||u'XX'v |s -\tau|\\
		&\quad{} \lesssim  \frac{2\bar{f}L_fL_\theta\|z\|_2}{\kappa_2(\infty)}\varphi_{\max}^{1/2}(p)\varphi_{\max}^{1/2}(s_v)\varphi_{\max}^{1/2}(2s_\theta) |s - \tau|.
	\end{align*}
	To conclude, simplify the bound.
\end{proof}
\begin{proof}[\textbf{Proof of Lemma~\ref{lemma:MaximaSumMulticonvexFunctions}}]
	We generalize the proof of Lemma~\ref{lemma:MaximaBiconvexFunction} from $K=2$ to arbitrary $K \in \mathbb{N}$.
	
	For each $x_j \in \mathrm{conv}(\mathcal{X}_j)$, $1 \leq j \leq K$, there exist $n_j \in \mathbb{N}$ and $\lambda_j^1, \ldots, \lambda_j^n \geq 0$, $\sum_{i_j=1}^{n_j} \lambda_j^{i_j} = 1$, such that $x_j = \sum_{i_j=1}^{n_j} \lambda_j^{i_j} x_j^{i_j}$ for some $x_j^{i_j} \in \mathcal{X}_j$. Thus, by the multi-convexity of the $f_i$'s and $N \times K$ applications of Jensen's inequality, for all $x_j \in \mathrm{conv}(\mathcal{X}_j)$, $ 1 \leq j \leq K$,
	\begin{align*}
		\sum_{\ell=1}^N f_\ell(x_1, \ldots, x_K) &\leq \sum_{j=1}^K\sum_{i_j=1}^{n_j} \lambda_1^{i_1}\cdots \lambda_K^{i_K} \left(\sum_{\ell=1}^N f_\ell(x_1^{i_1}, \ldots, x_K^{i_K}) \right)\\
		&\leq \sum_{j=1}^K\sum_{i_j=1}^{n_j} \lambda_1^{i_1}\cdots \lambda_K^{i_K} \left(\sup_{x_j \in \mathcal{X}_j, 1\leq j \leq K} \sum_{\ell=1}^N f_\ell(x_1, \ldots, x_K) \right)\\
		&\leq \sup_{x_j \in \mathcal{X}_j, 1\leq j \leq K} \sum_{\ell=1}^N f_\ell(x_1, \ldots, x_K).
	\end{align*}
	Thus, $\sup_{x_j \in \mathrm{conv}(\mathcal{X}_j), 1\leq j \leq K} \sum_{i=1}^N f_i(x_1, \ldots, x_K) \leq \sup_{x_j \in \mathcal{X}_j, 1\leq j \leq K} \sum_{i=1}^N f_i(x_1, \ldots, x_K)$. The reverse inequality holds trivially true since $\mathcal{X}_j \subseteq \mathrm{conv}(\mathcal{X}_j)$, $1 \leq j \leq K$. The same arguments hold if  $\sum_{i=1}^N f_i$ is replaced by $\left|\sum_{i=1}^Nf_i\right|$. This concludes the proof.
\end{proof}

\begin{proof}[\textbf{Proof of Lemma~\ref{lemma:GradientConeSquare}}]
	This lemma is proved as Lemma~\ref{lemma:GradientCone} (iii) with $\psi_{1/2}$ replaced with $\psi_{1/3}$.  We omit the repetitive details.
\end{proof}

\begin{proof}[\textbf{Proof of Lemma~\ref{lemma:LocalizedLossDual}}]
	To simplify notation, we write $K_k(\tau) = C^p_{q_k}(J_k(\tau), \vartheta_k) \cap B^p(0,1)$ for $k \in \{1, 2\}$ and $f_i(\tau) = f_{Y|X}(X_i'\theta_0(\tau)|X_i)$ for $1 \leq i \leq n$.
	
	\textbf{Proof of Case (i).} Let $\eta  \in (0,1)$ be arbitrary and $\mathcal{T}_\eta$ be an $\eta$-net with cardinality $\mathrm{card}(T_\eta) \leq 1 + 1/\eta$. We have the following decomposition:
	\begin{align}\label{eq:lemma:LocalizedLossDual-1}
		&\|\mathbb{G}_n\|_{\mathcal{G}} \nonumber\\
		&\quad{}\leq \sup_{s \in \mathcal{T}_\eta} \sup_{s' \in \mathcal{T}} \sup_{\substack{u_k \in K_k(s')\\ k \in \{1, 2\}}}\left|\frac{1}{\sqrt{n}} \sum_{i=1}^n f_i^2(s) (X_i'u_1) (X_i'u_2) - \mathbb{E}[f_i^2(s) (X_i'u_1) (X_i'u_2)  ]\right|\nonumber\\
		&\quad{} +  \sup_{s \in T_\eta} \sup_{\tau : |\tau - s| \leq \eta}  \sup_{s' \in \mathcal{T}}\sup_{\substack{u_k \in K_k(s') \\ k\in\{1,2\}}}\left|\frac{1}{\sqrt{n}} \sum_{i=1}^n \big( f_i^2(\tau) - f_i^2(s) \big) (X_i'u_1) (X_i'u_2) - \mathbb{E}\left[\big( f_i^2(\tau) - f_i^2(s)\big) (X_i'u_1) (X_i'u_2)\right]\right|\nonumber\\
		&\quad{} = \mathbf{I} + \mathbf{II}.
	\end{align}
	
	\textbf{Bound on $\mathbf{I}$.} By Lemma~\ref{lemma:MaxInequalityCovarianceCone} (ii) and the union bound over $s \in \mathcal{T}_\eta$, with probability at least $1 - \delta$,
	\begin{align}\label{eq:lemma:LocalizedLossDual-2}
		\mathbf{I} \lesssim  4(2 + \vartheta_1)(2+\vartheta_2) \bar{f}^2 \varphi_{\max}^{1/2}(s_1)\varphi_{\max}^{1/2}(s_2)  \pi_{n,1}^2\big(t_{s_1, s_2, \eta, \delta}\big),
	\end{align}
	where $t_{s_1, s_2, \eta, \delta} = s_1\log(ep/s_1) + s_2\log(ep/s_2)+ \log(1 +1/\eta)+ \log (1/\delta)$ and $\pi_{n,1}^2(z) = \sqrt{z/n} + z/n$ for $z \geq 0$.
	
	\textbf{Bound on $\mathbf{II}$.} We begin with deriving an upper bound on the supremum of an empirical processes indexed by an auxiliary function class. By Lemma~\ref{lemma:SizeConesDominatedCoordinates} there exist $\mathcal{M}_1, \mathcal{M}_2,\mathcal{W} \subset B^p(0,1)$ such that
	\begin{align*}
		&\mathrm{card}(\mathcal{M}_k) \leq \frac{3}{2}\left(\frac{5ep}{s_k}\right)^{s_k}, \quad{} \forall u \in \mathcal{M}_k: \: \|u\|_0 \leq s_k, \quad{} \forall \tau \in \mathcal{T}, \: k \in \{1, 2\} : \: K_k(\tau) \subset 2(2 + \vartheta_k) \mathrm{conv}(\mathcal{M}_k);\\
		&\mathrm{card}(\mathcal{W}) \leq \frac{3}{2}\left(\frac{5ep}{2s_{\theta}}\right)^{2s_{\theta}}, \quad{} \forall w \in \mathcal{W}: \: \|w\|_0 \leq 2s_\theta, \quad{}  \forall \tau, \tau' \in \mathcal{T}: \: C^p(T_\theta(\tau) \cup T_\theta(\tau'), 0) \cap B^p(0,1) \subset 4\mathrm{conv}(\mathcal{W}).
	\end{align*}
	Let $\{g_i\}_{i=1}^n$ be a sequence of i.i.d. standard normal random variables independent of $\{X_i\}_{i=1}^n$. Define
	\begin{align*}
		\mathcal{F} = \big\{ f(X, g) = g(X'u_1)(X'u_2)X'w : u_k \in \mathcal{M}_k, \: w \in \mathcal{W},\: k \in \{1,2\}\big\}
	\end{align*}
	Next, we verify that $\mathcal{F}$ satisfies the Lipschitz-type condition eq.~\eqref{eq:theorem:MaxInequalityBernsteinOrlicz-1}. For $\alpha \in (0,1)$ the $\psi_\alpha$-norm is a quasi-norm since $x \mapsto \exp(x^\alpha) -1$ is not convex for small values of $x$. However, by Assumption~\ref{assumption:SubGaussianity} and Lemma~\ref{lemma:ProductSubgaussian} there exists an absolute constant $C > 0$ such that for all $u_1, v_1 \in \mathcal{M}_1$, $u_2, v_2 \in \mathcal{M}_2$, and $w_1, w_2 \in \mathcal{W}$,
	\begin{align*}
		&\left\| g X'u_1X'v_1X'w_1 - \varepsilon X'u_2X'v_2X'w_2  \right\|_{P, \psi_{1/2}} \\
		&\quad{} \leq C \sup_{u \in \mathcal{M}_1}\sup_{v \in \mathcal{M}_2} \sup_{w \in \mathcal{W}}\|X'u\|_{P, \psi_2}\|X'u\|_{P, \psi_2}\|X'w\|_{P, \psi_2}\big( \|u_1 - u_2\|_2 + \|v_1 - v_2\|_2  + \|w_1 - w_2\|_2\big)\\
		&\quad{} \lesssim C\varphi_{\max}^{1/2}(2s_\theta)\varphi_{\max}^{1/2}(s_1)\varphi_{\max}^{1/2}(s_2)\big( \|u_1 - u_2\|_2 + \|v_1 - v_2\|_2  + \|w_1 - w_2\|_2\big).
	\end{align*}
	Whence, by Corollary~\ref{corollary:DeviationInequalityBernsteinOrlicz}, with probability at least $1-\delta$,
	\begin{align}\label{eq:lemma:LocalizedLossDual-4}
		\begin{split}
			\|\mathbb{G}_n\|_{\mathcal{F}} &\lesssim \varphi_{\max}^{1/2}(2s_\theta)\varphi_{\max}^{1/2}(s_1)\varphi_{\max}^{1/2}(s_2) \sqrt{s_\theta \log(ep/s_\theta) +  s_1 \log(ep/s_1) +  s_2 \log(ep/s_2) } \\
			&\quad{}+ \varphi_{\max}^{1/2}(2s_\theta)\varphi_{\max}^{1/2}(s_1)\varphi_{\max}^{1/2}(s_2)n^{-1/2}\Big(s_\theta \log(ep/s_\theta) + s_1 \log(ep/s_1) + s_2 \log(ep/s_2) \Big)^2 \\
			&\quad{} + \varphi_{\max}^{1/2}(2s_\theta)\varphi_{\max}^{1/2}(s_1)\varphi_{\max}^{1/2}(s_2) \left(\sqrt{ \log (1/\delta)} + n^{-1/2} \big(\log(1/\delta)\big)^2\right).
		\end{split}
	\end{align}
	
	We now turn the bound on $\|\mathbb{G}_n\|_{\mathcal{F}}$ into a bound on $\mathbf{II}$. For any increasing and convex function $F$ we have
	\begin{align}
		&\mathbb{E}\left[F\left(\sup_{\tau : |\tau - s| \leq \eta}\sup_{s' \in \mathcal{T}} \sup_{\substack{u_k \in K_k(s')\\ k \in \{1, 2\}}}\left|\frac{1}{\sqrt{n}} \sum_{i=1}^n \big( f_i^2(\tau) - f_i^2(s) \big) (X_i'u_1)(X_i'u_2) - \mathbb{E}\left[\big( f_i^2(\tau) - f_i^2(s)\big)(X_i'u_1)(X_i'u_2)\right]\right|\right)\right] \nonumber\\
		&\overset{(a)}{\leq} \mathbb{E}\left[F\left(\sqrt{2\pi}\sup_{\tau : |\tau - s| \leq \eta} \sup_{s' \in \mathcal{T}} \sup_{\substack{u_k \in K_k(s')\\ k \in \{1, 2\}}}\left|\frac{1}{\sqrt{n}} \sum_{i=1}^n g_i\big( f_i^2(\tau) - f_i^2(s) \big) (X_i'u_1)(X_i'u_2)\right|\right)\right]\nonumber\\
		&\overset{(b)}{\leq} \mathbb{E}\left[F\left(4\sqrt{2\pi}\bar{f}L_f\sup_{\tau : |\tau - s| \leq \eta} \sup_{s' \in \mathcal{T}} \sup_{\substack{u_k \in K_k(s')\\ k \in \{1, 2\}}}\left|\frac{1}{\sqrt{n}} \sum_{i=1}^n g_i\big(X_i'\theta_0(\tau) - X_i'\theta_0(s) \big) (X_i'u_1)(X_i'u_2)\right|\right)\right]\nonumber\\
		&\overset{(c)}{\leq} \mathbb{E}\left[F\left(32 \sqrt{2\pi}(2+ \vartheta_1)(2+ \vartheta_2)\bar{f}L_fL_\theta \eta  \|\mathbb{G}_n\|_{\mathcal{F}}\right)\right]\label{eq:lemma:LocalizedLossDual-5}
	\end{align}
	where $(a)$ holds by Lemma 6.3 followed by Lemma 4.5 in~\cite{ledoux1996probability}, $(b)$ holds by Lemma~\ref{lemma:Composite-GaussianContraction} (note that the map $z \mapsto f_{Y|X}^2(z - X'\theta_0(s)|X) - f_{Y|X}^2(X'\theta_0(s)|X)$ is Lipschitz-continuous and vanishes at 0), and $(c)$ holds by Assumption~\ref{assumption:LipschitzQRVector}, Lemma~\ref{lemma:MaximaSumMulticonvexFunctions}, and construction of $\mathcal{M}_1$, $\mathcal{M}_2$, and $\mathcal{W}$.
	
	By Lemma~\ref{lemma:Panchenko2003Handel} (with $\alpha = 1/2$ and $\varphi(t) = t^{1/4} + n^{-1/2}t$), eq.~\eqref{eq:lemma:LocalizedLossDual-4} and~\eqref{eq:lemma:LocalizedLossDual-5}, and the union bound over $\eta \in \mathcal{T}_\eta$, with probability at least $1-\delta$,
	\begin{align}\label{eq:lemma:LocalizedLossDual-6}
		\begin{split}
			\mathbf{II} &\lesssim (2+ \vartheta_1)(2+ \vartheta_2)\bar{f}L_fL_\theta \varphi_{\max}^{1/2}(2s_\theta)\varphi_{\max}^{1/2}(s_1)\varphi_{\max}^{1/2}(s_2) \\
			&\quad{} \times \left(\sqrt{s_\theta \log(ep/s_\theta) +  s_1 \log(ep/s_1) +  s_2 \log(ep/s_2) } \right.\\
			&\left. \quad{} \quad{} + n^{-1/2}\Big(s_\theta \log(ep/s_\theta) + s_1 \log(ep/s_1) + s_2 \log(ep/s_2) \Big)^2 \right. \\
			&\left. \quad{} \quad{} +  \sqrt{ \log (1/\delta)} + n^{-1/2} \big(\log(1/\delta)\big)^2\right) \\
			&\quad{} \times \eta.
		\end{split}
	\end{align}
	
	\textbf{Conclusion.} Since $\eta \in (0,1)$ is arbitrary, we can choose $\eta \asymp 1/(L_f L_\theta n) $. Combine the bounds in eq.~\eqref{eq:lemma:LocalizedLossDual-1},~\eqref{eq:lemma:LocalizedLossDual-2}, and~\eqref{eq:lemma:LocalizedLossDual-6}, adjust the constants, and conclude that with probability at least $1 - \delta$,	
	\begin{align*}
		\|\mathbb{G}_n\|_{\mathcal{G}} &\lesssim (2 + \vartheta_1)(2+\vartheta_2) \bar{f}^2 \varphi_{\max}^{1/2}(s_1)\varphi_{\max}^{1/2}(s_2) (1 + \varphi_{\max}^{1/2}(2s_\theta)) \psi_n\big(t_{s_\theta, s_1, s_2, n, \delta }\big),
	\end{align*}
	where $t_{s_1, s_2, s_\theta, n, \delta } = s_1 \log(ep/s_1) + s_2 \log(ep/s_2) + s_\theta \log(ep/s_\theta) + \log (nL_f L_\theta/\delta)$ and $\psi_n(z) = \sqrt{z}\big(1 + n^{-1/2} \sqrt{z} + n^{-3/2} z^{3/2}\big)$ for $z \geq 0$.
	
	\textbf{Proof of Case (ii).} Let $\mathcal{M}_k$ be as in the proof of case (i) and define $\widetilde{\mathcal{M}}_k = \{u \in \mathbb{R}^p: u \in \mathcal{M}_k \mathrm{\:or\:} -u \in \mathcal{M}_k\}$. Then, $\mathrm{card}(\widetilde{\mathcal{M}}_k) \leq 2 \mathrm{card}(\mathcal{M}_k)$, $\|u\|_0 \leq s_k$ for all $u \in\widetilde{\mathcal{M}}_k$ and $K(\tau) \subset 2(2+\vartheta_k) \mathrm{conv}(\widetilde{\mathcal{M}}_k)$. The claim follows now by the same arguments used to proof case (i).
	
	\textbf{Proof of Case (iii).} The claim is a simple consequence from the fact that the proofs of cases (i) and (ii) rely on an $\varepsilon$-net approximation of $s$-sparse sets. In fact, the proofs of these cases establish case (iii) and then use Lemma~\ref{lemma:SizeConesDominatedCoordinates} to deduce the case of $v \in C^p_{q_k}(J_k(\tau), \vartheta_k) \cap B^p_2(0,1)$. This completes the proof.
\end{proof}
\begin{proof}[\textbf{Proof of Lemma~\ref{lemma:GradientCone-2}}]
	The proof strategy is identical to the one of Lemma~\ref{lemma:LocalizedLossDual}. To simplify notation, we write $K(\tau) = C^p_q(J(\tau), \vartheta) \cap B^p(0,1)$ and $f_i(\tau) = f_{Y|X}(X_i'\theta_0(\tau)|X_i)$ for $1 \leq i \leq n$.
	
	\textbf{Proof of Case (i).} Let $\eta  \in (0,1)$ be arbitrary and $\mathcal{T}_\eta$ be an $\eta$-net with cardinality $\mathrm{card}(T_\eta) \leq 1 + 1/\eta$.  We have the following decomposition:
	\begin{align}\label{eq:lemma:GradientCone-2-1}
		\|\mathbb{G}_n\|_\mathcal{G} &\leq \sup_{\tau\in \mathcal{T}_\eta} \sup_{\tau' \in \mathcal{T}} \sup_{v  \in K(\tau')}  \left|\frac{1}{\sqrt{n}} \sum_{i=1}^n f_i(\tau)\big(\tau' - \mathbf{1}\{Y_i \leq X_i'\theta_0(\tau')\}\big)X_i'v\right| \nonumber\\
		&\quad{} + \sup_{\tau \in \mathcal{T}_\eta} \sup_{\tau' : |\tau' - \tau| \leq \eta} \sup_{\tau''\in \mathcal{T}}\sup_{v \in K(\tau'')} \left|\frac{1}{\sqrt{n}} \sum_{i=1}^n \big(f_i(\tau) - f_i(\tau')\big)\big(\tau'' - \mathbf{1}\{Y_i \leq X_i'\theta_0(\tau'')\}\big)X_i'v \right|\nonumber\\
		& = \mathbf{I} + \mathbf{II}.
	\end{align}
	
	\textbf{Bound on $\mathbf{I}$.} By Lemma~\ref{lemma:GradientCone} and the union bound over $\tau \in \mathcal{T}_\eta$, with probability at least $1 - \delta$,
	\begin{align}\label{eq:lemma:GradietnCone-2-2}
		\mathbf{I} \lesssim  \varphi_{\max}^{1/2}(s) \bar{f}\sqrt{t_{s, \eta, \delta}} \sqrt{1 + \pi_{n,1}^2(t_{s, \eta, \delta})},
	\end{align}
	where $t_{s, \eta, \delta} = s \log (ep/s) + \log(1 + 1/\eta) + \log(1/\delta)$ and $\pi_{n,1}^2(z) = \sqrt{z/n} + z/n$ for $ z \geq  0$.	
	
	\textbf{Bound on $\mathbf{II}$.}  We begin with deriving an upper bound on the supremum of an empirical processes indexed by an auxiliary function class. By Lemma~\ref{lemma:SizeConesDominatedCoordinates} there exist $\mathcal{M},\mathcal{W} \subset B^p(0,1)$ such that
	\begin{align*}
		&\mathrm{card}(\mathcal{M}) \leq \frac{3}{2}\left(\frac{5ep}{s}\right)^{s}, \quad{} \forall u \in \mathcal{M}: \: \|u\|_0 \leq s, \quad{} \forall \tau \in \mathcal{T} : \: K(\tau) \subset 2(2 + \vartheta) \mathrm{conv}(\mathcal{M});\\
		&\mathrm{card}(\mathcal{W}) \leq \frac{3}{2}\left(\frac{5ep}{2s_{\theta}}\right)^{2s_{\theta}}, \quad{} \forall w \in \mathcal{W}: \: \|w\|_0 \leq 2s_\theta, \quad{}  \forall \tau, \tau' \in \mathcal{T}: \: C^p(T_\theta(\tau) \cup T_\theta(\tau'), 0) \cap B^p(0,1) \subset 4\mathrm{conv}(\mathcal{W}).
	\end{align*}
	Let $\{g_i\}_{i=1}^n$ be a sequence of i.i.d. standard normal random variables independent of $\{(X_i, Y_i)\}_{i=1}^n$. Define
	\begin{align*}
		\mathcal{F} = \big\{ f(X, Y, g) = g (X'v)(X'w)\big(\tau - \mathbf{1}\{Y \leq X'\theta_0(\tau)\}\big) :\: v \in \mathcal{M}, \: w \in \mathcal{W}, \: \tau \in \mathcal{T} \big\}.
	\end{align*}
	Note the following: First, $\mathcal{F} =\{hj: h \in \mathcal{H}, \: j \in \mathcal{J}\} $, where $\mathcal{H} = \big\{h(X,Y) =  \big(\tau - \mathbf{1}\{F_{Y|X}(Y|X) \leq \tau \}\big) : \: \tau \in \mathcal{T} \big\}$ and $\mathcal{J} = \{j(X, g) = g(X'v) (X'w) :  v \in \mathcal{M}, \: w \in \mathcal{W}\}$. The set $\mathcal{H}$ is the difference of two VC-subgraph classes with VC-indices at most 2, respectively~\citep[][Lemma 2.6.15 and Example 2.6.1]{vaartwellner1996weak}. Thus, $\mathcal{H}$ is VC-subgraph class with VC-index at most $3$~\citep[][Lemma 2.6.18]{vaartwellner1996weak}. The function class $\mathcal{J}$ is finite with $\mathrm{card}(\mathcal{J}_\mathcal{M}) = \mathrm{card}(\mathcal{M}) \times \mathrm{card}(\mathcal{W}) $. By Assumption~\ref{assumption:SubGaussianity} and Lemma~\ref{lemma:ProductSubgaussian}, for $v_1, v_2 \in \mathcal{M}$, $w_1, w_2 \in \mathcal{W}$ arbitrary, 
	\begin{align*}
		&\left\| g^2(X'v_1)^2(X'w_1)^2 - \mathbb{E}[g^2(X'v_1)^2(X'w_1)^2] - \big(g^2(X'v_2)^2(X'w_2)^2 - \mathbb{E}[g^2(X'v_2)^2(X'w_2)^2]\big) \right\|_{P, \psi_{1/3}} \nonumber\\
		&\quad{} \lesssim \left\| g^2(X'v_1)^2(X'w_1)^2 - g^2(X'v_2)^2(X'w_2)^2\right\|_{P, \psi_{1/3}} \nonumber\\
		&\quad{} \lesssim \sup_{u_1, u_2}\left\|g^2\|_{\psi_1}\|(X'u_1)^2\|_{\psi_1}\|(X'u_2)^2 \right\|_{P, \psi_1} \left( \|v_1 - v_2\|_2 + \|w_1 - w_2\|_2 \right) \nonumber\\
		&\quad{} \lesssim \varphi_{\max}(s)\varphi_{\max}(2s_\theta)\left( \|v_1 - v_2\|_2 + \|w_1 - w_2\|_2 \right) ,
	\end{align*}
	where the supremum in the third line is taken over all $u_1, u_2$ such that $\|u_1\|_2, \|u_2\|_2 \leq 1$ and $\|u_1\|_0 \leq s$ and $\|u_2\|_0 \leq 2 s_\theta$. Therefore, for $j_v, j_u \in \mathcal{J}$,
	\begin{align*}
		\| (j_v^2 - Pj_v^2) - (j_u^2 - Pj_u^2) \|_{P, \psi_1} \lesssim \varphi_{\max}(s) \varphi_{\max}(2s_\theta)\|v - u\|_2.
	\end{align*}
	Thus, by Corollary~\ref{corollary:MaxInequalityBernsteinOrlicz-Gradient}, with probability at least $1- \delta$, 
	\begin{align}\label{eq:lemma:GradientCone-2-2}
		\|\mathbb{G}_n\|_{\mathcal{F}} \lesssim  \varphi_{\max}^{1/2}(s)\varphi_{\max}^{1/2}(2s_\theta)\sqrt{t_{s, s_\theta, \delta}}\sqrt{1 + \pi_{n,1/3}^2(t_{s, s_\theta, \delta})},
	\end{align}
	where $t_{s, s_\theta, \delta} = s\log(ep/s) + s_\theta\log(ep/s_\theta) + \log(1/\delta)$ and $\pi_{n,1/3}^2(z) = \sqrt{z/n} + z^3/n$ for $z \geq 0$.
	
	We now turn the bound on this process into a bound on $\mathbf{II}$. For any increasing and convex function $F$ we have, for $\tau \in \mathcal{T}_\eta$ arbitrary,
	\begin{align}\label{eq:lemma:GradientCone-2-3}
		&\mathbb{E}\left[F\left(\sup_{\tau' : |\tau' - \tau| \leq \eta} \sup_{\tau''\in \mathcal{T}}\sup_{v \in K(\tau'')} \left|\frac{1}{\sqrt{n}} \sum_{i=1}^n \big(f_i(\tau) - f_i(\tau')\big)\big(\tau'' - \mathbf{1}\{Y_i \leq X_i'\theta_0(\tau'')\}\big)X_i'v \right|\right)\right] \nonumber\\
		&\overset{(a)}{\leq} \mathbb{E}\left[F\left(\sqrt{2\pi} \sup_{\tau' : |\tau' - \tau| \leq \eta} \sup_{\tau''\in \mathcal{T}}\sup_{v \in K(\tau'')} \left|\frac{1}{\sqrt{n}} \sum_{i=1}^n  g_i\big(f_i(\tau) - f_i(\tau')\big)\big(\tau'' - \mathbf{1}\{Y_i \leq X_i'\theta_0(\tau'')\}\big)X_i'v \right|\right)\right] \nonumber\\
		&\overset{(b)}{\leq} \mathbb{E}\left[F\left(4\sqrt{2 \pi} L_f \sup_{\tau' : |\tau' - \tau| \leq \eta} \sup_{\tau''\in \mathcal{T}}\sup_{v \in K(\tau'')}  \left|\frac{1}{\sqrt{n}} \sum_{i=1}^n g_i\big(X_i'\theta_0(\tau) - X_i'\theta_0(\tau')\big)\big(\tau'' - \mathbf{1}\{Y_i \leq X_i'\theta_0(\tau'')\}\big)X_i'v\right|\right)\right]\nonumber\\
		&\overset{(c)}{\leq} \mathbb{E}\left[F\left(32 (2 + \vartheta)\sqrt{2\pi}L_fL_\theta \eta \|\mathbb{G}_n\|_\mathcal{F}\right)\right],
	\end{align}
	where $(a)$ holds by Lemma 6.3 followed by Lemma 4.5 in~\cite{ledoux1996probability}, $(b)$ holds by Lemma~\ref{lemma:Composite-GaussianContraction} (note that the map $z \mapsto f_{Y|X}(z - X'\theta_0(\tau)|X) - f_{Y|X}(X'\theta_0(\tau)|X)$ is Lipschitz-continuous and vanishes at 0), and $(c)$ holds by Assumption~\ref{assumption:LipschitzQRVector}, Lemma~\ref{lemma:MaximaBiconvexFunction}, and construction of $\mathcal{M}$ and $\mathcal{W}$.
	
	Set $t_{s, s_\theta, \eta, \delta} = s\log (5ep/s) + 2s_\theta\log (5ep/s_\theta) + \log(1 + 1/\eta) + \log(1/\delta)$. By Lemma~\ref{lemma:Panchenko2003Handel}, eq.~\eqref{eq:lemma:GradientCone-2-2} and~\eqref{eq:lemma:GradientCone-2-3}, and the union bound over $\tau \in \mathcal{T}_\eta$ we have, with probability at least $1- \delta$,
	\begin{align}\label{eq:lemma:GradientCone-2-4}
		\mathbf{II}\lesssim L_fL_\theta (2 + \vartheta) \varphi_{\max}^{1/2}(s)\varphi_{\max}^{1/2}(2s_\theta)\sqrt{t_{s, s_\theta, \eta, \delta}}\sqrt{1 + \pi_{n,1/3}^2(t_{s, s_\theta, \eta, \delta})}\eta.
	\end{align}
	
	\textbf{Conclusion.} Since $\eta \in (0,1)$ is arbitrary, we can choose $\eta \asymp 1/(L_f L_\theta n) $.
	Combine the bounds in eq.~\eqref{eq:lemma:GradientCone-2-1},~\eqref{eq:lemma:GradientCone-2-2}, and~\eqref{eq:lemma:GradientCone-2-4}, adjust the constants, and conclude that with probability at least $1 - \delta$,	
	\begin{align*}
		\|\mathbb{G}_n\|_{\mathcal{G}} &\lesssim (2 + \vartheta) \bar{f}\varphi_{\max}^{1/2}(s)\big(1 + \varphi_{\max}^{1/2}(2s_\theta)\big) \psi_n\big(t_{s, s_\theta, n, \delta }\big),
	\end{align*}
	where $t_{s, s_\theta, n, \delta } = s \log(ep/s) + s_\theta \log(ep/s_\theta) + \log (nL_f L_\theta/\delta)$ and $\psi_n(z) = \sqrt{z}\big(1 + n^{-1/2} \sqrt{z} + n^{-3/2} z^{3/2}\big)$ for $z \geq 0$.
	
	\textbf{Proof of Case (ii).} The claim is an immediate consequence from the fact that the proof of case (i) uses an $\varepsilon$-net approximation of $s$-sparse sets. In fact, the proof of this case establish case (ii) first, and then uses Lemma~\ref{lemma:SizeConesDominatedCoordinates} to deduce the case of $v \in C^p_q(J(\tau), \vartheta) \cap B^p_2(0,1)$. This completes the proof.
\end{proof}
\begin{proof}[\textbf{Proof of Lemma~\ref{lemma:GradientCone-3}}]
	The proof is an easy modification of the proof of Lemma~\ref{lemma:GradientCone-2}. We only sketch the details:
	
	\textbf{Proof of Case (i).} To simplify notation, we write $K_k(\tau) = C^p(J_k(\tau), \vartheta_k) \cap B^p(0,1)$ for $k \in \{1, 2\}$ and $f_i(\tau) = f_{Y|X}(X_i'\theta_0(\tau)|X_i)$ for $1 \leq i \leq n$. Let $\eta  \in (0,1)$ be arbitrary and $\mathcal{T}_\eta$ be an $\eta$-net with cardinality $\mathrm{card}(T_\eta) \leq 1 + 1/\eta$. We have the following decomposition:
	\begin{align}
		&\|\mathbb{G}_n\|_{\mathcal{G}} \nonumber\\
		&\quad{}\leq \sup_{s \in \mathcal{T}_\eta} \sup_{s' \in \mathcal{T}} \sup_{\substack{u_k \in K_k(s')\\ k \in \{1, 2\}}}\left|\frac{1}{\sqrt{n}} \sum_{i=1}^n f_i^2(s) \big(s' - \mathbf{1}\{Y_i \leq X_i'\theta_0(s')\}
		\big)(X_i'u_1) (X_i'u_2)\right|\nonumber\\
		&\quad{} +  \sup_{s \in T_\eta} \sup_{\tau : |\tau - s| \leq \eta}  \sup_{s' \in \mathcal{T}}\sup_{\substack{u_k \in K_k(s') \\ k\in\{1,2\}}}\left|\frac{1}{\sqrt{n}} \sum_{i=1}^n \big( f_i^2(\tau) - f_i^2(s) \big) \big(s' - \mathbf{1}\{Y_i \leq X_i'\theta_0(s')\}\big)(X_i'u_1) (X_i'u_2)\right|\nonumber\\
		&\quad{} = \mathbf{I} + \mathbf{II}.
	\end{align}
	
	A trivial modification of Lemma~\ref{lemma:GradientCone} and the union bound over $\tau \in \mathcal{T}_\eta$ yield, with probability at least $1 - \delta$,
	\begin{align*}
		\mathbf{I} \lesssim  (2 + \vartheta_1) (2 + \vartheta_2)\bar{f}^2 \varphi_{\max}^{1/2}(s_1)\varphi_{\max}^{1/2}(s_2)\sqrt{t_{s_1, s_2, \eta, \delta}} \sqrt{1 + \pi_{n,1/2}^2(t_{s_1, s_2, \eta, \delta})},
	\end{align*}
	where $t_{s_1, s_2, \eta, \delta} = s_1 \log (ep/s_1) + s_2 \log (ep/s_2) + \log(1 + 1/\eta) + \log(1/\delta)$ and $\pi_{n,1/2}^2(z) = \sqrt{z/n} + z^2/n$ for $ z \geq  0$.	
	
	By straightforward calculations as those leading up to eq.~\eqref{eq:lemma:GradientCone-2-2} we have, with probability at least $1- \delta$,
	\begin{align*}
		\mathbf{II} \lesssim  (2 + \vartheta_1) (2 + \vartheta_2)\varphi_{\max}^{1/2}(s_1)  \varphi_{\max}^{1/2}(s_2) \varphi_{\max}^{1/2}(2s_\theta)\sqrt{t_{s_1, s_2, s_\theta, \delta}}\sqrt{1 + \pi_{n,1/4}^2(t_{s_1, s_2, s_\theta, \delta})},
	\end{align*}
	where $t_{s_1, s_2, s_\theta, \delta} = s_1\log(ep/s_1) + s_2\log(ep/s_2) + s_\theta\log(ep/s_\theta) + \log(1/\delta)$ and $\pi_{n,1/4}^2(z) = \sqrt{z/n} + z^4/n$ for $z \geq 0$.	
	
	Note that $z \mapsto f_{Y|X}^2(z- X'\theta_0(\tau)|X) - f_{Y|X}^2(X'\theta_0(\tau)|X)$ has Lipschitz constant $2L_f \bar{f}$ and vanishes at 0. Hence, adapting the arguments in eq.~\eqref{eq:lemma:GradientCone-2-3} and eq.~\eqref{eq:lemma:GradientCone-2-4}, we conclude that, with probability at least $1- \delta$,
	\begin{align*}
		\mathbf{II} \lesssim L_fL_\theta \bar{f} (2 + \vartheta_1)(2 + \vartheta_2) \varphi_{\max}^{1/2}(s_1)\varphi_{\max}^{1/2}(s_2)\varphi_{\max}^{1/2}(2s_\theta)\sqrt{t_{s_1, s_2, s_\theta, \eta, \delta}}\sqrt{1 + \pi_{n,1/4}^2(t_{s_1, s_2, s_\theta, \eta, \delta})}\eta,
	\end{align*}
	where $t_{s_1, s_2, s_\theta, \eta, \delta} = s_1\log(ep/s_1) + s_2\log(ep/s_2) + s_\theta\log(ep/s_\theta) + \log(1/\eta) + \log(1/\delta)$. 
	
	Since $\eta \in (0,1)$ is arbitrary, we can choose $\eta \asymp 1/(L_f L_\theta n) $.
	Combine the bounds in on $\mathbf{I}$ and  $\mathbf{II}$ to conclude that with probability at least $1 - \delta$,	
	\begin{align*}
		\|\mathbb{G}_n\|_{\mathcal{G}} &\lesssim (2 + \vartheta_1)  (2 + \vartheta_2) \bar{f}^2 \varphi_{\max}^{1/2}(s_1) \varphi_{\max}^{1/2}(s_2)\varphi_{\max}^{1/2}(s_\theta)\big(1 + \varphi_{\max}^{1/2}(2s_\theta)\big) \psi_n\big(t_{s_1, s_2, s_\theta, n, \delta }\big),
	\end{align*}
	where $t_{s_1, s_2, s_\theta, n, \delta } = s_1 \log(ep/s_1) + s_2\log(ep/s_2) + s_\theta \log(ep/s_\theta) + \log (nL_f L_\theta/\delta)$ and $\psi_n(z) = \sqrt{z}\big(1 + n^{-1/2} \sqrt{z} + n^{-3/2} z^2\big)$ for $z \geq 0$.
	
	\textbf{Proof of Case (ii).} The claim is an immediate consequence from the fact that the proof of case (i) uses an $\varepsilon$-net approximation of $s$-sparse sets.
\end{proof}
\begin{proof}[\textbf{Proof of Lemma~\ref{lemma:LocalizedRankScoresCone}}]
	The proof strategy is similar to the one of Lemma~\ref{lemma:LocalizedLossDual}. However, instead of simply applying Corollary~\ref{corollary:DeviationInequalityBernsteinOrlicz}, we us a combination of Lemmas~\ref{lemma:MaxInequalityChernozhukov2014-VC-Class} and~\ref{lemma:MaxInequalityAdamczak2008}. This allows us to leverage the fact that $\|\theta - \theta_0(\tau)\|_2 \leq r_0$ for all $\tau \in \mathcal{T}$ at the cost of additional $(\log n)$-factors.
	\noindent
	
	\textbf{Proof of Case (i).} To simplify notation, we write $K(\tau) = C^p_q(J(\tau), \vartheta) \cap B^p_2(0,1)$ and $f_i(\tau = f_{Y|X}(X_i'\theta_0(\tau)|X_i)$ fo $1 \leq i \leq n$. By Lemma~\ref{lemma:SizeConesDominatedCoordinates} there exist $\mathcal{M} \subset B^p(0,1)$ such that
	\begin{align*}
		&\mathrm{card}(\mathcal{M}) \leq \frac{3}{2}\left(\frac{5ep}{s}\right)^s, \quad{} \forall v \in \mathcal{M}: \: \|v\|_0 \leq s, \quad{} \forall \tau \in \mathcal{T} : \: K(\tau) \subset 2(2 + \vartheta) \mathrm{conv}(\mathcal{M}).
	\end{align*}
	For $S \subseteq \{1, \ldots, p\}$, $v \in \mathcal{M}$, and $\tau \in \mathcal{T}$ we define the following function classes:
	\begin{align*}	
		\mathcal{H}_{S,v} &= \big\{ h(X, Y) = \left(\mathbf{1}\big\{Y \leq X'\theta \big\} - \mathbf{1}\big\{Y \leq X'\theta_0(\tau) \}\right)X'v, \\
		&\quad{}\quad{}\quad{}\theta \in \mathbb{R}^p, \: \mathrm{supp}(\theta) = S, \|\theta - \theta_0(\tau)\|_2 \leq r_0, \tau \in \mathcal{T}\big\},\\
		\mathcal{G}_{S, v, \tau} &= \big\{ g(X,Y) =  f_{Y|X}(X'\theta_0(\tau)|X)h(X,Y): h \in \mathcal{H}_{S,v}\}\\
		\mathcal{G}_{S, v} &= \bigcup_{\tau \in \mathcal{T}} \mathcal{G}_{S,v, \tau}.
	\end{align*}
	Each $g \in \mathcal{G}_{S,v, \tau}$ and $h \in \mathcal{H}_{S,v}$ is uniquely determined by the triplet $(v, \theta ,\tau)$. We therefore also write $g_{v, \theta, \tau}$ and $h_{v, \theta, \tau}$ whenever we need to identify a function via its parameters. Let $\eta  \in (0,1)$ be arbitrary and $\mathcal{T}_\eta$ be an $\eta$-net with cardinality $\mathrm{card}(T_\eta) \leq 1 + 1/\eta$. We have the following decomposition:
	\begin{align}\label{eq:lemma:LocalizedRankScoresCone-1}
		\|\mathbb{G}_n\|_{\mathcal{G}_{S, v} } &\leq \sup_{\tau\in \mathcal{T}_\eta} \|\mathbb{G}_n\|_{\mathcal{G}_{S,v, \tau}}\nonumber\\
		&\quad{} + \sup_{\tau \in \mathcal{T}_\eta} \sup_{\tau' : |\tau' - \tau| \leq \eta} \sup_{h \in \mathcal{H}_{S,v}}  \left|\frac{1}{\sqrt{n}} \sum_{i=1}^n \big(f_i(\tau') - f_i(\tau)\big)h(X_i, Y_i)  - \mathbb{E}\big[\big(f_i(\tau') - f_i(\tau)\big) h(X_i, Y_i) \big]\right|\nonumber\\
		& = \mathbf{I} + \mathbf{II}.
	\end{align}
	In the following we first derive an upper bound on $\|\mathbb{G}_n\|_{\mathcal{G}_{S, v} }$ and then assemble a bound on $\|\mathbb{G}_n\|_\mathcal{G}$.
	
	\textbf{Bound on $\mathbf{I}$.} It is standard to verify that $\mathcal{G}_{S,v,\tau}$ is a VC-subgraph class with VC-index at most a constant multiple of $|S| + 1$ (see proof of Lemma~\ref{lemma:LocalizedRankScores}). Also, $G_v(X) = \bar{f}|X'v|$ is an envelope of $\mathcal{G}_{S,v,\tau}$, $\|\max_{1 \leq i \leq n} G_v(X_i)\|_{\psi_1} \lesssim (\log n) \bar{f} \varphi_{\max}^{1/2}(s)$, and for all $g \in \mathcal{G}_{S,v,\tau}$,
	\begin{align*}
		Pg^2 &\leq 2\bar{f}^2\mathbb{E}\left[(X'v)^2\big|F_{Y|X}(X'\theta_0(\tau)|X) - F_{Y|X}(X'\theta|X)\big|\right] \\
		&\leq 2\bar{f}^3 \mathbb{E}\left[(X'v)^2 |X'(\theta_0(\tau) - \theta)|\right]\\
		& \lesssim r_0 \bar{f}^3\varphi_{\max}(s)\varphi_{\max}(|S| + s_\theta)\\
		& \lesssim r_0 \varphi_{\max}(s) \big(1 + \bar{f}^3\varphi_{\max}(|S| + s_\theta)\big).
	\end{align*}
	Thus, by Lemma~\ref{lemma:MaxInequalityAdamczak2008}, there exists an absolute constant $C_1 > 0$ (independent of $S, v, \tau$) such that
	\begin{align}\label{eq:lemma:LocalizedRankScoresCone-2}
		\mathbb{P}\left\{ \|\mathbb{G}_n\|_{\mathcal{G}_{S,v,\tau}} \geq  C_1 \mathbb{E}\|\mathbb{G}_n\|_{\mathcal{G}_{S,v,\tau}} + C_1\varphi_{\max}^{1/2}(s)\big(1 + \bar{f}^{3/2} \varphi_{\max}^{1/2}(|S| + s_\theta)\big) \upsilon_{r_0, n, 2}(t)\right\} \leq e^{-t},
	\end{align}
	and by Lemma~\ref{lemma:MaxInequalityChernozhukov2014-VC-Class}, there exists an absolute constant $C_2 > 1$ (independent of $S, v, \tau$) such that 
	\begin{align*}
		\mathbb{E}\|\mathbb{G}_n\|_{\mathcal{G}_{S,v,\tau}}
		&\leq C_2 \varphi_{\max}^{1/2}(s)\big(1 +  \bar{f}^{3/2}\varphi_{\max}^{1/2}(|S| + s_\theta)\big) \upsilon_{r_0, n, 2}\big(|S| \log (1/r_0) \big),
	\end{align*}
	where $\upsilon_{r_0, n, \gamma}(z) = \sqrt{r_0 z} + \sqrt{\frac{\log^\gamma n}{n} z^\gamma}$ for $z \geq 0$ and $\gamma > 0$. Set $t_{\eta, \delta} = \log (1 + 1/\eta) + \log(1/\delta)$. Now, eq.~\eqref{eq:lemma:LocalizedRankScoresCone-2} and the union bound over $\tau \in \mathcal{T}_\eta$ yield, with probability at least $1- \delta$, 
	\begin{align}\label{eq:lemma:LocalizedRankScoresCone-3}
		\mathbf{I} \lesssim \varphi_{\max}^{1/2}(s)\big(1 + \bar{f}^{3/2}\varphi_{\max}^{1/2}(|S| + s_\theta)\big) \big(\upsilon_{r_0, n, 2}\big(|S| \log (1/r_0) \big) + \upsilon_{r_0, n, 2}(t_{\eta, \delta})\big).
	\end{align}
	
	\textbf{Bound on $\mathbf{II}$.} Let $\{g_i\}_{i=1}^n$ be a sequence of i.i.d. standard normal random variables independent of $\{(X_i, Y_i)\}_{i=1}^n$. By Lemma~\ref{lemma:SizeConesDominatedCoordinates} there exist $\mathcal{W} \subset B^p(0,1)$ such that
	\begin{align*}
		&\mathrm{card}(\mathcal{W}) \leq \frac{3}{2}\left(\frac{5ep}{s_\theta}\right)^{s_\theta}, \quad{} \forall w \in \mathcal{W}: \: \|w\|_0 \leq s_\theta, \quad{} B^p(0,1) \subset 4 \mathrm{conv}(\mathcal{W}).
	\end{align*}	
	For $S \subseteq \{1, \ldots, p\}$, $v \in \mathcal{M}$, and $w \in \mathcal{W}$ consider
	\begin{align*}
		\mathcal{J}_{S, v, w} = \big\{j(X,Y,g) = gh(X,Y) X'w : h \in \mathcal{H}_{S,v}  \big\}.
	\end{align*}
	Again, we easily verify that $\mathcal{J}_{S,v,w}$ is a VC-subgraph class with VC-index at most a constant multiple of $|S| + 2$ (see proof of Lemma~\ref{lemma:LocalizedRankScores}). Also, $J_{v,w}(X, g) = |g||(X'v)(X'w)|$ is an envelope of $\mathcal{J}_{S,v,w}$, $\|\max_{1 \leq i \leq n} J_{v,w}(X_i, g_i)\|_{\psi_{2/3}} \lesssim (\log n)^{3/2} \varphi_{\max}^{1/2}(s)\varphi_{\max}^{1/2}(2s_\theta)$, and for all $j \in \mathcal{J}_{S,v,w}$,
	\begin{align*}
		Pj^2 &\leq 2\mathbb{E}\left[g^2(X'v)^2(X'w)^2\big|F_{Y|X}(X'\theta_0(\tau)|X) - F_{Y|X}(X'\theta|X)\big|\right] \\
		&\leq 2\bar{f}\mathbb{E}\left[g^2(X'v)^2(X'w)^2 |X'(\theta_0(\tau) - \theta)|\right]\\
		& \lesssim r_0 \bar{f}\varphi_{\max}(s)\varphi_{\max}(2s_\theta)\varphi_{\max}(|S| + s_\theta)\\
		& \lesssim r_0 \varphi_{\max}(s) \varphi_{\max}(2s_\theta)\big(1 + \bar{f}\varphi_{\max}(|S| + s_\theta)\big).
	\end{align*}
	Thus, by Lemma~\ref{lemma:MaxInequalityAdamczak2008}, there exists an absolute constant $C_3 > 0$ (independent of $S, v, w$) such that
	\begin{align}\label{eq:lemma:LocalizedRankScoresCone-4}
		\mathbb{P}\left\{ \|\mathbb{G}_n\|_{\mathcal{J}_{S,v,w}} \geq  C_3 \mathbb{E}\|\mathbb{G}_n\|_{\mathcal{J}_{S,v,w}} + C_3\varphi_{\max}^{1/2}(s)\varphi_{\max}^{1/2}(2s_\theta)\big(1 + \bar{f}^{1/2} \varphi_{\max}^{1/2}(|S| + s_\theta)\big) \upsilon_{r_0, n, 3}(t)\right\} \leq e^{-t},
	\end{align}
	and by Lemma~\ref{lemma:MaxInequalityChernozhukov2014-VC-Class}, there exists an absolute constant $C_4 > 1$ (independent of $S, v, w$) such that 
	\begin{align*}
		\mathbb{E}\|\mathbb{G}_n\|_{\mathcal{J}_{S,v,w}}
		&\leq C_4 \varphi_{\max}^{1/2}(s)\varphi_{\max}^{1/2}(2s_\theta)\big(1 +  \bar{f}^{1/2}\varphi_{\max}^{1/2}(|S| + s_\theta)\big) \upsilon_{r_0, n, 3}\big(|S| \log (1/r_0) \big),
	\end{align*}
	where $\upsilon_{r_0, n, \gamma}(z) = \sqrt{r_0 z} + \sqrt{\frac{\log^\gamma n}{n} z^\gamma}$ for $z \geq 0$ and $\gamma > 0$. Set $t_{s_\theta, \delta} = 2s_\theta\log (5ep/s_\theta) + \log(1/\delta)$. (Note that the upper bound on $\mathbb{E}\|\mathbb{G}_n\|_{\mathcal{G}_{S,v,\tau}}$ is not tight, but it is a convenient choice since it matches with the other terms in eq.~\eqref{eq:lemma:LocalizedRankScoresCone-4}.) Now, by eq.~\eqref{eq:lemma:LocalizedRankScoresCone-4} and the union bound over $w \in \mathcal{W}$ we have, with probability at least $1- \delta$,
	\begin{align}\label{eq:lemma:LocalizedRankScoresCone-5}
		\sup_{w \in\mathcal{W}} \|\mathbb{G}_n\|_{\mathcal{J}_{S,v, w}} \lesssim \varphi_{\max}^{1/2}(s) \varphi_{\max}^{1/2}(2s_\theta)\big(1 + \bar{f}^{1/2}\varphi_{\max}^{1/2}(|S| + s_\theta)\big) \big(\upsilon_{r_0, n, 3}\big(|S| \log (1/r_0) \big) + \upsilon_{r_0, n, 3}(t_{s_\theta, \delta})\big).
	\end{align}
	We now turn the bound on this process into a bound on $\mathbf{II}$. For any increasing and convex function $F$ we have, for $\tau \in \mathcal{T}_\eta$ arbitrary,
	\begin{align}\label{eq:lemma:LocalizedRankScoresCone-6}
		&\mathbb{E}\left[F\left(\sup_{\tau' : |\tau' - \tau| \leq \eta} \sup_{h \in \mathcal{H}_{S,v}}  \left|\frac{1}{\sqrt{n}} \sum_{i=1}^n \big(f_i(\tau') - f_i(\tau)\big)h(X_i, Y_i)  - \mathbb{E}\big[\big(f_i(\tau') - f_i(\tau)\big) h(X_i, Y_i) \big]\right|\right)\right] \nonumber\\
		&\overset{(a)}{\leq} \mathbb{E}\left[F\left(\sqrt{2\pi}\sup_{\tau' : |\tau' - \tau| \leq \eta} \sup_{h \in \mathcal{H}_{S,v}}  \left|\frac{1}{\sqrt{n}} \sum_{i=1}^n g_i\big(f_i(\tau') - f_i(\tau)\big)h(X_i, Y_i)\right|\right)\right]\nonumber\\
		&\overset{(b)}{\leq} \mathbb{E}\left[F\left(4\sqrt{2 \pi} L_f \sup_{\tau' : |\tau' - \tau| \leq \eta} \sup_{h \in \mathcal{H}_{S,v}}  \left|\frac{1}{\sqrt{n}} \sum_{i=1}^n g_i\big(X_i'\theta_0(\tau') - X_i'\theta_0(\tau)\big)h(X_i, Y_i)\right|\right)\right]\nonumber\\
		&\overset{(c)}{\leq} \mathbb{E}\left[F\left(16\sqrt{2\pi}L_fL_\theta \eta \sup_{w \in\mathcal{W}} \|\mathbb{G}_n\|_{\mathcal{J}_{S,v, w}}\right)\right],
	\end{align}
	where $(a)$ holds by Lemma 6.3 followed by Lemma 4.5 in~\cite{ledoux1996probability}, $(b)$ holds by Lemma~\ref{lemma:Composite-GaussianContraction} (note that the map $z \mapsto f_{Y|X}(z - X'\theta_0(\tau)|X) - f_{Y|X}(X'\theta_0(\tau)|X)$ is Lipschitz-continuous and vanishes at 0), and $(c)$ holds by Assumption~\ref{assumption:LipschitzQRVector}, Lemma~\ref{lemma:MaximaBiconvexFunction}, and construction of $\mathcal{W}$.
	
	Set $t_{s_\theta, \eta, \delta} = 2s_\theta\log (5ep/s_\theta) + \log(1 + 1/\eta) + \log(1/\delta)$. By Lemma~\ref{lemma:Panchenko2003Handel}, eq.~\eqref{eq:lemma:LocalizedRankScoresCone-5} and~\eqref{eq:lemma:LocalizedRankScoresCone-6}, and the union bound over $\tau \in \mathcal{T}_\eta$ we have, with probability at least $1- \delta$,
	\begin{align}\label{eq:lemma:LocalizedRankScoresCone-7}
		\mathbf{II}\lesssim L_fL_\theta \varphi_{\max}^{1/2}(s) \varphi_{\max}^{1/2}(2s_\theta)\big(1 + \bar{f}^{1/2}\varphi_{\max}^{1/2}(|S| + s_\theta)\big) \big(\upsilon_{r_0, n, 3}\big(|S| \log (1/r_0) \big) + \upsilon_{r_0, n, 3}(t_{s_\theta, \eta, \delta}\big)\eta.
	\end{align}
	
	\textbf{Conclusion.} Since $\eta \in (0,1)$ is arbitrary, we can choose $\eta \asymp 1/(L_f L_\theta n^{1/2}) $. Combining eq.~\eqref{eq:lemma:LocalizedRankScoresCone-1},~\eqref{eq:lemma:LocalizedRankScoresCone-3}, and~\eqref{eq:lemma:LocalizedRankScoresCone-7} yields, with probability at least $1-\delta$,
	\begin{align}\label{eq:lemma:LocalizedRankScoresCone-8}
		\begin{split}
			\|\mathbb{G}_n\|_{\mathcal{G}_{S, v} } \lesssim \varphi_{\max}^{1/2}(s) \big( 1 +  \varphi_{\max}^{1/2}(2s_\theta)\big)\big(1 + \bar{f}^{3/2}\varphi_{\max}^{1/2}(|S| + s_\theta)\big)\Big(\upsilon_{r_0, n}\big(|S| \log(1/r_0)\big) + \upsilon_{r_0, n}(t_{s_\theta, n, \delta})\Big),
		\end{split}
	\end{align}
	where $t_{s_\theta, n, \delta} = 2s_\theta\log (5ep/s_\theta) + \log(L_fL_\theta n) + \log(1/\delta)$ and  	$\upsilon_{r_0, n}(z) = \sqrt{z}\big(\sqrt{r_0} + n^{-1/2}(\log n) \sqrt{z} + n^{-1} (\log n)^{3/2} z\big)$ for $z \geq 0$.
	
	Next, observe that $\mathrm{card}\big(\{ S\subseteq \{1, \ldots, p\} : |S| \leq k \}\big) \leq \sum_{i=1}^k  { p \choose i} \leq \left(ep/k\right)^k$. Set $t_{s, k, s_\theta,n, \delta } =  s\log (5ep/s) + k \log (ep/k) + 2s_\theta\log (5ep/s_\theta) + \log(L_fL_\theta n) + \log(1/\delta)$. By eq.~\eqref{eq:lemma:LocalizedRankScoresCone-8} and the union bound over $v \in \mathcal{M}$ and $S \subset \{1, \ldots p\}$ with $\mathrm{card}(S) \leq k$ there exists an absolute constant $C_5 > 0$ such that
	\begin{align}\label{eq:lemma:LocalizedRankScoresCone-9}
		\begin{split}
			&\mathbb{P}\left\{  \sup_{v \in \mathcal{M}} \sup_{1 \leq k \leq n} \sup_{\mathrm{card}(S) \leq k} \frac{\|\mathbb{G}_n\|_{\mathcal{G}_{S, v}}}{	\Theta(s, k, s_\theta) \big(\upsilon_{r_0, n}\big(k \log(1/r_0)\big) + \upsilon_{r_0, n}(t_{s, k, s_\theta, n, \delta})\big)} > C_5 \right\}\\
			&\quad{}\leq \sum_{k=1}^n \left(\frac{ep}{k}\right)^k e^{- k \log (ep/k) - \log n - \log (1/\delta)} \\
			&\quad{} \leq \delta,
		\end{split}
	\end{align}	
	where $\Theta(s, k, s_\theta) = \bar{f}^{3/2}\varphi_{\max}^{1/2}(s) \big( 1 +  \varphi_{\max}^{1/2}(2s_\theta)\big)\big(1 + \varphi_{\max}^{1/2}(k + s_\theta)\big)$. For $S \subset \{1, \ldots, p\}$ define $\mathcal{G}_S = \left\{g \in \mathcal{G}_{S,v} : v \in C^p(J, \vartheta) \cap B^p(0,1)\right\}$. By Lemma~\ref{lemma:MaximaBiconvexFunction} and construction of $\mathcal{M}$, for all $S \subset \{1, \ldots, p\}$,
	\begin{align*}
		\|\mathbb{G}_n\|_{\mathcal{G}_S} \leq 2(2 + \vartheta) \sup_{v \in \mathcal{M}}\|\mathbb{G}_n\|_{\mathcal{G}_{S,v}},
	\end{align*}
	and
	\begin{align*}
		\mathcal{G} = \bigcup_{S \subset \{1, \ldots p\}, \:\mathrm{card}(S) \leq n} \mathcal{G}_S.
	\end{align*}	
	Hence, by eq.~\eqref{eq:lemma:LocalizedRankScoresCone-9}, with probability at least $1-\delta$, 
	\begin{align*}
		&\forall g_{v, \theta, \tau} \in \mathcal{G} : |\mathbb{G}_n(g_{v, \theta, \tau})| \lesssim 2(2 + \vartheta) \Theta\big(s, \|\theta\|_0, s_\theta \big) \big(\upsilon_{r_0, n}\big(\|\theta\|_0 \log(1/r_0)\big) + \upsilon_{r_0, n}(t_{s, \|\theta\|_0, s_\theta, n, \delta})\big).
	\end{align*}	
	
	\textbf{Proof of Case (ii).} Observe that $s \mapsto s \log (ep/s)$ and $s \mapsto \varphi_{\max}(s)$ are monotone increasing on $[1, p]$. Thus, the bound of case (i) for $g_{v, \theta, \tau} \in \mathcal{G}$ with $\|\theta\|_0 = m$ holds also for all $g_{v, \theta', \tau} \in \mathcal{G}$ with $\|\theta'\|_0 \leq m$. To conclude, adjust some constants.
	
	\textbf{Proof of Case (iii).} Note that 
	\begin{align*}
		\|\mathbb{G}_n\|_{|\mathcal{G}|} \leq \|\mathbb{G}_n\|_{\mathcal{G}^A} + \|\mathbb{G}_n\|_{\mathcal{G}^B},
	\end{align*}
	where
	\begin{align*}
		\mathcal{G}^A &= \left\{g: \mathbb{R}^{p+1} \rightarrow \mathbb{R}: g(X, Y) = f_{Y|X}(X'\theta_0(\tau)|X) \mathbf{1}\big\{ X'\theta < Y \leq X'\theta_0(\tau)\}|X'u|,  \: \theta \in \mathbb{R}^p, \right.\\
		&\left.\quad{}\quad{}\quad{}\quad{}\quad{}\quad{}\quad{} \quad{}\quad{}\|\theta\|_0 \leq n, \: \|\theta - \theta_0(\tau)\|_2 \leq r_0, \: u \in C^p_{q}(J(\tau), \vartheta) \cap B^p(0,1), \:\tau \in \mathcal{T}\right\},\\
		\mathcal{G}^B & \left\{g: \mathbb{R}^{p+1} \rightarrow \mathbb{R}: g(X, Y) = f_{Y|X}(X'\theta_0(\tau)|X) \mathbf{1}\big\{ X'\theta_0(\tau) < Y \leq X'\theta\}|X'u|,  \: \theta \in \mathbb{R}^p, \right.\\
		&\left.\quad{}\quad{}\quad{}\quad{}\quad{}\quad{}\quad{}\quad{} \|\theta\|_0 \leq n, \: \|\theta - \theta_0(\tau)\|_2 \leq r_0, \: u \in C^p_{q}(J(\tau), \vartheta) \cap B^p(0,1), \:\tau \in \mathcal{T}\right\}.
	\end{align*}
	Using the notation from case (i), we define, for $S \subseteq \{1, \ldots, p\}$, $u \in \mathcal{M}$, and $\tau \in \mathcal{T}$, the following function classes:
	\begin{align*}	
		\mathcal{H}^A_{S,u} &= \big\{ h(X, Y) = \mathbf{1}\big\{ X'\theta < Y \leq X'\theta_0(\tau)\}|X'u|, \\
		&\quad{}\quad{}\quad{}\quad{}\quad{}\quad{}\quad{}\quad{}\theta \in \mathbb{R}^p, \: \mathrm{supp}(\theta) = S, \|\theta - \theta_0(\tau)\|_2 \leq r_0, \tau \in \mathcal{T}\big\},\\
		\mathcal{G}^A_{S, u, \tau} &= \big\{ g(X,Y) =  f_{Y|X}^2(X'\theta_0(\tau)|X)h(X,Y): h \in \mathcal{H}^A_{S, u}\},\\
		\mathcal{G}^A_{S, u} &= \bigcup_{\tau \in \mathcal{T}} \mathcal{G}^A_{S,u, \tau},
	\end{align*}
	and analogously $\mathcal{H}^B_{S,u},\mathcal{G}^B_{S, u, \tau}, \mathcal{G}^B_{S, u}$.
	The proof of the statement now follows by applying the same arguments as in the cases (i) and (ii) to these function classes. We only need to modify the argument that $\mathcal{H}^A_{S,u}$ and $\mathcal{H}^B_{S,u}$ are VC-subgraph classes of functions with VC-index at most a constant multiple of $|S| + 2$:
	
	The indicator $\mathbf{1}\big\{ X'\theta < Y \leq X'\theta_0(\tau)\}$ can be written as the indicator of the set difference of $\{Y \leq X'\theta\}$ and $\{Y \leq X'\theta_0(\tau)\}$. Both these sets are VC-classes of sets with VC-indices at most $|S| + 3$~\citep[][Lemma 2.6.15]{vaartwellner1996weak} and 2~\citep[][Theorem 4.10 (a)]{dudley2014uniform}. By Lemma 2.6.17 (i) and (ii) in~\cite{vaartwellner1996weak} the set difference of these VC-classes of sets is again a VC-class of sets with VC-index at most $|S| + 3 + 2 -1 = |S| + 4$. The same argument applies to the indicator $\mathbf{1}\big\{ X'\theta_0(\tau) < Y \leq X'\theta\}$.

	\textbf{Proof of Case (iv).} This trivially follows from the fact that the proofs of cases (i) and (ii) rely on an $\varepsilon$-net approximation of $s$-sparse sets. In fact, the proofs of these cases establish case (iv) and then invoke Lemma~\ref{lemma:SizeConesDominatedCoordinates} to deduce the case of $v \in C^p_q(J(\tau), \vartheta) \cap B^p_2(0,1)$. This completes the proof.
\end{proof}

\subsection{Proofs of Section~\ref{subsec:AuxResults-EP}}
\begin{proof}[\textbf{Proof of Corollary~\ref{corollary:MaxInequalityBernsteinOrlicz-Gradient}}]
	We first derive a bound on the symmetrized process $\|\mathbb{G}_n^\circ\|_{\mathcal{F}}$. Then, using Lemma~\ref{lemma:Panchenko2003Handel}, we deduce a bound for the original process $\|\mathbb{G}_n\|_{\mathcal{F}}$. In the following, we slightly abuse notation and write $\big\|\|g\|_{P,2}\big\|_{\mathcal{G}}$ for $\sup_{g \in \mathcal{G}} \sqrt{Pg^2}$. For $t,s > 0$ we have the following decomposition:
	\begin{align}\label{eq:corollary:MaxInequalityBernsteinOrlicz-Gradient-1}
		&\mathbb{P}\left\{\|\mathbb{G}_n^\circ\|_\mathcal{F} > t + \sqrt{s}t \right\}\nonumber\\
		&\quad{}\leq \mathbb{P}\left\{  \|\mathbb{G}_n^\circ\|_\mathcal{F} \frac{\big\|\|g\|_{P,2}\big\|_{\mathcal{G}}}{\big\|\|g\|_{P_n,2}\big\|_{\mathcal{G}}}\left|\frac{\big\|\|g\|_{P_n,2}\big\|_{\mathcal{G}}}{\big\|\|g\|_{P,2}\big\|_{\mathcal{G}}}-1\right| + \|\mathbb{G}_n^\circ\|_\mathcal{F} \frac{\big\|\|g\|_{P,2}\big\|_{\mathcal{G}}}{\big\|\|g\|_{P_n,2}\big\|_{\mathcal{G}}} > t + \sqrt{s}t \right\} \nonumber\\
		&\quad{}\leq 2\mathbb{P}\left\{ \big\|\|g\|_{P,2}\big\|_{\mathcal{G}}  \frac{\|\mathbb{G}_n^\circ\|_\mathcal{F}}{\big\|\|g\|_{P_n,2}\big\|_{\mathcal{G}}} > t \right\} +  \mathbb{P}\left\{\left|\frac{\big\|\|g\|_{P_n,2}\big\|_{\mathcal{G}}}{\big\|\|g\|_{P,2}\big\|_{\mathcal{G}}}-1\right| > \sqrt{s} \right\}  \nonumber\\
		&\quad{}\leq  2\mathbb{P}\left\{\frac{\|\mathbb{G}_n^\circ\|_\mathcal{F}}{\big\|\|g\|_{P_n,2}\big\|_{\mathcal{G}}} > \frac{t}{ \big\|\|g\|_{P,2}\big\|_{\mathcal{G}} } \right\} +  \mathbb{P}\Big\{\big\|\|g\|_{P_n,2}^2- \|g\|_{P,2}^2\big\|_{\mathcal{G}} > s\big\|\|g\|_{P,2}^2\big\|_{\mathcal{G}} \Big\} \nonumber\\
		&\quad{} = \mathbf{I} + \mathbf{II},
	\end{align}
	where the second inequality holds since for all $a,b \in \mathbb{R}$ and $s, t > 0$,
	\begin{align*}
		|ab| > st \quad{}  \implies \quad{} |a| > s \quad{} \text{or} \quad{} |b| > t,\\
		|a + b| > s + t \quad{} \implies  \quad{} |a| > s \quad{} \text{or} \quad{} |b| > t,
	\end{align*}
	and the third inequality holds since $|\sqrt{a} - \sqrt{b}| \leq \sqrt{|a - b|}$ for all $a, b \geq 0$.
	
	\textbf{Bound on $\mathbf{I}$.}  Define the following (data-dependent) classes:
	\begin{align*}
		\mathcal{F}_g = \left\{h g/\|g\|_{P_n,2} : h \in \mathcal{H}\right\}, \quad{} g \in \mathcal{G}.
	\end{align*}
	
	For all $f \in \mathcal{F}_g$ and any $e_1, e_2 \in \{-1,1\}$ we have $|fe_1 - fe_2| \leq 2|g|/\|g\|_{P_n,2}$. Thus, conditionally on $\{X_i\}_{i=1}^n$, the map $(\varepsilon_1, \ldots, \varepsilon_n) \mapsto \|\mathbb{G}_n^\circ\|_{\mathcal{F}_g}$ is a function of bounded differences with constant $c^2 = 4 n^{-1}\sum_{i=1}^n g^2(X_i)/\|g\|_{P_n,2}^2 = 4$. Hence, for all $u \geq 0$,
	\begin{align}\label{eq:corollary:MaxInequalityBernsteinOrlicz-Gradient-2}
		\mathbb{P}_\varepsilon \left\{\|\mathbb{G}_n^\circ\|_{\mathcal{F}_g} \geq \mathbb{E}_\varepsilon \|\mathbb{G}_n^\circ\|_{\mathcal{F}_g} + \sqrt{u}\right\} \leq e^{-2u/c^2} = e^{-u/2}.
	\end{align}
	
	By construction of $\mathcal{F}_g$, the envelope $F_g = \sup_{f \in \mathcal{F}_g}|f|$ has $L_2(P_n)$-semi-norm bounded by one and is VC subgraph with VC-index $V(\mathcal{H})$. Thus, by Dudley's maximal inequality applied conditionally on $\{X_i\}_{i=1}^n$ and Theorem 2.6.7 in~\cite{vaartwellner1996weak},
	\begin{align}\label{eq:corollary:MaxInequalityBernsteinOrlicz-Gradient-3}
		\mathbb{E}_\varepsilon \|\mathbb{G}_n^\circ\|_{\mathcal{F}_g} \lesssim \left\|F_g\right\|_{P_n, 2} \int_0^1\sqrt{\log N\big(\varepsilon\left\|F_g\right\|_{P_n, 2}, \mathcal{F}_g, L_2(P_n)\big)} d\varepsilon \lesssim \sqrt{V(\mathcal{H})}.
	\end{align}
	
	Combine eq.~\eqref{eq:corollary:MaxInequalityBernsteinOrlicz-Gradient-2} and~\eqref{eq:corollary:MaxInequalityBernsteinOrlicz-Gradient-3} with the union bound over $g \in \mathcal{G}$ to conclude that there exists an absolute constant $C_1 > 0$ (independent of $\{X_i\}_{i=1}^n, n, \mathcal{F}_g, \mathcal{G}$) such that for all $u \geq 0$,
	\begin{align*}
		\mathbb{P}_\varepsilon \left\{\sup_{g \in \mathcal{G}} \|\mathbb{G}_n^\circ\|_{\mathcal{F}_g} \geq C_1 \sqrt{V(\mathcal{H})} + \sqrt{u + \log\mathrm{card}(\mathcal{G})}\right\} \leq e^{-u/2}.
	\end{align*}
	Clearly, $\sup_{g \in \mathcal{G}} \|\mathbb{G}_n^\circ\|_{\mathcal{F}_g} \geq \|\mathbb{G}_n^\circ\|_\mathcal{F}/\big\|\|g\|_{P_n,2}\big\|_{\mathcal{G}}$. Thus, in above display, take expectation with respect to $\{X_i\}_{i=1}^n$, and conclude that, for all $u \geq 0$,
	\begin{align}\label{eq:corollary:MaxInequalityBernsteinOrlicz-Gradient-4}
		\mathbb{P} \left\{\frac{\|\mathbb{G}_n^\circ\|_\mathcal{F}}{\big\|\|g\|_{P_n,2}\big\|_{\mathcal{G}}} \geq C_1 \sqrt{V(\mathcal{H})}  + \sqrt{\log\mathrm{card}(\mathcal{G})} + \sqrt{u}\right\}  \leq e^{-u/2}.
	\end{align}
	
	\textbf{Bound on $\mathbf{II}$.} Recall that $\mathcal{G}^2 = \{g^2 : g \in \mathcal{G}\}$. Hence, $\|\mathbb{G}_n\|_{\mathcal{G}^2} = \sqrt{n}\big\|\|g\|_{P_n,2}^2- \|g\|_{P,2}^2\big\|_{\mathcal{G}}$. Therefore, by eq.~\eqref{eq:corollary:MaxInequalityBernsteinOrlicz-Gradient-0} and Corollary~\ref{corollary:DeviationInequalityBernsteinOrlicz}, there exists an absolute constant $C_2 > 0$ such that, for all $u \geq 0$, 
	\begin{align}\label{eq:corollary:MaxInequalityBernsteinOrlicz-Gradient-5}
		\mathbb{P}\left\{\big\|\|g\|_{P_n,2}^2- \|g\|_{P,2}^2\big\|_{\mathcal{G}} > C_2K \delta \pi_{n,\alpha}^2\big(\log\mathrm{card}(\mathcal{G})\big)+ C_2 K \delta \pi_{n,\alpha}^2(u)\right\} \leq e^{-u},
	\end{align}
	where $\delta= \sup_{g_1, g_2 \in \mathcal{G}^2}\rho(g_1, g_2)$ and $\pi_{n,\alpha}^2(z) = \sqrt{z/n} + z^{1/\alpha}/n$ for $ z \geq  0$.
	
	\textbf{Conclusion.} Set
	\begin{align*}
		t = \left(\sqrt{V(\mathcal{H})} + \sqrt{\log\mathrm{card}(\mathcal{G})} + \sqrt{u}\right) \big\|\|g\|_{P,2}\big\|_{\mathcal{G}} \quad{} \mathrm{and}\quad{} s = \left(K \delta \pi_{n,\alpha}^2\big(\log\mathrm{card}(\mathcal{G})\big)+ K \delta \pi_{n,\alpha}^2(u)	\right) \big\|\|g\|_{P,2}\big\|_{\mathcal{G}}^{-2}.	\end{align*}
	Now, combine eq.~\eqref{eq:corollary:MaxInequalityBernsteinOrlicz-Gradient-1},~\eqref{eq:corollary:MaxInequalityBernsteinOrlicz-Gradient-4}, and~\eqref{eq:corollary:MaxInequalityBernsteinOrlicz-Gradient-5} and conclude that there exists an absolute constant $C_3 > 0$ such that, for all $u>0$,
	\begin{align}\label{eq:corollary:MaxInequalityBernsteinOrlicz-Gradient-6}
		\begin{split}
			\mathbb{P}\left\{\|\mathbb{G}_n^\circ\|_\mathcal{F} > C_3 \left(\sqrt{V(\mathcal{H})} + \sqrt{\log\mathrm{card}(\mathcal{G})} + \sqrt{u}\right) \left( \sqrt{ \sigma^2 + K \delta \pi_{n,\alpha}^2\big(\log\mathrm{card}(\mathcal{G})\big)} + \sqrt{K \delta \pi_{n,\alpha}^2(u)} \right)  \right\}\\
			\leq 2e^{-u} e^{-u}.	
		\end{split}
	\end{align}
	Since for any convex and increasing function $F: \mathbb{R}_+ \rightarrow \mathbb{R}_+$,
	\begin{align*}
		\mathbb{E}\left[F\left(\|\mathbb{G}_n\|_{\mathcal{F}}\right) \right]\leq \mathbb{E}\left[F\left(2\|\mathbb{G}_n^\circ\|_{\mathcal{F}}\right) \right],
	\end{align*}
	Lemma~\ref{lemma:Panchenko2003Handel} and eq.~\eqref{eq:corollary:MaxInequalityBernsteinOrlicz-Gradient-6} imply that, there exist cosntants $c, c' \geq 0$ (depending on $\alpha$) such that, with probability at least $1 - c e^{-c't}$,
	\begin{align*}
		\|\mathbb{G}_n\|_{\mathcal{F}} \lesssim \left(\sqrt{V(\mathcal{H})} + \sqrt{\log\mathrm{card}(\mathcal{G})} + \sqrt{t}\right) \left( \sqrt{ \sigma^2 + K \delta \pi_{n,\alpha}^2\big(\log\mathrm{card}(\mathcal{G})\big)} + \sqrt{K \delta \pi_{n,\alpha}^2(t)} \right) .
	\end{align*}
	To conclude, adjust some absolute constants.
\end{proof}

\begin{proof}[\textbf{Proof of Lemma~\ref{lemma:ProductSubgaussian}}]
	Recall Young's inequality: $\prod_{i=1}^K x_i^{\lambda_i} \leq \sum_{i=1}^K \lambda_i x_i$ for all $x_i, \lambda_i \geq 0$, $i=1, \ldots, K$ with $\sum_{i=1}^K \lambda_i = 1$. Without loss of generality, we can assume that $\|X_i\|_{\psi_\alpha} = 1$ for all $i=1, \ldots, K$. Thus, the claim of the lemma follows if we can show the following: If $\mathbb{E}\left[\exp(|X_i|^\alpha)\right] \leq 2$ for all $i=1, \ldots, K$, then $\mathbb{E}[\exp(\prod_{i=1}^K|X_i|^{\alpha/K} )] \leq 2$. This assertion follows from straightforward calculations:
	\begin{align*}
		&\mathbb{E}\left[\psi_{\alpha/K}\left(\prod_{i=1}^KX_i\right)\right] + 1 = \mathbb{E}\left[\exp\left(\prod_{i=1}^K |X_i|^{\alpha/K}\right)\right] \leq \mathbb{E}\left[\exp\left(\frac{1}{K}\sum_{i=1}^K |X_i|^\alpha\right)\right]\\ &\quad{}=\mathbb{E}\left[\prod_{i=1}^K\exp\left(\frac{1}{K} |X_i|^\alpha\right)\right]\leq \frac{1}{K} \left(\sum_{i=1}^K\mathbb{E}\left[\exp\left(|X_i|^\alpha\right)\right] \right)\leq 2,
	\end{align*}
	where in the first and second inequalities we have used Young's inequality.
\end{proof}
\begin{proof}[\textbf{Proof of Lemma~\ref{lemma:MomentsToTails}}]
	Let $t \geq 0$ be arbitrary. Set $A : = \left\{ \omega \in \Omega: \xi(\omega) > \phi(t)\right\}$. By the premise, $\phi(t) \leq	\int_A \xi d\mathbb{P} \leq \phi\left(\varphi^{-1}\left(1/\mathbb{P}\{A\}\right)\right)$. Thus, $\phi(t) \leq \phi\left(\varphi^{-1}\left(1/\mathbb{P}\{A\}\right)\right)$. Solving for $\mathbb{P}\{A\}$ yields the claim.
\end{proof}
\begin{proof}[\textbf{Proof of Lemma~\ref{lemma:Panchenko2003Handel}}]
	We split the proof into two cases with similar yet distinct proofs.
	
	\textbf{Case $\alpha \in (0,1)$.} Denote by $\varphi^{-1}$ the inverse of $\varphi$ and note that $\varphi^{-1}$ is convex and increasing. Therefore, $F(z) = \max\{\varphi^{-1}(z) - t, 0\}$  is also convex and increasing. Hence, by assumption, for all $t \geq 0$,
	\begin{align}\label{eq:lemma:Panchenko2003Handel-1}
		\begin{split}
			&\int_t^\infty \mathbb{P}\left\{\varphi^{-1}(X) \geq s\right\} ds  = \mathbb{E}[F(X)] \leq \mathbb{E}[F(Y)] = \int_t^\infty \mathbb{P}\left\{\varphi^{-1}(Y) \geq s\right\} ds\\
			& \leq \int_t^\infty c_1 e^{-c_2s^\alpha}ds =  \frac{c_1}{\alpha c_2^{1/\alpha} }\int_{c_2 t^\alpha}^\infty  u^{1/\alpha -1}e^{-u} du  = \frac{c_1}{\alpha c_2^{1/\alpha} }\Gamma\left(1/\alpha, c_2t^\alpha\right),
		\end{split}
	\end{align}
	where $\Gamma(a, z) = \int_z^\infty u^{a-1} e^{-u} du$ is the incomplete Gamma function.
	
	By Theorem 1.1 and Proposition 2.10 in~\cite{pinelis2020exact}, for all $z \geq 0$,
	\begin{align}\label{eq:lemma:Panchenko2003Handel-2}
		\Gamma(a,z) \leq 
		\begin{cases}
			z^{a-1} e^{-x} + (a-1) G_{a-1}(z) & 1 < a < 2\\
			G_a(z) & a \geq 2,
		\end{cases}
	\end{align}
	where $G_a(z) = \frac{(z + b_a)^a - z^a}{ab_a} e^{-z}$ and $b_a = \Gamma(a + 1)^{1/(a-1)}$. Thus, by eq.~\eqref{eq:lemma:Panchenko2003Handel-2} there exist constants $c, c', c'', c''' > 0$ (depending on $a > 1$ only) such that, for all $z \geq 0$,
	\begin{align}\label{eq:lemma:Panchenko2003Handel-3}
		\Gamma(a,z) \leq (c z^a + c'z^{a-1} + c'')e^{-z} \leq c'''e^{-z/a}.
	\end{align}
	Combine eq.~\eqref{eq:lemma:Panchenko2003Handel-3} and~\eqref{eq:lemma:Panchenko2003Handel-1} to conclude that there exists a constant $c_3 > 1$ (depending on $\alpha, c_1$ only) such that for all $t \geq 0$ and all $u \leq t$,
	\begin{align*}
		\mathbb{P}\left\{\varphi^{-1}(X) \geq t\right\} \leq \frac{1}{u} \int_{t-u}^t \mathbb{P}\left\{\varphi^{-1}(X) \geq s\right\} ds \leq \frac{c_3e^{\alpha c_2u^\alpha}}{c_2^{1/\alpha}u} e^{-\alpha c_2 t^\alpha}.
	\end{align*}
	Optimizing over $u$ yields $u^* = (1/\alpha^2 c_2)^{1/\alpha}$ and, hence, for all $t \geq (1/\alpha^2 c_2)^{1/\alpha}$,
	\begin{align*}
		\mathbb{P}\left\{\varphi^{-1}(X) \geq t\right\} \leq c_3 e^{1/\alpha} e^{-\alpha c_2t^\alpha}.
	\end{align*}
	Since $c_3 e^{1/\alpha}e^{-\alpha c_2t^\alpha} \geq 1$ for $t \leq (1/ c_2)^{1/\alpha}$, this bound holds also for all $0 \leq t < (1/ c_2)^{1/\alpha}$.
	
	\textbf{Case $\alpha \in [1, \infty)$. } The proof strategy is the same as for case (i), but the calculations are simpler. Denote by $\varphi^{-1}$ the inverse of $\varphi$ and note that $z \mapsto \left(\varphi^{-1}(z)\right)^\alpha$ is convex and increasing. Therefore, $F(z) = \max\left\{ \left(\varphi^{-1}(z)\right)^\alpha - t, 0\right\}$  is also convex and increasing.  We have, for all $t \geq 0$,
	\begin{align*}
		\int_t^\infty \mathbb{P}\left\{\left(\varphi^{-1}(X)\right)^\alpha \geq s\right\} ds  = \mathbb{E}[F(X)] \leq \mathbb{E}[F(Y)] = \int_t^\infty \mathbb{P}\left\{\left(\varphi^{-1}(Y)\right)^\alpha \geq s\right\} ds 
		\leq \frac{c_1}{c_2} e^{-c_2t}.
	\end{align*}
	Thus, we have, for all $t \geq 0$ and all $u \leq t$,
	\begin{align*}
		\mathbb{P}\left\{\left(\varphi^{-1}(X)\right)^\alpha \geq t\right\} \leq \frac{1}{u} \int_{t-u}^t \mathbb{P}\left\{\left(\varphi^{-1}(X)\right)^\alpha\geq s\right\} ds \leq \frac{c_1e^{c_2u}}{c_2u} e^{-c_2t}.
	\end{align*}
	Optimizing over $u$ yields, for all $t \geq 1/c_2$,
	\begin{align*}
		\mathbb{P}\left\{\left(\varphi^{-1}(X)\right)^\alpha \geq t\right\} \leq c_1 e^{1-c_2t}.
	\end{align*}
	Again, since $c_1 e^{1-c_2t} \geq 1$ for $t \leq 1/c_2$, this bound holds also for all $0 \leq t < 1/c_2$.
\end{proof}

\begin{proof}[\textbf{Proof of Lemma~\ref{lemma:Composite-GaussianContraction}}]
	The proof is a straightforward modification of the classical proof of Corollary 3.17 in~\cite{ledoux1996probability}. Let $\tilde{f} \in \mathcal{F}$ be arbitrary. We have the following:
	\begin{align*}
		&\mathbb{E} \left[ F\left(\frac{1}{2} \sup_{f \in \mathcal{F}} \sup_{h \in \mathcal{H}} \left|\sum_{i=1}^n g_i \varphi_i\big(f(X_i)\big)h(X_i) \right|\right)\right] \\
		&\overset{(a)}{\leq} \frac{1}{2} \mathbb{E} \left[ F\left(\sup_{f \in \mathcal{F}} \sup_{h \in \mathcal{H}} \left|\sum_{i=1}^n g_i \Big(\varphi_i\big(f(X_i)\big) - \varphi_i\big(\tilde{f}(X_i)\big)\Big)h(X_i) \right|\right)\right] + \frac{1}{2} \mathbb{E} \left[ F\left(\sup_{h \in \mathcal{H}} \left|\sum_{i=1}^n g_i \varphi_i\big(\tilde{f}(X_i)\big)h(X_i) \right|\right)\right]\\
		&\overset{(b)}{\leq}  \mathbb{E} \left[ F\left(\sup_{f, f' \in \mathcal{F}} \sup_{h \in \mathcal{H}} \left|\sum_{i=1}^n g_i \Big(\varphi_i\big(f(X_i)\big) - \varphi_i\big(f'(X_i)\big)\Big)h(X_i) \right|\right)\right],
	\end{align*}
	where (a) holds by convexity of $F$ and (b) holds since $\varphi_i(0) = 0$ and symmetry of $g_i$, $1 \leq i \leq n$. The expression in above display can be further upper bounded by
	\begin{align}\label{eq:lemma:Composite-GaussianContraction-1}
		\mathbb{E} \left[ F\left(\sup_{f,f' \in \mathcal{F}} \sup_{h \in \mathcal{H}} \left|\sum_{i=1}^n g_i \Big(f(X_i) - f'(X_i)\Big)h(X_i) \right|\right)\right],
	\end{align}
	since for all $f, f' \in \mathcal{F}$ and $h \in \mathcal{H}$,
	\begin{align*}
		\sum_{i=1}^n \Big(\varphi_i\big(f(X_i)\big) - \varphi_i\big(f'(X_i)\big)\Big) \Big)^2 h^2(X_i) \leq \sum_{i=1}^n \Big(f(X_i) - f'(X_i) \Big)^2 h^2(X_i),
	\end{align*}	
	and hence the the Gaussian contraction theorem~\citep[e.g.][Theorem 3.15]{ledoux1996probability} applies. To conclude the proof, upper bound eq.~\eqref{eq:lemma:Composite-GaussianContraction-1} by
	\begin{align*}
		\mathbb{E} \left[ F\left(2\sup_{f \in \mathcal{F}} \sup_{h \in \mathcal{H}} \left|\sum_{i=1}^n g_i f(X_i) h(X_i) \right|\right)\right].
	\end{align*}
\end{proof}





\end{document}